\documentclass[iop]{emulateapj}
\usepackage{enumerate}
\usepackage{morefloats}
\def\arcsec{$\,^{\prime\prime}$~}

\def\kms{km~s$^{-1}$}
\def\sers{$n$~}
\newcommand{\lsim }{{\lower0.8ex\hbox{$\buildrel <\over\sim$}}}
\newcommand{\gsim }{{\lower0.8ex\hbox{$\buildrel >\over\sim$}}}

\def\simge{\mathrel{%
   \rlap{\raise 0.511ex \hbox{$>$}}{\lower 0.511ex \hbox{$\sim$}}}}
\def\simle{\mathrel{
   \rlap{\raise 0.511ex \hbox{$<$}}{\lower 0.511ex \hbox{$\sim$}}}}

\newcommand{\Msun}{\ifmmode {M_{\odot}}\else${M_{\odot}}$\fi}
\newcommand{\Lsun}{\ifmmode {L_{\odot}}\else${L_{\odot}}$\fi}
\newcommand{\Rsun}{\ifmmode {R_{\odot}}\else${R_{\odot}}$\fi}

\shorttitle{Massive Ellipticals at Large Radius}
\shortauthors{RASKUTTI, ET AL.}

\begin{document}
\title{The Stellar Haloes of Massive Elliptical Galaxies III: Kinematics at Large Radius}  

\author{Sudhir Raskutti \altaffilmark{1}, Jenny E. Greene\altaffilmark{1,2}, 
Jeremy D. Murphy\altaffilmark{1,2}}

\altaffiltext{1}{Department of Astrophysics, Princeton University, }
\altaffiltext{2}{Department of Astronomy, UT Austin, 1 University Station C1400, 
Austin, TX 71712}

%\slugcomment{}

\begin{abstract}

  We present a 2D kinematic analysis out to $\sim 2 - 5$ effective
  radii ($R_e$) of 33 massive elliptical galaxies with stellar
  velocity dispersions $\sigma > 150$~\kms.  Our observations
  were taken using the Mitchell Spectrograph (formerly VIRUS-P), a spectrograph with a large 
  $107 \times 107$ arcsec$^2$ field-of-view that allows us to 
  construct robust, spatially resolved kinematic maps of 
  $V$ and $\sigma$ for each galaxy extending to at least 2
  $R_e$. Using these maps we study the radial dependence of the stellar 
  angular momentum and other kinematic properties. We see
  the familiar division between slow and fast rotators persisting out
  to large radius in our sample. Centrally slow rotating galaxies, which are
  almost universally characterised by some form of kinematic
  decoupling or misalignment, remain slowly rotating in their halos. 
  The majority of fast rotating galaxies show either increases in
  specific angular momentum outwards or no change beyond $R_e$. The 
  generally triaxial nature of the slow rotators suggests that they 
  formed through mergers, consistent with a ``two-phase'' picture
  of elliptical galaxy formation. However, we do not observe the sharp 
  transitions in kinematics proposed in the literature as a signpost of moving from central
  dissipationally-formed components to outer accretion-dominated haloes.

\end{abstract}

\maketitle
%----------------------------------------------------------------------

\section{Introduction}
\label{Sec:Introduction}

\begin{figure*}[!htb]
\begin{center}
\includegraphics[width=0.45\textwidth,angle=0,clip]{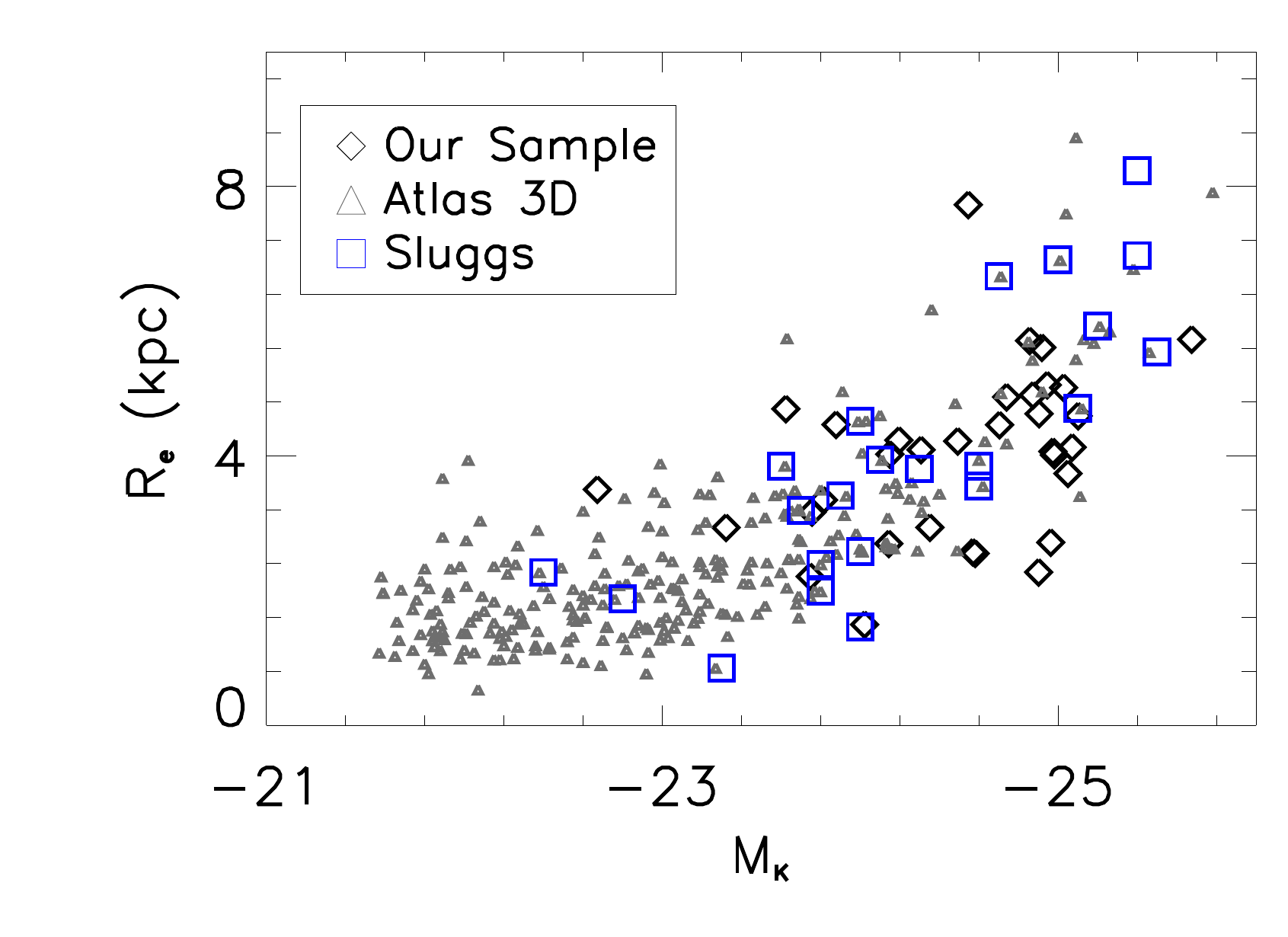}
\includegraphics[width=0.45\textwidth,angle=0,clip]{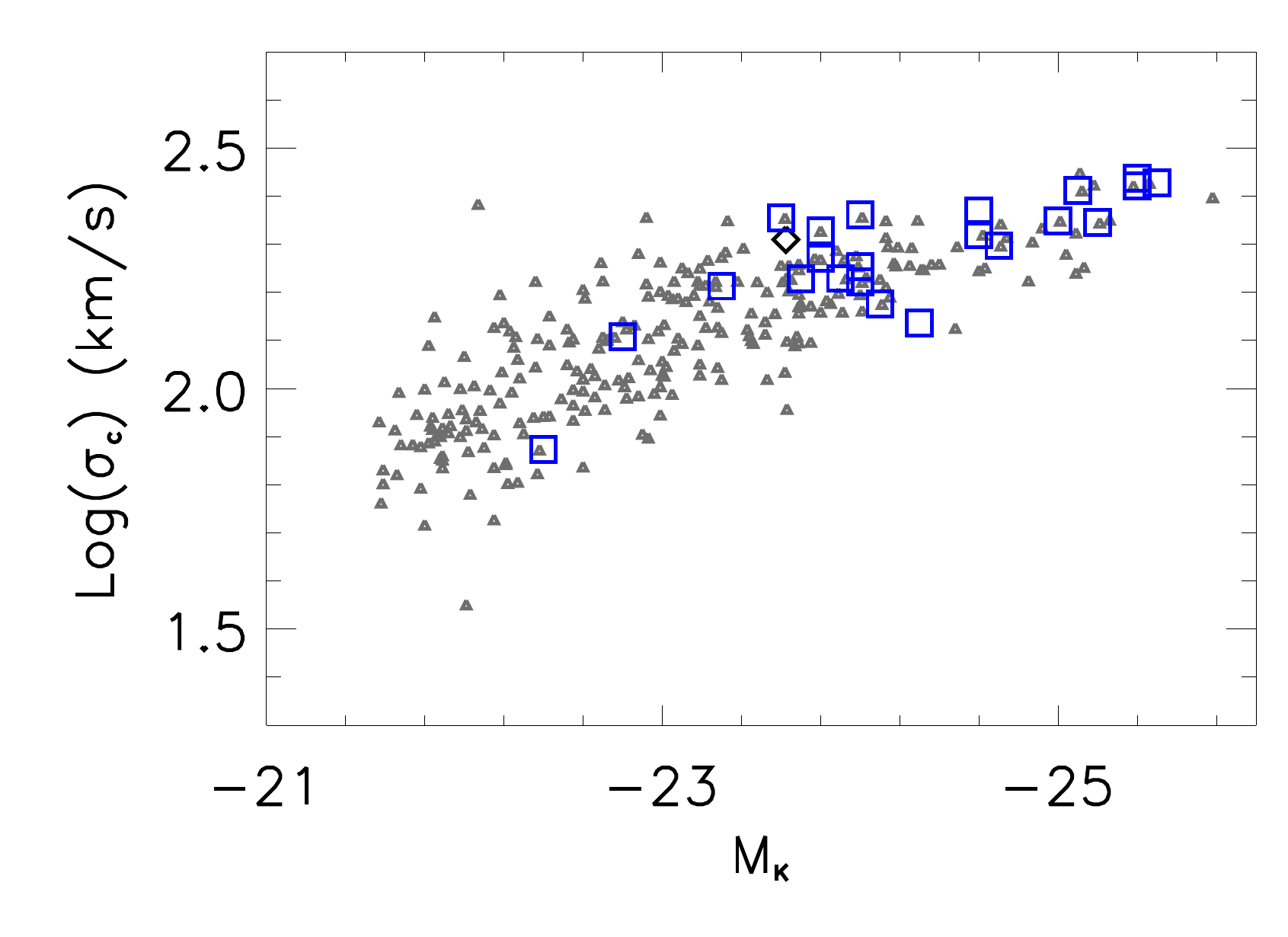}
\includegraphics[width=0.45\textwidth,angle=0,clip]{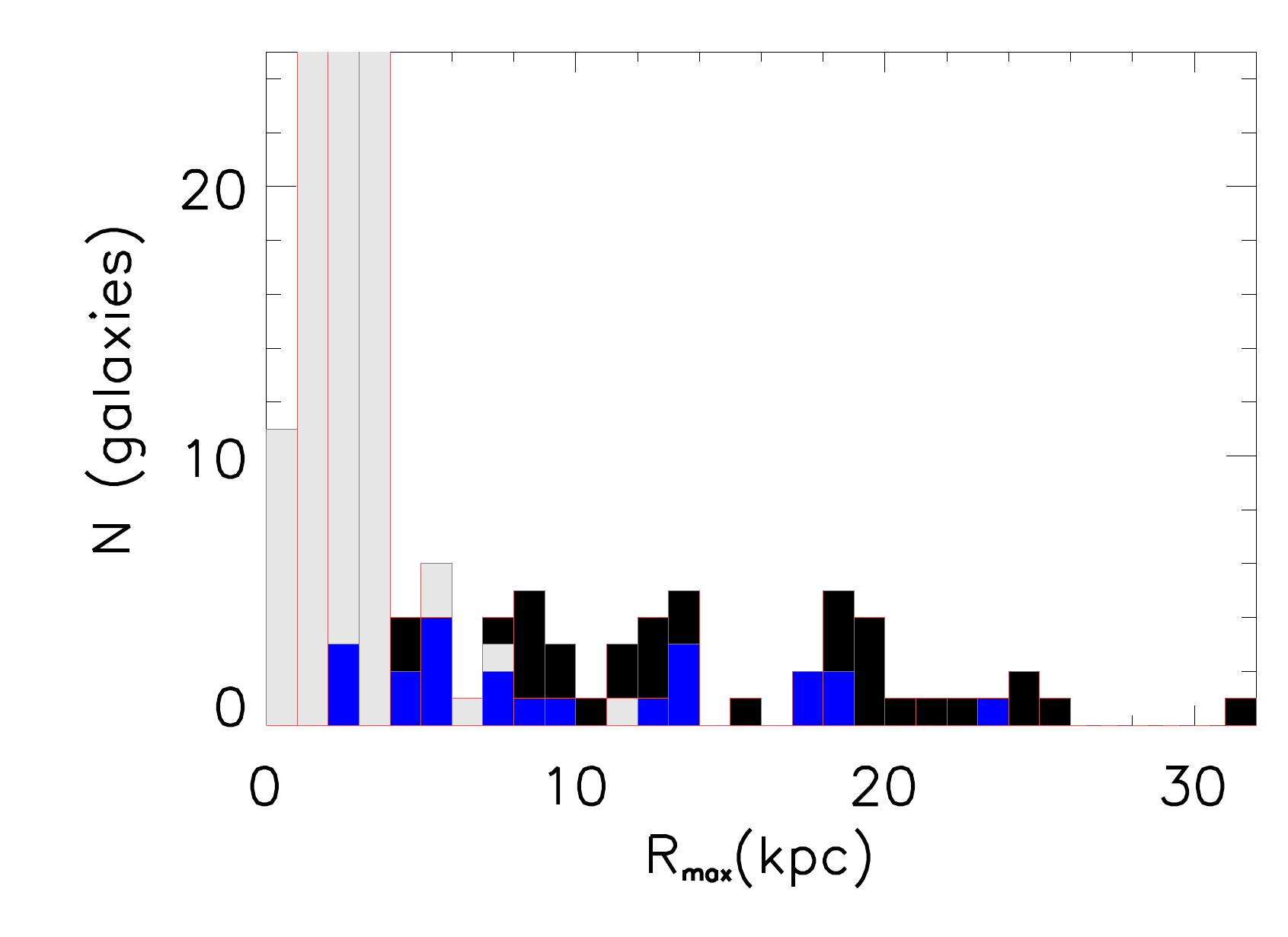}
\includegraphics[width=0.45\textwidth,angle=0,clip]{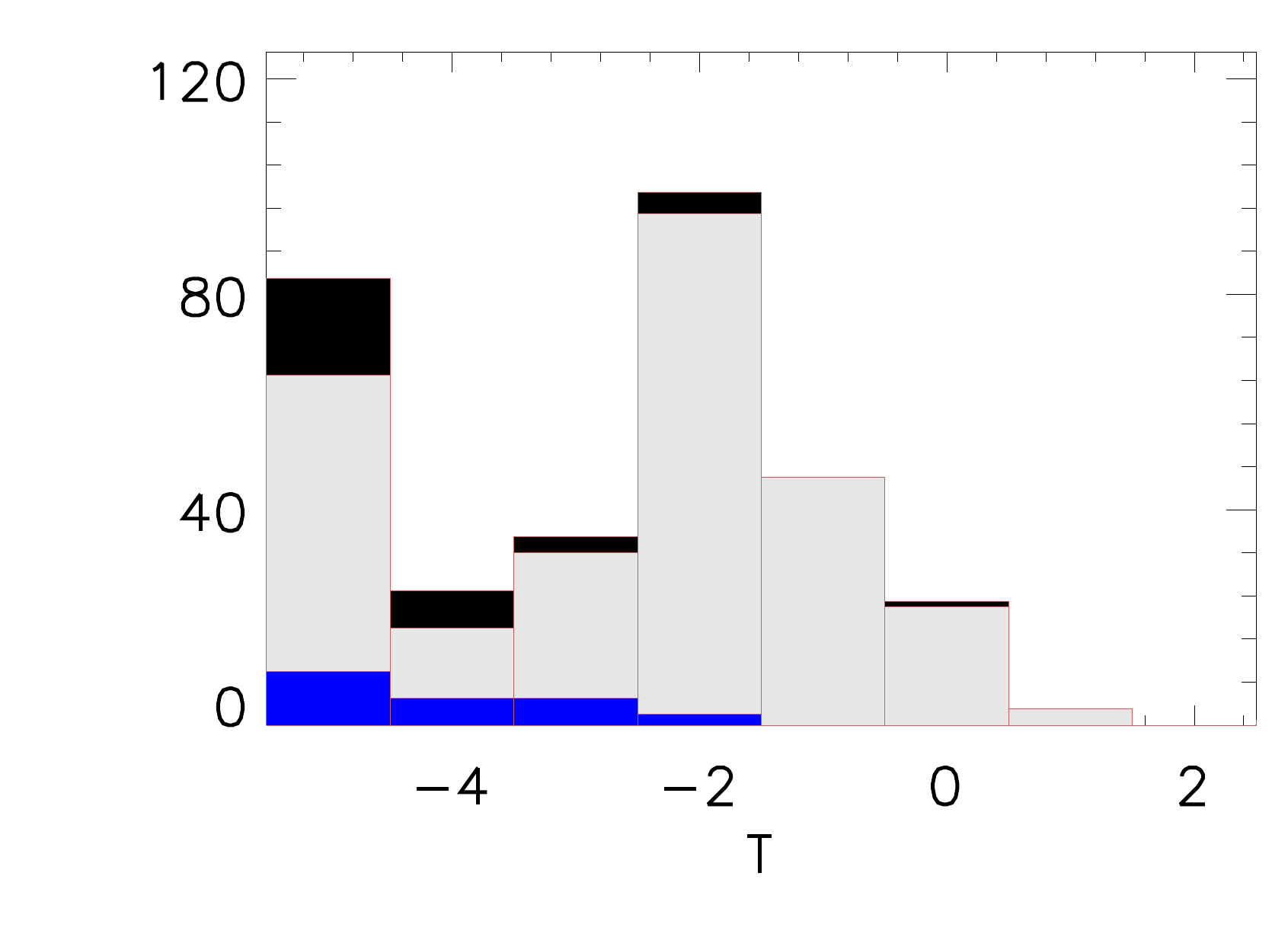}
\caption{
  Characteristics of our galaxy sample (black) as compared to the
  volume-limited ATLAS$^{\rm 3D}$ survey of ETG's \citep[grey][]{Cappellari2011} and
  the 22 massive galaxies in the SLUGGS survey \citep[blue][]{Arnold2013}.  We
  show the K-band magnitude and half-light radii (top left panel), the
  distribution of central dispersions ($\sigma_c$) as a function of luminosity (top
  right panel), the distribution of maximum observed radii (bottom
  left) and the distribution of Hubble Types (bottom right).  We note that the 
radii are measured by the SDSS for our galaxies using a deVaucouleurs fit to the 
light profile, while they are based on RC3 for the ATLAS$^{\rm 3D}$ and SLUGGS galaxies. 
For clarity, we have truncated the histogram at bottom 
left. In the truncated bins, there are 120, 95 and 30 ATLAS$^{\rm 3D}$ galaxies 
respectively.}
\label{Fig:Sample}
\end{center}
\end{figure*}

Much attention has been paid recently to the formation and evolution
of Early-Type Galaxies [ETGs, including both elliptical (E) and
lenticular (S0) galaxies], driven in large part by the discovery that
ETG's at $z \sim 2$ are $\sim 2 - 4$ times smaller at fixed mass 
than their present day counterparts \citep{vanderWel2005,
  diSeregoAlighieri2005, Daddi2005, Trujillo2006, Longhetti2007,
  Toft2007, vanDokkum2008, Cimatti2008, Buitrago2008, vanderWel2008,
  Franx2008, vanDokkum2008, Damjanov2008, CenarroTrujillo2009, Bezanson2009,
  vanDokkum2010, vandeSande2011, Whitaker2012}. 
To explain the rapid size evolution from $z \sim 2$ until today, a
two-phase picture of ETG growth has emerged. At early times, ETG's
form in a highly dissipative environment, with rapid star formation
creating massive, compact cores, where most of the stars formed in
situ \citep{Keres2005, KhochfarSilk2006, DeLucia2006, Krick2006,
  Naab2007, Naab2009, Joung2009, Dekel2009, Keres2009, Oser2010,
  Feldmann2010, DominguezSanchez2011, Feldmann2011, Oser2012}. The second 
 phase, dry accretion, is dominated by collisionless
dynamics during which star formation is suppressed and most of the
stellar mass increase occurs in the galactic outskirts
\citep{Hopkins2009, vanDokkum2010, Szomoru2012, Saracco2012}.

While such a two-phase picture is generally compelling, it is
uncertain precisely how and when mass is added (e.g., the balance of
major to minor mergers).  Simple virial arguments \citep{Cole2000,
  Naab2009, Bezanson2009} as well as recent cosmological simulations
\citep{Hilz2012, OogiHabe2013, Hilz2013} suggest that major and minor
mergers have very different effects. Violent relaxation in
major mergers generally results in moderate, factor of $\sim 2-3$, increases in the 
half-mass radius for every merger event. Meanwhile, mass build-up via minor mergers 
deposits more mass in the outskirts, resulting in $\sim 5-$fold increases in the radius for similar
growth in mass \citep{Hilz2012}.  Simulations therefore currently favor a $1:5$ mass
ratio in mergers \citep{Oser2012, Lackner2012, GaborDave2012}. However, 
incomplete modelling of feedback processes (e.g., AGN and supernovae
winds) makes these results uncertain.

Kinematic observations of local ellipticals also contain important
information. It has long been known that ETG's are well separated into
those that rotate and those that do not 
\citep[e.g.,][]{bertolacapaccioli1975,Illingworth1977,Davies1983}. 
The former tend to have lower stellar mass, disky isophotes and
cuspy light profiles, while the latter are triaxial, cored, and
massive \citep[e.g.,][]{Bender1989,KormendyBender1996, deZeeuw1985, 
Franx1991, deZeeuwFranx1991, vandenBosch2008}. Modern
integral-field studies have provided strong confirmation of this
general bimodal picture with excellent statistics
\citep[e.g.,][]{Emsellem2004, Cappellari2007, Emsellem2007, Krajnovic2011, Cappellari2011,
  Emsellem2011} and have made interesting comparisons
with cosmological simulations \citep{Khochfar2011, Davis2011,
  Serra2014}. 

In the context of two-phase assembly, it is thought that the global properties of each family can be
linked to their formation history. Slow Rotators (SRs) are thought to
accrete most of their mass in minor dry mergers with up to $\sim 3$
major mergers \citep{Khochfar2011}. This explains both their low net
rotation and their preponderance of kinematically decoupled cores that
are likely long-lived remnants of mergers 
\citep[KDC, e.g.,][]{Kormendy1984, Forbes1994, Carollo1997, 
Emsellem2004, Emsellem2007, Krajnovic2008}. 
In contrast Fast Rotators (FRs) likely grew
predominantly through cold gas accretion with at most one major merger 
\citep{Bois2011,Khochfar2011, Davis2011, Serra2012}, and 
thus have high rotation velocities. 
%% Meanwhile, observations of the
%% local Mg-V$_{\rm esc}$ relation linking stellar populations to
%% kinematics \citep{Scott2012} suggest that most of the stellar mass in the halo 
%% was formed in situ, with a limit of $\lesssim 1.5$ major mergers.

However, this picture remains uncertain since most observations are limited to within the half-light radius
of the galaxy. In contrast, if late-stage growth occurs through dry
accretion, then most of the dynamical changes occur beyond the
half-light radius, where stars have longer relaxation times and so
carry a record of the merger history \citep{vanDokkum2005, Duc2011,
  RomanowskyFall2012}.  It is also only in the outer regions that observations 
become sensitive to dark matter, for which there are concrete
predictions from cosmological simulations. Therefore, wide-field kinematic 
data are required to provide more direct signatures of two-phase growth. 

A number of kinematic measurements of ETG's out to large radius have
been made using spatially sparse measurements of planetary nebulae (PNe) and
globular clusters (GCs) 
\citep{Mendez2001, Coccato2009, Strader2011, McNeil-Moylan2012,
  Arnold2011, Pota2013}. Most recently, \citet{Arnold2013} presented
spatially well-sampled measurements of 22 massive ETG's out to $\sim 4
R_e$ as part of the SLUGGS survey. They showed that a significant
fraction of their galaxies (particularly Es) show a transition from
rotation to dispersion-dominated beyond $\sim R_e$.  They
interpreted this as a transition between a central dissipational component, 
formed at early times, and an outer halo-dominated region formed through later
dry merging.

However, without full 2D kinematic coverage from
integral-field spectroscopic (IFS) studies of stellar continua, these
results alone can be difficult to interpret.  Thus far, at large
radius, most studies of stellar kinematics either utilize one or two long-slit positions
\citep[][]{CarolloDanziger1994, thomas2011}, or focus on individual
objects with IFS \citep[e.g.,][]{Weijmans2009, Proctor2009,
  Coccato2010,Murphy2011}. By contrast, \cite{Greene2012, Greene2013} assembled a sample
of 33 massive, local ETG's with observations extending over $\sim 2 - 4
R_e$. They studied stellar population gradients, finding that most
stars in the outskirts were comparatively old and metal-poor,
consistent with accretion from much smaller galaxies. While they 
were able to constrain when the stars at large radius formed, dynamical
studies are much better suited to revealing where they were formed and
how they were assembled.

In this study, we therefore extend the Greene et al.\ survey by
studying the stellar kinematics in conjunction with the stellar
populations.  We begin in \S\ref{Sec:Sample} by briefly discussing the 
galaxy sample, before describing in \S\ref{Sec:Observations} our
observations, reduction methods and dynamical modelling. In
\S\ref{Sec:Observed Kinematics} we discuss the basic kinematic
characteristics of our galaxies at large radius, with particular
reference to the Slow and Fast Rotator paradigm.  We then go on to
explore, in \S\ref{Sec:Analysis}, the possible theoretical implications 
of our results before concluding in\S\ref{Sec:Conclusions}.

%----------------------------------------------------------------------

\section{The Galaxy Sample, Observations, and Data Reduction}
\label{Sec:Sample}

\begin{table*}
\vspace{0.5 cm}
\caption{Galaxy Sample}
\centering
\tiny
\begin{tabular}{ccccccccccc}
\hline \hline
Galaxy & RA & Dec & z & Mag & Morph. & PA & $\epsilon$ & $R_e$ & $\sigma_{c}$ & Env \\
(1) & (2) & (3) & (4) & (5) & (6) & (7) & (8) & (9) & (10) & (11) \\
\hline
NGC~219 & 00:42:11.3 & +00:54:16.3 & 0.018 & 15.0 & $-$5.0 & 172.0 & 0.11 & 4.4 & 184 & F \\
NGC~426 & 01:12:48.6 & -00:17:24.6 & 0.018 & 14.0 & $-$2.5 & 150.0 & 0.32 & 8.3 & 285 & F \\
NGC~474 & 01:20:06.6 & +03:24:55.8 & 0.008 & 12.4 & $-$2.17 & 4.9 & 0.19 & 18.1 & 163 & F \\
CGCG~390-096 & 03:30:17.1 & $-$00:55:12.6 & 0.021 & 14.7 & $-$5.0 & 61.8 & 0.13 & 7.8 & 204 & F \\
NGC~661 & 01:44:14.6 & +28:42:21.1 & 0.013 & 13.2 & $-$4.0 & 50.1 & 0.36 & 19.9 & 190 & G \\
NGC~677 & 01:49:14.0 & +13:03:19.1 & 0.017 & 13.7 & $-$5.0 & 0.6 & 0.13 & 9.6 & 257 & G \\
UGC~1382 & 01:54:41.0 & $-$00:08:36.0 & 0.019 & 14.3 & $-$5.0 & 65.0 & 0.25 & 9.9 & 195 & F \\
NGC~774 & 01:59:34.7 & +14:00:29.5 & 0.015 & 13.8 & $-$2.2 & 165.4 & 0.24 & 20.9 & 165 & F \\
IC~301 & 03:14:47.7 & +42:13:21.6 & 0.016 & 14.2 & $-$4.7 & 148.0 & 0.21 & 12.6 & 159 & C \\
NGC~1286 & 03:17:48.5 & $-$07:37:00.6 & 0.014 & 14.1 & $-$4.0 & 151.5 & 0.19 & 18.1 & 163 & F \\
IC~312 & 03:18:08.4 & +41:45:15.6 & 0.017 & 14.4 & $-$4.0 & 124.9 & 0.49 & 18.1 & 218 & C \\
NGC~1267 & 03:18:44.7 & +41:28:02.8 & 0.018 & 15.4 & $-$3.3 & 51.7 & 0.15 & 6.4 & 236 & C \\
NGC~1270 & 03:18:58.1 & +41:28:12.4 & 0.017 & 14.3 & $-$5.0 & 179.1 & 0.20 & 6.4 & 373 & C \\
NVSS & 03:20:50.7 & +41:36:01.5 & 0.018 & 15.5 & $-$5.0 & 131.7 & 0.18 & 4.5 & 274 & C \\
UGC~4051 & 07:51:17.6 & +50:10:45.4 & 0.021 & 14.2 & $-$5.0 & 13.1 & 0.19 & 8.6 & 300 & G \\
NGC~3837 & 11:43:56.4 & +19:53:40.4 & 0.021 & 14.4 & $-$5.0 & 109.9 & 0.26 & 8.1 & 265 & C \\
NGC~3482 & 11:44:02.1 & +19:56:59.3 & 0.021 & 13.5 & $-$5.0 & 176.9 & 0.19 & 20.5 & 284 & C \\
NGC~4065 & 12:04:06.1 & +20:14:06.2 & 0.021 & 13.6 & $-$5.0 & 108.4 & 0.17 & 12.5 & 278 & C \\
IC~834 & 12:56:18.5 & +26:21:32.0 & 0.021 & 14.6 & $-$4.3 & 97.7 & 0.35 & 7.3 & 255 & F \\
NGC~4908 & 13:00:54.4 & +28:00:27.4 & 0.017 & 14.1 & $-$4.0 & 102.2 & 0.31 & 18.5 & 236 & F \\
NGC~4952 & 13:04:58.3 & +29:07:20.0 & 0.020 & 13.6 & $-$4.1 & 21.5 & 0.36 & 12.1 & 292 & F \\
NGC~5080 & 13:19:19.2 & +08:25:44.9 & 0.022 & 14.6 & $-$2.0 & 93.4 & 0.09 & 7.8 & 269 & F \\
NGC~5127 & 13:23:45.0 & +31:33:57.0 & 0.016 & 13.3 & $-$4.8 & 71.2 & 0.27 & 22.9 & 275 & F \\
NGC~5423 & 14:02:48.6 & +09:20:29.0 & 0.020 & 13.7 & $-$3.3 & 75.9 & 0.33 & 10.9 & 263 & G \\
NGC~5982 & 15:38:39.8 & +59:21:21.0 & 0.010 & 12.1 & $-$5.0 & 102.5 & 0.30 & 17.9 & 239 & F \\
IC~1152 & 15:56:43.3 & +48:05:42.0 & 0.020 & 13.9 & $-$5.0 & 28.0 & 0.17 & 7.7 & 258 & G \\
IC~1153 & 15:57:03.0 & +48:10:06.1 & 0.020 & 13.6 & $-$2.8 & 165.4 & 0.19 & 9.8 & 241 & G \\
CGCG~137-019 & 16:02:30.4 & +21:07:14.5 & 0.015 & 14.2 & $-$4.0 & 18.3 & 0.15 & 8.7 & 174 & F \\
NGC~6127 & 16:19:11.5 & +57:59:02.8 & 0.016 & 13.0 & $-$5.0 & 33.8 & 0.03 & 11.2 & 247 & F \\
NGC~6482 & 17:51:48.8 & +23:04:19.0 & 0.013 & 12.4 & $-$5.0 & 65.0 & 0.15 & 9.7 & 292 & G \\
NGC~6964 & 20:47:24.3 & +00:18:02.9 & 0.013 & 13.8 & $-$4.9 & 23.2 & 0.13 & 17.0 & 188 & G \\
NGC~7509 & 23:12:21.4 & +14:36:33.8 & 0.016 & 14.1 & $-$5.0 & 175.8 & 0.07 & 9.0 & 180 & F \\
NGC~7684 & 23:30:32.0 & +00:04:51.8 & 0.017 & 13.7 & 0.25 & 153.5 & 0.47 & 15.8 & 169 & F \\
\hline
\end{tabular}
\label{Tab:Sample}
\\[0.5 cm]
Notes: Col. (1): Galaxy Name. Col. (2): RA (hrs) in J2000. 
Col. (3): Dec (deg) in J2000. Col. (4): Redshift from the SDSS. Col. 
(5) $g-$band magnitude (mag) from the SDSS. 
Col. (6): morphological T type from HyperLeda. E: $T \le -3.5$; S0: $-3.5 < T \le -0.5$. 
Col. (7): SDSS photometric position angle (deg). 
Col. (8): SDSS photometric ellipticity. 
Col. (9): SDSS major axis half-light radius ("). 
Col. (10): SDSS stellar velocity dispersion ($\rm km s^{-1}$). 
Col. (11): We sort galaxies into ’F’ield, ’G’roup, and ’C’luster based on the 
number of group members in the \cite{Yang2007} catalogue as 
described in \cite{Greene2013}. Field galaxies have 
$N_{\rm group} < 5$, group indicates $5 < N_{\rm group} < 50$, 
and cluster indicates richer than 50 group members. 
\\[0.5 cm]
\end{table*}

The observations analyzed here were taken with the George and Cynthia
Mitchell Spectrograph \citep[the Mitchell Spectrograph, formerly
VIRUS-P;][]{Hill2008} on the 2.7m Harlan J. Smith telescope at
McDonald Observatory. The Mitchell Spectrograph is an integral-field
spectrograph composed of 246 fibers covering a
107\arcsec$\times$107\arcsec\ field of view with a one-third filling
factor. Each of the 246 fibers subtends $4\farcs2$ and they are
assembled in an array similar to Densepak \citep{Barden1998}. The
Mitchell Spectrograph has performed a very successful search for
Ly$\alpha$ emitters
\citep{Adams2011,Finkelstein2011,Blanc2011} and has become
a highly productive tool to study spatially resolved kinematics and
stellar populations in nearby galaxies
\citep{Blanc2009,Yoachim2010,Murphy2011,Adams2012}.

We use the low-resolution (R $\approx$ 850) blue setting of the
Mitchell Spectrograph. Our wavelength range spans 3550-5850~\AA\ with
an average spectral resolution of 5~\AA\ FWHM. This resolution
delivers a dispersion of $\sim$~1.1~\AA\ pixel$^{-1}$ and corresponds
to $\sigma~\approx 150$~\kms\ at 4300~\AA, our bluest Lick index.
Each galaxy was observed for a total of $\sim 2$ hours on source with
one-third of the time spent at each of three dither positions to fill
the field of view.  Initial data reduction is accomplished using the
custom code Vaccine \citep{Adams2011,Murphy2011}, which
performs basic bias subtraction, wavelength calibration, cosmic-ray
rejection, sky subtraction, and spectral extraction.  Final processing
and flux calibration is performed using code developed for the VENGA
project \citep{Blanc2009,Blanc2013}. The details of our data
reduction are described in \citet{Murphy2011},
\citet{Greene2012} and \citet{Murphy2013}.

Properties of the entire sample of massive galaxies are shown in
Table~\ref{Tab:Sample}. The sample, selected from the SDSS 
\citep{York2000}, is identical to that presented in
\cite{Greene2013}, and details of the selection criteria can be
found there. However, briefly, galaxies were chosen to have central stellar velocity 
dispersions $\sigma_{c} > 150~{\rm km s^{-1}}$, to be observable
in a single 107\arcsec$\times $107\arcsec\ pointing, and to 
have $u - r > 2.2$ \citep{Strateva2001}. We 
then removed {spiral} galaxies by hand. Finally, only
galaxies with half-light radii at least twice the fiber diameter of
$4\farcs2$ were included.

Figure~\ref{Fig:Sample} shows some of the key characteristics of our
sample compared to that of the volume-limited ATLAS$^{\rm 3D}$ survey
as well as the more recent SLUGGS survey. By focusing on
high stellar velocity dispersion, we have deliberately selected a
population of more massive and more distant ellipticals than the ATLAS$^{\rm 3D}$
and SLUGGS samples. As a result of their distance they also tend to
be more compact on the sky.

We also show the distribution of maximum radii $R_{\rm max}$, defined as
the exterior radius of the outermost spatial bins
(Figure~\ref{Fig:Sample}). Beyond $R_{\rm max}$ we cannot achieve our
limiting signal-to-noise ratio (S/N) of 15 even in bins extending over half
the face of the galaxy and with a width of $R_e$. Our maximum radii extend
well beyond the ATLAS$^{\rm 3D}$ sample in both kpc and $R_e$, and
achieve comparable depth to the SLUGGS sample. However, we mention two
caveats.  First, as in our prior papers we adopt the
SDSS ``model'' radius \citep[based mostly on the de Vaucouleurs fit;][]
{deVaucouleurs1948, Graham2005} as the effective radius ($R_e$).  In principle, galaxy profile
shape is a function of mass \citep[e.g.,][]{caon1993,kormendy2009},
but fitting the galaxies with a fixed Sersic (\sers\ ) index of four has the
benefit that we are less sensitive to both sky subtraction errors
\citep{mandelbaum2005,bernardi2007} and to the detailed shape of the
light profile in the very faint wings
\citep[e.g.,][]{LacknerGunn2012}.  In the effort to have a uniform
analysis, we have therefore adopted the effective radii published by
the SDSS, which tend to be small compared to literature values.
Furthermore, our outermost radii correspond to measurements over wide
bins in both the radial ($R_e$) and azimuthal ($\pi$) directions, and
so it is worth bearing in mind that while we reach large radius we do
so at low spatial resolution.

Finally, we will often examine properties of our sample as a function
of stellar mass.  Stellar masses are based on the stellar population
synthesis presented in \citet{Greene2013} based on Lick index modeling
within $\sim R_e$ \citep{GravesSchiavon2008}.  Using the
$\alpha$-enhanced models of \cite{Schiavon2007}, {SDSS r-band photometry}, and assuming a
Salpeter IMF, we derive a luminosity-weighted global $M/L$. The inferred
stellar masses are subject to systematics from emission-line
contamination, which primarily impacts H$\beta$ and therefore the
stellar ages \citep[e.g.,][]{Graves2007}.  As a sanity check, we
extract $K$-band luminosities from the 2MASS Extended Source 
Catalog \citep[XSC, available online\footnote{http://www.ipac.caltech.edu/2mass.}]
[]{Huchra2012}, and use the
empirical scaling of \citet{Cappellari2013} based on kinematics to
calculate an independent stellar mass. We find agreement within $\sim
30 \%$ in all cases. Given the order of magnitude range probed by our
galactic sample, we are therefore able to robustly separate galaxies into mass bins.

\section{Analysis: Kinematic Modelling}
\label{Sec:Observations}

We briefly outline here the higher-level analysis (source masking and
binning) involved in preparing the data for kinematic measurements.
We then describe the extraction of kinematic parameters using the
penalised PiXel Fitting (pPXF) technique of
\citet{CappellariEmsellem2003}.

\subsection{Source Masking}
\label{SubSec:Source Masking}

Before coadding or fitting any spectra, we first mask out any fibers
on the IFU that are dominated by foreground sources. Close to half of our
galaxies (NGC~219, NGC~661, NGC~677, IC~301, NGC~1286, IC~312, NGC~1267,
NGC~3837, NGC~3842, NGC~4065, NGC~4952, NGC~6127, NGC~6964, NGC~7509,
NGC~7684) require masking of some sort beyond $R_e$, though in most
cases the external sources are not extended and so do not affect more
than one or two fibers.

Two galaxies, NGC~1267 and NGC~6482, also have bright stars 
between us and the galaxy center, which contaminate 4 or 5 fibers in the
core. Kinematic measurements are therefore not able to probe to radii
within $\sim 0.5 R_e$ for these galaxies. Finally, we note that a
small number of galaxies, most notably NGC~426 and NGC~7509, have
central fibers dominated by strong emission lines characteristic of
AGN, which can affect measurements of dispersion. However, we choose
not to mask out these fibers, but instead deal with the emission lines
in our fitting procedure (see Appendix).

\subsection{Spatial Binning}
\label{SubSec:Spatial Binning}

\begin{figure}
\begin{center}
\includegraphics[width=0.8\columnwidth,angle=0,clip]{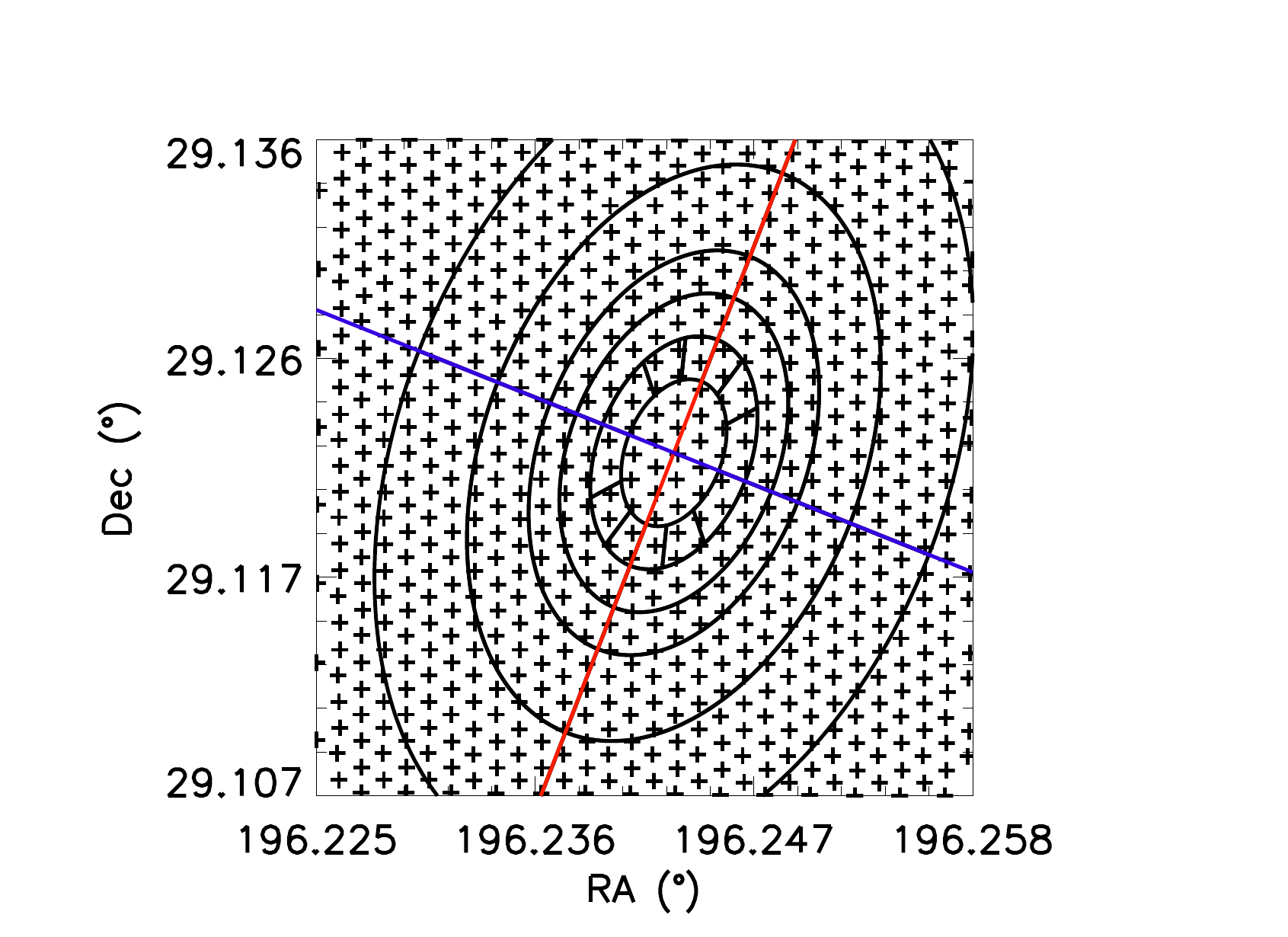}
\includegraphics[width=0.8\columnwidth,angle=0,clip]{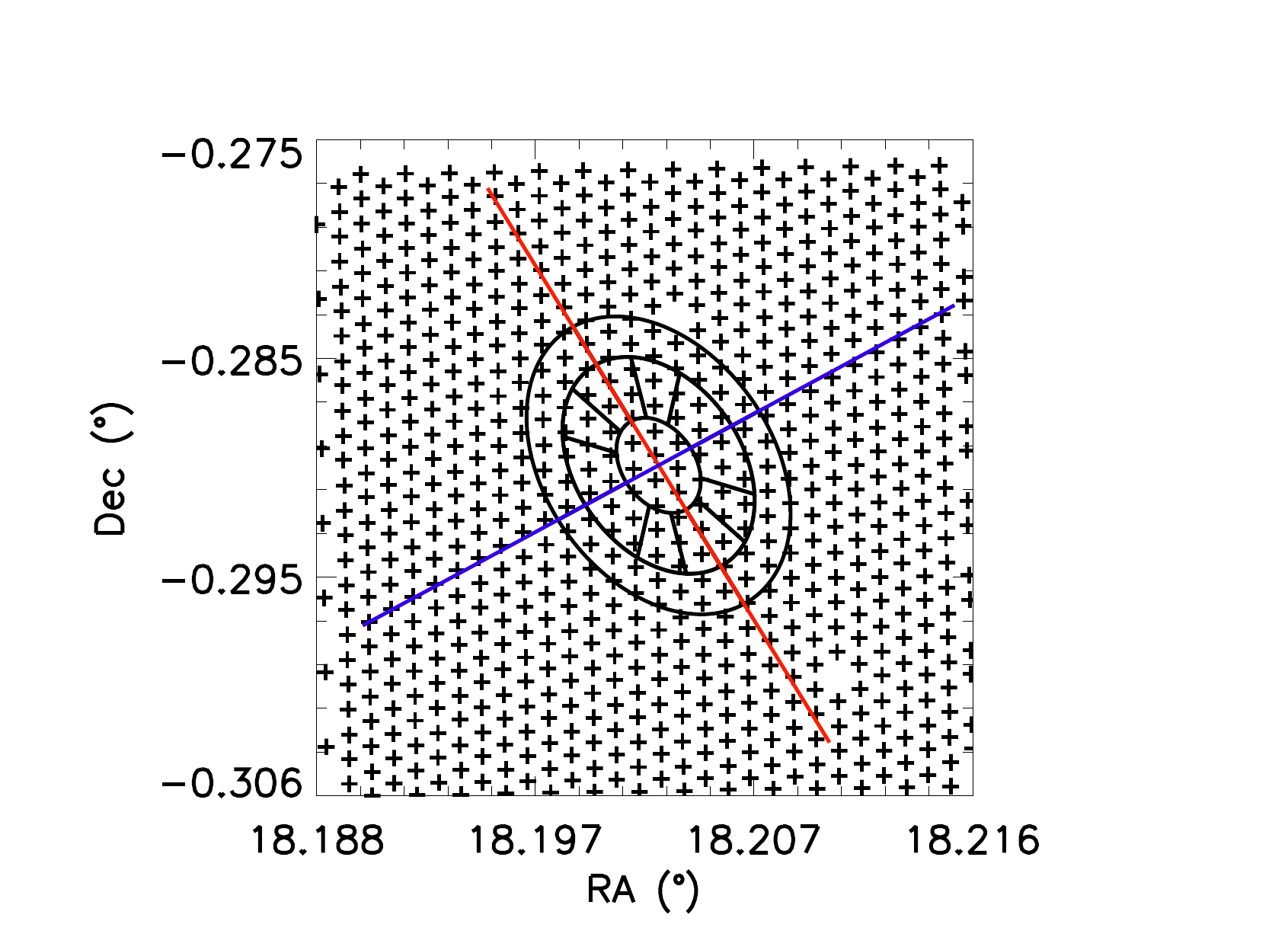}
\caption{
  Locations of bins for galaxies NGC~4952 (top) and NGC~426
  (bottom).  We show fiber positions for one dither (crosses), bin locations, and 
the major (red) and minor(blue) axes.  Radial bins vary in size
  between 0.5 and 1 $R_e$. We note that a typical observation involves
  3 dithers of the IFU so that sky coverage is 3 times denser than
  shown.}
\label{Fig:Binning}
\end{center}
\end{figure}

We require a minimum S/N of 15 (justified in the Appendix) to extract
robust kinematics, but only the very central individual fibers have
such high S/N.  Therefore, we must perform our analysis on radially
binned spectra. All spectra are resampled onto a common wavelength
grid over the range $4000$~\AA $< \lambda < 5420$~\AA.  Both spectra and
errors are then weighted by their flux, coadded, and
renormalised. We use flux-weighted addition with
iterative sigma-clipping, which provides a simple and reliable
estimate of the coadded errors. The spectra combined in this manner
are nearly identical to those derived from the biweight estimator
\citep{Beers1990} employed in \citet{Murphy2011}.

The size of each spatial bin is set by the minimum S/N requirement.
The innermost fibers that pass this threshold are analysed without
further coaddition. We then bin in elliptical annuli set by the axial
ratio measured from SDSS. Each radial bin begins with a width of
$0.5~R_e$ along the major axis and is further separated into $5$
angular bins in each quadrant, each with equal width in $\rm sin
\theta$. Spectra are folded across the minor axis, so that we are left
with $10$ angular bins in total \citep{Gebhardt2000, Gebhardt2003,
  McConnell2012}. When an angular bin falls below the S/N threshold,
it is merged with its nearest neighbors until we are left with only
two angular bins, one on each side of the major axis, at a given
radius. If more binning is required, then the size of the bin is
increased radially by $0.1~R_e$ until we cross the S/N threshold or
the edge of the integral-field unit (IFU) is reached.  With
this procedure we are typically able to probe out to $\sim 4~R_e$ with
full angular information available only for the inner bins.

Figure~\ref{Fig:Binning} demonstrates our binning scheme. In the case
of the spatially extended galaxy NGC~4952, shown in the top panel of 
Figure~\ref{Fig:Binning}, we are able to retain
many individual fibers and then eventually end with five purely
radial bins of varying width. Even here, we can only resolve angular
structure out to $\sim 1.5~R_e$, while the outermost bin runs over the
edge of the IFU.  NGC~426 (bottom), while still quite massive, is one of our
smallest galaxies on the sky.  It has a much smaller inner region, and
only 3 radial bins.  Nevertheless, because the galaxy is so compact,
we are able to probe out to $\sim 3~R_e$.

\subsection{Kinematics}
\label{SubSec:Kinematics}

\begin{figure*}
\begin{center}
\includegraphics[width=0.8\columnwidth,angle=0,clip]{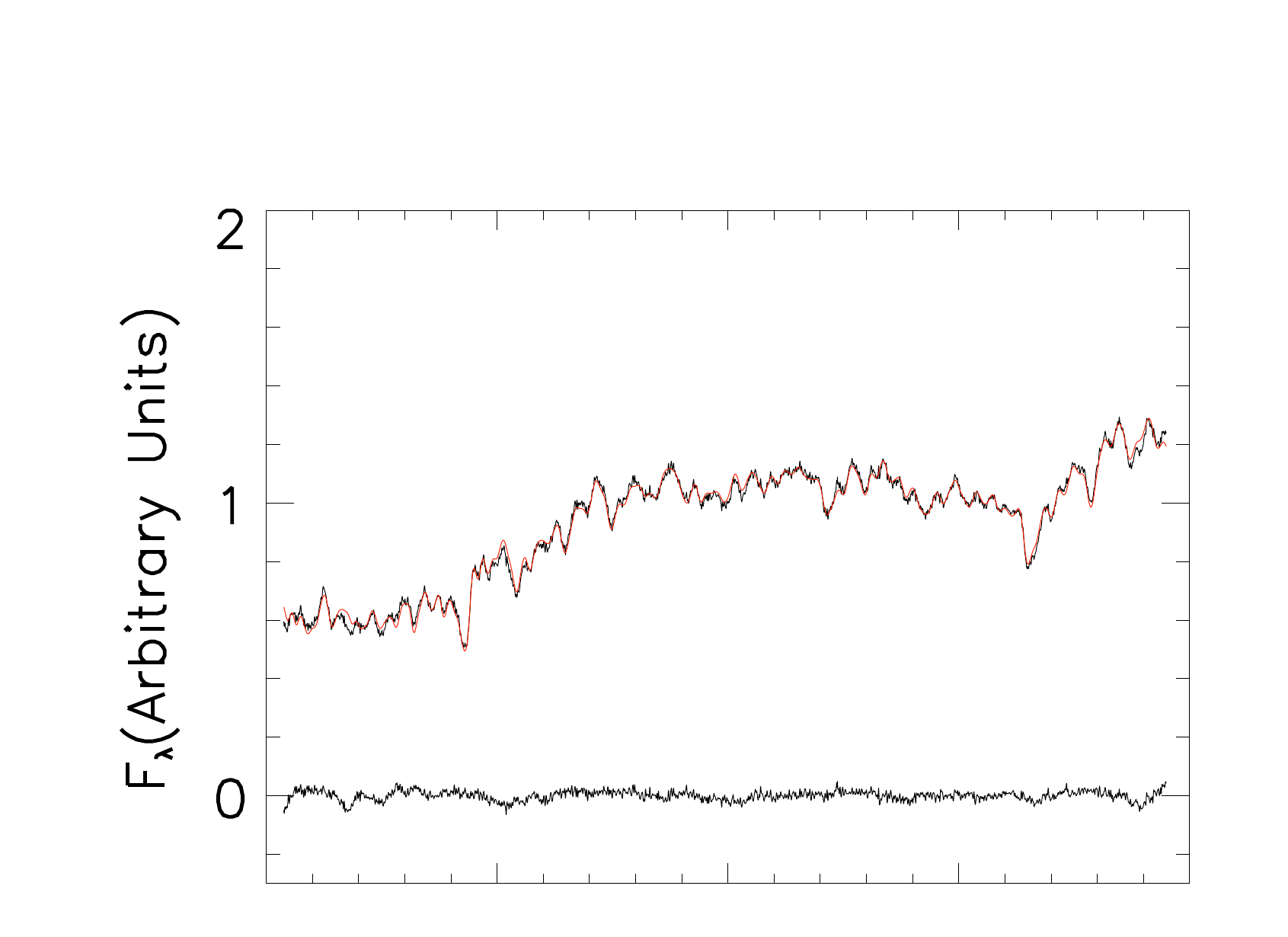}
\includegraphics[width=0.8\columnwidth,angle=0,clip]{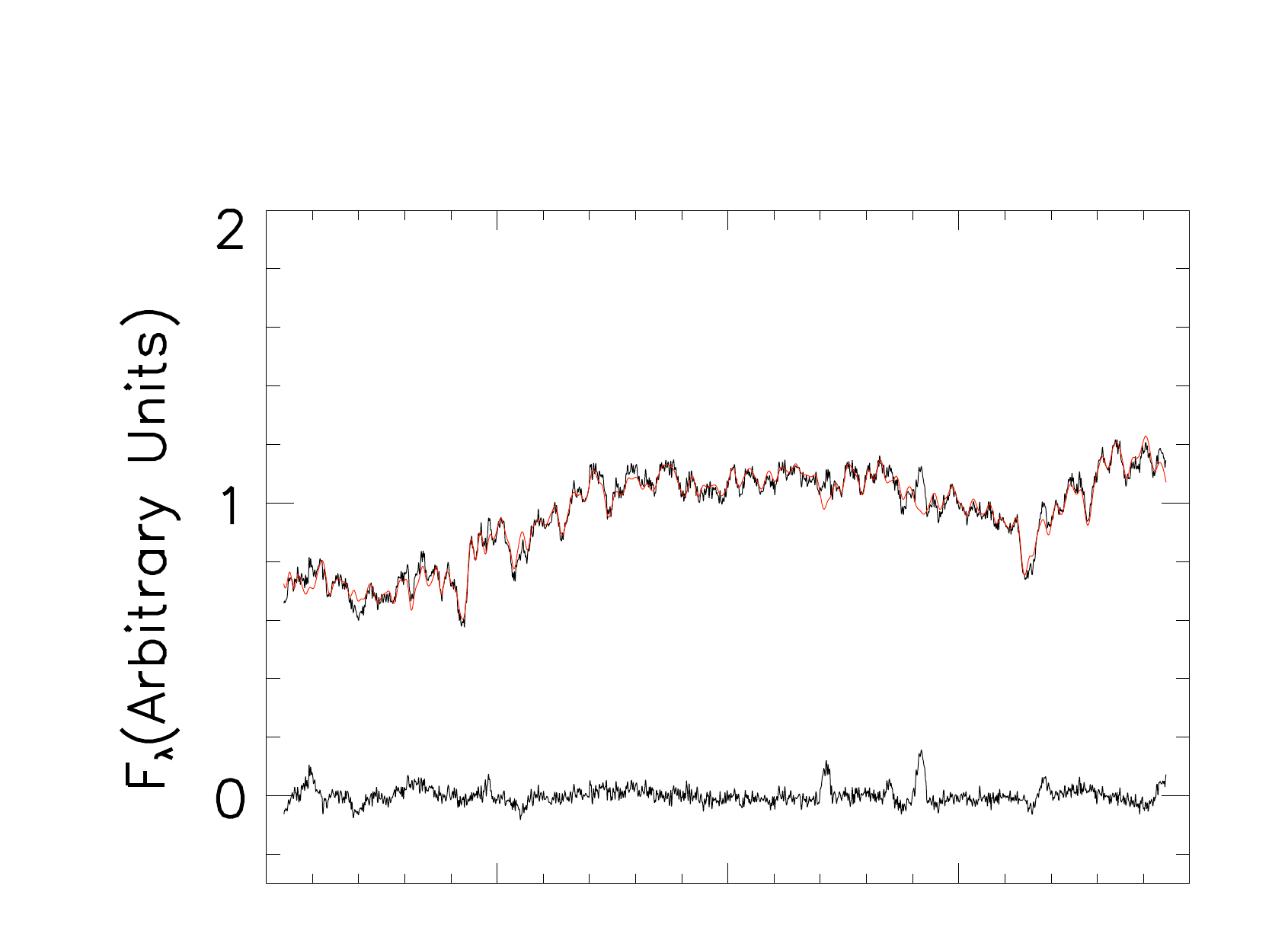}
\includegraphics[width=0.8\columnwidth,angle=0,clip]{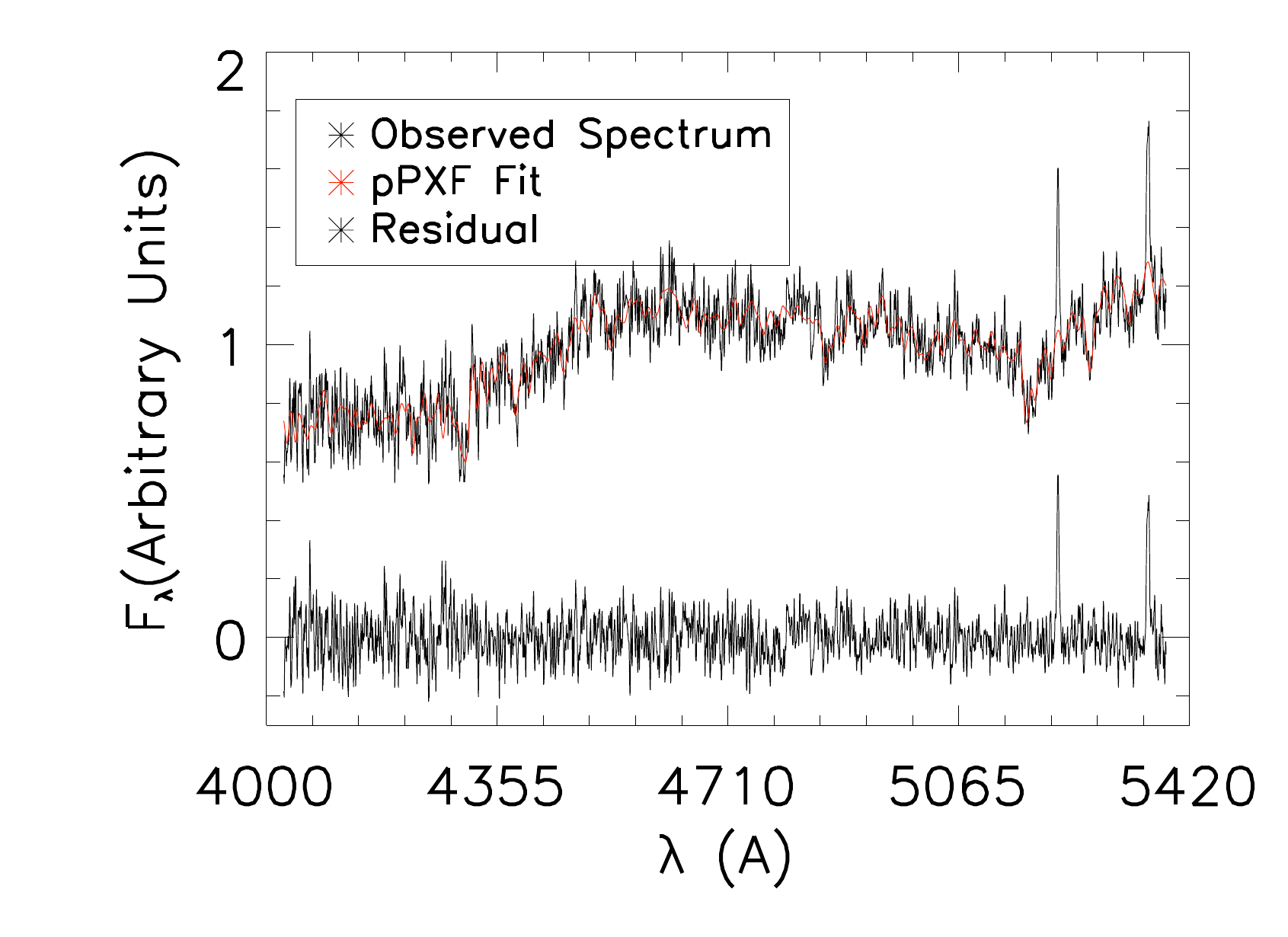}
\includegraphics[width=0.8\columnwidth,angle=0,clip]{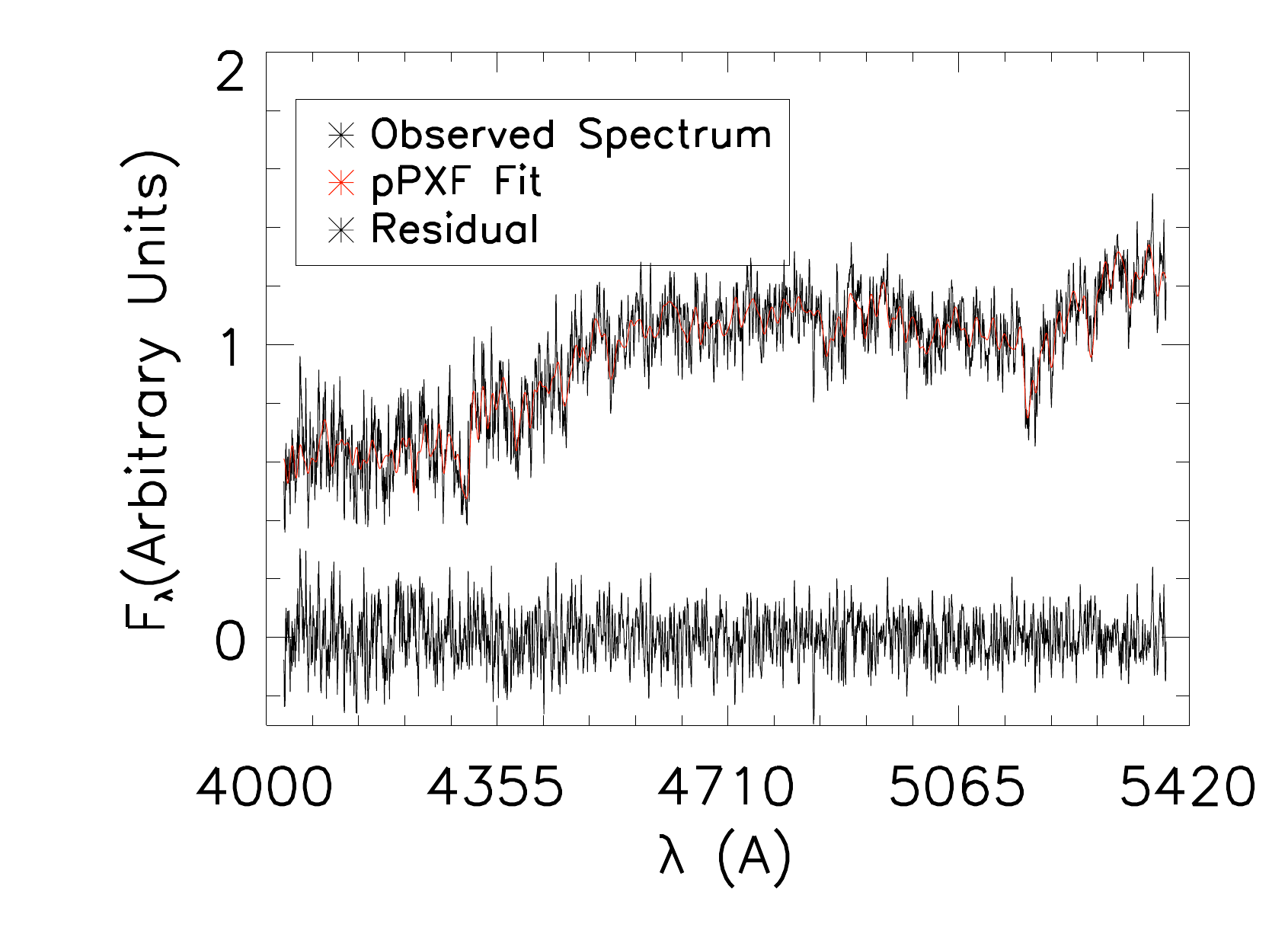}
\caption{Spectra and pPXF fits for the central (top) and outermost (bottom) 
bins in NGC~4952 (left) and 
NGC~426 (right) over our fiducial wavelength range. 
The fit for NGC~4952 especially suffers no issues 
with template mismatch even at low S/N.}
\label{Fig:PPXF1}
\end{center}
\end{figure*}

In principle, stellar velocity dispersions can be measured in a number
of ways, including Fourier techniques such as the cross-correlation
\citep{TonryDavis1979} and the Fourier quotient \citep{Simkin1974,
  Sargent1977}. However, now that computational costs are no
limitation, direct pixel-by-pixel fitting
\citep{Burbidge1961, RixWhite1992} allows for masking of emission
lines and does not suffer from windowing problems. We employ the
direct fitting pPXF technique of \citet{CappellariEmsellem2003} to
calculate our stellar kinematics. In brief, pPXF convolves a library 
of stellar templates with a line-of-sight velocity distribution (LOSVD) 
function that is modeled as a Gauss-Hermite series 
\citep{MarelFranx1993,Gerhard1993}:
\begin{equation}
\mathcal{L}(v)=\frac{e^{-(1/2)y^2}}{\sigma\sqrt{2\pi}}
    \left[ 1 + \sum_{m=3}^M h_m H_m(y) \right]
    \label{Eq:Losvd}
\end{equation}
where $y=(v-V)/\sigma$ and the $H_m$ are the Hermite polynomials. The
free LOSVD parameters $\{V, \sigma, h_i\}$ are fit by minimizing over
a given objective function using the Levenberg-Marquardt method for
non-linear least squares problems. The objective function itself is
regularised to favour gaussian profiles so that
\begin{eqnarray}
	\chi_{\rm p}^2\;=\;\chi^2\; (1+\lambda^2 \mathcal{D}^2), \nonumber \\
	\mathcal{D}^2\approx\sum_{m=3}^{M} h_m^2.
\label{Eq:Chi2p}
\end{eqnarray}
with $\lambda \sim 0.7$ found to work well empirically
\citep{Cappellari2011}.  We find no significant deviations from
Gaussian LOSVDs in our data, so we simply set $\lambda = 0$, {or equivalently 
fix all hermite moments to zero}, and the
problem reduces to pixel-fitting by standard $\chi^2$ minimisation
with a Gaussian LOSVD. We fit over a large wavelength range that
starts just redward of the $4000$~\AA\ break and extends to the Fe
lines at $5420$~\AA. Over this region the continuum is well-modelled
with multiplicative Legendre polynomials of order $10$, and the
presence of emission lines has little systematic effect on the derived
kinematics. A more detailed justification for our choice of wavelength
region, continuum polynomial degree, and LOSVD can be found in the
Appendix.

Errors for our Gaussian fits were estimated using a Monte-Carlo
method. We started with the noiseless fit to each spectrum and
added Gaussian-distributed noise to each pixel according to its error
array.  The new noisy spectrum was then fit using pPXF. When repeated
over many realizations of the added noise, this produces an estimate
of the errors in our fits to $V$ and $\sigma$. {In a small number of galaxies, 
measured dispersions fell well below the instrumental dispersion ($\sigma \lesssim 100$~\kms), 
and measured errors approached $\sim 20 \%$. These measurements were 
deemed unreliable and ignored in all further analysis.}

Our stellar templates were chosen from stellar population synthesis
(SPS) spectra, as generated by the Flexible SPS (FSPS) code
\citep{Conroy2009} calibrated to the observational data in
\cite{ConroyGunn2010}, with an intrinsic resolution of $2.5$~\AA\
FWHM. Since pPXF is sensitive to the completeness of the stellar
library \citep{Cappellari2011} we adopted a wide range of ages, $3~{\rm
  Gyr} < t < 13.5~{\rm Gyr}$ and {alpha-enhancements}, $0.0 < [\alpha/Fe] < 0.4$, 
alongside a Chabrier IMF for our template library. We then allowed each binned
spectrum to fit to a weighted sum of these templates. We used 
SPS models rather than stars to gain some additional insight into the 
stellar populations of our galaxies. However, we did cross-compare with 
stellar templates, and examine our sensitivity to template mismatch 
in detail, with resulting systematic errors in our kinematic
estimates of $\sim 10 - 20~{\rm kms^{-1}}$.

\subsection{Robustness of the Kinematic Fits}
\label{subsec:Robustness}

\begin{figure}
\begin{center}
\includegraphics[width=0.9\columnwidth,angle=0,clip]{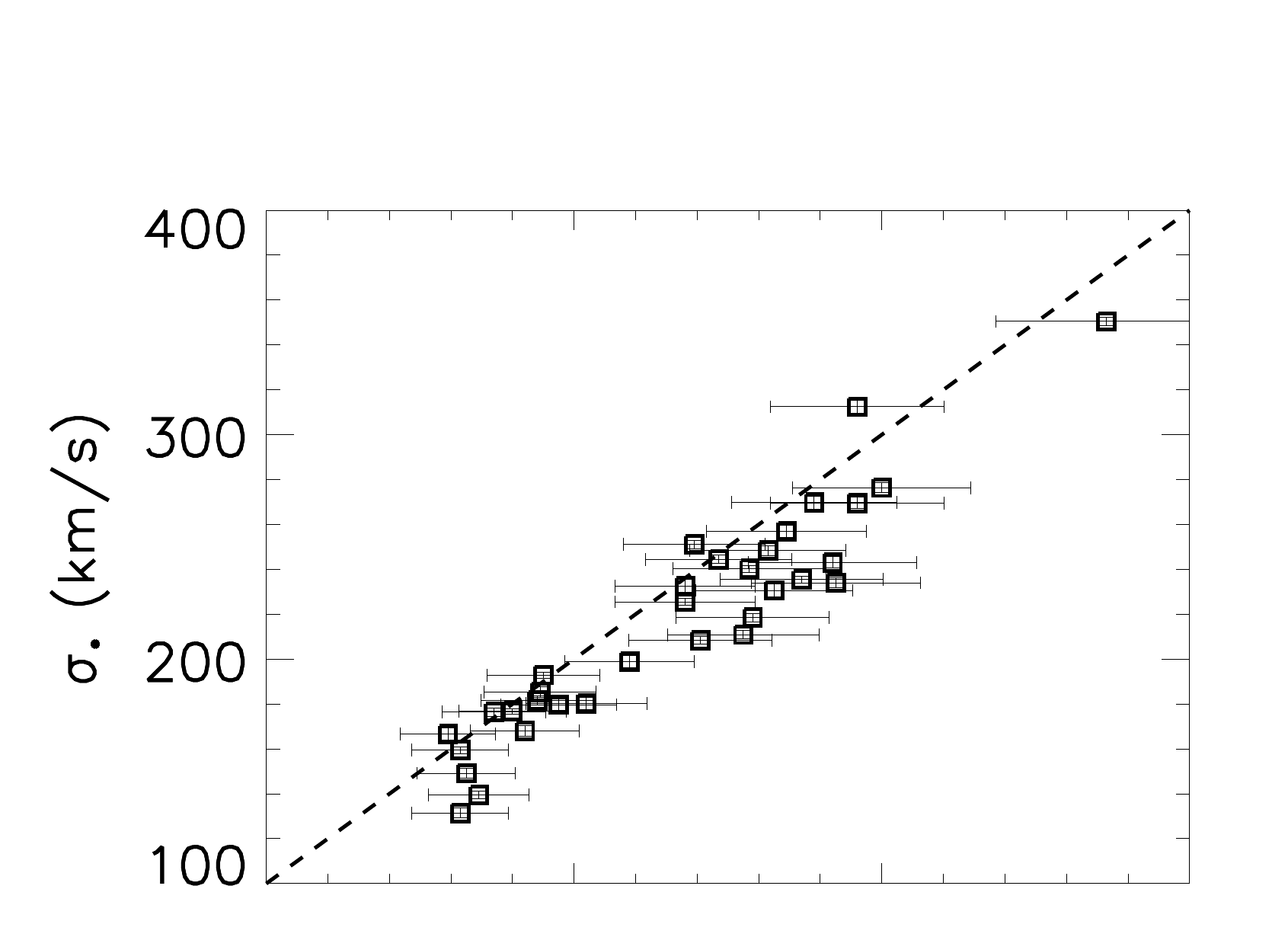}
\includegraphics[width=0.9\columnwidth,angle=0,clip]{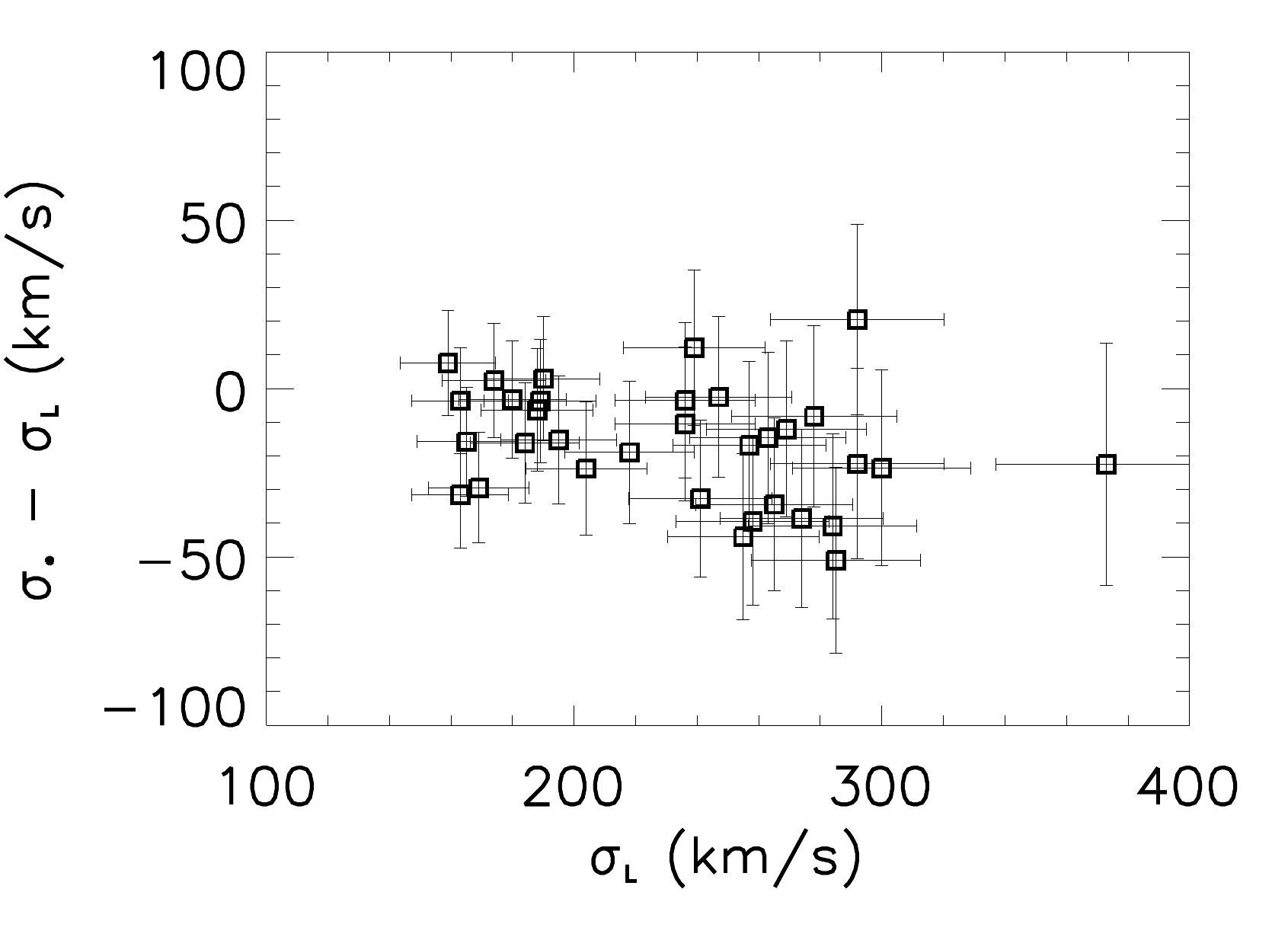}
\caption{
Velocity dispersions in the central fiber $\sigma_{c}$ of our 
galaxies as compared with values taken from the literature $\sigma_{\rm L}$. We 
show both the dispersions themselves (top) and the offset 
between literature and our measurement (bottom). The literature values 
have been taken from 
the SDSS catalog \citep{Blanton2005} as well as several collated data sets 
of galactic dispersions \citep{Whitmore1985, Mcelroy1995, Ho2009}. The 
$\sim$20~\kms\ offset likely reflects our larger aperture, {although our 
aperture does match relatively well to the SDSS.}}
\label{Fig:DispvsL}
\end{center}
\end{figure}

Our extracted kinematics are subject to both statistical 
errors and systematic uncertainties in the continuum, the degree of
non-gaussianity in our LOSVD, the choice of stellar templates, and the
presence of emission lines in certain galaxies. We therefore tested
the robustness of our kinematic fits to all of these uncertainties in
Appendix A. As a summary, we show in Figure~\ref{Fig:PPXF1} example 
fits to the central and outermost fibers in NGC~4952 and NGC~426. We see
that the effects of template mismatch are relatively small, though not
entirely negligible, especially in the case of NGC~426.

We also show comparisons of our central velocity dispersions to literature
data in Figure~\ref{Fig:DispvsL}. Almost all of our galaxies have dispersions in the
SDSS catalog \citep{Blanton2005}, 
and we supplement these with the
lists compiled by \cite{Whitmore1985}, 
\cite{Mcelroy1995} and most recently \cite{Ho2009}. {We match the SDSS aperture 
relatively well, while the \cite{Ho2009} measurements had a similar 2\arcsec by 2 \arcsec 
aperture, and the \cite{Whitmore1985} and \cite{Mcelroy1995} compilations were placed 
on a 2\arcsec by 4\arcsec system}. As can be seen
from Figure~\ref{Fig:DispvsL}, we find a small systematic
negative bias in our dispersion calculations of around 20~\kms\ 
on average. The low resolution of our fibers means that we 
blend dispersions out to larger radius than SDSS and so expect to find 
systematically smaller dispersions. Furthermore, this bias is typically less than 
$\sim 2 \sigma$ and is much smaller than the 
large measured dispersions of $\gtrsim$~150~\kms.

\subsection{Tracing Angular Momentum}
\label{subsec: Angular Momentum}

In moving from an inner disk or bulge-like component to an outer
stellar halo we may expect changes in how much of the galaxy
angular momentum is stored in random as opposed to ordered motion.
With long-slit spectroscopy, typically only the ratio of maximum
observed rotational velocity ($V_{\rm max}$) to central dispersion
\citep[$\sigma_c$; ][]{Illingworth1977, Binney1978, Davies1983} can be 
measured. With IFS, which provides full 2D kinematic information, we
can construct a more robust measure incorporating both radial and
spatial variations.  \cite{Binney2005} introduced the ratio of the
luminosity-weighted integrated quantites $\langle V^2 \rangle$ and
$\langle \sigma^2 \rangle$. However, weighting by the surface
brightness alone tends to overestimate the importance of central
regions and conflate very different kinematic structures, such as
organized rotation versus small central KDCs. To counteract this
limitation, the SAURON survey introduced
the parameter $\lambda_R$ \citep{Emsellem2007}, which weights by radius as well as flux and
therefore measures the projected baryonic specific angular momentum
\begin{equation}
\lambda_R \equiv \frac{\langle R \, \left| V \right| \rangle }{\langle R \, \sqrt{V^2 + \sigma^2} \rangle}
\,
\end{equation}
where the brackets indicate a flux-weighted sum within an ellipse of
mean radius $R$.  In calculating $\lambda_R$, we adopt the approach of
\cite{Wu2014} and slightly modify the above definition. Instead of taking
the absolute value of the velocity, we sum over the actual velocity
separately on both sides of the rotation axis. This allows positive and negative
noise terms to cancel, resulting in a smoother profile.

Errors on $\lambda_R$ are estimated using the formal fit errors to $V$
and $\sigma$. While systematic errors in the pPXF fits are likely 
an important factor, particularly in the inner regions \citep{Emsellem2011},  
they will tend to have roughly the same effect on all galaxies. Therefore, 
these error estimates are at least indicative of the relative differences between 
galaxies.

Using $\lambda_R$, \cite{Emsellem2007} found that SRs and FRs could
be quite robustly separated, with SRs being specified by
$\lambda(R_e) \le 0.1$ and FRs by $\lambda(R_e) \ge
0.1$. \cite{Emsellem2011} and \cite{Krajnovic2011}, looking at the
full ATLAS$^{\rm 3D}$ sample, found that this simple picture was slightly
blurred by inclination effects since a flattened system viewed
face-on would have a similar profile to a nearly spherical galaxy seen
at a large inclination angle. However, with a modified definition of
FRs, $\lambda(R_e) \ge 0.31 \times \epsilon_e$, the same picture of
two dichotomous families separated by their kinematics and formation
processes still holds \citep[e.g.,][]{Bender1989,KormendyBender1996,
deZeeuw1985,Franx1991,deZeeuwFranx1991, vandenBosch2008}.

We calculate $\lambda_R$ evaluated at the effective radius for each
galaxy and classify them as slow (SRs) or fast rotators (FRs) based on
the \cite{Emsellem2011} definition. Twenty-one of our galaxies are
FRs, and the remaining 12 are SRs. As we will see below, there are
some borderline cases, galaxies classified as FR that are right at the
boundary within $\sim R_e$ but then do not rotate at all in their
outer parts.  Even sticking to the strict definition, we have a much
higher percentage of SRs than the $14 \%$ found in the ATLAS$^{\rm
  3D}$ sample. We also note that
these classifications are entirely derived from the kinematics within
$R_e$. In \S~\ref{subsec:Changing Kinematics} we will explore changes
in $\lambda_R$ with radius.

\subsection{Kinemetry}
\label{subsec: Kinemetry}

Beyond just the angular momentum content with radius, we 
can also investigate different kinematic components in the galaxies.
We turn to {\it kinemetry} \citep{Krajnovic2006} to help identify 
and characterize substructures in these maps. Kinemetry
extends the basic assumption of photometry, that the surface
brightness of ETG's is constant to within $\lesssim 1 \%$ along
best-fit ellipses, to higher order moments of the LOSVD. 
It assumes that symmetric (even) moments, such as the
dispersion, are constant along ellipses while antisymmetric (odd) moments,
such as the velocity, satisfy a simple cosine law to first order.
Therefore, by fitting these kinematic parameters in elliptical annuli
we may obtain simple radial profiles of the velocity and dispersion
showing the scales at which important kinematic transitions occur.

In particular, application of kinemetry to our velocity maps allows
us to derive radial profiles of the kinematic position angle
PA$_{\rm kin}$ and flattening $q_{\rm kin}$ of our best-fit ellipses. We may
furthermore decompose the velocity along these ellipses into Fourier
coefficients
\begin{equation}
	V(a, \phi) = V_0 + \sum_{n=1}^{\infty}A_n(a){\rm sin}(n \phi) + B_n(a){\rm cos}(n \phi)
\end{equation}
where $a$ is the ellipse radius and $V_0$ is the systemic velocity,
set to a constant across all radii.  The
simple cosine law decomposition is dominated by the
first order term, $k_1 = \sqrt{A_1^2 + B_1^2}$, which captures the
rotation curve.  By choosing best-fit ellipses we
effectively minimise all higher order terms up to $k_5$. Radial
profiles of these coefficients therefore also provide information on
the rotation curve ($k_1$) and any deviations from the assumption of
ellipticity ($k_5$), which tend to occur in transitions between
components rotating at different velocities and/or PA.

This kinemetric analysis is essentially identical to that performed
for the SAURON galaxies in \cite{Krajnovic2008} and the ATLAS$^{\rm
  3D}$ sample in \cite{Krajnovic2011}, but because our data are so
different we uncover different kinds of structures.  Our galaxies
have on average about $40$ binned data points out to $\sim 3-4 R_e$ compared
to the thousands of points available to ATLAS$^{\rm 3D}$ galaxies within
$R_e$. We cannot resolve classic kinematically decoupled components
but we cover much larger-scale features.

Our sample characteristics are also different; a large fraction
display little-to-no net rotation.  These galaxies must be treated
differently, since the determination of best-fitting ellipses for a
velocity map close to zero everywhere is highly degenerate. Therefore,
for cases where we have very low velocities $k_1 < 15$~\kms,
we follow the approach of \cite{Krajnovic2008} and rerun the
kinemetric analysis assuming $q_{\rm kin} = 1$ (where $q$ is 
the axis ratio).  Deviations from a
cosine profile, seen in the $k_5$ term, will be artificially inflated
since a circular profile does not necessarily match
the isovelocity contours and the PA$_{\rm kin}$ is poorly
defined. However, they both provide some indication of the extent of
the non-rotating component.

A final caveat is that our S/N, particularly at large radius, tends to
be lower than for the ATLAS$^{\rm 3D}$ galaxies. For instance, whereas the
average error on $k_5 / k_1$ for the SAURON sample is 0.015
\citep{Krajnovic2008}, our galaxies typically have errors in this
ratio that approach 0.04. This means that as the S/N deteriorates, 
we are only sensitive to relatively large changes in  PA$_{\rm kin}$, $q_{\rm kin}$ 
and $k_5 / k_1$, around 10\degr, 0.2 and 0.1 respectively.

Nevertheless, with these caveats in mind, we may attempt to identify
kinematic substructures on each map using a similar classification
scheme to \cite{Krajnovic2008} but with more conservative limits.  In
particular, we first differentiate between galaxies that are
\begin{enumerate}
	\item 
          Single Component (SC): Constant or slowly varying PA and
          $q$, with changes in the former restricted to less than
          $10^{\circ}$. We do not constrain $q$ well enough to take
          changes in $q$ alone as signs of a transition between
          components.

	\item 
          Multiple Component (MC): Abrupt change in $\Delta {\rm PA}
          > 10^{\circ}$, or a double hump in $k_1$ with a corresponding local
          minimum or peak in $k_5$ with $k_5 / k_1 > 0.05$
\end{enumerate}
Individual subcomponents of each galaxy can then be identified as
\begin{enumerate}
	\item Disk-like Rotation (DR): $k_5 / k_1 < 0.05$, while variation in PA is less than $10^{\circ}$.
	\item Low-level Velocity (LV): Maximum of $k_1$ is less than $15$~\kms.
	\item Kinematic Twist (KT): Smooth variation in PA of more than $10^{\circ}$.
	\item Kinematically Distinct (KD) components: 
          Abrupt change of larger than $20^{\circ}$ between adjacent
          components or an outer LV component in which the derivation
          of PA is uncertain.  Note we will not call these ``kinematically decoupled 
          cores'' because we are only sensitive to structures larger than one kpc, while 
          classic KDs can have scales of 100s of pc \citep{Kormendy1984, Forbes1994, Carollo1997}.
	  {They are however quite similar in scale to the ``kinematically distinct haloes" discussed 
	  in \cite{Foster2013}.}
\end{enumerate}
We will classify each galaxy and discuss the characteristics of the 
different subclasses in \S~\ref{Sec:Observed Kinematics}.

\section{Results: LOS Kinematics at Large Radius}
\label{Sec:Observed Kinematics}

\subsection{Angular Momentum Classification}
\label{subsec: Angular Momentum Classification}

\begin{table*}
\vspace{0.5 cm}
\caption{Kinematic Properties of the Galaxy Sample}
\centering
\tiny

\begin{tabular}{ccccccccccr}
\hline \hline
Galaxy & $V_{\rm m}$ & $R_{\rm max}$ & $\sigma(R_e)$ & $\sigma(R_{\rm m})$ & $\lambda(R_e)$ & 
$\lambda(R_{\rm m})$ & Rot & Structure & $R_t$ & $\Psi$ \\
(1) & (2) & (3) & (4) & (5) & (6) & (7) & (8) & (9) & (10) & (11) \\
\hline
NGC~219 & $76.5 \pm 5.7$ & 1.42 - 2.87 & $159 \pm 11$ & $154 \pm 15$ & 0.20 & 0.21 & FR & SC (KT) & --- & $27 \pm 25$ \\
NGC~426 & $213.6 \pm 5.7$ & 2.35 - 3.23 & $166.7 \pm 3.6$ & $121 \pm 41$ & 0.54 & 0.64 & FR & SC (DR) & --- & $-10 \pm 12$ \\
NGC~474 & $50.0 \pm 3.0$ & 2.50 - 3.29 & $125.7 \pm 2.4$ & $98.1 \pm 7.4$ & 0.16 & 0.17 & FR & MC (KD/DR, DR) & 2.2 & $-3 \pm 29$ \\
CGCG~390-096 & $99.2 \pm 4.5$ & 1.91 - 4.11 & $159.5 \pm 2.7$ & $151 \pm 12$ & 0.35 & 0.38 & FR & SC (DR) & --- & $53 \pm 34$ \\
NGC~661 & $67.1 \pm 5.3$ & 2.13 - 2.57 & $143.2 \pm 3.1$ & $92 \pm 14$ & 0.19 & 0.23 & FR & MC (LV, DR) & 1.9 & $3.0 \pm 6.6$ \\
NGC~677 & $54.7 \pm 3.8$ & 4.57 - 5.14 & $209.4 \pm 3.3$ & $185 \pm 19$ & 0.11 & 0.057 & FR & MC (KD/KT, DR) & 4.3 & $0 \pm 11$ \\
UGC~1382 & $104.4 \pm 6.7$ & 1.53 - 2.23 & $153.1 \pm 2.9$ & $167 \pm 16$ & 0.30 & 0.35 & FR & SC (KT) & --- & $25 \pm 40$ \\
NGC~774 & $141.2 \pm 5.4$ & 1.91 - 2.53 & $115.4 \pm 2.4$ & $130 \pm 24$ & 0.51 & 0.47 & FR & SC (DR) & --- & $-2 \pm 24$ \\
IC~301 & $32.7 \pm 5.7$ & 2.52 - 4.13 & $151.3 \pm 3.3$ & $81 \pm 51$ & 0.089 & 0.072 & FR & MC (KD/DR, LV) & 2.6 & $7.4 \pm 8.3$ \\
NGC~1286 & $77.7 \pm 3.6$ & 1.58 - 2.08 & $63.0 \pm 2.5$ & ---* & 0.37 & 0.57 & FR & SC (DR) & --- & $-2 \pm 34$ \\
IC312 & $199.7 \pm 5.1$ & 1.11 - 2.15 & $152.8 \pm 5.3$ & $141 \pm 10$ & 0.59 & 0.61 & FR & SC (DR) & --- & $-43 \pm 22$ \\
NGC~1267 & $64.4 \pm 9.8$ & 3.65 - 5.41 & $211.7 \pm 7.8$ & $196 \pm 29$ & 0.067 & 0.15 & SR & MC (KT, DR) & 3.8 & $35.9 \pm 4.9$ \\
NGC~1270 & $145.7 \pm 2.8$ & 4.09 - 5.53 & $233.7 \pm 1.7$ & $183 \pm 48$ & 0.24 & 0.33 & FR & SC (DR) & --- & $-28 \pm 42$ \\
NVSS & $103.4 \pm 3.7$ & 1.97 - 3.10 & $167.0 \pm 3.6$ & $112 \pm 27$ & 0.24 & 0.53 & FR & SC (DR) & --- & $-28 \pm 32$ \\
UGC~4051 & $135 \pm 14$ & 3.01 - 4.69 & $244.4 \pm 5.4$ & $234 \pm 16$ & 0.23 & 0.22 & FR & SC (DR) & --- & $15 \pm 32$ \\
NGC~3837 & $40.9 \pm 3.4$ & 2.78 - 4.29 & $199.9 \pm 2.6$ & $143 \pm 12$ & 0.11 & 0.087 & FR & MC (KD/KT, LV) & 7.1 & $-42 \pm 27$ \\
NGC~3842 & $24.5 \pm 5.8$ & 2.13 - 2.77 & $219.6 \pm 2.9$ & $216 \pm 13$ & 0.039 & 0.037 & SR & MC (KD/DR, LV) & 3.2 & $31 \pm 40$ \\
NGC~4065 & $41 \pm 11$ & 2.04 - 2.83 & $244.7 \pm 4.3$ & $218 \pm 28$ & 0.053 & 0.043 & SR & MC (KD/KT, LV) & 5.4 & $-4.8 \pm 2.8$ \\
IC~834 & $143.5 \pm 4.1$ & 2.32 - 4.31 & $143.7 \pm 5.0$ & $119 \pm 17$ & 0.42 & 0.58 & FR & MC (DR, DR) & 3.2 & $-20.3 \pm 3.4$ \\
NGC~4908 & $53 \pm 17$ & 1.64 - 2.52 & $254.4 \pm 7.8$ & $257 \pm 49$ & 0.11 & 0.11 & SR & MC (KD/DR, LV) & 5.7 & $-53 \pm 33$ \\
NGC~4952 & $100.4 \pm 7.6$ & 2.74 - 3.91 & $213.0 \pm 2.7$ & $159 \pm 15$ & 0.21 & 0.30 & FR & MC (DR, DR, DR) & 3.1, 8.9 & $-1.3 \pm 0.4$ \\
NGC~5080 & $57.7 \pm 9.2$ & 3.52 - 5.58 & $239.6 \pm 4.9$ & $146 \pm 16$ & 0.08 & 0.13 & SR & MC (LV, DR) & 2.6 & $-67 \pm 47$ \\
NGC~5127 & $25.3 \pm 4.7$ & 1.90 - 2.36 & $162.5 \pm 2.8$ & $170 \pm 16$ & 0.041 & 0.038 & SR & MC (KD/DR, LV) & 4.7 & $73 \pm 44$ \\
NGC~5423 & $101.2 \pm 6.0$ & 3.60 - 4.78 & $196.5 \pm 3.6$ & $225 \pm 49$ & 0.27 & 0.30 & FR & SC (DR) & --- & $17 \pm 26$ \\
NGC~5982 & $65.0 \pm 9.8$ & 3.34 - 4.01 & $206.7 \pm 1.6$ & $204 \pm 9 $ & 0.076 & 0.13 & SR & MC (LV, DR, DR) & 1.6, 5.2 & $-10 \pm 32$ \\
IC~1152 & $20.3 \pm 6.0$ & 3.61 - 4.71 & $181.9 \pm 4.0$ & $155 \pm 16$ & 0.043 & 0.038 & SR & SC (LV) & --- & $-15 \pm 27$ \\
IC~1153 & $155.5 \pm 3.2$ & 3.37 - 4.89 & $172.7 \pm 2.9$ & $104 \pm 13$ & 0.43 & 0.41 & FR & SC (DR) & --- & $-1 \pm 30$ \\
CGCG~137-019 & $34.2 \pm 7.3$ & 2.32 - 4.18 & $155.4 \pm 3.0$ & $114 \pm 15$ & 0.091 & 0.11 & SR & MC (LV, DR) & 2.4 & $23.5 \pm 9.3$ \\
NGC~6125 & $38.8 \pm 8.7$ & 3.85 - 4.51 & $207.2 \pm 3.2$ & $204 \pm 41$ & 0.067 & 0.080 & SR & MC (LV, DR) & 4.9 & $16 \pm 39$ \\
NGC~6482 & $114.8 \pm 9.2$ & 4.25 - 4.97 & $304.3 \pm 4.2$ & $225 \pm 22$ & 0.10 & 0.18 & SR & MC (DR, DR) & 1.6 & $51 \pm 15$ \\
NGC~6964 & $99.9 \pm 6.2$ & 1.54 - 2.03 & $140.7 \pm 4.5$ & $88 \pm 17$ & 0.29 & 0.35 & FR & MC (KT, DR) & 4.0 & $27.5 \pm 5.5$ \\
NGC~7509 & $41.3 \pm 5.2$ & 3.56 - 4.33 & $140.2 \pm 3.0$ & $148 \pm 31$ & 0.11 & 0.098 & FR & MC (LV, DR) & 3.6 & $-41 \pm 36$ \\
NGC~7684 & $158.2 \pm 4.7$ & 1.97 - 4.07 & $114.3 \pm 4.7$ & $41 \pm 22$ & 0.60 & 0.70 & FR & MC (KD/DR, DR) & 4.0 & $-29 \pm 17$ \\
\hline
\end{tabular}
\label{Tab:SampleKinematics}
\\[0.5 cm]
Notes: 
Col. (1): Galaxy Name. 
Col. (2): Maximum rotational velocity measured as the maximum 
absolute deviation from the systematic recessional velocity. 
Col. (3): Maximum radius with robust kinematic results (in units of $R_e$). 
We show the minimum and maximum 
radius of the outermost spatial bin.
Col. (4) Dispersion at the half-light radius.
Col. (5) Dispersion at the maximum radius. 
*No measurement is possible for NGC~1286 as we are below the 
resolution limit.
Col. (6): $\lambda$ parameter at $R_e$ as defined in \cite{Emsellem2007}. 
Col. (7): $\lambda$ parameter at $R_{\rm max}$
Col. (8): Classification as fast (FR) or slow (SR) rotator based on 
$\lambda(R_e)$ and the definitions in 
\cite{Cappellari2011}.
Col. (9): Structural Classification, based on \cite{Krajnovic2008}. The classes are DR (Disk-like Rotation), 
LV (Low-Level Velocity), KT (Kinematic Twist) and KD (Kinematically {Distinct}).
Col. (10): Transition radius in kpc between first and second component, 
where relevant.
Col. (11): Global Misalignment angle $\Psi = \langle {\rm PA}_{\rm kin} - {\rm PA}_{\rm phot}\rangle$ 
in degrees for the outermost component of each galaxy 
\\[0.5 cm]
\end{table*}

\begin{figure}
\begin{center}
\includegraphics[width=0.9\columnwidth,angle=0,clip]{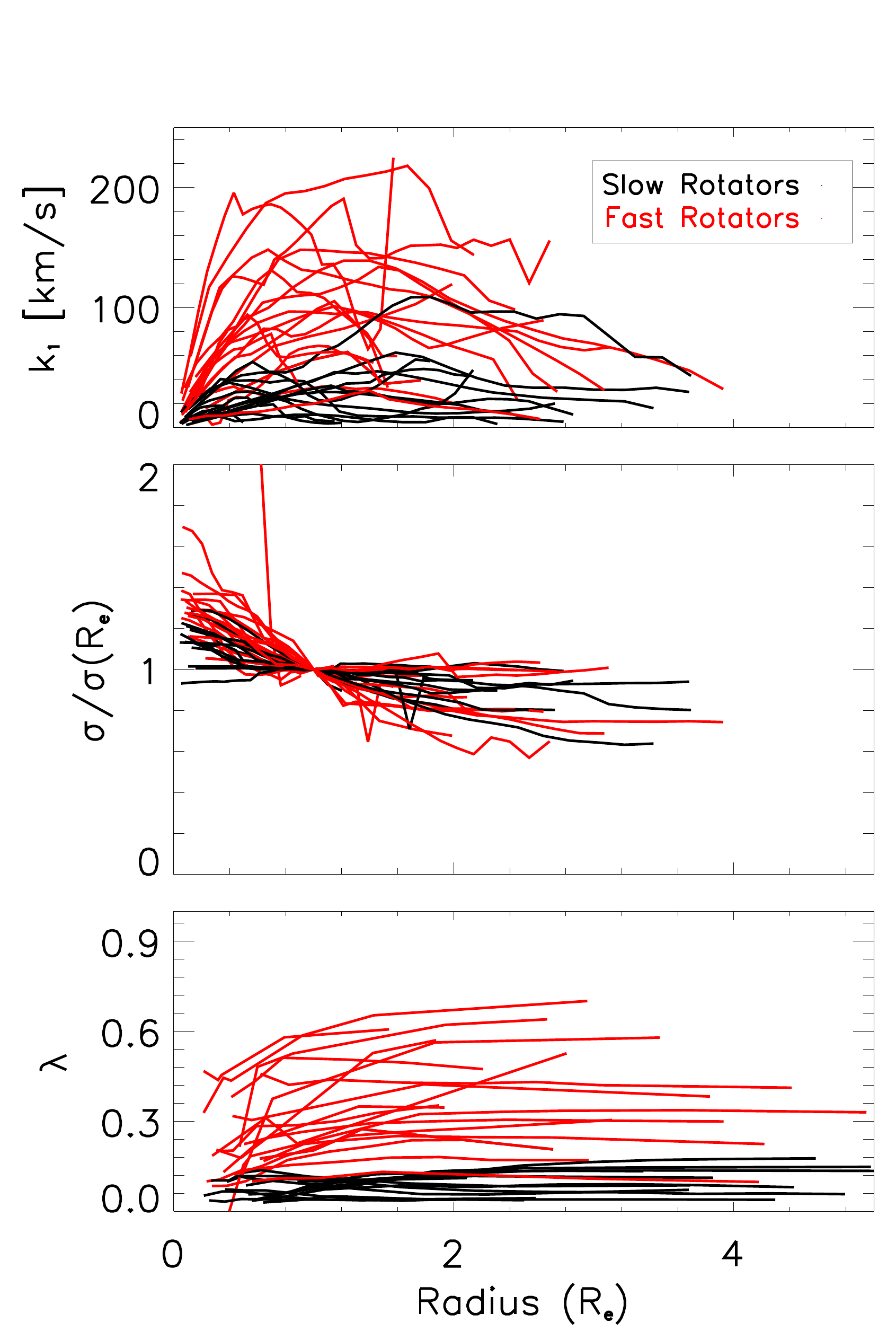}
\caption{Velocity, dispersion and angular momentum profiles for all 
galaxies in our sample based on kinemetry. We show 
the first order harmonic velocity term $k_1$ (top), the velocity dispersion 
in elliptical annuli (middle), and 
the cumulative measure $\lambda_R$ as a function of radius (bottom) for both SRs (Black) and FRs (Red).}
\label{Fig:LambdaProfile}
\end{center}
\end{figure}

\begin{figure*}
\begin{center}
\includegraphics[width=0.8\columnwidth,angle=0,clip]{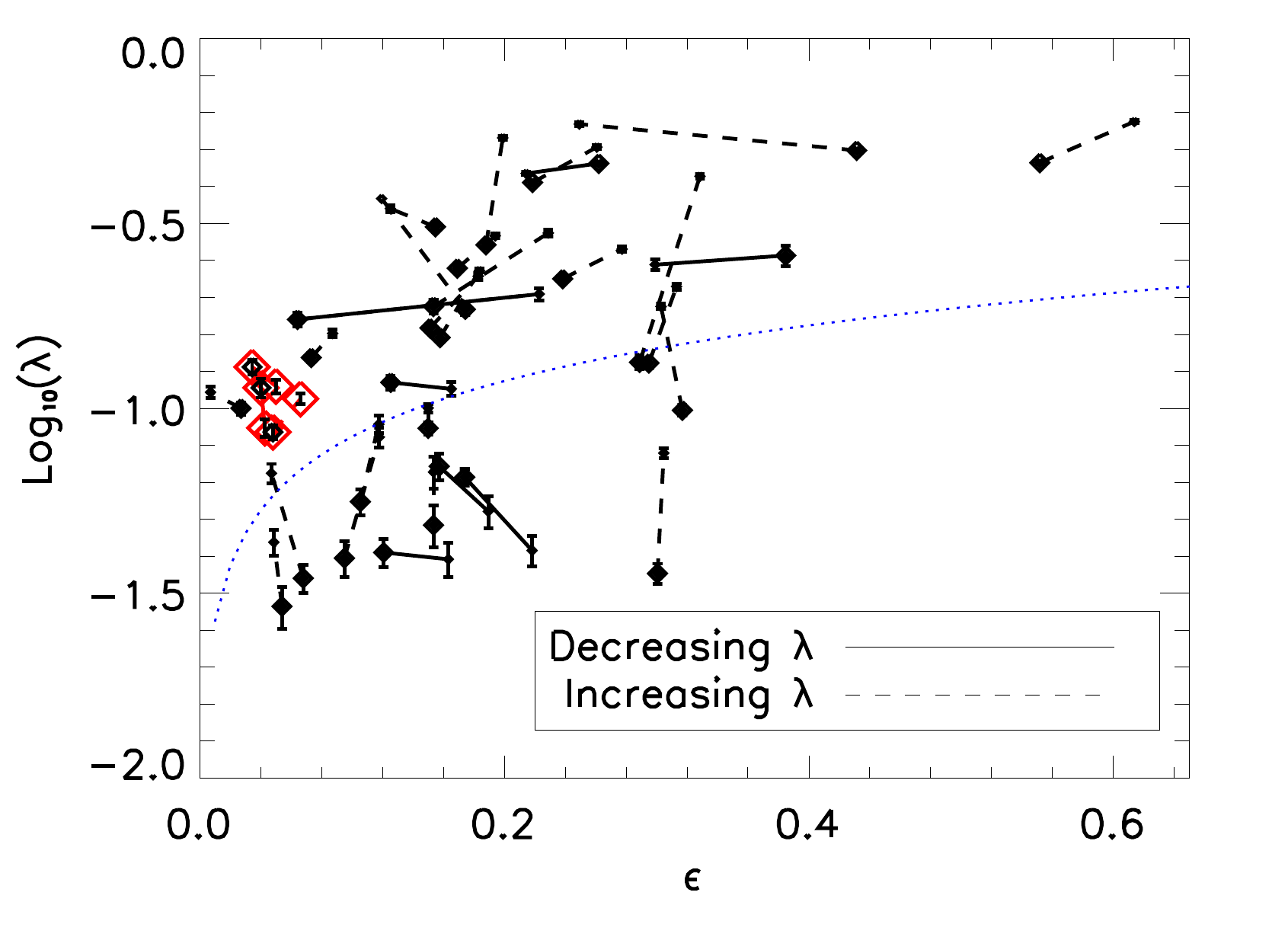}
\includegraphics[width=0.8\columnwidth,angle=0,clip]{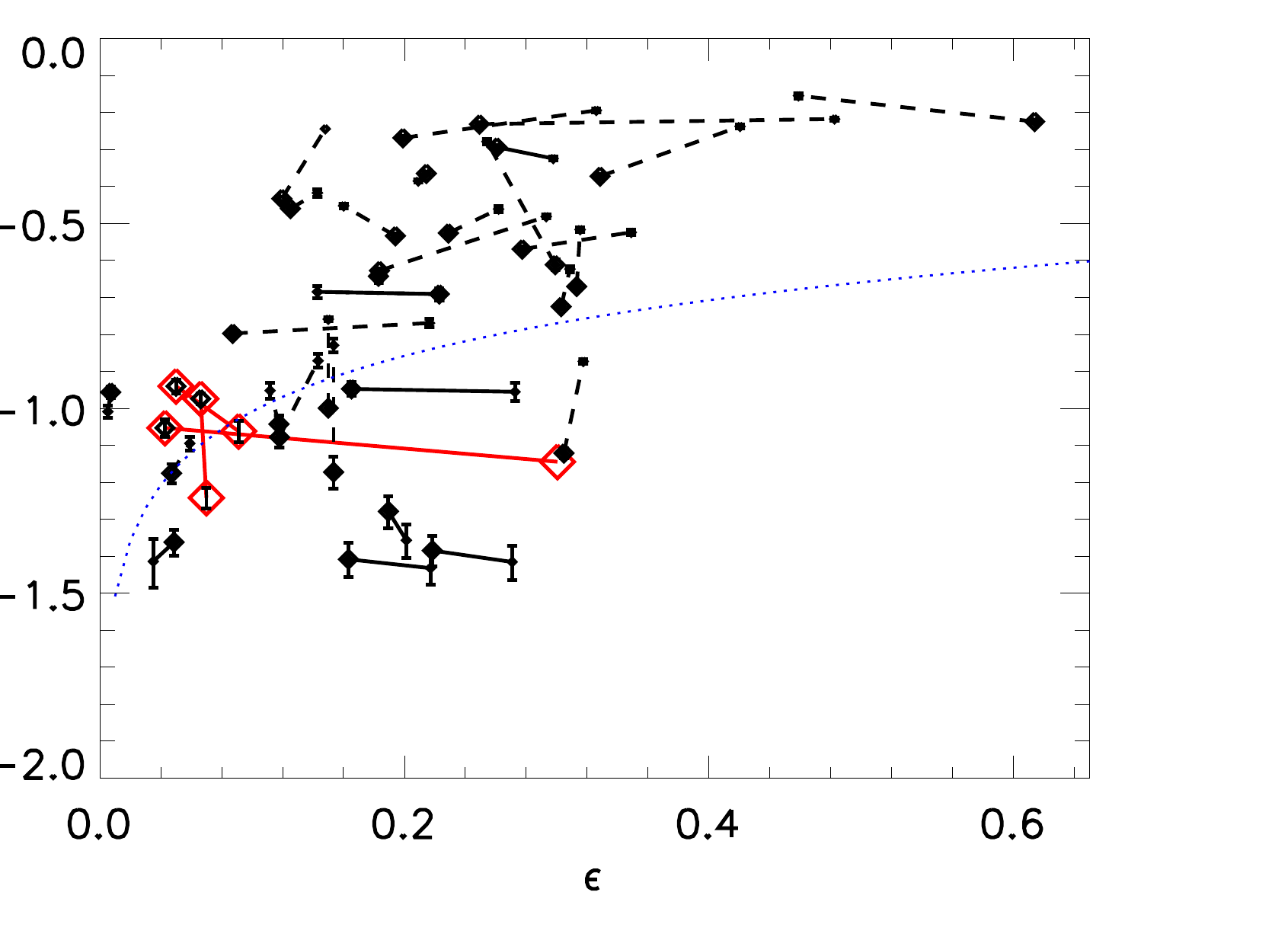}
\caption{We show the cumulative measure $\lambda_R$ as a function of 
ellipticity at both $R_e / 2$ (left) 
and the outermost radius (right). In both cases, values at $R_e$ 
are also plotted, with measurements at each radius connected by 
a solid line when $\lambda_R$ increases outwards and a dashed line
for decreasing $\lambda_R$. {In a small number of cases, the ellipticity 
decreases moving outwards, leading to values of $\lambda$ that decrease with radius, but 
increase with $\epsilon$.} For reference, the function $\lambda = 0.31 \sqrt{\epsilon}$, which divides SRs and FRs, is overplotted (blue dotted line) 
though we note that the scaling of 
this function is radius-dependent. Galaxies NGC~677, IC~301 and NGC~3837, 
intermediate between SR and FR as discussed in the text, are
shown with large diamonds. All three are low ellipticity galaxies with $\lambda(R_e) \sim 0.1$.}
\label{Fig:LambdaTransition}
\end{center}
\end{figure*}

\begin{figure*}
\begin{center}
\includegraphics[width=0.95\columnwidth,angle=0,clip]{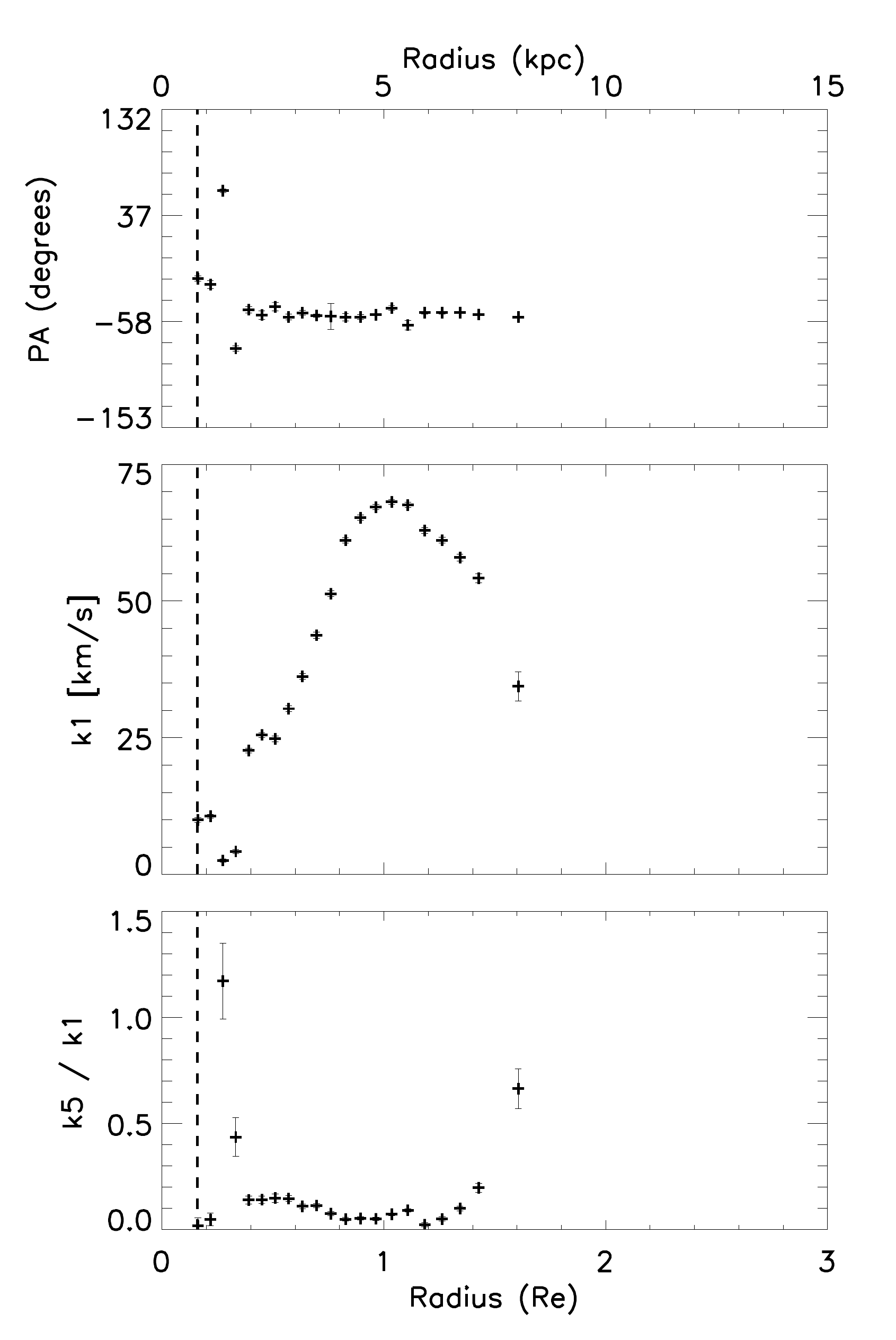}
\includegraphics[width=0.95\columnwidth,angle=0,clip]{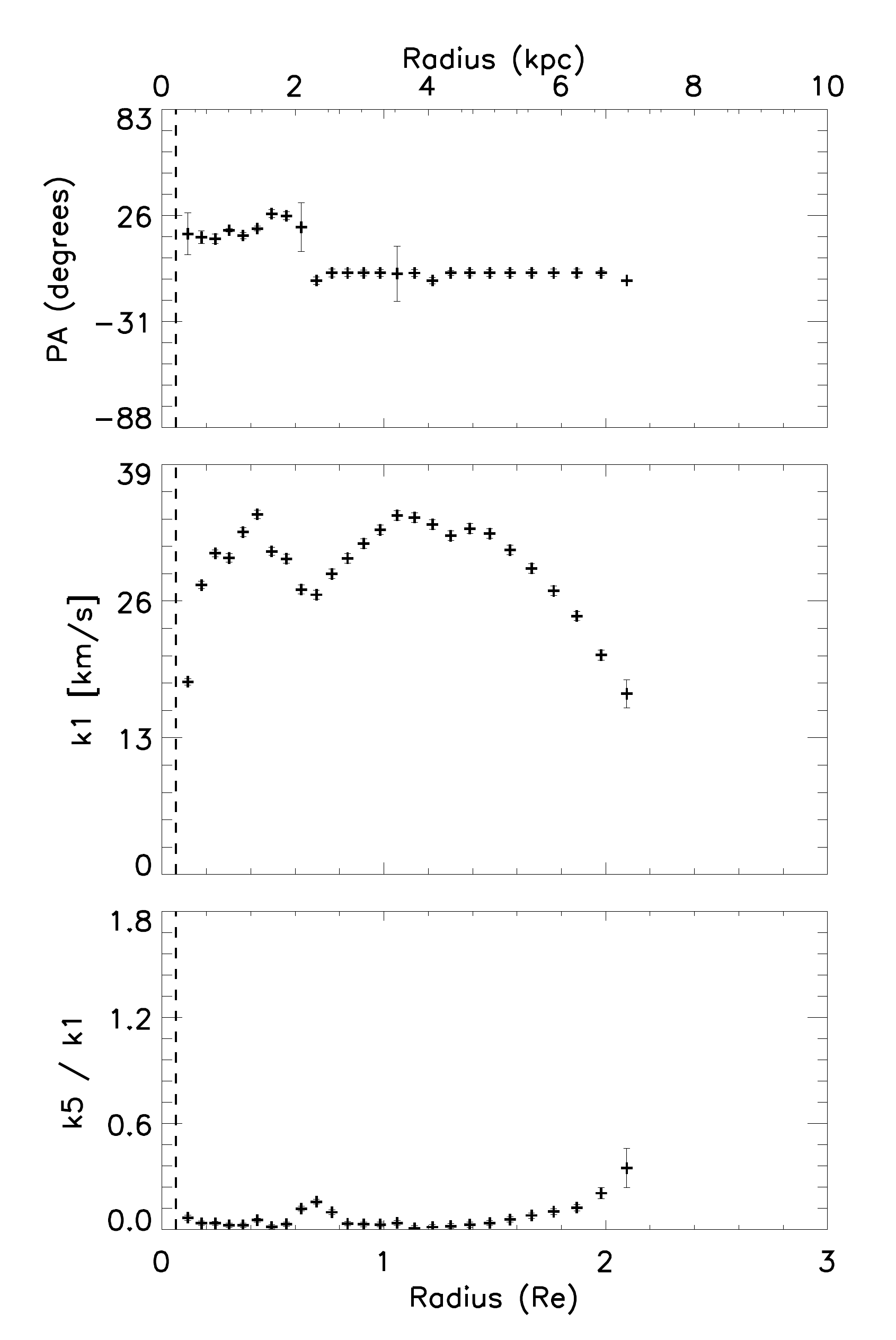}
\includegraphics[width=0.95\columnwidth,angle=0,clip]{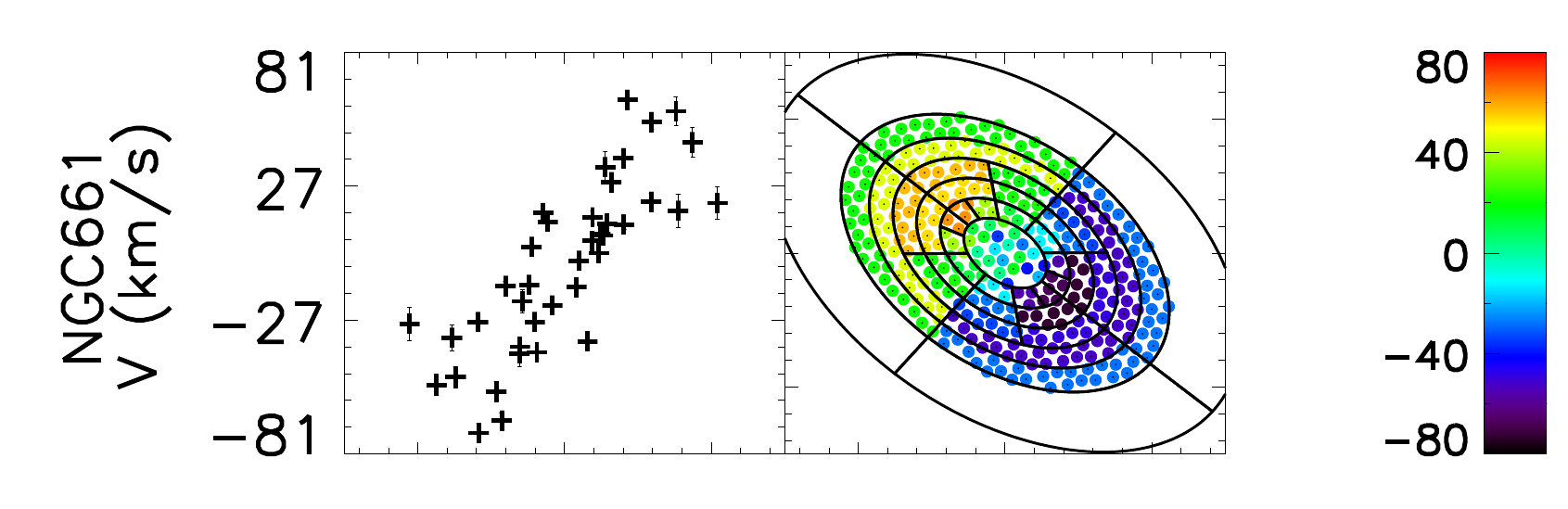}
\includegraphics[width=0.95\columnwidth,angle=0,clip]{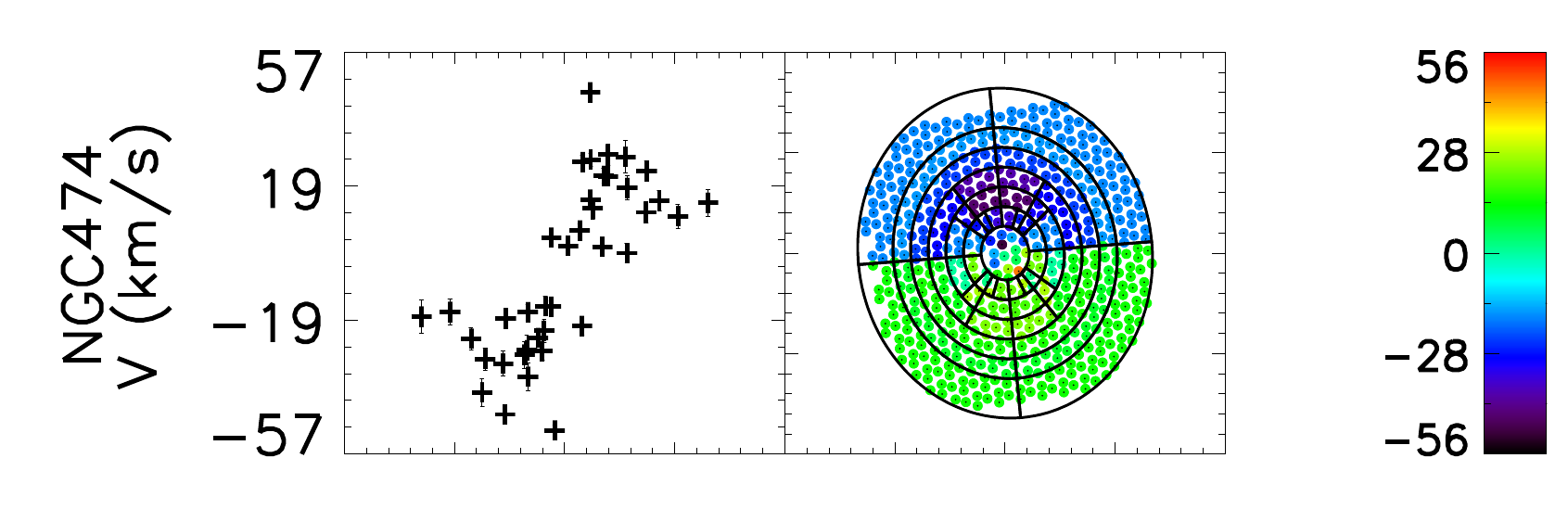}
\includegraphics[width=0.95\columnwidth,angle=0,clip]{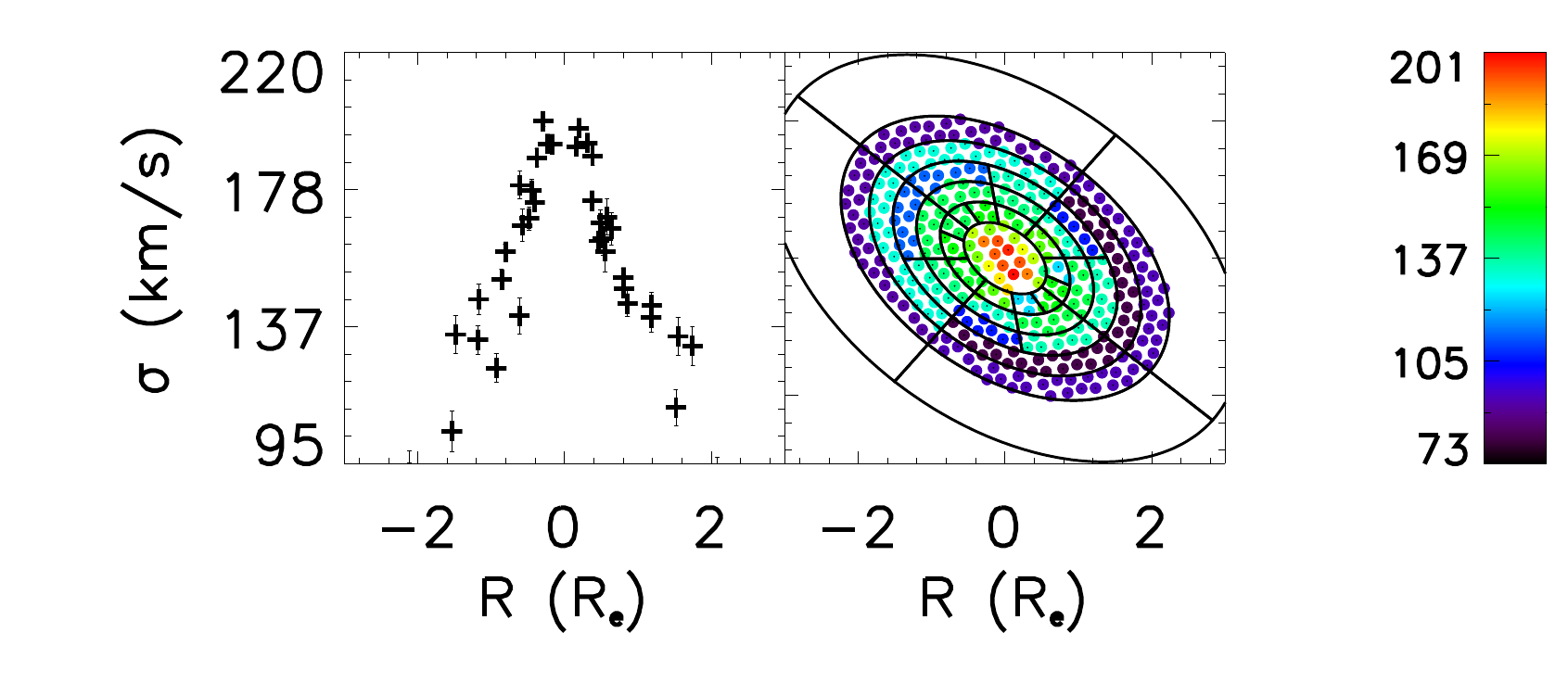}
\includegraphics[width=0.95\columnwidth,angle=0,clip]{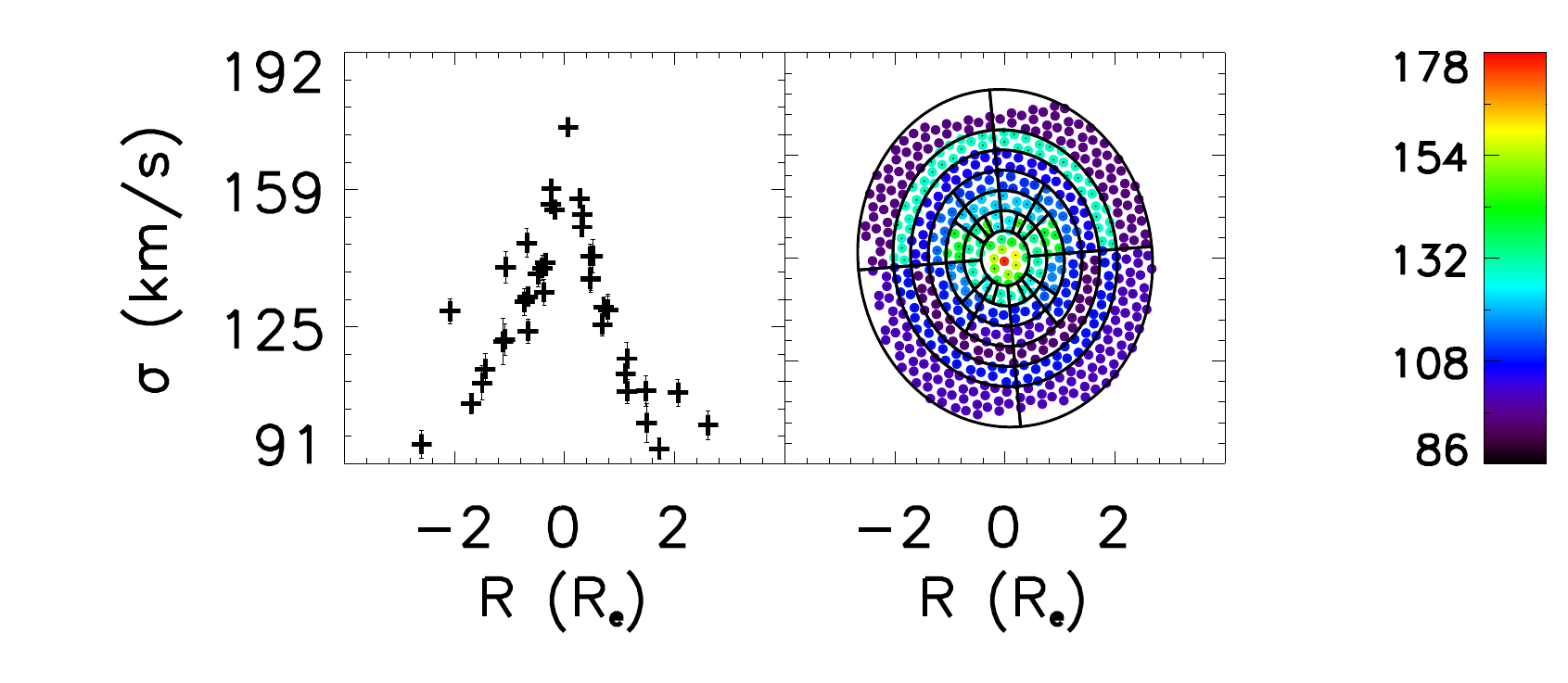}
\caption{
  Kinematic and kinemetric maps for galaxies NGC~661 (left) and NGC~474
  (right). In each case we show 1D radial kinemetric profiles
  including the PA (top), first coefficient in the harmonic dispersion
  expansion $k_1$ (the rotation curve; top middle) and $k_5 / k_1$ (middle). 
  Below are 2D maps of the velocity (bottom middle) and velocity dispersion
  (bottom) for each galaxy. NGC~661 is very similar to most of the SC
  FRs in our sample, with flat $k_5 / k_1$ and PA, and a rotation
  profile that flattens or decreases slightly beyond $\sim R_e$. It
  also possesses a more pronounced LV region in its inner kpc,
  which we believe is characteristic of an unresolved KD
  component. NGC~661 is a more typical MC FR, with a clear transition
  between two disk-like components at 2 kpc.}
\label{Fig:KinematicsSetGala}
\end{center}
\end{figure*}

\begin{figure*}
\begin{center}
\includegraphics[width=0.95\columnwidth,angle=0,clip]{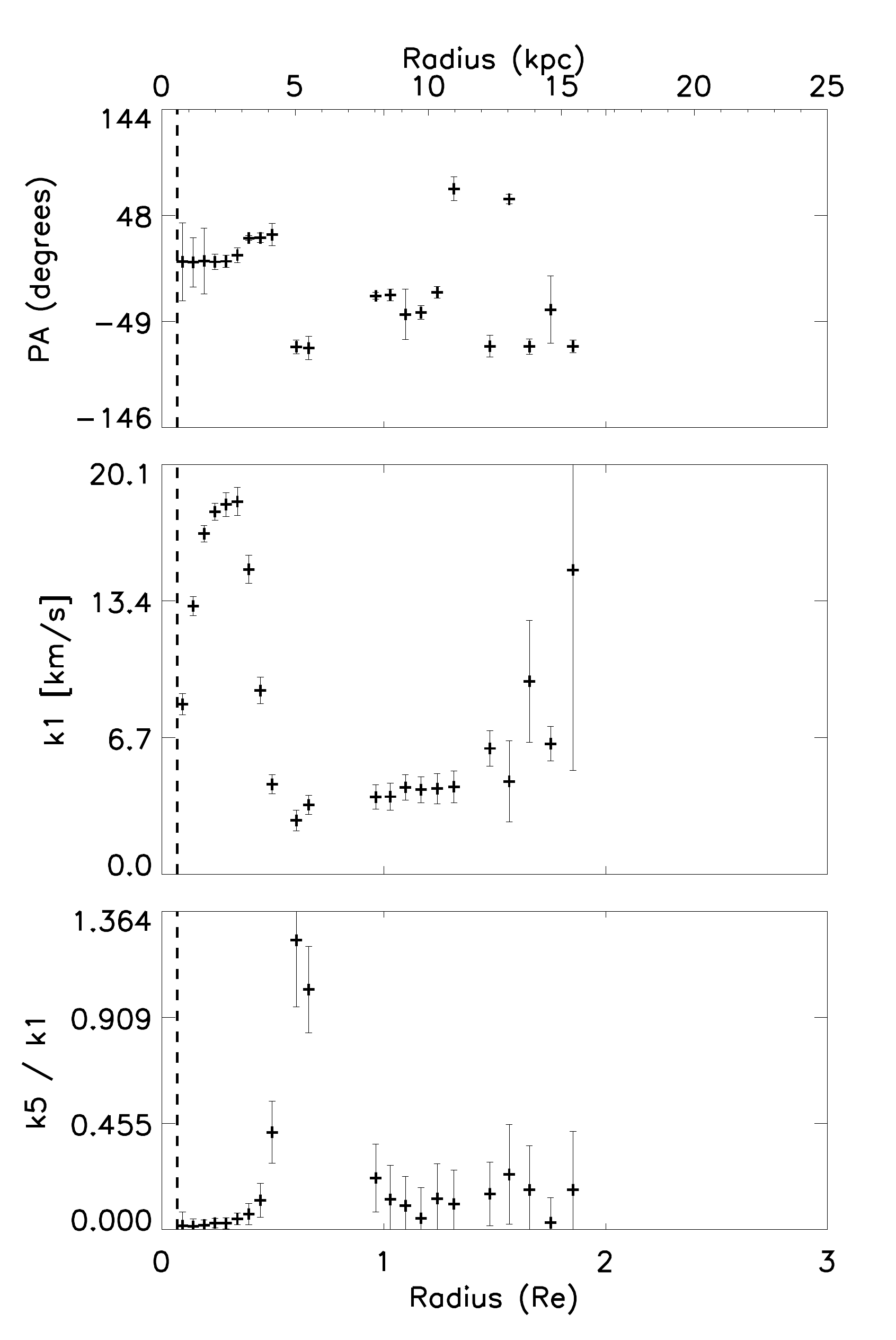}
\includegraphics[width=0.95\columnwidth,angle=0,clip]{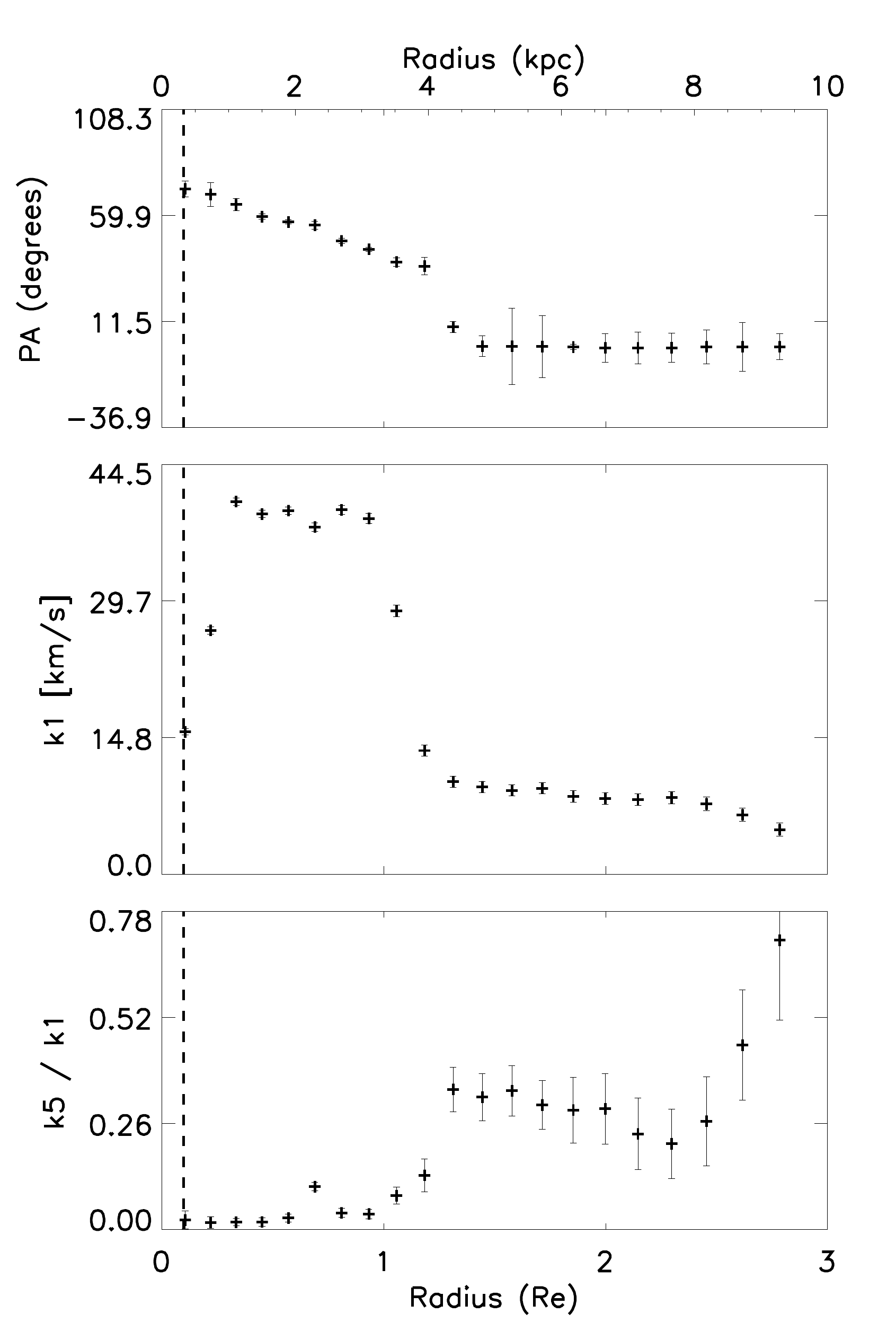}
\includegraphics[width=0.95\columnwidth,angle=0,clip]{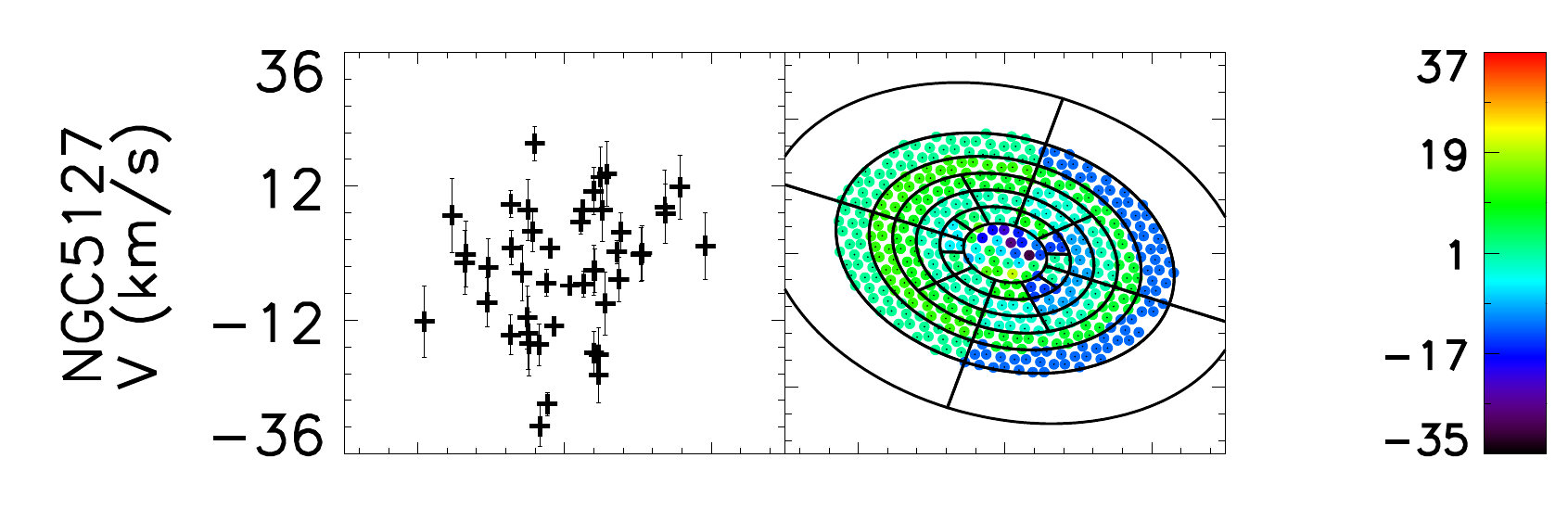}
\includegraphics[width=0.95\columnwidth,angle=0,clip]{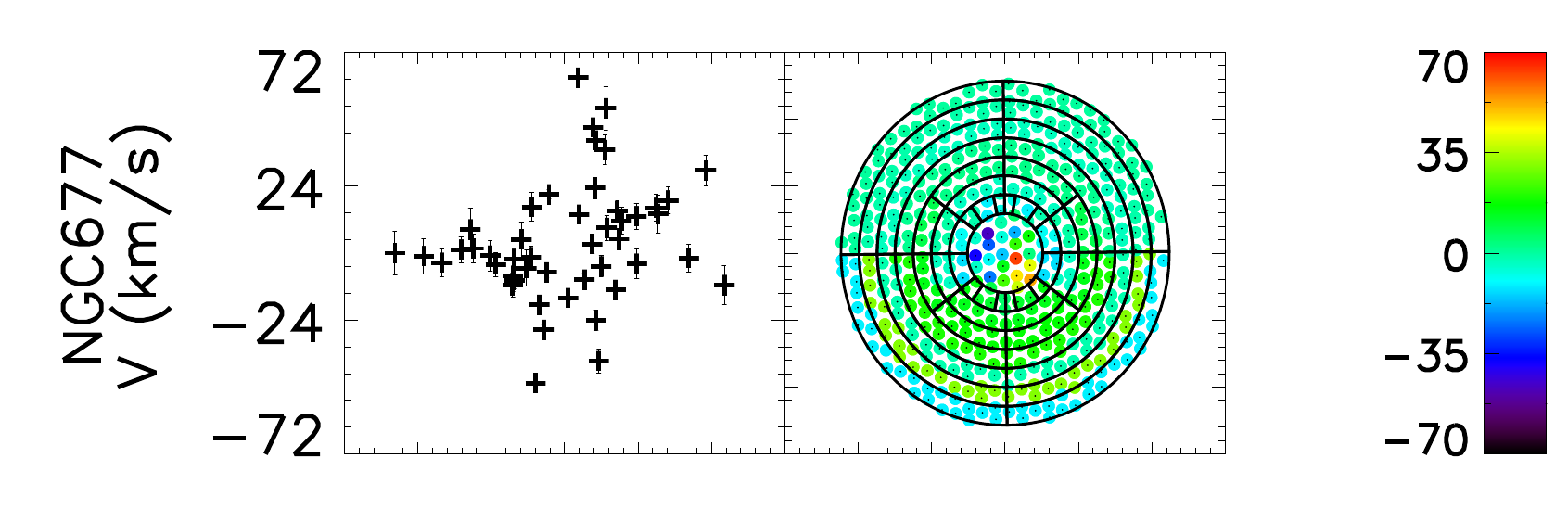}
\includegraphics[width=0.95\columnwidth,angle=0,clip]{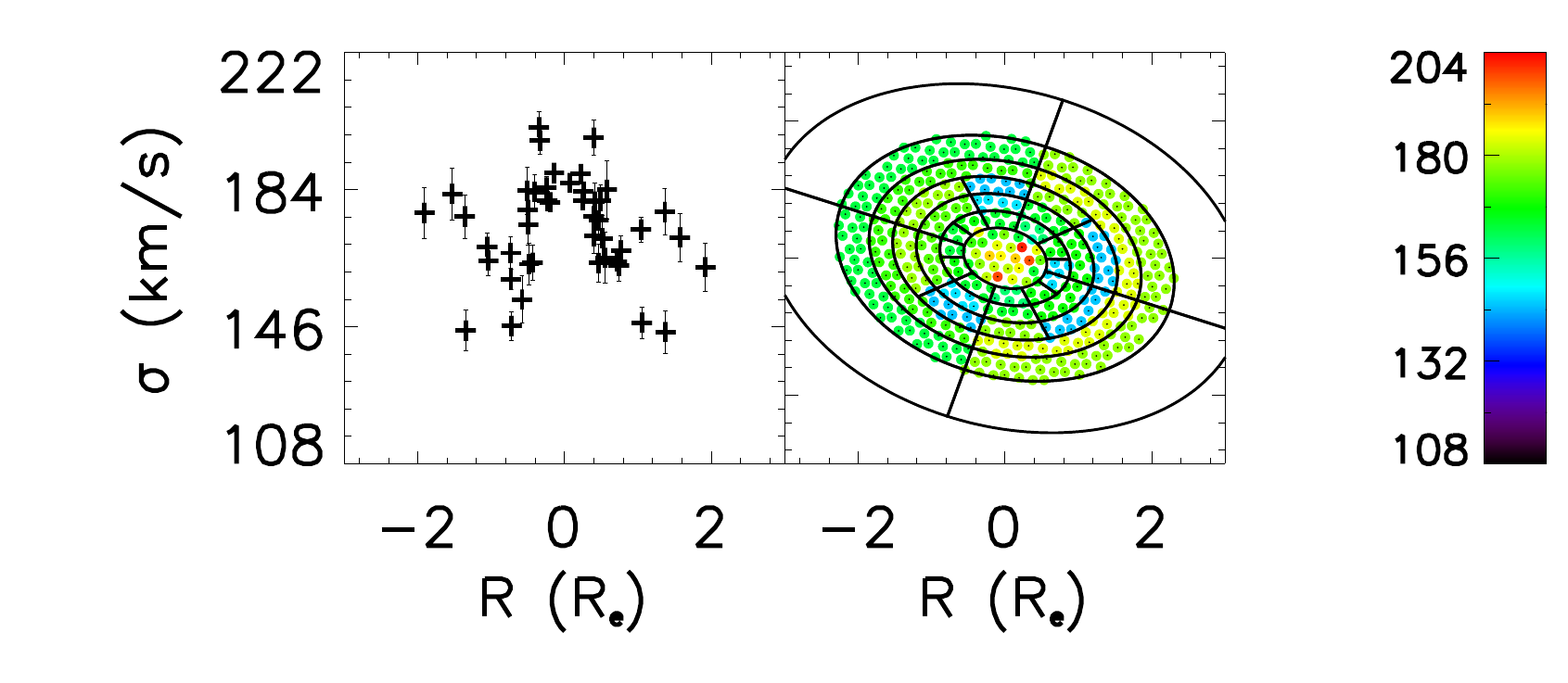}
\includegraphics[width=0.95\columnwidth,angle=0,clip]{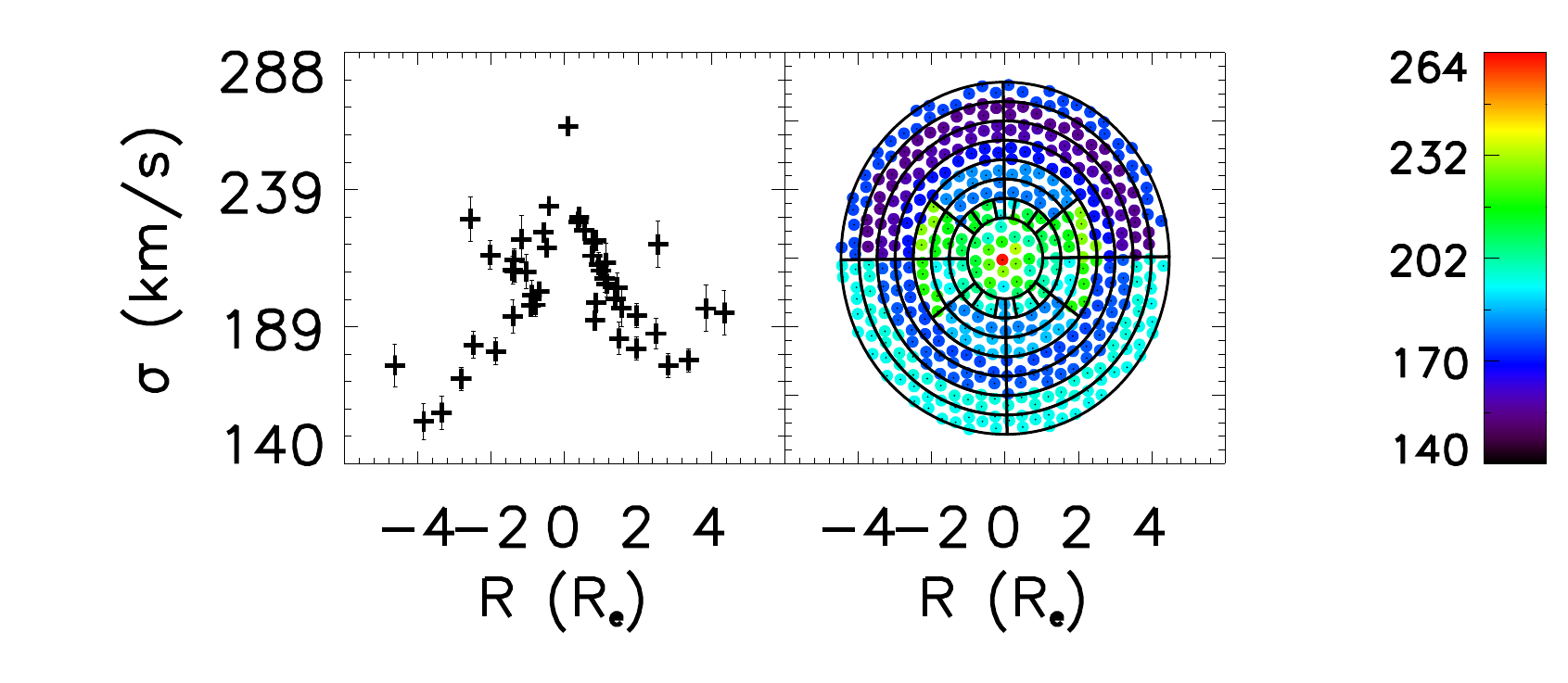}
\caption{
  Identical to Figure~\ref{Fig:KinematicsSetGala} for galaxies NGC~5127
  (left) and NGC~677 (right).  NGC~5127 is a typical example of an SR
  with an extended disk-like KD component at its center. At 5 kpc,
  there is a sharp transition in PA, and a hump in $k_5 / k_1$, as the
  galaxy transitions to an outer LV component. On the other hand,
  NGC~677 is an example of a galaxy classified as an FR based on
  $\lambda(R_e)$, but displaying all the same characteristics as
  NGC~5127. It is only the extended nature of the inner disk-like component that leads
  to its classification as an FR.}
\label{Fig:KinematicsSetGalb}
\end{center}
\end{figure*}

\begin{figure}
\begin{center}
\includegraphics[width=0.8\columnwidth,angle=0,clip]{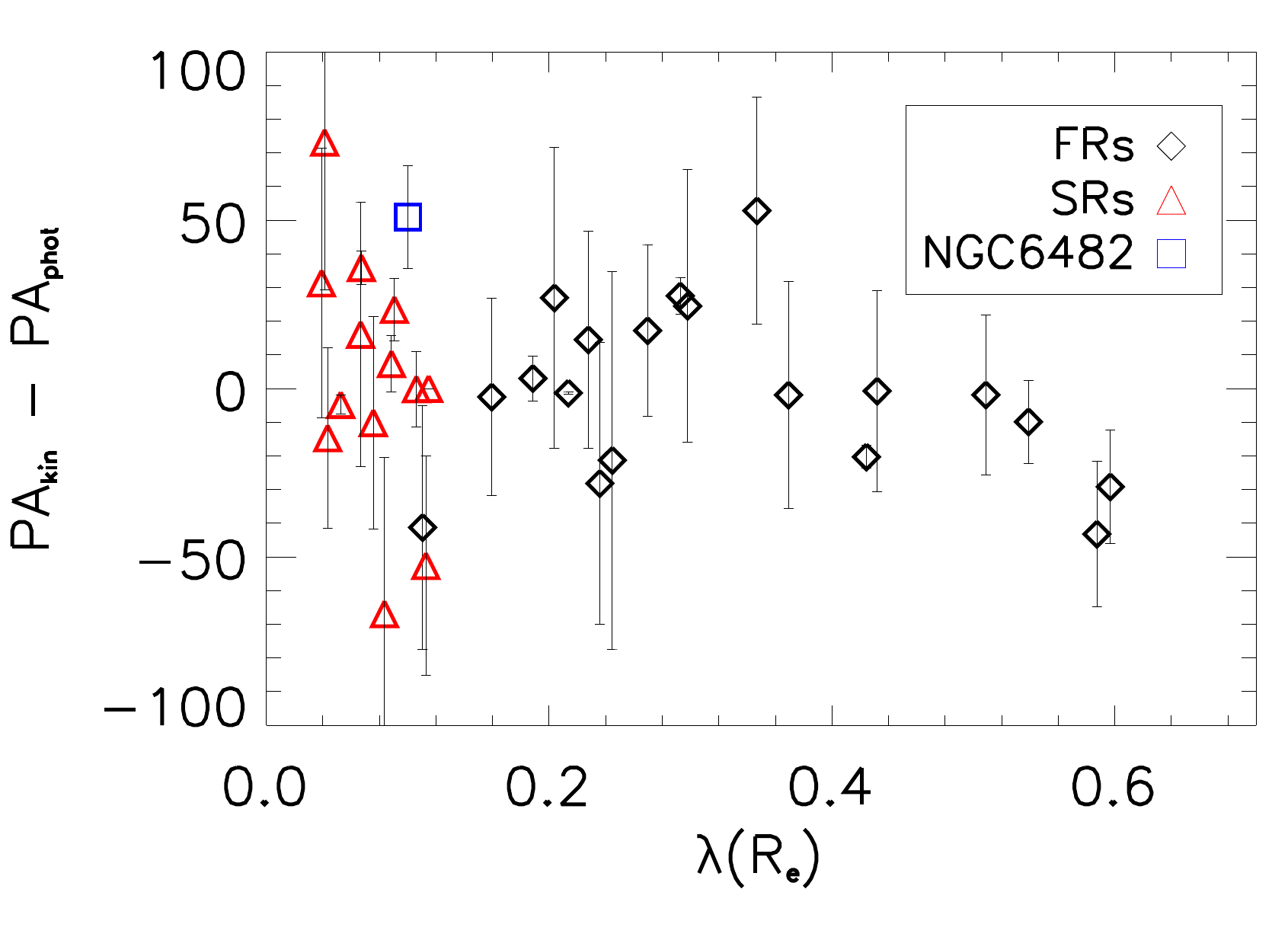}
\caption{Global misalignment angle $\langle {\rm PA}_{\rm kin} - {\rm PA}_{\rm phot}\rangle$ as a 
function of $\lambda(R_e)$ for the outermost kinemetric component of all galaxies. 
{${\rm PA}_{\rm kin}$ and ${\rm PA}_{\rm phot}$ are both 
calculated in best-fit elliptical annuli and the average misalignment 
angle of this component is then found by taking a luminosity-weighted 
average over all radii in the region}. We choose the 
outermost component to avoid LV regions at the centres of most galaxies where 
${\rm PA}_{\rm kin}$ is poorly defined, and typically cover between $\sim 0.5 - 2.5 R_e$. 
Error bars are the 1-$\sigma$ standard deviations when we average over all radii. 
We observe a general trend of greater misalignment amongst 
SRs, including NGC~677, IC~301 and NGC~3837, and highlight NGC~6482 (blue), 
which has the largest well-defined misalignment.}
\label{Fig:Misalignment}
\end{center}
\end{figure}

Full 2D profiles of the velocity and velocity dispersion of all galaxies, as
well as 1D kinemetric profiles, can be found in
Figures~\ref{Fig:AllKinematicsa}-\ref{Fig:AllKinematicsf} of Appendix
B. In addition, some key kinematic characteristics out to large radius
are listed in Table~\ref{Tab:SampleKinematics}.  We include the
traditional measures of maximum velocity, the dispersion
within an effective radius, and the kinematic position angle. In
addition, as we are largely interested in the evolution of kinematic
properties at large radius, we include the radius
of our maximum robust measurement and the dispersion at this radius.

For almost all galaxies our observations extend
to beyond $2~R_e$, or around 8 kpc.  The only exception is
NGC~1286, which has a relatively low dispersion and large $R_e$, as
well as two nearby sources that contaminate our outermost bin and
force us to discard it.  For six galaxies, we are able to observe to
$\gtrsim 5~R_e$ (close to 20 kpc), although with wide
bins ($\sim R_e$). Even accounting for our small $R_e$ estimates, we
probe the stellar kinematics out to distances well beyond most
existing samples.

In Figure~\ref{Fig:LambdaProfile}, we show the velocity, velocity dispersion, 
and $\lambda_R$ profiles for all galaxies. In general, we find that
$\sigma$ is declining gently ($\sim 23\%$ per decade in radius
on average), as has been observed many times at smaller radius
\citep[e.g.,][]{Jorgensen1997}, and has also been
reproduced in simulations \citep[e.g.,][]{Remus2013, Wu2014}.  
While some galaxies appear to show rising dispersion profiles 
at large radius, most are consistent with remaining flat beyond $R_e$ 
within the measurement uncertainties.
%In fact, a
%small number of galaxies (UGC4051, UGC1382, NGC774, IC312, CGCG390-096, 
%NGC677 and NGC4908) appear to actually show small increases 
%in dispersion beyond $R_e$.  Generally, these tend to be 
%low mass galaxies with relatively larger errors in their outskirts, 
%so that any increases are not significant. 
However, the two most 
massive galaxies in our sample, NGC~677 and UGC~4051, do show significant, 
albeit small ($\sim 5\%$ above $\sigma_{R_e}$) rises in dispersion. 
The velocity dispersion profiles also appear to show a rough split
between FRs and SRs. Most SRs have comparatively flat dispersions,
particularly within $R_e$. By contrast, FRs tend to show
more rapidly declining dispersion profiles out to $R_e$. 

Given the integral nature of $\lambda_R$, in most cases galaxies reach
a limiting value of $\lambda_R$ at $1 - 2~R_e$. In this large radius limit, most SRs have 
$\lambda_R \lesssim 0.1$, while most FRs occupy a much broader
continuum above $\lambda_R \sim 0.2$.  Classifications based on
kinematics within $R_e$ are therefore quite accurate out to much
larger radii. The main exceptions are the galaxies NGC~677,
IC~301 and NGC~3837, which are classified as FRs based on
$\lambda(R_e)$, but at large radius show $\lambda_R$ characteristic of
minimal rotation.

This is also clear if we consider the behaviour of $\lambda$ as a function of 
observed ellipticity \citep{Binney1978}. In Figure~\ref{Fig:LambdaTransition}, 
we show galaxies on the $(\lambda-\epsilon)$ plane at $R_e / 2$, $R_e$ and $R_{\rm max}$. 
Very few galaxies show significant changes in $\lambda$ even
beyond $R_e / 2$. Almost all changes are restricted to increases in
ordered rotation amongst FRs, which tend to continue rotating out to
radii well beyond $R_e$. The only galaxies that show significant
enough decreases to transition from fast rotation within $R_e$ to slow
rotation in their outskirts are the three galaxies identified
in the previous paragraph, which uniformly also have higher ellipticity in their outskirts.

\subsection{Kinemetric Classification}
\label{subsec: Kinemetric Classification}

We now turn to identifying substructures based on kinemetry.  
We first classify each galaxy as single (SC) or multi-component (MC),
and then classify each component as either disk-like rotation (DR), low
velocity (LV), kinematic twist (KT) or kinematically decoupled
(KD). 13 (40$\%$) of our galaxies are SC systems, a slightly
higher percentage than the 31\% in the SAURON sample. {We do not necessarily 
expect these numbers to align since the two samples cover different mass 
ranges and additionally, we are unable to resolve central structures with sizes $\lesssim 1$
kpc}.  Of the SCs, only one (IC~1152) is an SR, and the remaining 12 are
SC FRs. Almost universally, the single component in these cases shows
disk-like rotation and close to perfect alignment between kinematic and 
photometric PA. Furthermore, the $\lambda_R$ profiles tend to be flat or
rising for the SC FRs, as shown in Figure~\ref{Fig:LambdaProfile}. 
There are one or two examples with small kinematic
twisting (NGC~219 and UGC~1382), as well as two disk-like rotators 
(NVSS~J032053+413629 and IC~312) that show strong {kinematic misalignment 
in the sense that ${\rm PA}_{\rm kin}$ deviates from the photometric PA by close to 
$90^{\circ}$ at large radius}.

Of the MC galaxies, 33$\%$ contain an LV component in the
center with the rotation curve increasing outward.  An additional 28$\%$ 
have a kinematic twist.  The remaining 39$\%$ contain a KD
ranging in size from 1-7 kpc.  We now describe the properties of these
subclasses.

There are ten MC FRs, which come in a few varieties. Three of them
have disky kinematics everywhere except in a small LV region in the
center. Two such examples of these, NGC~474 and NGC~661, are identified
in Figure~\ref{Fig:KinematicsSetGala}, which shows radial profiles of
$k_1$, ${\rm PA}_{\rm kin}$ and $k_5 / k_1$ as well as 2D kinematic maps for these two
galaxies.  Very likely these rapid drops in velocity and $\lambda_R$
point to small rotating components that are unresolved by our
observations. In particular, our observations of NGC~474 and NGC~5982
show low velocities in their central fibers whereas higher resolution
SAURON kinematics identify KDC's rotating at $\sim 50$~\kms\ within 1
kpc. The velocities also appear to drop in the outer parts of the FRs, 
but this is likely due to our large spatial and angular bins at large radius.

Another four of the MC FRs are typified by NGC~474, shown in the
bottom panel of Figure~\ref{Fig:KinematicsSetGala}. These relatively
slowly rotating FRs contain two distinct disk-like components, in the
case of NGC~474 separated by a PA shift of about $15^{\circ}$ at 2
kpc. The transition is also marked by a clear hump in $k_5$ and a
minimum of $k_1$. SAURON kinematics of NGC~474 also show an additional
inner disk-like component at 0.8 kpc \citep{Krajnovic2008} that we do
not detect, empirically determining that structures with sizes $<
7$\arcsec\ are not discernible in our data.

There are three remaining MC FRs (NGC~677, shown in the top of
Figure~\ref{Fig:KinematicsSetGalb}, NGC~3837 and IC~301), and these are
the most difficult to classify.  While technically they satisfy the FR
criterion, their $\lambda_R$ values are borderline between FRs and
SRs, and they do not rotate at all beyond $\sim~1.5-2~R_e$. From
Figure~\ref{Fig:LambdaTransition}, we see that all of these are low
ellipticity galaxies, which never rotate particularly fast in the
sense that $\lambda(R_e) \lesssim 0.15$. NGC~677 exemplifies this
class. NGC~677 also has an interesting outer dispersion profile that
appears to start to rise again at around $3~R_e$ (10 kpc) on both
sides of the galaxy along the minor axis, as has been seen in a few 
central galaxies \citep[see e.g.][for other examples of dispersion 
profiles rising outwards]{Dressler1979, Kelson2002, Loubser2008, Jimmy2013, Murphy2013}

In fact, these three borderline cases have very similar kinemetric
profiles to five of the MC SRs. Specifically, they all have a central component with
low-amplitude rotation and large angular differences between photometric and kinematic 
PAs, which then transitions to an LV component at larger radius. NGC~5127, top left of
Figure~\ref{Fig:KinematicsSetGala}, is one good example. The three
borderline cases were classified as FRs only because the scale of the
inner rotating component is a bit larger than for the typical MC
SRs. Therefore, while we technically count these three galaxies as
FRs, perhaps they are better considered as SRs with a very extended KD
in the center. These central KD components, with sizes of 2.2 to 7 kpc, 
are considerably larger in size than KDCs \citep[0.2 to 1.8 kpc in
SAURON][]{Kormendy1984, Forbes1994, Carollo1997, Krajnovic2008}, but 
are otherwise kinematically quite similar.

Finally, beyond the one SC SR, and the five MC SRs with KD
components, we see one more type of MC SR, comprising five galaxies, that
show steadily increasing $\lambda_R$ profiles in
Figure~\ref{Fig:LambdaProfile}. As can be seen from
Figure~\ref{Fig:LambdaTransition}, none of them increase their
rotation enough to become FRs in their outskirts. All of the five galaxies, 
and particularly NGC~6482, also show evidence of 
misalignment between their photometric and kinematic 
PAs. 

We see this in Figure~\ref{Fig:Misalignment}, which plots the misalignment angle 
$\langle {\rm PA}_{\rm kin} - {\rm PA}_{\rm phot}\rangle$ of the outermost 
component in each galaxy, generally covering radii between 
$\sim 0.5 - 2.5 R_e$. {Here, ${\rm PA}_{\rm kin}$ and ${\rm PA}_{\rm phot}$ are 
calculated independently in best-fit elliptical annuli, using kinemetry applied 
to our kinematic data and r-band photometry respectively. The average misalignment 
angle for a given galaxy or component is then found by taking a luminosity-weighted 
average over all radii in that component. Since we restrict 
our attention to the range $[-90^{\circ}, 90^{\circ}]$, 
this is comparable to the kinematic misalignment angle 
${\rm sin}\Psi = |{\rm sin}{\rm PA}_{\rm kin} - {\rm PA}_{\rm phot}|$ defined by \cite{Franx1991}.  
We choose to only show the outer component to avoid any strong 
luminosity bias towards central regions and to avoid the rapid transitions in 
${\rm PA}_{\rm kin}$ seen between different components.}

In the FRs, ${\rm PA}_{\rm kin}$ is well-defined, and in 
almost all cases is aligned with ${\rm PA}_{\rm phot}$.  It is more difficult to determine the 
situation for the SRs since they are nearly round and have poorly defined ${\rm PA}_{\rm phot}$.  
However, the five MC SRs with increasing $\lambda_R$ profiles have 
better defined and quite large misalignment angles. NGC~6482 specifically, highlighted 
in Figure~\ref{Fig:Misalignment}, is misaligned by $\sim 60^{\circ}$. Such misalignment 
is typically interpreted as evidence of triaxiality, since projection effects in triaxial 
galaxies can lead to observed differences between the angular momentum vector and 
the major axis \citep{Statler1991}. Therefore, we suggest that at least half of our SR 
galaxies probably have triaxial structure. This fits in with the existing picture of SRs 
within $R_e$, which show signs, based on their photometry, of being mildly triaxial 
\citep{Binney1978b,TremblayMerritt1995,TremblayMerritt1996,Krajnovic2008, Krajnovic2011}.

\section{Discussion}
\label{Sec:Analysis}

\subsection{Expectations for Large Radius Kinematics}
\label{subsec:Expectations}

Before we examine the kinematics of our galaxy sample at large radii,
we begin by reviewing the possible formation paths for ETG's and the
results we may expect from any given formation scenario. The so-called
two-phase picture of elliptical galaxy formation \citep{vanderWel2008, Naab2009,
  Oser2010, Khochfar2011, vandeSande2013} posits that the central $\sim 1 - 5$ kpc of
galaxies are initially formed by a fast, dissipational phase, 
which leaves behind a compact stellar disk with relatively high
rotational support $\lambda \sim 0.5$ \citep{Elmegreen2008, Dekel2009,
  Ceverino2010, Khochfar2011}.  At later times dry merging expands the
galaxy's outskirts in a manner that reduces $\lambda$ and leaves
behind rounder and kinematically hotter remnants 
{\citep[e.g.,][]{Naab2009, Hilz2013, Taranu2013}}.

The two-phase picture predicts that ETG's are inherently
multi-component systems, with rotationally supported disks comprised
primarily of in situ stars at their center and much rounder halos made
up of accreted material. However, observations at large radius remain
limited.  While KDCs on small scales are interpreted as evidence of 
prior dissipational merging, most observed ETG's are FRs for
which no evidence of such transition has been found, e.g., by ATLAS$^{\rm 3D}$.

We thus focus on the MC galaxies 
discussed in Section~\ref{Sec:Observed Kinematics} and whether or not
the transitions we observe beyond $R_e$ fit into the two-phase
formation picture. We consider kinemetric transitions between rotation-supported 
and dispersion-supported regions, how similar they are to the KDCs of
\cite{Krajnovic2011} and whether they are accompanied by any similar
transitions in $\lambda_R$. Finally, we consider the stellar
populations associated with each subgroup, and whether they are
characteristic of a move from in situ to accreted stars.

We are also interested in comparing to the
picture presented by \cite{Arnold2013}, who were able to use the
SLUGGS survey to measure kinematics out to $\sim 5~R_e$.  They
reported falling profiles {in local angular momentum}, perhaps
reflecting transitions in some FRs from an inner disk to an
outer halo at $\sim 5$ kpc, most dramatically in NGC~3377. They also
found that S0s with more extended disks are most likely to show
rising $\lambda$ profiles at large radius while elliptical galaxies
are most likely to have falling $\lambda$ profiles. Finally they reported 
signs of PA alignment between inner disk and outer halo. Together these 
were used to argue for the two-phase picture and against the formation 
of disks by late-time major mergers \citep{Hoffman2009}, since
{1:1 mergers} result in significant kinematic decoupling between
the inner disk and outer halo \citep{Hoffman2010}. {However, 
we note that \cite{Naab2013} present a more nuanced view of the origin of
SRs and FRs, in which either class can emerge from either a recent major merger,
or a series of minor mergers, depending on the fraction of in-situ star formation and
gas-richness of the last major merger.}

We also compare with the simulations of \cite{Wu2014}. This work
derives galaxy kinematics at large radii from cosmological simulations
of galaxy formation.  They focus on a lower-mass sample (stellar
masses of $\sim 3 - 5 \times 10^{10} M_{\odot}$ compared to our $\sim
2 - 20 \times 10^{10} M_{\odot}$) with kinematics that extend out to $\sim 6~R_e$. However, 
they present simulated rotation and angular momentum profiles that correspond 
quite well with our observations.

\subsection{Galaxies with Changing Kinematics}
\label{subsec:Changing Kinematics}

In order to emphasize radial changes, \cite{Arnold2013} consider a
spatially varying specific angular momentum $\Lambda$, defined in
elliptical annuli rather than full elliptical apertures. Since a local determination
largely removes the effect of radial weighting, $\Lambda$ is very similar to
the flux-weighted ratio of velocity to dispersion, $\langle V^2
\rangle / \langle \sigma^2 \rangle$ used by \cite{Binney2005} and
\cite{Wu2014}. Our elliptical annuli in the central regions are
calculated using 5\arcsec windows, and outside of this region are aligned
with our previously described spatial bins. Additionally, instead of
flux-weighting, which does not vary much in each elliptical bin, we
weight by the measurement errors. Since S/N is
correlated with flux, the two methods do not differ much, but our approach 
is more robust to outlying measurements.

Figure~\ref{Fig:LambdaAllC} shows rotation curves ($k_1$),
normalized velocity dispersions, and $\Lambda$ profiles for galaxies
split into SRs and FRs. To highlight the different kinematic transitions 
observed, we further subdivide our sample into SC systems, 
MC galaxies with KD's, and other MC galaxies. In all cases, the 
local measure $\Lambda$ naturally shows much greater variation 
than $\lambda$ out to large radius. Partly this is due to the lower quality spectra
in these regions, which means that errors increase outwards, rather
than decreasing as in the cumulative case. However, we
truncate the $\Lambda$ profiles where the errors exceed $\pm 0.025$,
while the changes we observe in $\Lambda$ are larger than this, and 
thus likely real.

\subsubsection{FRs at Large Radius}
\label{subsubsec:FRR}

\begin{figure*}
\begin{center}
\includegraphics[width=0.3\textwidth,angle=0,clip]{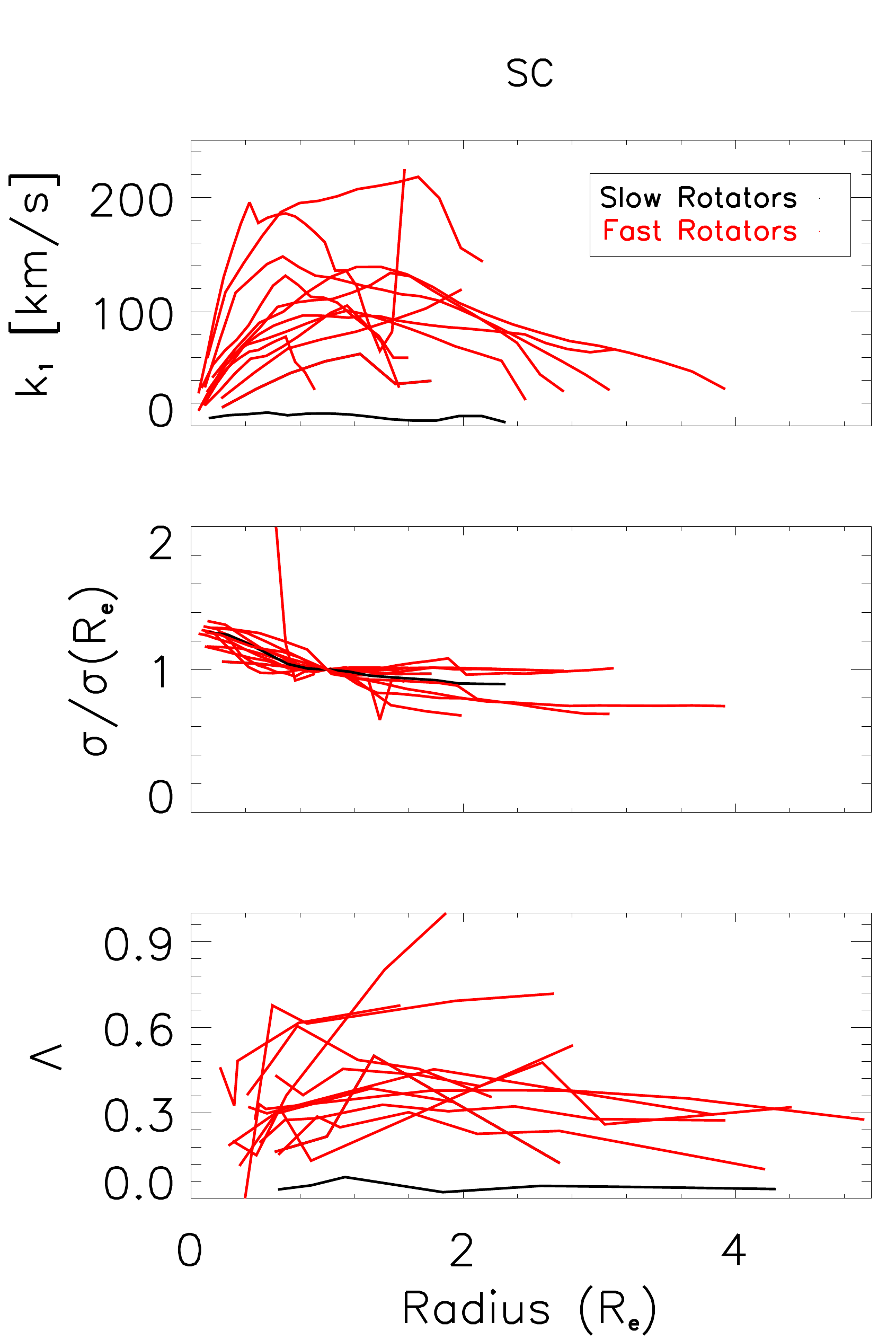}
\includegraphics[width=0.3\textwidth,angle=0,clip]{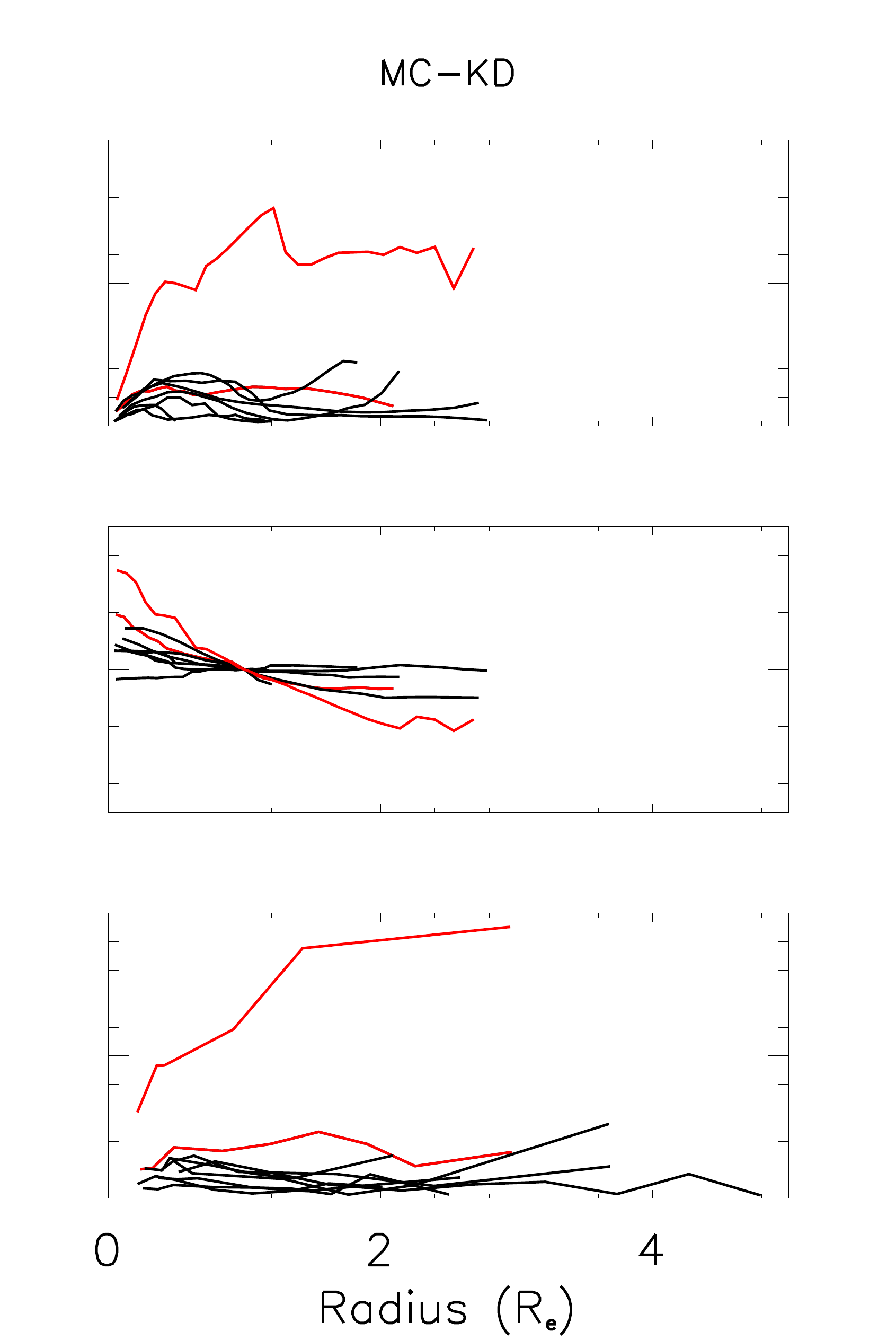}
\includegraphics[width=0.3\textwidth,angle=0,clip]{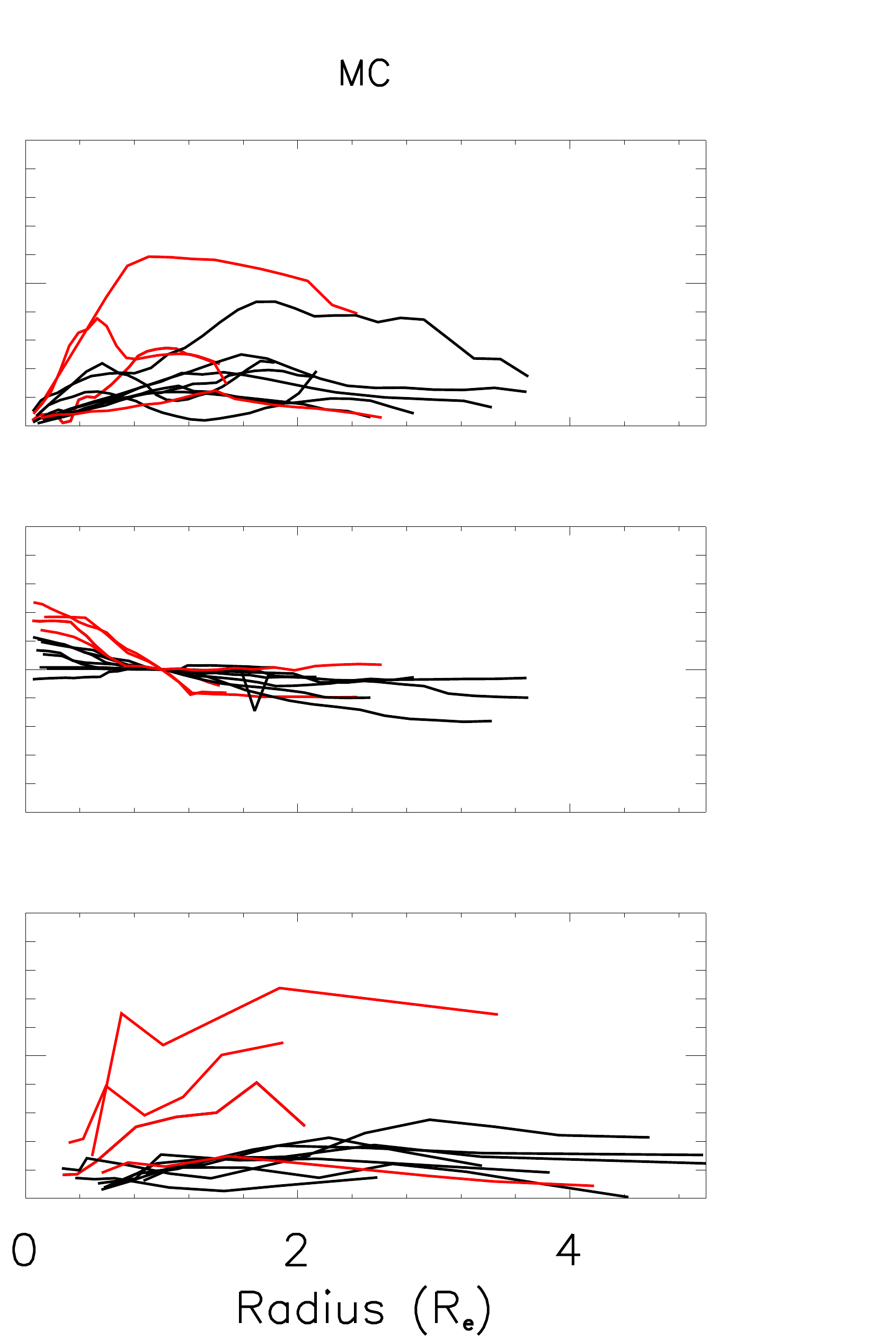}
\caption{
  Velocity, normalized velocity dispersion and angular momentum
  profiles for all galaxies in our sample. We show the first order
  harmonic velocity term $k_1$ (the rotation curve, top), the velocity dispersion in
  elliptical annuli (middle), and the local measure of angular momentum, 
  $\Lambda_R$ as a
  function of radius (bottom) for both SRs (Black) and FRs
  (Red). Results are additionally divided into SC galaxies (left), MC
  galaxies with KD's (middle) and other MC galaxies (right). For
  simplicity, we classify the 3 galaxies NGC~677, IC~301 and NGC~3837 as
  SRs with KDs, since their kinematic behaviour is most similar to this class.}
\label{Fig:LambdaAllC}
\end{center}
\end{figure*}

\begin{figure*}
\begin{center}
\includegraphics[width=0.8\columnwidth,angle=0,clip]{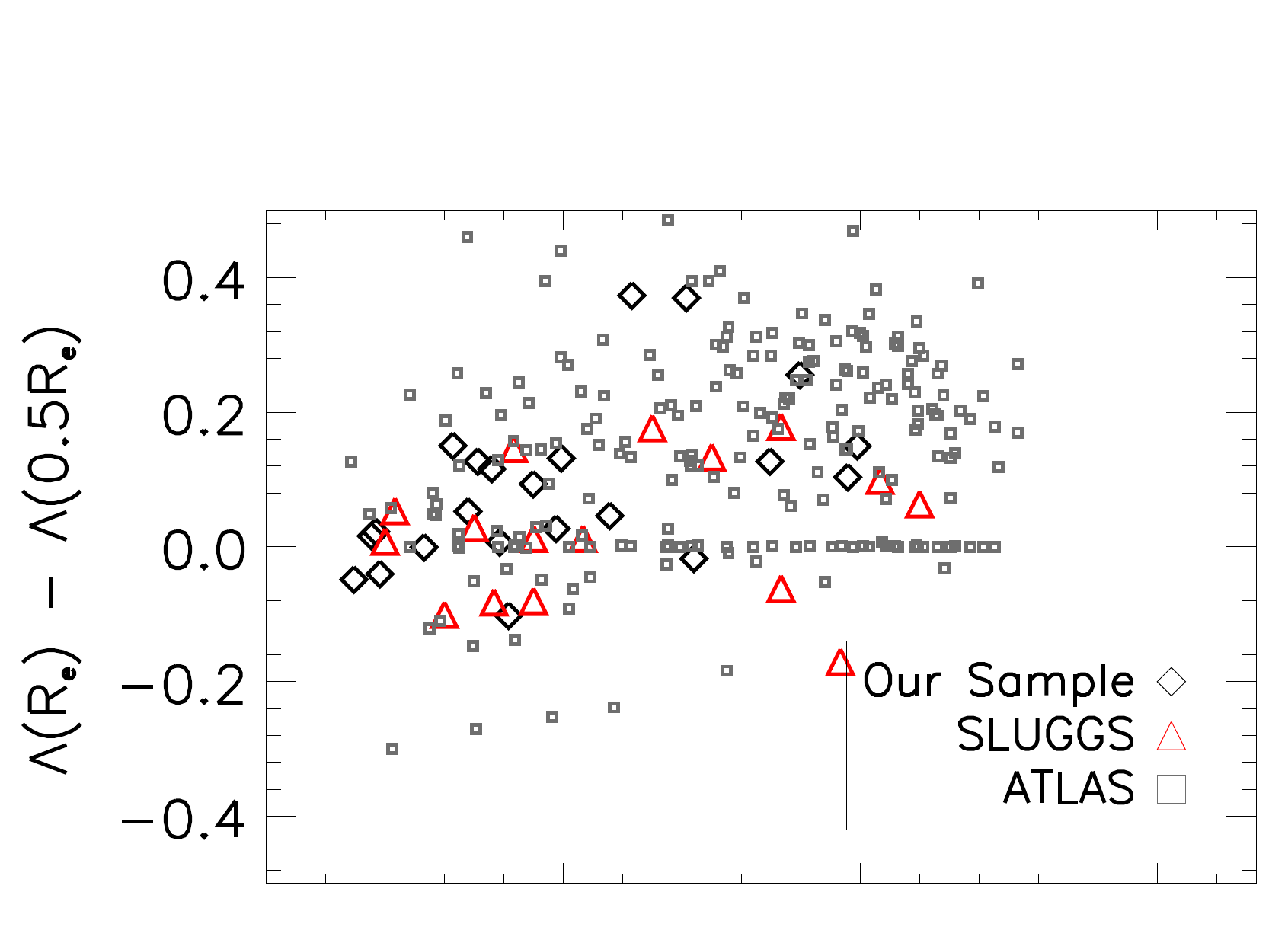}
\includegraphics[width=0.8\columnwidth,angle=0,clip]{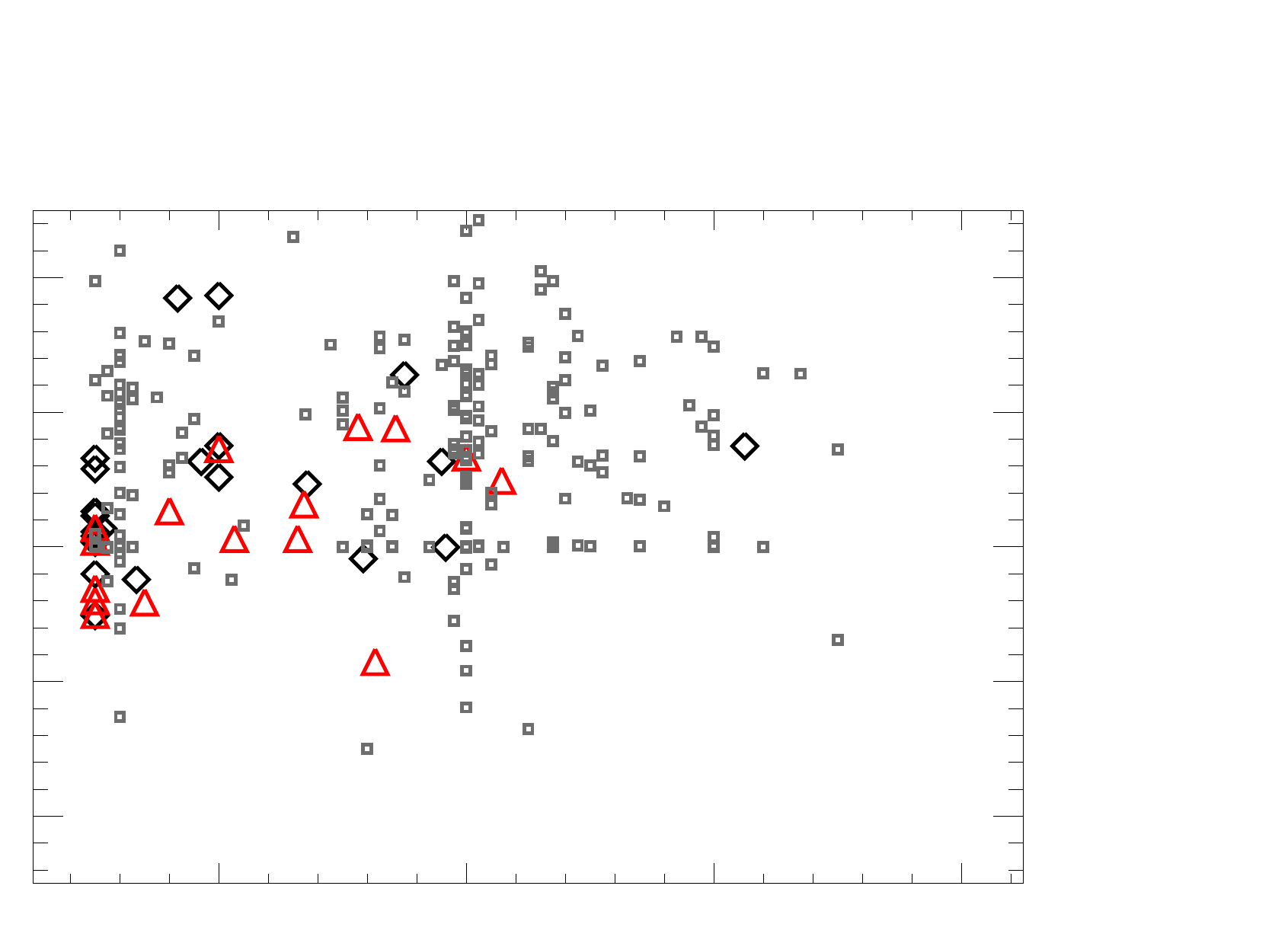}
\includegraphics[width=0.8\columnwidth,angle=0,clip]{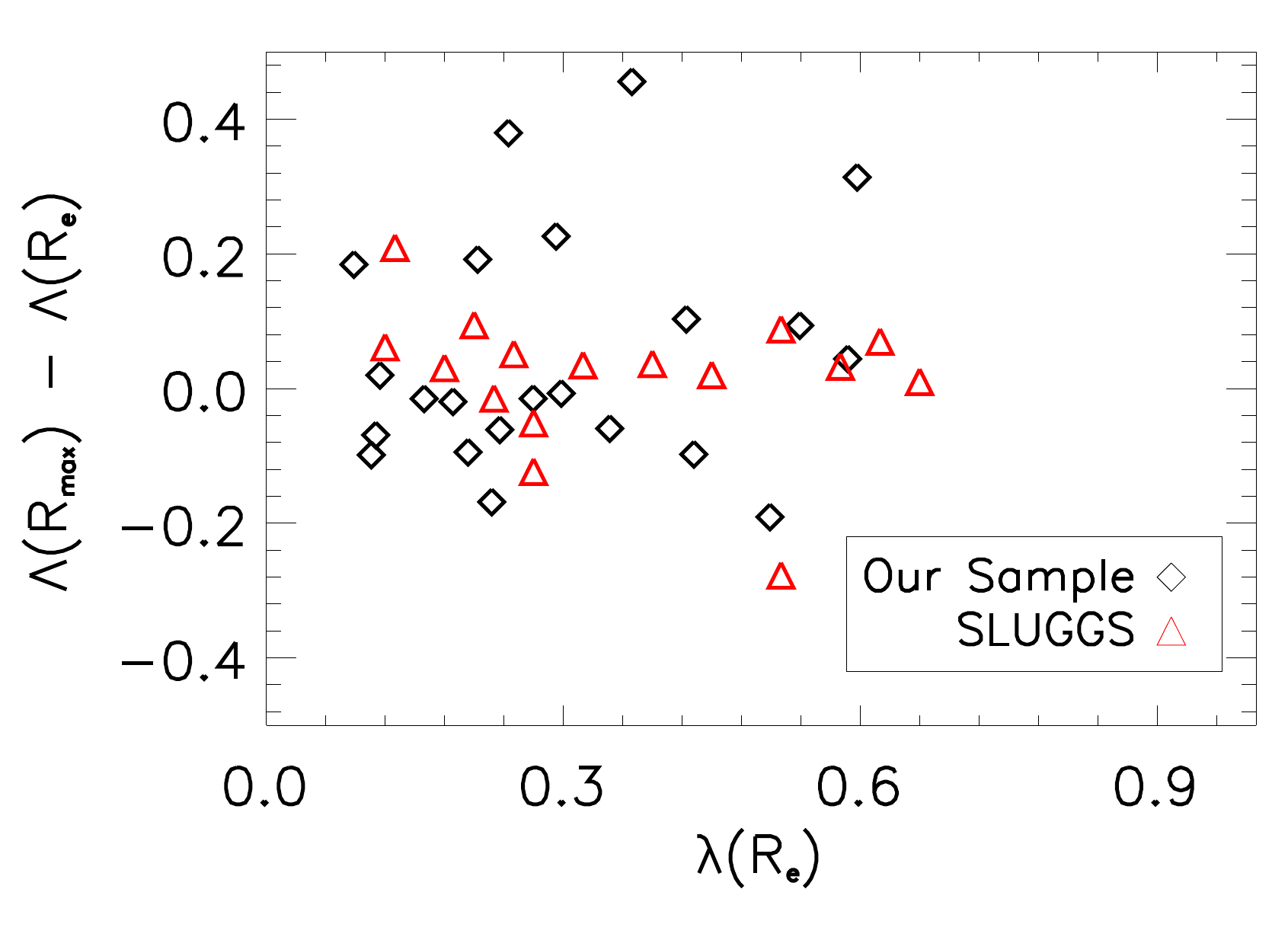}
\includegraphics[width=0.8\columnwidth,angle=0,clip]{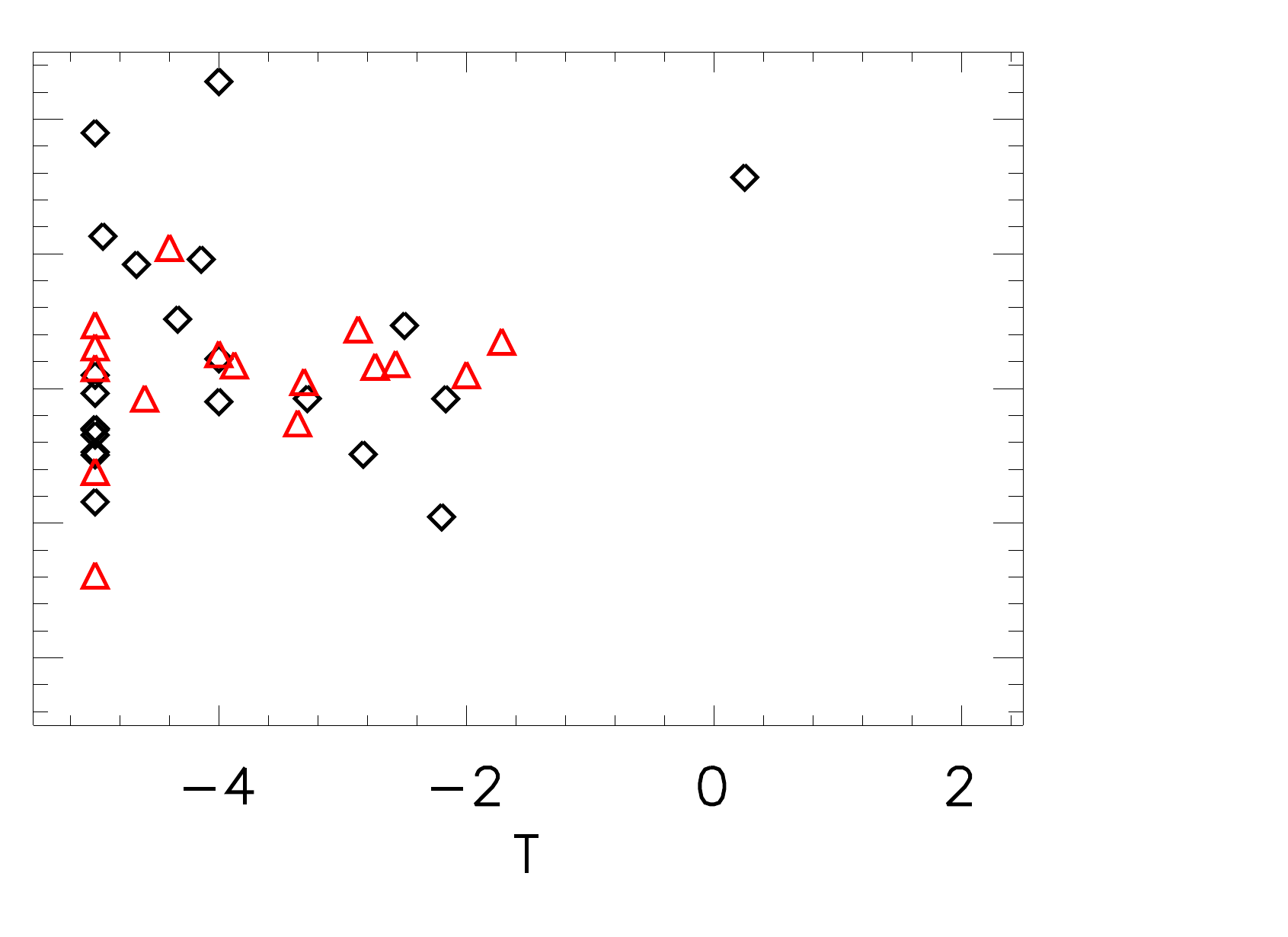}
\caption{
  Radial gradient in the angular momentum between $R_e$ and
  0.5$R_e$ (top) and the outermost measured radius $R_{\rm max}$ 
  and $R_e$ (bottom). We show the gradient vs. both $\lambda(R_e)$ and
  the morphological T-type number, where $T > -3.5$ indicates a
  lenticular galaxy.  In all cases, we show only FRs from our sample
  (black diamonds), SLUGGS (red triangles) and where appropriate 
  ATLAS$^{\rm 3D}$ (grey squares). The ATLAS$^{\rm 3D}$
  values were calculated from their published stellar kinematics,
  while for SLUGGS, the relevant values were drawn from
  \cite{Arnold2013}. For the ATLAS$^{\rm 3D}$ sample, the large subset of
  galaxies clustered around zero gradient arises from those galaxies
  observed with relatively small apertures.}
\label{Fig:LambdaGrad}
\end{center}
\end{figure*}

\begin{figure}
\begin{center}
\includegraphics[width=0.95\columnwidth,angle=0,clip]{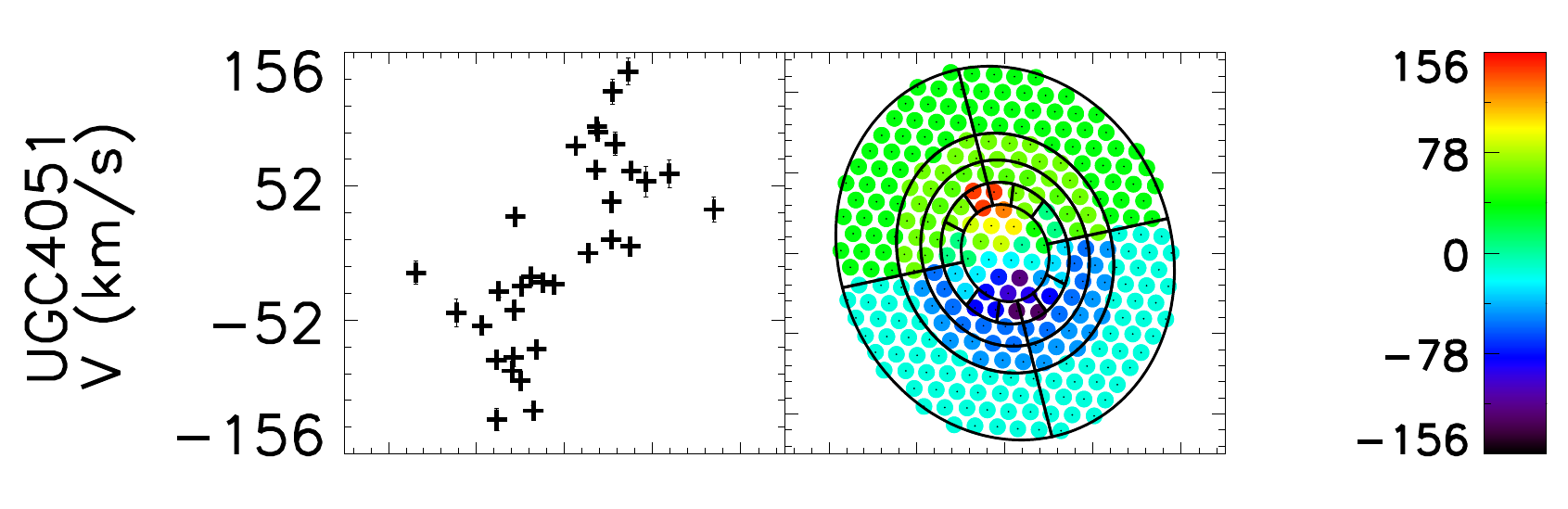}
\includegraphics[width=0.95\columnwidth,angle=0,clip]{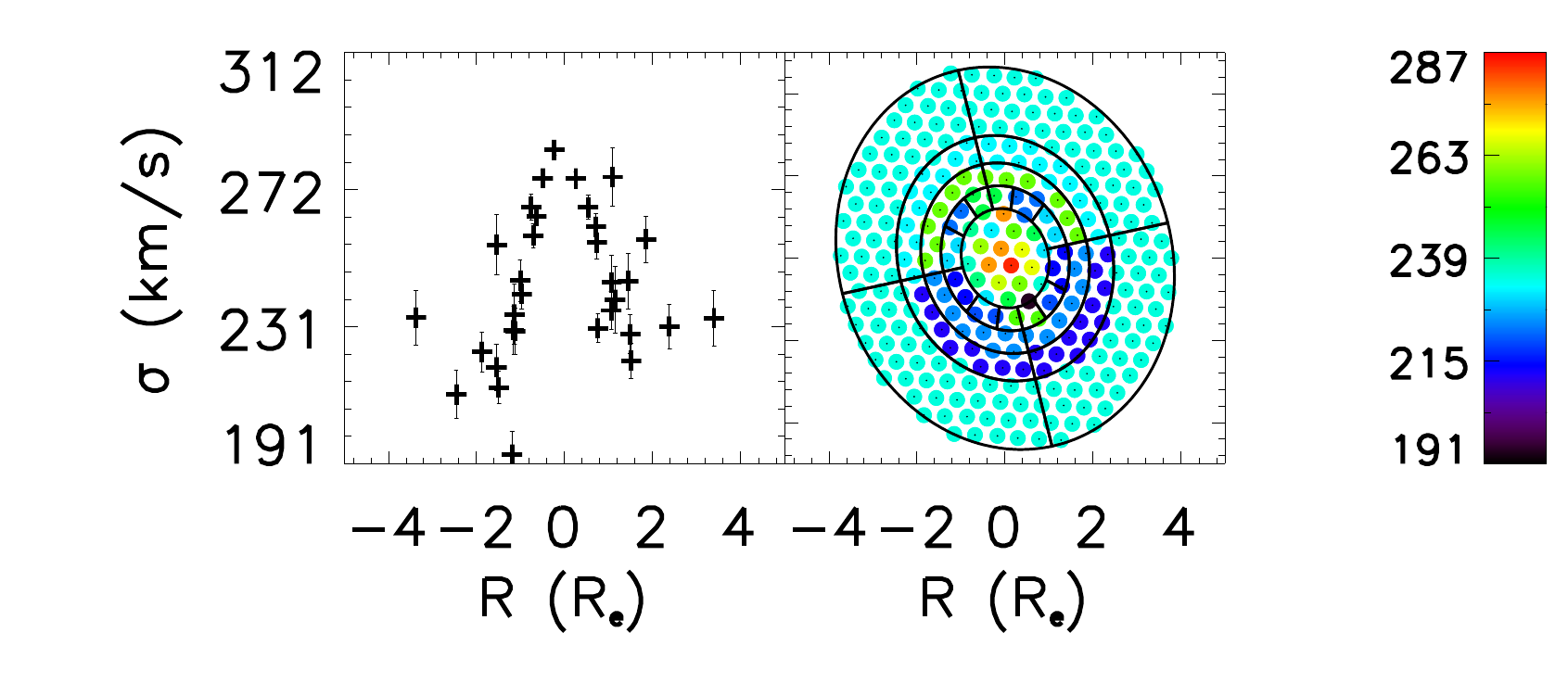}
\caption{2D velocity (Top) and dispersion (Bottom) maps for UGC~4051.}
\label{Fig:KinematicsUGC4051}
\end{center}
\end{figure}

For FRs, the distribution of $\Lambda$ looks qualitatively similar to
the 22 galaxies observed by the SLUGGS survey and the numerical
results of \cite{Wu2014}.  In higher mass FRs, $\Lambda$ tends to
decline slightly or remain flat, while the majority of the FRs with
lower mass tend to have rising $\Lambda$ profiles. Since we can only
reach $\sim 2 R_e$ in these lower-mass galaxies, it is possible that
we simply have not reached a large enough radius to see the $\Lambda$
profile flatten/fall.

The galaxies with declining $\Lambda$ profiles are predominantly SC disk-like
FRs, as can be seen from the leftmost panel of
Figure~\ref{Fig:LambdaAllC}. This subset includes the galaxies with
the sharpest declines in $\Lambda$: NGC~774 at low mass and UGC~4051 at
high mass, both of which have $\delta \Lambda \sim 0.1 - 0.2$. These SC
FRs also almost all show a decline in the rotation curve beyond $\sim
R_e$, which is typically accompanied by increases in $k_5 / k_1$ to
$\sim 0.3$. The declining S/N and large spatial
bins also contribute to the large $k_5 / k_1$ values. However, given
the rough correspondence between drops in $\Lambda$ and $k_1$, both of
which are calculated independently, it seems unlikely that S/N alone is
behind the radial changes.

We now ask whether these galaxies are showing signs of the transition
from inner disk to outer halo detected by \cite{Arnold2013}. As
mentioned earlier, none show nearly as rapid a decline in $\Lambda$ as
that seen in NGC~3377, so it is not clear on a galaxy-by-galaxy basis
that we are seeing this transition. However, statistically, we may ask
whether we see the correlation between angular momentum gradients and
Hubble Type seen by \cite{Arnold2013}. In Figure~\ref{Fig:LambdaGrad}
we show the radial variations in $\Lambda$ between $R_{\rm max}$ and
$R_e$ (top) and also between $R_e$ and 0.5$R_e$ (bottom) for all FRs
as a function of the Hubble Type and of $\lambda(R_e)$. In this case, we
omit the two galaxies with $R_{\rm max} \lesssim 2R_e$ as these lack
sufficient data for a robust measurement of $\Lambda(>R_e)$. We also
show the corresponding measurements for the SLUGGS, and where possible,
ATLAS$^{\rm 3D}$, surveys. The former are taken directly from \cite{Arnold2013},
while for the ATLAS$^{\rm 3D}$ survey, values of $\Lambda$ within $R_e$ are
calculated from the full 2D stellar kinematics. In both
cases, Hubble types are taken from the HyperLeda\footnote{http://leda.univ-lyon1.fr.} 
database \citep[][]{Paturel2003}.

As described in \cite{Arnold2013}, the fastest declining SLUGGS
galaxies tend to be elliptical, while most of those that rotate more
outwards are S0's. However, we do not notice any such trend for our
sample. If anything the reverse holds true, with our S0s having the 
fastest declining $\Lambda$ profiles while the ellipticals show the 
largest $\Lambda$ increases. Hubble type is not a continuous quantity, 
but if we naively fit lines to the radial gradient $\Lambda(R_{\rm max}) -
\Lambda(R_e)$ as a function of $T$, then we obtain a positive
Pearson correlation coefficient of $r = 0.45$ {($p = 0.036$)} for the SLUGGS sample as opposed
to $r = -0.18$ {($p = 0.32$)} for ours and $r = -0.01$ {($p = 0.94$)} for the joint sample. This
seems to point to a lack of correlation between declining $\Lambda$
and disky galaxies.

We lack the statistical significance to make
any strong statement about correlations between morphology and large
scale kinematics. However, as an interesting exercise, we may ask the
same question of the entire ATLAS$^{\rm 3D}$ sample, as shown in the top right
panel of Figure~\ref{Fig:LambdaGrad}. Naturally, in this case we are
restricted to $<R_e$, but even within this smaller aperture, we already
see gradients comparable to, or exceeding, the changes out to $\sim
4R_e$. Equally, within this much larger sample, we see no evidence of
any difference in $\Lambda$ gradients between the E and S0 galaxies, and a
simple fit gives a correlation coefficient of $r = 0.01$, entirely
consistent with zero.

As a final comparison between the two samples, we may consider the
kinematics and morphology of our fastest declining FR,
UGC~4051. Figure~\ref{Fig:KinematicsUGC4051} shows the velocity and
dispersion maps for this galaxy. If there were an embedded disk we may
expect that along the major axis, where the disk is located, there
would be lower velocity dispersion with respect to the minor axis,
which contains mostly halo stars. We see no such evidence of
such a feature. Kinematic maps of other rapidly declining galaxies
(particularly NGC~774) also show no such behaviour, {although this effect 
may only be pronounced if the galaxy were edge-on. Given also  
our low kinematic resolution this does not necessarily preclude the 
presence of stellar disks in these systems.}

There are a number of key differences between our sample and SLUGGS
that may explain the differences in our results. Firstly from a methodological 
perspective, our galaxies are binned at much lower spatial resolution, particularly 
at large radius. However, it seems unlikely that this could explain our  
dearth of galaxies with pronounced declines in $\Lambda$ as compared 
to SLUGGS. If anything, averaging over large spatial bins 
would tend to artificially lower the measured velocity and thus also $\Lambda$.

More physically, our sample covers more massive galaxies, which may tend 
to have smaller $\Lambda$ gradients. For instance, the simulated 
galaxies in \cite{Wu2014} show  a trend with stellar mass, in the sense that the low-mass FRs are 
more likely to show declining $\Lambda$ profiles.  Perhaps we need to probe even 
larger radii to see the transition to a halo component in these more massive galaxies.
Thus, our differences with the SLUGGS sample may be simply explained by the bias towards higher 
mass in our sample. 

\subsubsection{SRs at Large Radius}
\label{subsubsec:SRR}

For at least half the SRs, the picture is comparatively simple. Aside
from the completely non-rotating SC SR IC~1152, five SRs show central
kinematically decoupled components characteristic of a transition from
an inner disky structure to an outer halo.  The decoupled components
seem to be similar to the KDCs described in \cite{Krajnovic2011},
which were interpreted as remnants of old, wet, major
mergers.  

If these kinematic transitions actually signal 
a component with a different formation history, then we could 
be seeing the remnant of an early dissipational component transitioning 
to an outer halo \citep{Arnold2013}.  On the other hand, these components 
are large (1-7 kpc) and have low amplitude rotation 
($\Lambda \lesssim 0.2$ as compared to $\Lambda \sim 0.6$).  Furthermore, the 
kinematic and photometric position angles are generally misaligned.  For all 
of these reasons, we believe we are instead seeing signs of triaxiality 
\citep[e.g.,][]{Statler1991}. This 
triaxiality also likes results from merging, as pointed out for NGC~5982 by 
\citet{Oosterloo1994}. In fact, simulations suggest that triaxiality is strongly 
correlated with the box orbits that result from specifically dry major mergers 
\citep{Jesseit2005, Jesseit2007, Hoffman2009}.

In the same way, we have argued that the SRs with rising $\Lambda$ profiles also show 
clear signs of triaxiality (as typified by NGC~5982 and NGC~6482). They
generally show some evidence of a central LV component that
transitions to slow disk-like rotation. In addition, the PA tends to be misaligned with the 
photometric axis in the central regions. NGC~6482 particularly shows strong kinematic 
misalignment of between $20^{\circ}$ and $50^{\circ}$ out to at least $\sim 2 R_e$. 
Based on their complicated kinematics, both galaxies have been put forward as recent 
merger remnants \citep{Statler1991, Oosterloo1994, DelBurgo2008}. 
While more detailed comparisons are needed, 
it seems likely from simulations that a series of minor mergers are needed to reproduce 
both the low $\lambda$ and generic triaxial properties of the MC SRs 
\citep[e.g.,][]{Bois2011}.

\subsection{Correlations with Stellar Populations}
\label{subsec:Changing SSPs}

\begin{figure*}
\begin{center}
\includegraphics[width=0.8\textwidth,angle=270,clip]{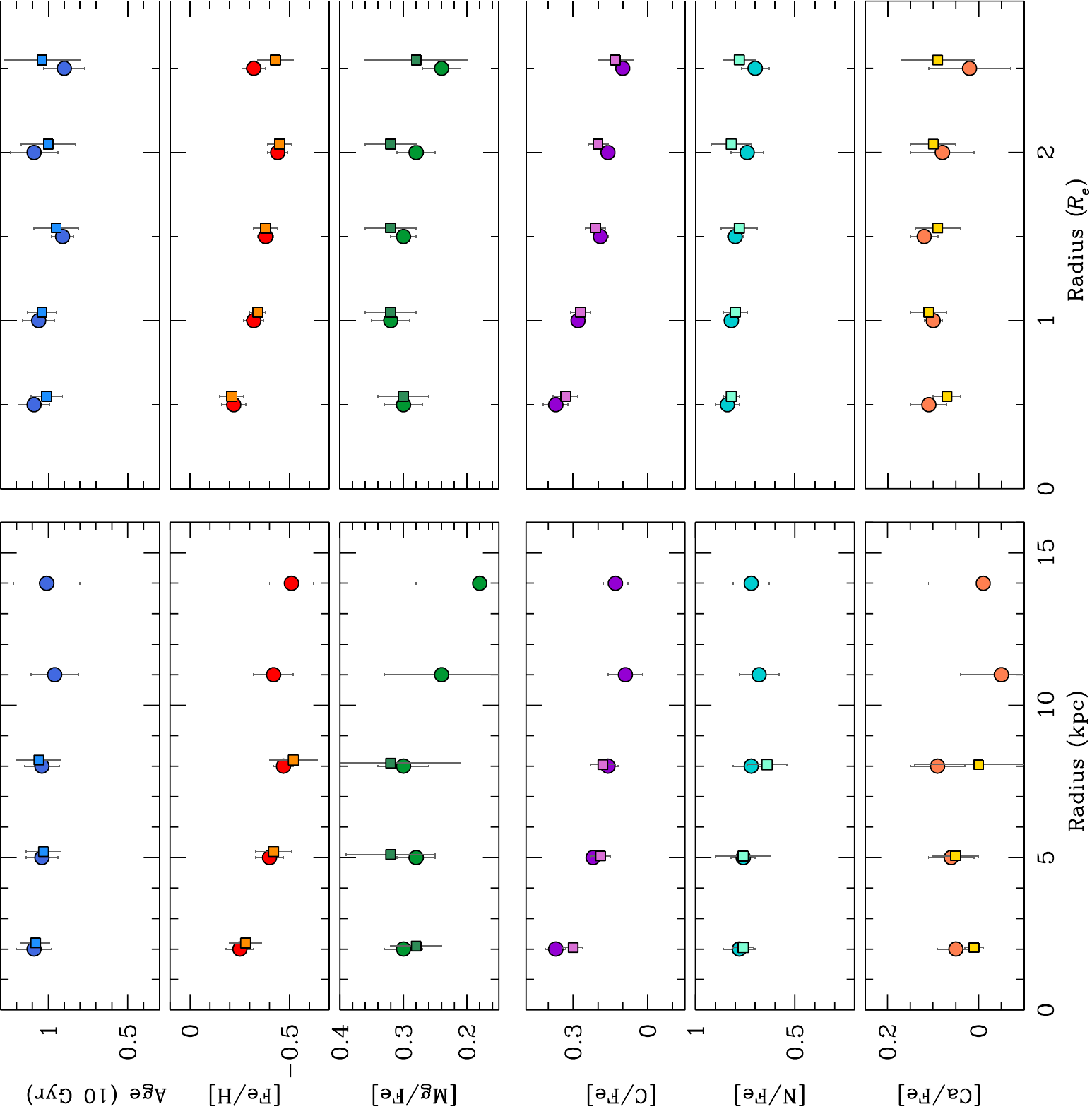}
\caption{Radial gradients in age, [Fe/H], [Mg/Fe], [C/Fe], [N/Fe], and [Ca/Fe] as calculated by 
{\it EZ\_Ages} from the Lick indices measured in the composite spectra. We show both the measurements 
for SR (circles) and FR (squares) galaxies as a function of R in kpc (left) or ${\rm R} / {\rm R_e}$ (right).}
\label{Fig:SSP}
\end{center}
\end{figure*}

We now ask whether there are any differences in the stellar
populations of our sample as a function of $\lambda$. For instance, if
high $\lambda$ is a signpost of dissipational formation, we might
expect younger, more metal-rich stellar populations in the outer parts
of FRs. Following \cite{Greene2013}, we construct composite spectra
as a function of radius, dividing the sample into FRs and SRs. {To try
and mitigate the strong impact of $\sigma$, we 
restrict our attention to galaxies with central stellar velocity dispersion 
$\sigma_{c}$, as measured by the SDSS, greater than 200~\kms}. There are 10 SRs and 12 FRs 
included in our stacked spectra.

We construct composite spectra as described in \cite{Greene2013}.
In brief, we first substract emission-lines iteratively using
continuum fits \citep[e.g.,][]{Graves2007}. Then, we divide each
spectrum by a heavily smoothed version of itself to remove the
continuum, and combine them using the biweight estimator
\citep{Beers1990}.  We then measure the Lick indices, and invert them
to infer the ages, metallicities, and abundance ratios at each radial
bin for the SRs and FRs, using {\it EZ\_Ages}
\citep{GravesSchiavon2008}. In addition to stellar age, [Fe/H], and
[$\alpha$/Fe] abundance ratios, the code also iteratively solves for
the [C/Fe] and [N/Fe] abundance ratios, the former based mostly on the
C$_2~\lambda 4668$ Swann band, and the latter on the blue CN bands.

We note that the absolute values of [C/Fe] and [N/Fe] are uncertain
because they depend directly on the oxygen abundance.  Oxygen, as the
most abundant heavy element, has a large indirect impact on the
spectra but as there are no broad-band O indices, we must assume a
value for [O/Fe].  Here we assume that it tracks the other $\alpha$
elements.  Because the C gets bound up in CO molecules, the assumed
oxygen abundance has a significant effect on the modeled [C/Fe] and
therefore [N/Fe] \citep{Graves2007, Greene2013}. Specifically, if we
lowered the assumed [O/Fe] to a solar value, the [C/Fe] and 
the [N/Fe] would fall, while their relative trend is robust 
\citep[see discussion in][]{Greene2013}.

The radial profiles of our measured stellar population properties are
shown in Figure \ref{Fig:SSP}.  There are no significant differences
between SRs and FRs.  However, there are some intriguing hints.
First of all, the FRs appear to have a slight tendency to get older in
the outermost bins.  In fact, we see a weak trend for positive age
gradients as well when we consider individual galaxies, but it 
is not statistically significant.  If true, 
  we may be seeing the transition from stellar disk to stellar halo in
  the FRs.  Over the past year, we have gathered data
for twice as many galaxies, which will allow us to bin in both
$\sigma_{c}$ and $\lambda$.

We are left with a slightly ambiguous picture of how our galaxy sample
ties into two-phase galaxy formation. Our observed FRs may
show signs of a transition from inner disk to outer halo through 
small drops in the net rotation. However, these are typically not accompanied by the
significant drops in angular momentum reported in
\citet{Arnold2013} or any significant change in stellar populations. 
Nor are the observed drops in angular momentum
correlated with E galaxies, as we might expect if S0's were
characterised by more extended disks.  Perhaps this is entirely a
function of mass, since simulations of two-phase galaxy assembly by
\cite{Wu2014}, with which our observations seem to agree quite well,
show fewer angular momentum transitions as we move to higher mass.

\section{Summary and Future Work}
\label{Sec:Conclusions}

We have presented wide-field 2D kinematic LOSVD's (out to radii
between 2 and 5 $R_e$) for a sample of 33 massive elliptical galaxies
previously described in \citet{Greene2013}. Our sample comprises 12
Slow (SRs) and 21 Fast Rotators (FRs), with classifications based on kinematic
information out to $R_e$. By design, this is a higher fraction of SRs
than the volume-limited ATLAS$^{\rm 3D}$ and SLUGGS samples.

Despite covering a broad range of central dispersions, sizes and
environments, we find that most of the galaxies can be well classified
on the basis of their kinematic and kinemetric information. A majority
of the FRs are single component disk-like rotators, with any decoupled components
usually being in the form of central low velocity regions that likely signal unresolved 
rotation. They do show some tentative indications of transitioning from
stellar disk to halo beyond $R_e$, but we observe no galaxies with the dramatic drops in
angular momentum reported in \citet{Arnold2013}. Generally, our FR
galaxies continue rotating as far out as we observe. A majority of
the SRs meanwhile show kinematically distinct components with disk-like properties at
their centre, but these typically rotate more slowly and are much
larger than the KDCs found in SAURON.

Our work, along with SLUGGS, represents an early effort to classify the outer parts of
massive elliptical galaxies based on their kinematic properties.  If we interpret the
ubiquitous multi-component nature of the SRs as evidence for 
different formation histories, then we see some evidence for an ``inner'' and ``outer'' component,
with a transition $\sim 5$ kpc.  However, without more concrete comparisons with theory, it is hard
to say whether we are seeing different phases of elliptical galaxy growth, or just triaxial galaxies
in projection.

Furthermore, many of the trends we observe are relatively uncertain due to
the small size, and incomplete selection, of our sample.  We are
currently in the process of doubling our sample, which combined with
more detailed dynamical modelling of our galaxies, should be able to
provide more insight into elliptical galaxy formation and particularly
two-phase assembly. Specifically, dynamical studies, which reveal the
DM fraction, velocity anisotropy and gravitational potential, should,
in combination with stellar population results, be able to
constrain both when and where outer halo stars were assembled. By
offering a more direct comparison with cosmological-scale simulations
of elliptical galaxy formation, we may then be able to comment more
conclusively on exactly when and how the most massive galaxies were
formed.

\acknowledgements
We would like to thank the referee for a prompt and helpful review. We also thank D. Krajnovic, 
A. Romanowky, and J. Brodie for their feedback, which substantially improved this manuscript. 
We acknowledge the usage of the HyperLeda database (http://leda.univ-lyon1.fr)

%----------------------------------------------------------------------

\bibliographystyle{apj}
\bibliography{Paper1}

\begin{thebibliography}{155}
\expandafter\ifx\csname natexlab\endcsname\relax\def\natexlab#1{#1}\fi

\bibitem[{Adams {et~al.}(2012)Adams, Gebhardt, Blanc, Fabricius, Hill, Murphy,
  van~den Bosch, \& van~de Ven}]{Adams2012}
Adams, J.~J., Gebhardt, K., Blanc, G.~A., {et~al.} 2012, The Astrophysical
  Journal, 745, 92

\bibitem[{Adams {et~al.}(2011)Adams, Blanc, Hill, Gebhardt, Drory, Hao, Bender,
  Byun, Ciardullo, Cornell, Finkelstein, Fry, Gawiser, Gronwall, Hopp, Jeong,
  Kelz, Kelzenberg, Komatsu, MacQueen, Murphy, Odoms, Roth, Schneider, Tufts,
  \& Wilkinson}]{Adams2011}
Adams, J.~J., Blanc, G.~A., Hill, G.~J., {et~al.} 2011, The Astrophysical
  Journal Supplement, 192, 5

\bibitem[{Arnold {et~al.}(2011)Arnold, Romanowsky, Brodie, Chomiuk, Spitler,
  Strader, Benson, \& Forbes}]{Arnold2011}
Arnold, J.~A., Romanowsky, A.~J., Brodie, J.~P., {et~al.} 2011, The
  Astrophysical Journal Letters, 736, L26

\bibitem[{{Arnold} {et~al.}(2013){Arnold}, {Romanowsky}, {Brodie}, {Forbes},
  {Strader}, {Spitler}, {Foster}, {Kartha}, {Pastorello}, {Pota}, {Usher}, \&
  {Woodley}}]{Arnold2013}
{Arnold}, J.~A., {Romanowsky}, A.~J., {Brodie}, J.~P., {et~al.} 2013, ArXiv
  e-prints

\bibitem[{Barden {et~al.}(1998)Barden, Sawyer, \& Honeycutt}]{Barden1998}
Barden, S.~C., Sawyer, D.~G., \& Honeycutt, R.~K. 1998, Proc. SPIE Vol. 3355,
  3355, 892

\bibitem[{{Barth} {et~al.}(2002){Barth}, {Ho}, \& {Sargent}}]{Barth2002}
{Barth}, A.~J., {Ho}, L.~C., \& {Sargent}, W.~L.~W. 2002, \aj, 124, 2607

\bibitem[{Beers {et~al.}(1990)Beers, Flynn, \& Gebhardt}]{Beers1990}
Beers, T.~C., Flynn, K., \& Gebhardt, K. 1990, Astronomical Journal (ISSN
  0004-6256), 100, 32

\bibitem[{{Bender} {et~al.}(1989){Bender}, {Surma}, {Doebereiner},
  {Moellenhoff}, \& {Madejsky}}]{Bender1989}
{Bender}, R., {Surma}, P., {Doebereiner}, S., {Moellenhoff}, C., \& {Madejsky},
  R. 1989, \aap, 217, 35

\bibitem[{{Bernardi} {et~al.}(2007){Bernardi}, {Hyde}, {Sheth}, {Miller}, \&
  {Nichol}}]{bernardi2007}
{Bernardi}, M., {Hyde}, J.~B., {Sheth}, R.~K., {Miller}, C.~J., \& {Nichol},
  R.~C. 2007, \aj, 133, 1741

\bibitem[{{Bertola} \& {Capaccioli}(1975)}]{bertolacapaccioli1975}
{Bertola}, F., \& {Capaccioli}, M. 1975, \apj, 200, 439

\bibitem[{{Bezanson} {et~al.}(2009){Bezanson}, {van Dokkum}, {Tal},
  {Marchesini}, {Kriek}, {Franx}, \& {Coppi}}]{Bezanson2009}
{Bezanson}, R., {van Dokkum}, P.~G., {Tal}, T., {et~al.} 2009, \apj, 697, 1290

\bibitem[{{Binney}(1978{\natexlab{a}})}]{Binney1978b}
{Binney}, J. 1978{\natexlab{a}}, Comments on Astrophysics, 8, 27

\bibitem[{{Binney}(1978{\natexlab{b}})}]{Binney1978}
---. 1978{\natexlab{b}}, \mnras, 183, 501

\bibitem[{{Binney}(2005)}]{Binney2005}
---. 2005, \mnras, 363, 937

\bibitem[{Blanc {et~al.}(2009)Blanc, Heiderman, Gebhardt, Evans, \&
  Adams}]{Blanc2009}
Blanc, G.~A., Heiderman, A., Gebhardt, K., Evans, N. J.~I., \& Adams, J. 2009,
  \apj, 704, 842

\bibitem[{{Blanc} {et~al.}(2011){Blanc}, {Adams}, {Gebhardt}, {Hill}, {Drory},
  {Hao}, {Bender}, {Ciardullo}, {Finkelstein}, {Fry}, {Gawiser}, {Gronwall},
  {Hopp}, {Jeong}, {Kelzenberg}, {Komatsu}, {MacQueen}, {Murphy}, {Roth},
  {Schneider}, \& {Tufts}}]{Blanc2011}
{Blanc}, G.~A., {Adams}, J.~J., {Gebhardt}, K., {et~al.} 2011, \apj, 736, 31

\bibitem[{Blanc {et~al.}(2013)Blanc, Weinzirl, Song, Heiderman, Gebhardt,
  Jogee, Evans, van~den Bosch, Luo, Drory, Fabricius, Fisher, Hao, Kaplan,
  Marinova, Vutisalchavakul, \& Yoachim}]{Blanc2013}
Blanc, G.~A., Weinzirl, T., Song, M., {et~al.} 2013, \aj, 145, 138

\bibitem[{{Blanton} {et~al.}(2005){Blanton}, {Eisenstein}, {Hogg}, {Schlegel},
  \& {Brinkmann}}]{Blanton2005}
{Blanton}, M.~R., {Eisenstein}, D., {Hogg}, D.~W., {Schlegel}, D.~J., \&
  {Brinkmann}, J. 2005, \apj, 629, 143

\bibitem[{Bois {et~al.}(2011)Bois, Emsellem, Bournaud, Alatalo, Blitz, Bureau,
  Cappellari, Davies, Davis, de~Zeeuw, Duc, Khochfar, Krajnovi{\'c},
  Kuntschner, Lablanche, McDermid, Morganti, Naab, Oosterloo, Sarzi, Scott,
  Serra, Weijmans, \& Young}]{Bois2011}
Bois, M., Emsellem, E., Bournaud, F., {et~al.} 2011, \mnras, 416, 1654

\bibitem[{{Bruzual} \& {Charlot}(2003)}]{BruzualCharlot2003}
{Bruzual}, G., \& {Charlot}, S. 2003, \mnras, 344, 1000

\bibitem[{{Buitrago} {et~al.}(2008){Buitrago}, {Trujillo}, {Conselice},
  {Bouwens}, {Dickinson}, \& {Yan}}]{Buitrago2008}
{Buitrago}, F., {Trujillo}, I., {Conselice}, C.~J., {et~al.} 2008, \apjl, 687,
  L61

\bibitem[{{Burbidge} {et~al.}(1961){Burbidge}, {Burbidge}, \&
  {Fish}}]{Burbidge1961}
{Burbidge}, E.~M., {Burbidge}, G.~R., \& {Fish}, R.~A. 1961, \apj, 133, 393

\bibitem[{{Caon} {et~al.}(1993){Caon}, {Capaccioli}, \& {D'Onofrio}}]{caon1993}
{Caon}, N., {Capaccioli}, M., \& {D'Onofrio}, M. 1993, \mnras, 265, 1013

\bibitem[{{Cappellari}(2013)}]{Cappellari2013}
{Cappellari}, M. 2013, \apjl, 778, L2

\bibitem[{{Cappellari} \& {Emsellem}(2004)}]{CappellariEmsellem2003}
{Cappellari}, M., \& {Emsellem}, E. 2004, \pasp, 116, 138

\bibitem[{{Cappellari} {et~al.}(2007){Cappellari}, {Emsellem}, {Bacon},
  {Bureau}, {Davies}, {de Zeeuw}, {Falc{\'o}n-Barroso}, {Krajnovi{\'c}},
  {Kuntschner}, {McDermid}, {Peletier}, {Sarzi}, {van den Bosch}, \& {van de
  Ven}}]{Cappellari2007}
{Cappellari}, M., {Emsellem}, E., {Bacon}, R., {et~al.} 2007, \mnras, 379, 418

\bibitem[{Cappellari {et~al.}(2011)Cappellari, Emsellem, Krajnovic, McDermid,
  Scott, Verdoes~Kleijn, Young, Alatalo, Bacon, Blitz, Bois, Bournaud, Bureau,
  Davies, Davis, de~Zeeuw, Duc, Khochfar, Kuntschner, Lablanche, Morganti,
  Naab, Oosterloo, Sarzi, Serra, \& Weijmans}]{Cappellari2011}
Cappellari, M., Emsellem, E., Krajnovic, D., {et~al.} 2011, \mnras, 413, 813

\bibitem[{Carollo \& Danziger(1994)}]{CarolloDanziger1994}
Carollo, C.~M., \& Danziger, I.~J. 1994, \mnras, 270, 523

\bibitem[{Carollo {et~al.}(1997)Carollo, Danziger, Rich, \& Chen}]{Carollo1997}
Carollo, C.~M., Danziger, I.~J., Rich, R.~M., \& Chen, X. 1997, \apj, 491, 545

\bibitem[{{Cenarro} \& {Trujillo}(2009)}]{CenarroTrujillo2009}
{Cenarro}, A.~J., \& {Trujillo}, I. 2009, \apjl, 696, L43

\bibitem[{Ceverino {et~al.}(2010)Ceverino, Dekel, \& Bournaud}]{Ceverino2010}
Ceverino, D., Dekel, A., \& Bournaud, F. 2010, \mnras, 404, 2151

\bibitem[{{Cimatti} {et~al.}(2008){Cimatti}, {Cassata}, {Pozzetti}, {Kurk},
  {Mignoli}, {Renzini}, {Daddi}, {Bolzonella}, {Brusa}, {Rodighiero},
  {Dickinson}, {Franceschini}, {Zamorani}, {Berta}, {Rosati}, \&
  {Halliday}}]{Cimatti2008}
{Cimatti}, A., {Cassata}, P., {Pozzetti}, L., {et~al.} 2008, \aap, 482, 21

\bibitem[{{Coccato} {et~al.}(2010){Coccato}, {Arnaboldi}, {Gerhard}, {Freeman},
  {Ventimiglia}, \& {Yasuda}}]{Coccato2010}
{Coccato}, L., {Arnaboldi}, M., {Gerhard}, O., {et~al.} 2010, \aap, 519, A95

\bibitem[{Coccato {et~al.}(2009)Coccato, Gerhard, Arnaboldi, Das, Douglas,
  Kuijken, Merrifield, Napolitano, Noordermeer, Romanowsky, Capaccioli,
  Cortesi, de~Lorenzi, \& Freeman}]{Coccato2009}
Coccato, L., Gerhard, O., Arnaboldi, M., {et~al.} 2009, \mnras, 394, 1249

\bibitem[{{Cole} {et~al.}(2000){Cole}, {Lacey}, {Baugh}, \& {Frenk}}]{Cole2000}
{Cole}, S., {Lacey}, C.~G., {Baugh}, C.~M., \& {Frenk}, C.~S. 2000, \mnras,
  319, 168

\bibitem[{{Conroy} \& {Gunn}(2010)}]{ConroyGunn2010}
{Conroy}, C., \& {Gunn}, J.~E. 2010, \apj, 712, 833

\bibitem[{{Conroy} {et~al.}(2009){Conroy}, {Gunn}, \& {White}}]{Conroy2009}
{Conroy}, C., {Gunn}, J.~E., \& {White}, M. 2009, \apj, 699, 486

\bibitem[{{Daddi} {et~al.}(2005){Daddi}, {Renzini}, {Pirzkal}, {Cimatti},
  {Malhotra}, {Stiavelli}, {Xu}, {Pasquali}, {Rhoads}, {Brusa}, {di Serego
  Alighieri}, {Ferguson}, {Koekemoer}, {Moustakas}, {Panagia}, \&
  {Windhorst}}]{Daddi2005}
{Daddi}, E., {Renzini}, A., {Pirzkal}, N., {et~al.} 2005, \apj, 626, 680

\bibitem[{{Damjanov} {et~al.}(2009){Damjanov}, {McCarthy}, {Abraham},
  {Glazebrook}, {Yan}, {Mentuch}, {Le Borgne}, {Savaglio}, {Crampton},
  {Murowinski}, {Juneau}, {Carlberg}, {J{\o}rgensen}, {Roth}, {Chen}, \&
  {Marzke}}]{Damjanov2008}
{Damjanov}, I., {McCarthy}, P.~J., {Abraham}, R.~G., {et~al.} 2009, \apj, 695,
  101

\bibitem[{{Davies} {et~al.}(1983){Davies}, {Efstathiou}, {Fall}, {Illingworth},
  \& {Schechter}}]{Davies1983}
{Davies}, R.~L., {Efstathiou}, G., {Fall}, S.~M., {Illingworth}, G., \&
  {Schechter}, P.~L. 1983, \apj, 266, 41

\bibitem[{{Davis} {et~al.}(2011){Davis}, {Alatalo}, {Sarzi}, {Bureau}, {Young},
  {Blitz}, {Serra}, {Crocker}, {Krajnovi{\'c}}, {McDermid}, {Bois}, {Bournaud},
  {Cappellari}, {Davies}, {Duc}, {de Zeeuw}, {Emsellem}, {Khochfar},
  {Kuntschner}, {Lablanche}, {Morganti}, {Naab}, {Oosterloo}, {Scott}, \&
  {Weijmans}}]{Davis2011}
{Davis}, T.~A., {Alatalo}, K., {Sarzi}, M., {et~al.} 2011, \mnras, 417, 882

\bibitem[{{De Lucia} {et~al.}(2006){De Lucia}, {Springel}, {White}, {Croton},
  \& {Kauffmann}}]{DeLucia2006}
{De Lucia}, G., {Springel}, V., {White}, S.~D.~M., {Croton}, D., \&
  {Kauffmann}, G. 2006, \mnras, 366, 499

\bibitem[{{de Vaucouleurs}(1948)}]{deVaucouleurs1948}
{de Vaucouleurs}, G. 1948, Annales d'Astrophysique, 11, 247

\bibitem[{{de Zeeuw}(1985)}]{deZeeuw1985}
{de Zeeuw}, T. 1985, \mnras, 215, 731

\bibitem[{{de Zeeuw} \& {Franx}(1991)}]{deZeeuwFranx1991}
{de Zeeuw}, T., \& {Franx}, M. 1991, \araa, 29, 239

\bibitem[{{Dekel} {et~al.}(2009){Dekel}, {Sari}, \& {Ceverino}}]{Dekel2009}
{Dekel}, A., {Sari}, R., \& {Ceverino}, D. 2009, \apj, 703, 785

\bibitem[{Del~Burgo {et~al.}(2008)Del~Burgo, Carter, \& Sikkema}]{DelBurgo2008}
Del~Burgo, C., Carter, D., \& Sikkema, G. 2008, \aap, 477, 105

\bibitem[{{di Serego Alighieri} {et~al.}(2005){di Serego Alighieri}, {Vernet},
  {Cimatti}, {Lanzoni}, {Cassata}, {Ciotti}, {Daddi}, {Mignoli}, {Pignatelli},
  {Pozzetti}, {Renzini}, {Rettura}, \& {Zamorani}}]{diSeregoAlighieri2005}
{di Serego Alighieri}, S., {Vernet}, J., {Cimatti}, A., {et~al.} 2005, \aap,
  442, 125

\bibitem[{Dom{\'\i}nguez~S{\'a}nchez {et~al.}(2011)Dom{\'\i}nguez~S{\'a}nchez,
  Pozzi, Gruppioni, Cimatti, Ilbert, Pozzetti, McCracken, Capak, Le~Floch,
  Salvato, Zamorani, Carollo, Contini, Kneib, Le~F{\`e}vre, Lilly, Mainieri,
  Renzini, Scodeggio, Bardelli, Bolzonella, Bongiorno, Caputi, Coppa, Cucciati,
  de~la Torre, de~Ravel, Franzetti, Garilli, Iovino, Kampczyk, Knobel, Kova{\v
  c}, Lamareille, Le~Borgne, Le~Brun, Maier, Mignoli, Pell{\'o}, Peng,
  Perez-Montero, Ricciardelli, Silverman, Tanaka, Tasca, Tresse, Vergani, \&
  Zucca}]{DominguezSanchez2011}
Dom{\'\i}nguez~S{\'a}nchez, H., Pozzi, F., Gruppioni, C., {et~al.} 2011,
  \mnras, 417, 900

\bibitem[{{Dressler}(1979)}]{Dressler1979}
{Dressler}, A. 1979, \apj, 231, 659

\bibitem[{Duc {et~al.}(2011)Duc, Cuillandre, Serra, Michel-Dansac, Ferriere,
  Alatalo, Blitz, Bois, Bournaud, Bureau, Cappellari, Davies, Davis, de~Zeeuw,
  Emsellem, Khochfar, Krajnovi{\'c}, Kuntschner, Lablanche, McDermid, Morganti,
  Naab, Oosterloo, Sarzi, Scott, Weijmans, \& Young}]{Duc2011}
Duc, P.-A., Cuillandre, J.-C., Serra, P., {et~al.} 2011, \mnras, 417, 863

\bibitem[{{Elmegreen}(2009)}]{Elmegreen2008}
{Elmegreen}, B.~G. 2009, 254, 289

\bibitem[{{Emsellem} {et~al.}(2004){Emsellem}, {Cappellari}, {Peletier},
  {McDermid}, {Bacon}, {Bureau}, {Copin}, {Davies}, {Krajnovi{\'c}},
  {Kuntschner}, {Miller}, \& {de Zeeuw}}]{Emsellem2004}
{Emsellem}, E., {Cappellari}, M., {Peletier}, R.~F., {et~al.} 2004, \mnras,
  352, 721

\bibitem[{Emsellem {et~al.}(2007)Emsellem, Cappellari, Krajnovic, van~de Ven,
  Bacon, Bureau, Davies, de~Zeeuw, Falc{\'o}n-Barroso, Kuntschner, McDermid,
  Peletier, \& Sarzi}]{Emsellem2007}
Emsellem, E., Cappellari, M., Krajnovic, D., {et~al.} 2007, \mnras, 379, 401

\bibitem[{{Emsellem} {et~al.}(2011){Emsellem}, {Cappellari}, {Krajnovi{\'c}},
  {Alatalo}, {Blitz}, {Bois}, {Bournaud}, {Bureau}, {Davies}, {Davis}, {de
  Zeeuw}, {Khochfar}, {Kuntschner}, {Lablanche}, {McDermid}, {Morganti},
  {Naab}, {Oosterloo}, {Sarzi}, {Scott}, {Serra}, {van de Ven}, {Weijmans}, \&
  {Young}}]{Emsellem2011}
{Emsellem}, E., {Cappellari}, M., {Krajnovi{\'c}}, D., {et~al.} 2011, \mnras,
  414, 888

\bibitem[{{Feldmann} {et~al.}(2011){Feldmann}, {Carollo}, \&
  {Mayer}}]{Feldmann2011}
{Feldmann}, R., {Carollo}, C.~M., \& {Mayer}, L. 2011, \apj, 736, 88

\bibitem[{{Feldmann} {et~al.}(2010){Feldmann}, {Carollo}, {Mayer}, {Renzini},
  {Lake}, {Quinn}, {Stinson}, \& {Yepes}}]{Feldmann2010}
{Feldmann}, R., {Carollo}, C.~M., {Mayer}, L., {et~al.} 2010, \apj, 709, 218

\bibitem[{Finkelstein {et~al.}(2011)Finkelstein, Hill, Gebhardt, Adams, Blanc,
  Papovich, Ciardullo, Drory, Gawiser, Gronwall, Schneider, \&
  Tran}]{Finkelstein2011}
Finkelstein, S.~L., Hill, G.~J., Gebhardt, K., {et~al.} 2011, \aj, 729, 140

\bibitem[{Forbes {et~al.}(1994)Forbes, Franx, \& Illingworth}]{Forbes1994}
Forbes, D.~A., Franx, M., \& Illingworth, G.~D. 1994, \apj, 428, L49

\bibitem[{{Foster} {et~al.}(2013){Foster}, {Arnold}, {Forbes}, {Pastorello},
  {Romanowsky}, {Spitler}, {Strader}, \& {Brodie}}]{Foster2013}
{Foster}, C., {Arnold}, J.~A., {Forbes}, D.~A., {et~al.} 2013, \mnras, 435,
  3587

\bibitem[{{Franx} {et~al.}(1991){Franx}, {Illingworth}, \& {de
  Zeeuw}}]{Franx1991}
{Franx}, M., {Illingworth}, G., \& {de Zeeuw}, T. 1991, \apj, 383, 112

\bibitem[{{Franx} {et~al.}(2008){Franx}, {van Dokkum}, {Schreiber}, {Wuyts},
  {Labb{\'e}}, \& {Toft}}]{Franx2008}
{Franx}, M., {van Dokkum}, P.~G., {Schreiber}, N.~M.~F., {et~al.} 2008, \apj,
  688, 770

\bibitem[{Gabor \& Dav{\'e}(2012)}]{GaborDave2012}
Gabor, J.~M., \& Dav{\'e}, R. 2012, \mnras, 427, 1816

\bibitem[{{Gebhardt} {et~al.}(2000){Gebhardt}, {Richstone}, {Kormendy},
  {Lauer}, {Ajhar}, {Bender}, {Dressler}, {Faber}, {Grillmair}, {Magorrian}, \&
  {Tremaine}}]{Gebhardt2000}
{Gebhardt}, K., {Richstone}, D., {Kormendy}, J., {et~al.} 2000, \aj, 119, 1157

\bibitem[{{Gebhardt} {et~al.}(2003){Gebhardt}, {Richstone}, {Tremaine},
  {Lauer}, {Bender}, {Bower}, {Dressler}, {Faber}, {Filippenko}, {Green},
  {Grillmair}, {Ho}, {Kormendy}, {Magorrian}, \& {Pinkney}}]{Gebhardt2003}
{Gebhardt}, K., {Richstone}, D., {Tremaine}, S., {et~al.} 2003, \apj, 583, 92

\bibitem[{{Gerhard}(1993)}]{Gerhard1993}
{Gerhard}, O.~E. 1993, \mnras, 265, 213

\bibitem[{{Graham} {et~al.}(2005){Graham}, {Driver}, {Petrosian}, {Conselice},
  {Bershady}, {Crawford}, \& {Goto}}]{Graham2005}
{Graham}, A.~W., {Driver}, S.~P., {Petrosian}, V., {et~al.} 2005, \aj, 130,
  1535

\bibitem[{{Graves} {et~al.}(2007){Graves}, {Faber}, {Schiavon}, \&
  {Yan}}]{Graves2007}
{Graves}, G.~J., {Faber}, S.~M., {Schiavon}, R.~P., \& {Yan}, R. 2007, \apj,
  671, 243

\bibitem[{{Graves} \& {Schiavon}(2008)}]{GravesSchiavon2008}
{Graves}, G.~J., \& {Schiavon}, R.~P. 2008, \apjs, 177, 446

\bibitem[{{Greene} \& {Ho}(2006)}]{Greene2005}
{Greene}, J.~E., \& {Ho}, L.~C. 2006, \apj, 641, 117

\bibitem[{{Greene} {et~al.}(2012){Greene}, {Murphy}, {Comerford}, {Gebhardt},
  \& {Adams}}]{Greene2012}
{Greene}, J.~E., {Murphy}, J.~D., {Comerford}, J.~M., {Gebhardt}, K., \&
  {Adams}, J.~J. 2012, \apj, 750, 32

\bibitem[{Greene {et~al.}(2013)Greene, Murphy, Graves, Gunn, Raskutti,
  Comerford, \& Gebhardt}]{Greene2013}
Greene, J.~E., Murphy, J.~D., Graves, G.~J., {et~al.} 2013, \apj, 776, 64

\bibitem[{{Hill} {et~al.}(2008){Hill}, {MacQueen}, {Smith}, {Tufts}, {Roth},
  {Kelz}, {Adams}, {Drory}, {Grupp}, {Barnes}, {Blanc}, {Murphy}, {Altmann},
  {Wesley}, {Segura}, {Good}, {Booth}, {Bauer}, {Popow}, {Goertz}, {Edmonston},
  \& {Wilkinson}}]{Hill2008}
{Hill}, G.~J., {MacQueen}, P.~J., {Smith}, M.~P., {et~al.} 2008, in Society of
  Photo-Optical Instrumentation Engineers (SPIE) Conference Series, Vol. 7014,
  Society of Photo-Optical Instrumentation Engineers (SPIE) Conference Series

\bibitem[{Hilz {et~al.}(2013)Hilz, Naab, \& Ostriker}]{Hilz2013}
Hilz, M., Naab, T., \& Ostriker, J.~P. 2013, \apj, 429, 2924

\bibitem[{Hilz {et~al.}(2012)Hilz, Naab, Ostriker, Thomas, Burkert, \&
  Jesseit}]{Hilz2012}
Hilz, M., Naab, T., Ostriker, J.~P., {et~al.} 2012, \apj, 425, 3119

\bibitem[{{Ho} {et~al.}(2009){Ho}, {Greene}, {Filippenko}, \&
  {Sargent}}]{Ho2009}
{Ho}, L.~C., {Greene}, J.~E., {Filippenko}, A.~V., \& {Sargent}, W.~L.~W. 2009,
  \apjs, 183, 1

\bibitem[{{Hoffman} {et~al.}(2009){Hoffman}, {Cox}, {Dutta}, \&
  {Hernquist}}]{Hoffman2009}
{Hoffman}, L., {Cox}, T.~J., {Dutta}, S., \& {Hernquist}, L. 2009, \apj, 705,
  920

\bibitem[{Hoffman {et~al.}(2010)Hoffman, Cox, Dutta, \&
  Hernquist}]{Hoffman2010}
Hoffman, L., Cox, T.~J., Dutta, S., \& Hernquist, L. 2010, \apj, 723, 818

\bibitem[{Hopkins {et~al.}(2009)Hopkins, Bundy, Murray, Quataert, Lauer, \&
  Ma}]{Hopkins2009}
Hopkins, P.~F., Bundy, K., Murray, N., {et~al.} 2009, \mnras, 398, 898

\bibitem[{{Huchra} {et~al.}(2012){Huchra}, {Macri}, {Masters}, {Jarrett},
  {Berlind}, {Calkins}, {Crook}, {Cutri}, {Erdo{\v g}du}, {Falco}, {George},
  {Hutcheson}, {Lahav}, {Mader}, {Mink}, {Martimbeau}, {Schneider},
  {Skrutskie}, {Tokarz}, \& {Westover}}]{Huchra2012}
{Huchra}, J.~P., {Macri}, L.~M., {Masters}, K.~L., {et~al.} 2012, \apjs, 199,
  26

\bibitem[{{Illingworth}(1977)}]{Illingworth1977}
{Illingworth}, G. 1977, \apjl, 218, L43

\bibitem[{{Jesseit} {et~al.}(2005){Jesseit}, {Naab}, \&
  {Burkert}}]{Jesseit2005}
{Jesseit}, R., {Naab}, T., \& {Burkert}, A. 2005, \mnras, 360, 1185

\bibitem[{{Jesseit} {et~al.}(2007){Jesseit}, {Naab}, {Peletier}, \&
  {Burkert}}]{Jesseit2007}
{Jesseit}, R., {Naab}, T., {Peletier}, R.~F., \& {Burkert}, A. 2007, \mnras,
  376, 997

\bibitem[{{Jimmy} {et~al.}(2013){Jimmy}, Tran, Brough, Gebhardt, von~der
  Linden, Couch, \& Sharp}]{Jimmy2013}
{Jimmy}, Tran, K.-V., Brough, S., {et~al.} 2013, \apj, 778, 171

\bibitem[{Jorgensen {et~al.}(1997)Jorgensen, Hjorth, Franx, \& van
  Dokkum}]{Jorgensen1997}
Jorgensen, I., Hjorth, J., Franx, M., \& van Dokkum, P. 1997, \aas, 190, 780

\bibitem[{Joung {et~al.}(2009)Joung, Cen, \& Bryan}]{Joung2009}
Joung, M.~R., Cen, R., \& Bryan, G.~L. 2009, \apjl, 692, L1

\bibitem[{Kelson {et~al.}(2002)Kelson, Zabludoff, Williams, Trager, Mulchaey,
  \& Bolte}]{Kelson2002}
Kelson, D.~D., Zabludoff, A.~I., Williams, K.~A., {et~al.} 2002, \apj, 576, 720

\bibitem[{{Kere{\v s}} {et~al.}(2009){Kere{\v s}}, {Katz}, {Fardal},
  {Dav{\'e}}, \& {Weinberg}}]{Keres2009}
{Kere{\v s}}, D., {Katz}, N., {Fardal}, M., {Dav{\'e}}, R., \& {Weinberg},
  D.~H. 2009, \mnras, 395, 160

\bibitem[{{Kere{\v s}} {et~al.}(2005){Kere{\v s}}, {Katz}, {Weinberg}, \&
  {Dav{\'e}}}]{Keres2005}
{Kere{\v s}}, D., {Katz}, N., {Weinberg}, D.~H., \& {Dav{\'e}}, R. 2005,
  \mnras, 363, 2

\bibitem[{{Khochfar} \& {Silk}(2006)}]{KhochfarSilk2006}
{Khochfar}, S., \& {Silk}, J. 2006, \apjl, 648, L21

\bibitem[{Khochfar {et~al.}(2011)Khochfar, Emsellem, Serra, Bois, Alatalo,
  Bacon, Blitz, Bournaud, Bureau, Cappellari, Davies, Davis, de~Zeeuw, Duc,
  Krajnovic, Kuntschner, Lablanche, McDermid, Morganti, Naab, Oosterloo, Sarzi,
  Scott, Weijmans, \& Young}]{Khochfar2011}
Khochfar, S., Emsellem, E., Serra, P., {et~al.} 2011, \mnras, 417, 845

\bibitem[{{Kormendy}(1984)}]{Kormendy1984}
{Kormendy}, J. 1984, \apj, 287, 577

\bibitem[{{Kormendy} \& {Bender}(1996)}]{KormendyBender1996}
{Kormendy}, J., \& {Bender}, R. 1996, \apjl, 464, L119

\bibitem[{{Kormendy} {et~al.}(2009){Kormendy}, {Fisher}, {Cornell}, \&
  {Bender}}]{kormendy2009}
{Kormendy}, J., {Fisher}, D.~B., {Cornell}, M.~E., \& {Bender}, R. 2009, \apjs,
  182, 216

\bibitem[{{Krajnovi{\'c}} {et~al.}(2006){Krajnovi{\'c}}, {Cappellari}, {de
  Zeeuw}, \& {Copin}}]{Krajnovic2006}
{Krajnovi{\'c}}, D., {Cappellari}, M., {de Zeeuw}, P.~T., \& {Copin}, Y. 2006,
  \mnras, 366, 787

\bibitem[{{Krajnovi{\'c}} {et~al.}(2008){Krajnovi{\'c}}, {Bacon}, {Cappellari},
  {Davies}, {de Zeeuw}, {Emsellem}, {Falc{\'o}n-Barroso}, {Kuntschner},
  {McDermid}, {Peletier}, {Sarzi}, {van den Bosch}, \& {van de
  Ven}}]{Krajnovic2008}
{Krajnovi{\'c}}, D., {Bacon}, R., {Cappellari}, M., {et~al.} 2008, \mnras, 390,
  93

\bibitem[{Krajnovic {et~al.}(2011)Krajnovic, Emsellem, Cappellari, Alatalo,
  Blitz, Bois, Bournaud, Bureau, Davies, Davis, de~Zeeuw, Khochfar, Kuntschner,
  Lablanche, McDermid, Morganti, Naab, Oosterloo, Sarzi, Scott, Serra,
  Weijmans, \& Young}]{Krajnovic2011}
Krajnovic, D., Emsellem, E., Cappellari, M., {et~al.} 2011, \mnras, 414, 2923

\bibitem[{{Krick} {et~al.}(2006){Krick}, {Bernstein}, \&
  {Pimbblet}}]{Krick2006}
{Krick}, J.~E., {Bernstein}, R.~A., \& {Pimbblet}, K.~A. 2006, \aj, 131, 168

\bibitem[{{Lackner} {et~al.}(2012){Lackner}, {Cen}, {Ostriker}, \&
  {Joung}}]{Lackner2012}
{Lackner}, C.~N., {Cen}, R., {Ostriker}, J.~P., \& {Joung}, M.~R. 2012, \mnras,
  425, 641

\bibitem[{{Lackner} \& {Gunn}(2012)}]{LacknerGunn2012}
{Lackner}, C.~N., \& {Gunn}, J.~E. 2012, \mnras, 421, 2277

\bibitem[{{Longhetti} {et~al.}(2007){Longhetti}, {Saracco}, {Severgnini},
  {Della Ceca}, {Mannucci}, {Bender}, {Drory}, {Feulner}, \&
  {Hopp}}]{Longhetti2007}
{Longhetti}, M., {Saracco}, P., {Severgnini}, P., {et~al.} 2007, \mnras, 374,
  614

\bibitem[{Loubser {et~al.}(2009)Loubser, Sansom, Sanchez-Blazquez, Soechting,
  \& Bromage}]{Loubser2008}
Loubser, S.~I., Sansom, A.~E., Sanchez-Blazquez, P., Soechting, I.~K., \&
  Bromage, G.~E. 2009, VizieR On-line Data Catalog, 739, 11009

\bibitem[{{Mandelbaum} {et~al.}(2005)}]{mandelbaum2005}
{Mandelbaum}, R., {et~al.} 2005, \mnras, 361, 1287

\bibitem[{{McConnell} {et~al.}(2012){McConnell}, {Ma}, {Murphy}, {Gebhardt},
  {Lauer}, {Graham}, {Wright}, \& {Richstone}}]{McConnell2012}
{McConnell}, N.~J., {Ma}, C.-P., {Murphy}, J.~D., {et~al.} 2012, \apj, 756, 179

\bibitem[{{McElroy}(1998)}]{Mcelroy1995}
{McElroy}, D.~B. 1998, VizieR Online Data Catalog, 210, 105

\bibitem[{McNeil-Moylan {et~al.}(2012)McNeil-Moylan, Freeman, Arnaboldi, \&
  Gerhard}]{McNeil-Moylan2012}
McNeil-Moylan, E.~K., Freeman, K.~C., Arnaboldi, M., \& Gerhard, O.~E. 2012,
  \aap, 539, 11

\bibitem[{{M{\'e}ndez} {et~al.}(2001){M{\'e}ndez}, {Riffeser}, {Kudritzki},
  {Matthias}, {Freeman}, {Arnaboldi}, {Capaccioli}, \& {Gerhard}}]{Mendez2001}
{M{\'e}ndez}, R.~H., {Riffeser}, A., {Kudritzki}, R.-P., {et~al.} 2001, \apj,
  563, 135

\bibitem[{Murphy {et~al.}(2013)Murphy, Gebhardt, Greene, \&
  Graves}]{Murphy2013}
Murphy, J., Gebhardt, K., Greene, J.~E., \& Graves, G. 2013, Probes of Dark
  Matter on Galaxy Scales, 50103

\bibitem[{Murphy {et~al.}(2011)Murphy, Gebhardt, \& Adams}]{Murphy2011}
Murphy, J.~D., Gebhardt, K., \& Adams, J.~J. 2011, \apj, 729, 129

\bibitem[{{Naab} {et~al.}(2009){Naab}, {Johansson}, \& {Ostriker}}]{Naab2009}
{Naab}, T., {Johansson}, P.~H., \& {Ostriker}, J.~P. 2009, \apjl, 699, L178

\bibitem[{Naab {et~al.}(2007)Naab, Johansson, Ostriker, \&
  Efstathiou}]{Naab2007}
Naab, T., Johansson, P.~H., Ostriker, J.~P., \& Efstathiou, G. 2007, \apj, 658,
  710

\bibitem[{{Naab} {et~al.}(2013){Naab}, {Oser}, {Emsellem}, {Cappellari},
  {Krajnovic}, {McDermid}, {Alatalo}, {Bayet}, {Blitz}, {Bois}, {Bournaud},
  {Bureau}, {Crocker}, {Davies}, {Davis}, {de Zeeuw}, {Duc}, {Hirschmann},
  {Johansson}, {Khochfar}, {Kuntschner}, {Morganti}, {Oosterloo}, {Sarzi},
  {Scott}, {Serra}, {van de Ven}, {Weijmans}, \& {Young}}]{Naab2013}
{Naab}, T., {Oser}, L., {Emsellem}, E., {et~al.} 2013, ArXiv e-prints

\bibitem[{Oogi \& Habe(2013)}]{OogiHabe2013}
Oogi, T., \& Habe, A. 2013, Monthly Notices of the Royal Astronomical Society,
  428, 641

\bibitem[{Oosterloo {et~al.}(1994)Oosterloo, Balcells, \&
  Carter}]{Oosterloo1994}
Oosterloo, T., Balcells, M., \& Carter, D. 1994, \mnras, 266, L10

\bibitem[{Oser {et~al.}(2012)Oser, Naab, Ostriker, \& Johansson}]{Oser2012}
Oser, L., Naab, T., Ostriker, J.~P., \& Johansson, P.~H. 2012, \apj, 744, 63

\bibitem[{{Oser} {et~al.}(2010){Oser}, {Ostriker}, {Naab}, {Johansson}, \&
  {Burkert}}]{Oser2010}
{Oser}, L., {Ostriker}, J.~P., {Naab}, T., {Johansson}, P.~H., \& {Burkert}, A.
  2010, \apj, 725, 2312

\bibitem[{{Paturel} {et~al.}(2003){Paturel}, {Petit}, {Prugniel}, {Theureau},
  {Rousseau}, {Brouty}, {Dubois}, \& {Cambr{\'e}sy}}]{Paturel2003}
{Paturel}, G., {Petit}, C., {Prugniel}, P., {et~al.} 2003, \aap, 412, 45

\bibitem[{{Pota} {et~al.}(2013){Pota}, {Graham}, {Forbes}, {Romanowsky},
  {Brodie}, \& {Strader}}]{Pota2013}
{Pota}, V., {Graham}, A.~W., {Forbes}, D.~A., {et~al.} 2013, \mnras, 433, 235

\bibitem[{Proctor {et~al.}(2009)Proctor, Forbes, Romanowsky, Brodie, Strader,
  Spolaor, Mendel, \& Spitler}]{Proctor2009}
Proctor, R.~N., Forbes, D.~A., Romanowsky, A.~J., {et~al.} 2009, \mnras, 398,
  91

\bibitem[{{Remus} {et~al.}(2013){Remus}, {Burkert}, {Dolag}, {Johansson},
  {Naab}, {Oser}, \& {Thomas}}]{Remus2013}
{Remus}, R.-S., {Burkert}, A., {Dolag}, K., {et~al.} 2013, \apj, 766, 71

\bibitem[{{Rix} \& {White}(1992)}]{RixWhite1992}
{Rix}, H.-W., \& {White}, S.~D.~M. 1992, \mnras, 254, 389

\bibitem[{{Romanowsky} \& {Fall}(2012)}]{RomanowskyFall2012}
{Romanowsky}, A.~J., \& {Fall}, S.~M. 2012, \apjs, 203, 17

\bibitem[{Saracco {et~al.}(2012)Saracco, Gargiulo, \& Longhetti}]{Saracco2012}
Saracco, P., Gargiulo, A., \& Longhetti, M. 2012, \mnras, 422, 3107

\bibitem[{Sargent {et~al.}(1977)Sargent, Schechter, Boksenberg, \&
  Shortridge}]{Sargent1977}
Sargent, W. L.~W., Schechter, P.~L., Boksenberg, A., \& Shortridge, K. 1977,
  \apj, 212, 326

\bibitem[{Schiavon(2007)}]{Schiavon2007}
Schiavon, R.~P. 2007, \apjs, 171, 146

\bibitem[{{Serra} {et~al.}(2012){Serra}, {Oosterloo}, {Morganti}, {Alatalo},
  {Blitz}, {Bois}, {Bournaud}, {Bureau}, {Cappellari}, {Crocker}, {Davies},
  {Davis}, {de Zeeuw}, {Duc}, {Emsellem}, {Khochfar}, {Krajnovi{\'c}},
  {Kuntschner}, {Lablanche}, {McDermid}, {Naab}, {Sarzi}, {Scott}, {Trager},
  {Weijmans}, \& {Young}}]{Serra2012}
{Serra}, P., {Oosterloo}, T., {Morganti}, R., {et~al.} 2012, \mnras, 422, 1835

\bibitem[{Serra {et~al.}(2014)Serra, Oser, Krajnovic, Naab, Oosterloo,
  Morganti, Cappellari, Emsellem, Young, Blitz, Davis, Duc, Hirschmann,
  Weijmans, Alatalo, Bayet, Bois, Bournaud, Bureau, Davies, de~Zeeuw, Khochfar,
  Kuntschner, Lablanche, McDermid, Sarzi, \& Scott}]{Serra2014}
Serra, P., Oser, L., Krajnovic, D., {et~al.} 2014, arXiv.org, 3180

\bibitem[{Simkin(1974)}]{Simkin1974}
Simkin, S.~M. 1974, \aap, 31, 129

\bibitem[{Statler(1991)}]{Statler1991}
Statler, T.~S. 1991, \aj, 102, 882

\bibitem[{Strader {et~al.}(2011)Strader, Romanowsky, Brodie, Spitler, Beasley,
  Arnold, Tamura, Sharples, \& Arimoto}]{Strader2011}
Strader, J., Romanowsky, A.~J., Brodie, J.~P., {et~al.} 2011, \apjs, 197, 33

\bibitem[{{Strateva} {et~al.}(2001){Strateva}, {Ivezi{\'c}}, {Knapp},
  {Narayanan}, {Strauss}, {Gunn}, {Lupton}, {Schlegel}, {Bahcall}, {Brinkmann},
  {Brunner}, {Budav{\'a}ri}, {Csabai}, {Castander}, {Doi}, {Fukugita}, {Gy{\H
  o}ry}, {Hamabe}, {Hennessy}, {Ichikawa}, {Kunszt}, {Lamb}, {McKay},
  {Okamura}, {Racusin}, {Sekiguchi}, {Schneider}, {Shimasaku}, \&
  {York}}]{Strateva2001}
{Strateva}, I., {Ivezi{\'c}}, {\v Z}., {Knapp}, G.~R., {et~al.} 2001, \aj, 122,
  1861

\bibitem[{Szomoru {et~al.}(2012)Szomoru, Franx, \& van Dokkum}]{Szomoru2012}
Szomoru, D., Franx, M., \& van Dokkum, P.~G. 2012, \apj, 749, 121

\bibitem[{{Taranu} {et~al.}(2013){Taranu}, {Dubinski}, \& {Yee}}]{Taranu2013}
{Taranu}, D.~S., {Dubinski}, J., \& {Yee}, H.~K.~C. 2013, \apj, 778, 61

\bibitem[{{Thomas} {et~al.}(2011){Thomas}, {Saglia}, {Bender}, {Thomas},
  {Gebhardt}, {Magorrian}, {Corsini}, {Wegner}, \& {Seitz}}]{thomas2011}
{Thomas}, J., {Saglia}, R.~P., {Bender}, R., {et~al.} 2011, \mnras, 415, 545

\bibitem[{{Toft} {et~al.}(2007){Toft}, {van Dokkum}, {Franx}, {Labbe},
  {F{\"o}rster Schreiber}, {Wuyts}, {Webb}, {Rudnick}, {Zirm}, {Kriek}, {van
  der Werf}, {Blakeslee}, {Illingworth}, {Rix}, {Papovich}, \&
  {Moorwood}}]{Toft2007}
{Toft}, S., {van Dokkum}, P., {Franx}, M., {et~al.} 2007, \apj, 671, 285

\bibitem[{Tonry \& Davis(1979)}]{TonryDavis1979}
Tonry, J., \& Davis, M. 1979, \aj, 84, 1511

\bibitem[{{Tremblay} \& {Merritt}(1995)}]{TremblayMerritt1995}
{Tremblay}, B., \& {Merritt}, D. 1995, \aj, 110, 1039

\bibitem[{{Tremblay} \& {Merritt}(1996)}]{TremblayMerritt1996}
---. 1996, \aj, 111, 2243

\bibitem[{{Trujillo} {et~al.}(2006){Trujillo}, {Feulner}, {Goranova}, {Hopp},
  {Longhetti}, {Saracco}, {Bender}, {Braito}, {Della Ceca}, {Drory},
  {Mannucci}, \& {Severgnini}}]{Trujillo2006}
{Trujillo}, I., {Feulner}, G., {Goranova}, Y., {et~al.} 2006, \mnras, 373, L36

\bibitem[{{van de Sande} {et~al.}(2011){van de Sande}, {Kriek}, {Franx}, {van
  Dokkum}, {Bezanson}, {Whitaker}, {Brammer}, {Labb{\'e}}, {Groot}, \&
  {Kaper}}]{vandeSande2011}
{van de Sande}, J., {Kriek}, M., {Franx}, M., {et~al.} 2011, \apjl, 736, L9

\bibitem[{van~de Sande {et~al.}(2013)van~de Sande, Kriek, Franx, van Dokkum,
  Bezanson, Bouwens, Quadri, Rix, \& Skelton}]{vandeSande2013}
van~de Sande, J., Kriek, M., Franx, M., {et~al.} 2013, \apj, 771, 85

\bibitem[{{van den Bosch} {et~al.}(2008){van den Bosch}, {van de Ven},
  {Verolme}, {Cappellari}, \& {de Zeeuw}}]{vandenBosch2008}
{van den Bosch}, R.~C.~E., {van de Ven}, G., {Verolme}, E.~K., {Cappellari},
  M., \& {de Zeeuw}, P.~T. 2008, \mnras, 385, 647

\bibitem[{van~der Marel \& Franx(1993)}]{MarelFranx1993}
van~der Marel, R.~P., \& Franx, M. 1993, \apj, 407, 525

\bibitem[{{van der Wel} {et~al.}(2006){van der Wel}, {Franx}, {van Dokkum},
  {Huang}, {Rix}, \& {Illingworth}}]{vanderWel2005}
{van der Wel}, A., {Franx}, M., {van Dokkum}, P.~G., {et~al.} 2006, \apjl, 636,
  L21

\bibitem[{{van der Wel} {et~al.}(2008){van der Wel}, {Holden}, {Zirm}, {Franx},
  {Rettura}, {Illingworth}, \& {Ford}}]{vanderWel2008}
{van der Wel}, A., {Holden}, B.~P., {Zirm}, A.~W., {et~al.} 2008, \apj, 688, 48

\bibitem[{{van Dokkum}(2005)}]{vanDokkum2005}
{van Dokkum}, P.~G. 2005, \aj, 130, 2647

\bibitem[{{van Dokkum} {et~al.}(2008){van Dokkum}, {Franx}, {Kriek}, {Holden},
  {Illingworth}, {Magee}, {Bouwens}, {Marchesini}, {Quadri}, {Rudnick},
  {Taylor}, \& {Toft}}]{vanDokkum2008}
{van Dokkum}, P.~G., {Franx}, M., {Kriek}, M., {et~al.} 2008, \apjl, 677, L5

\bibitem[{{van Dokkum} {et~al.}(2010){van Dokkum}, {Whitaker}, {Brammer},
  {Franx}, {Kriek}, {Labb{\'e}}, {Marchesini}, {Quadri}, {Bezanson},
  {Illingworth}, {Muzzin}, {Rudnick}, {Tal}, \& {Wake}}]{vanDokkum2010}
{van Dokkum}, P.~G., {Whitaker}, K.~E., {Brammer}, G., {et~al.} 2010, \apj,
  709, 1018

\bibitem[{Weijmans {et~al.}(2009)Weijmans, Cappellari, Bacon, de~Zeeuw,
  Emsellem, Falc{\'o}n-Barroso, Kuntschner, McDermid, van~den Bosch, \& van~de
  Ven}]{Weijmans2009}
Weijmans, A.-M., Cappellari, M., Bacon, R., {et~al.} 2009, \mnras, 398, 561

\bibitem[{{Whitaker} {et~al.}(2012){Whitaker}, {Kriek}, {van Dokkum},
  {Bezanson}, {Brammer}, {Franx}, \& {Labb{\'e}}}]{Whitaker2012}
{Whitaker}, K.~E., {Kriek}, M., {van Dokkum}, P.~G., {et~al.} 2012, \apj, 745,
  179

\bibitem[{{Whitmore} {et~al.}(1985){Whitmore}, {McElroy}, \&
  {Tonry}}]{Whitmore1985}
{Whitmore}, B.~C., {McElroy}, D.~B., \& {Tonry}, J.~L. 1985, \apjs, 59, 1

\bibitem[{{Wu} {et~al.}(2014){Wu}, {Gerhard}, {Naab}, {Oser},
  {Martinez-Valpuesta}, {Hilz}, {Churazov}, \& {Lyskova}}]{Wu2014}
{Wu}, X., {Gerhard}, O., {Naab}, T., {et~al.} 2014, \mnras

\bibitem[{{Yang} {et~al.}(2007){Yang}, {Mo}, {van den Bosch}, {Pasquali}, {Li},
  \& {Barden}}]{Yang2007}
{Yang}, X., {Mo}, H.~J., {van den Bosch}, F.~C., {et~al.} 2007, \apj, 671, 153

\bibitem[{Yoachim {et~al.}(2010)Yoachim, Ro{\v s}kar, \&
  Debattista}]{Yoachim2010}
Yoachim, P., Ro{\v s}kar, R., \& Debattista, V.~P. 2010, \apjl, 716, L4

\bibitem[{{York} {et~al.}(2000){York}, {Adelman}, {Anderson}, {Anderson},
  {Annis}, {Bahcall}, {Bakken}, {Barkhouser}, {Bastian}, {Berman}, {Boroski},
  {Bracker}, {Briegel}, {Briggs}, {Brinkmann}, {Brunner}, {Burles}, {Carey},
  {Carr}, {Castander}, {Chen}, {Colestock}, {Connolly}, {Crocker}, {Csabai},
  {Czarapata}, {Davis}, {Doi}, {Dombeck}, {Eisenstein}, {Ellman}, {Elms},
  {Evans}, {Fan}, {Federwitz}, {Fiscelli}, {Friedman}, {Frieman}, {Fukugita},
  {Gillespie}, {Gunn}, {Gurbani}, {de Haas}, {Haldeman}, {Harris}, {Hayes},
  {Heckman}, {Hennessy}, {Hindsley}, {Holm}, {Holmgren}, {Huang}, {Hull},
  {Husby}, {Ichikawa}, {Ichikawa}, {Ivezi{\'c}}, {Kent}, {Kim}, {Kinney},
  {Klaene}, {Kleinman}, {Kleinman}, {Knapp}, {Korienek}, {Kron}, {Kunszt},
  {Lamb}, {Lee}, {Leger}, {Limmongkol}, {Lindenmeyer}, {Long}, {Loomis},
  {Loveday}, {Lucinio}, {Lupton}, {MacKinnon}, {Mannery}, {Mantsch}, {Margon},
  {McGehee}, {McKay}, {Meiksin}, {Merelli}, {Monet}, {Munn}, {Narayanan},
  {Nash}, {Neilsen}, {Neswold}, {Newberg}, {Nichol}, {Nicinski}, {Nonino},
  {Okada}, {Okamura}, {Ostriker}, {Owen}, {Pauls}, {Peoples}, {Peterson},
  {Petravick}, {Pier}, {Pope}, {Pordes}, {Prosapio}, {Rechenmacher}, {Quinn},
  {Richards}, {Richmond}, {Rivetta}, {Rockosi}, {Ruthmansdorfer}, {Sandford},
  {Schlegel}, {Schneider}, {Sekiguchi}, {Sergey}, {Shimasaku}, {Siegmund},
  {Smee}, {Smith}, {Snedden}, {Stone}, {Stoughton}, {Strauss}, {Stubbs},
  {SubbaRao}, {Szalay}, {Szapudi}, {Szokoly}, {Thakar}, {Tremonti}, {Tucker},
  {Uomoto}, {Vanden Berk}, {Vogeley}, {Waddell}, {Wang}, {Watanabe},
  {Weinberg}, {Yanny}, {Yasuda}, \& {SDSS Collaboration}}]{York2000}
{York}, D.~G., {Adelman}, J., {Anderson}, Jr., J.~E., {et~al.} 2000, \aj, 120,
  1579

\end{thebibliography}

%----------------------------------------------------------------------

\newpage

\appendix
\section{Tests for Robustness of the Kinematic Fits}
\label{KinematicAppendix}

\setcounter{figure}{0} \renewcommand{\thefigure}{A.\arabic{figure}}

We show here the results of a series of tests conducted to
characterize the robustness of our fits to S/N, continuum fitting,
shape of the LOSVD, template mismatch and masking of emission
lines. These results were used to motivate our fiducial choice of
wavelength region $(4000$~\AA$-5420$~\AA), our minimum S/N threshold,
the degree of our continuum polynomial fit and the assumption of a
Gaussian LOSVD.

\subsection{S/N Thresholds}
\label{subsec: S/N Thresholds}

\begin{figure}[!htb]
\begin{center}
\includegraphics[width=0.4\columnwidth,angle=0,clip]{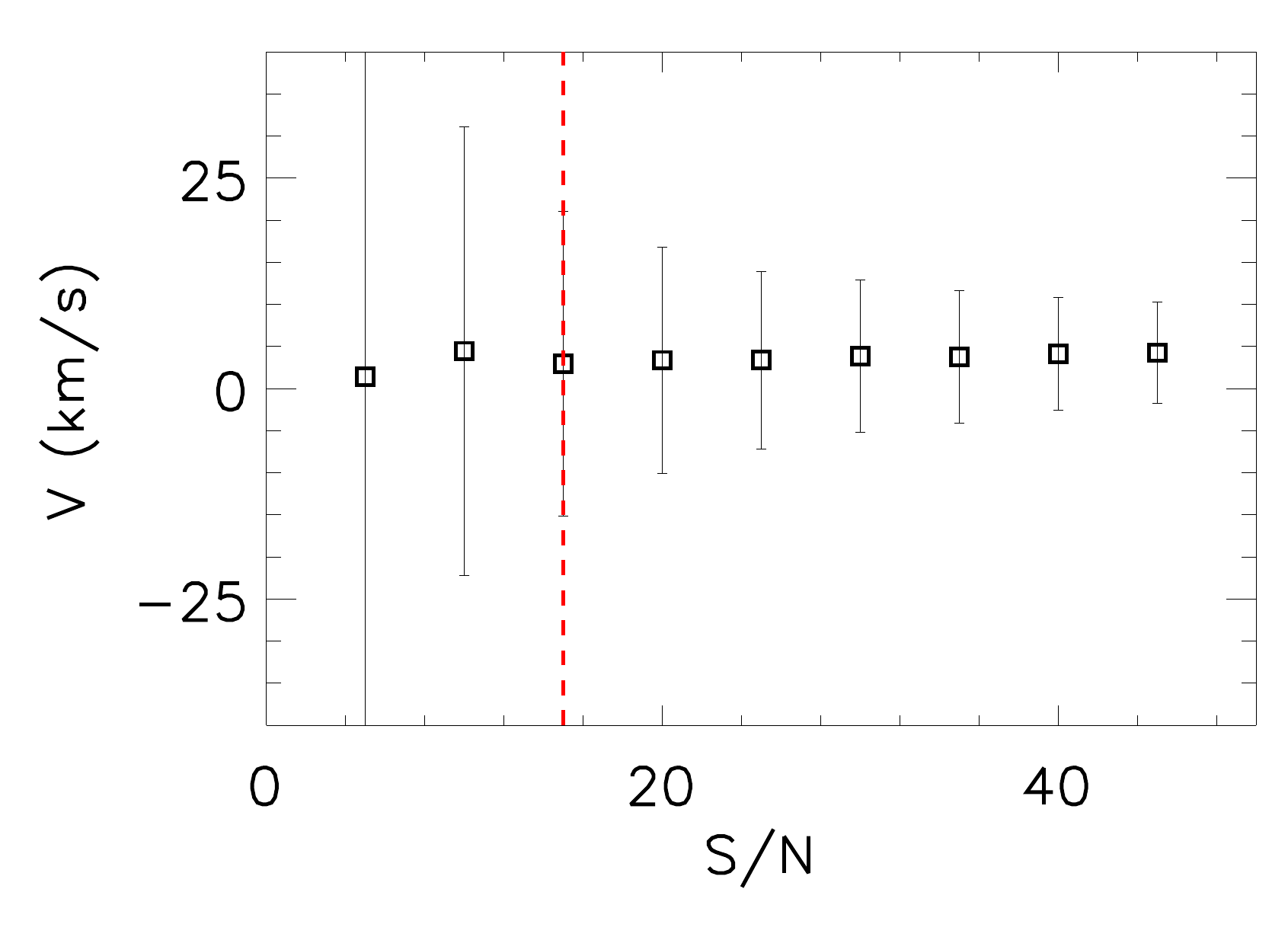}
\includegraphics[width=0.4\columnwidth,angle=0,clip]{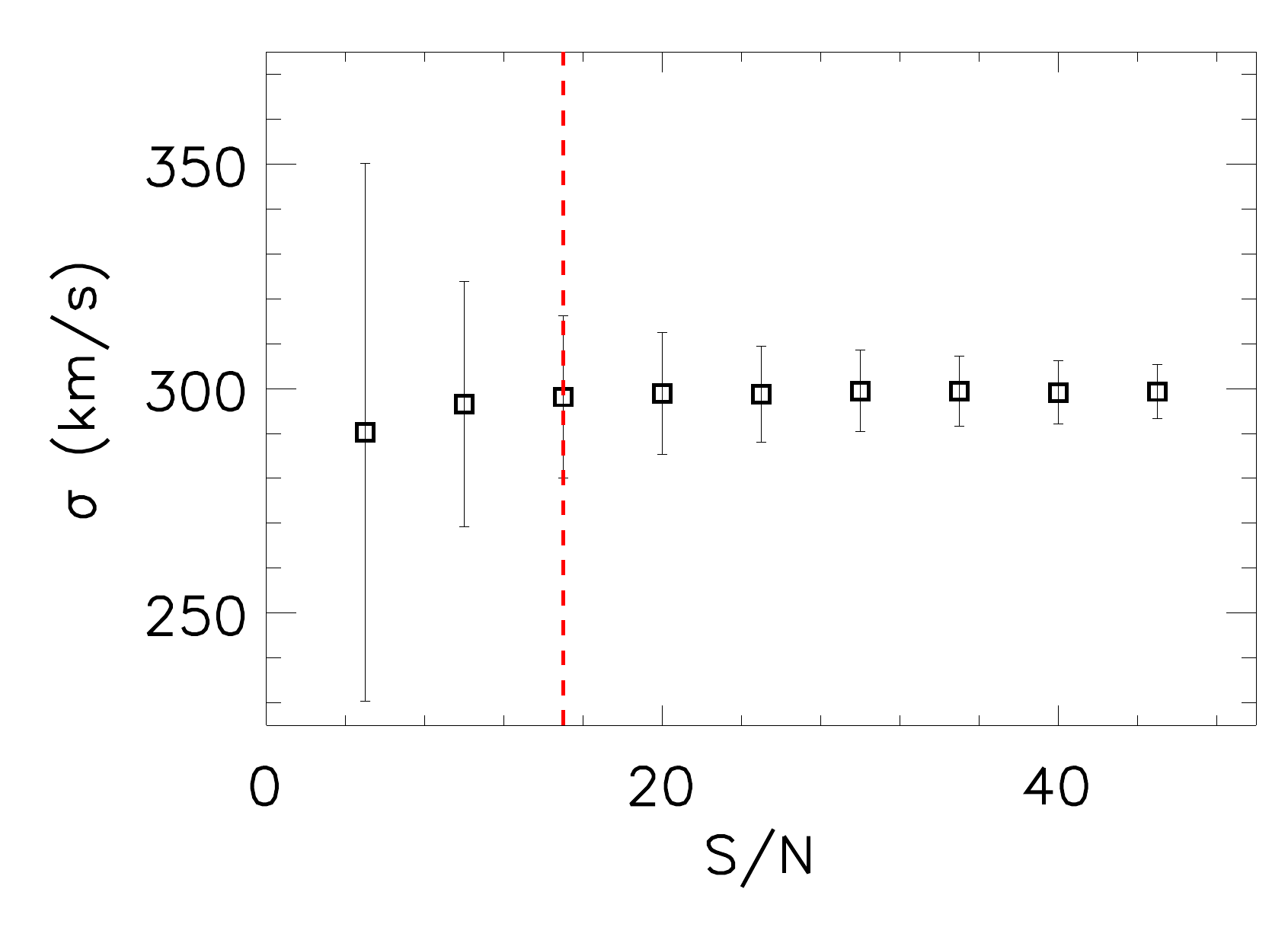}
\caption{ S/N vs. measured kinematics for a Gaussian LOSVD of mean
  zero and dispersion 300~\kms.  Error bars show 1-$\sigma$
  uncertainties on the derived values estimated over 100 noise
  realisations.  Values are averaged over the whole library of
  templates. Our chosen threshold of S/N = 15 is also shown 
  (red dashed line).}
\label{Fig:KinematicNoise}
\end{center}
\end{figure}

\begin{figure}[!htb]
\begin{center}
\includegraphics[width=0.4\columnwidth,angle=0,clip]{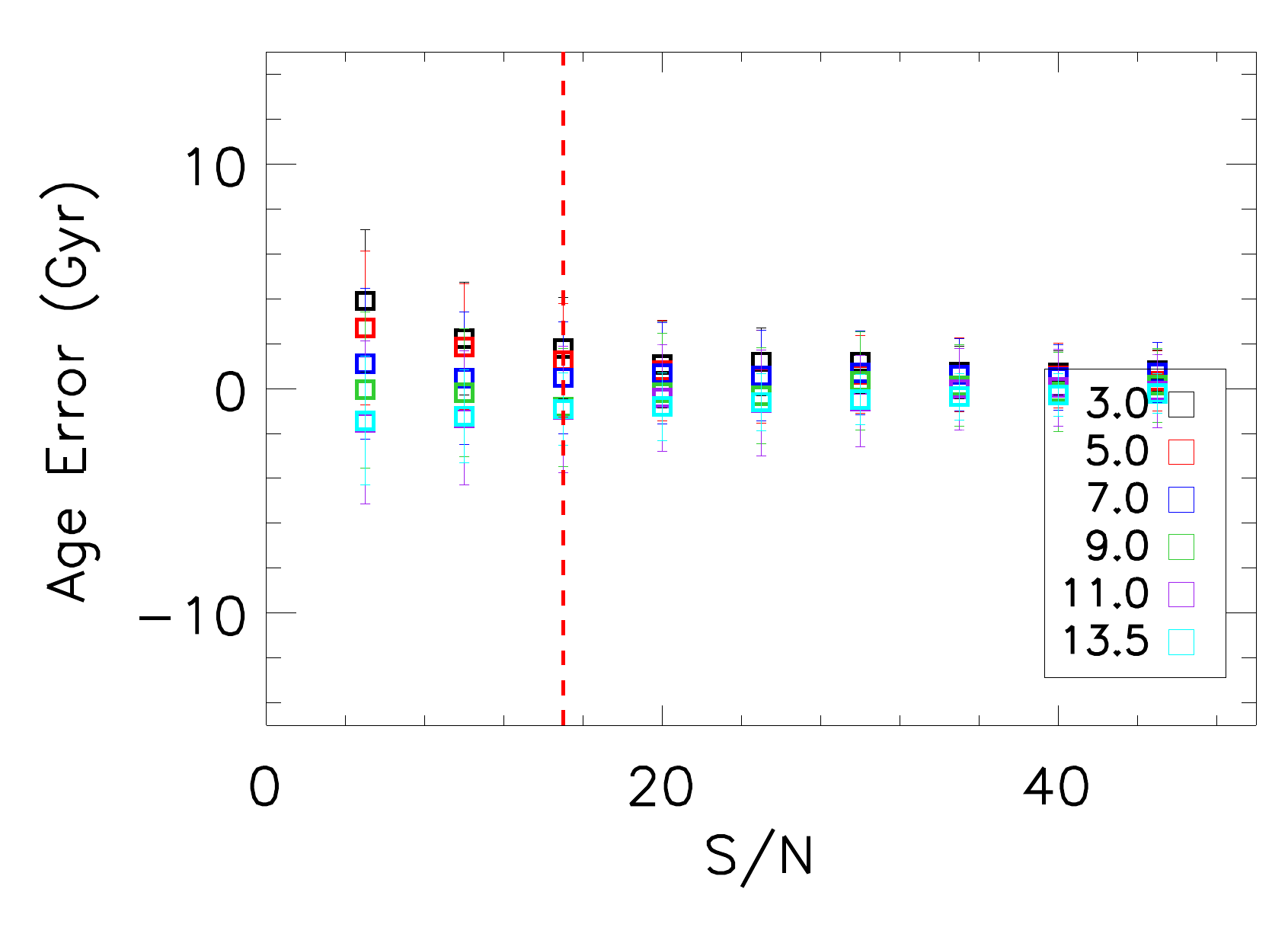}
\includegraphics[width=0.4\columnwidth,angle=0,clip]{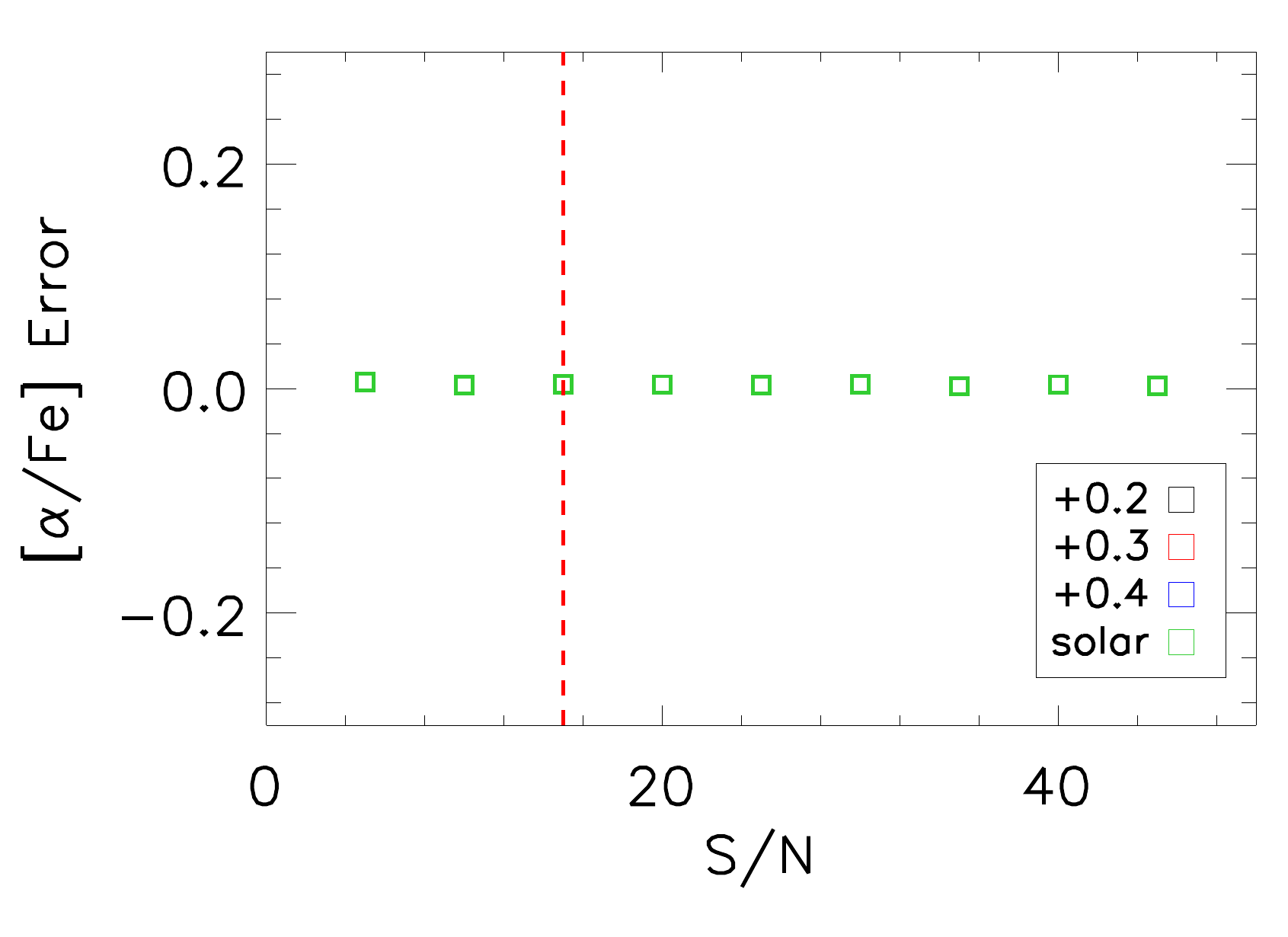}
\caption{
  S/N vs. galaxy characteristics inferred from the best fit
  templates. We show errors in the inferred age as a function of
  template age in Gyr (right) and errors in the inferred metallicity as a
  function of template metallicity  in [$\alpha$/Fe] (left). Error bars again show
  1-$\sigma$ uncertainties, and our chosen S/N threshold is overplotted in red.}
\label{Fig:CharacteristicNoise}
\end{center}
\end{figure}

We begin by briefly testing the response of pPXF to varying S/N, in
order to justify our minimum S/N threshold of 15. This is done by
fitting each of the templates in our library, convolved with a
Gaussian LOSVD with mean zero and dispersion $300$~\kms, over a
variety of S/N. A single realisation, at a given S/N, is found
by adding Gaussian random noise to each pixel with a standard
deviation set by the pixel flux divided by S/N. When this is repeated
over 100 Monte Carlo iterations at each S/N we characterise the
bias and scatter in our kinematic fits as a function of S/N. Fits are
made using the whole library of templates so that we don't
underestimate the scatter, and the process is repeated for each
template to test for any systematic effects with galaxy
templates. We test over our fiducial range of $4000$~\AA$-5420
$~\AA and since the degree of the continuum polynomial is unimportant
for this test, we assume a low order polynomial of degree $4$.

The results, shown in Figure~\ref{Fig:KinematicNoise}, firstly show no
bias in the derived dispersions as a function of S/N. There is perhaps
a slight bias in the inferred velocity (of $\sim 5$~\kms), however
this is significantly smaller than both the velocity uncertainty and
the pixel separation. The uncertainty in
the derived dispersions rise fairly dramatically for S/N $\lesssim
10$. We therefore conservatively adopt a minimum S/N threshold of 15,
where uncertainties are around $\pm 15~$\kms.

As a sanity check, we also consider how accurately our best fit
templates represent the characteristics of the input template. We show
in Figure~\ref{Fig:CharacteristicNoise} the errors in our derived
ages, and metallicities. The inferred ages, and in particular
metallicities, of our fits tend to be quite accurate. At low S/N,
older galaxies tend to be biased downwards by $\sim 1~{\rm Gyr}$ and
the reverse holds true for the youngest galaxies. This bias, 
combined with the spread of $\sim 2~{\rm Gyr}$ at
a S/N of 15, means that any estimates of galactic age based on our
fits are likely to be uncertain by at least $3~{\rm Gyr}$.

\subsection{Degree of the Hermite Polynomial}
\label{subsec: Hermite Degree}

\begin{figure}[!htb]
\begin{center}
\includegraphics[width=0.45\columnwidth,angle=0,clip]{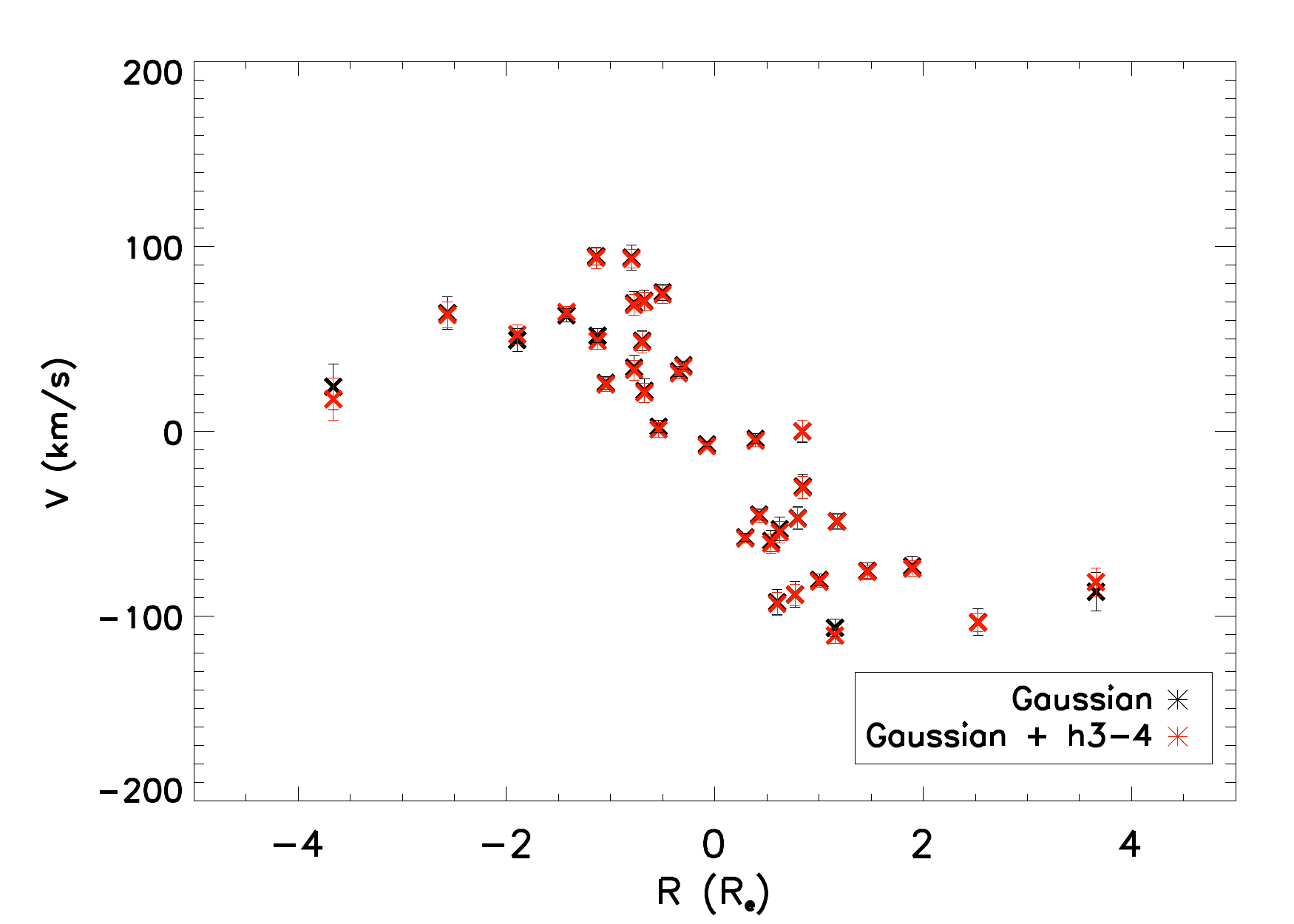}
\includegraphics[width=0.45\columnwidth,angle=0,clip]{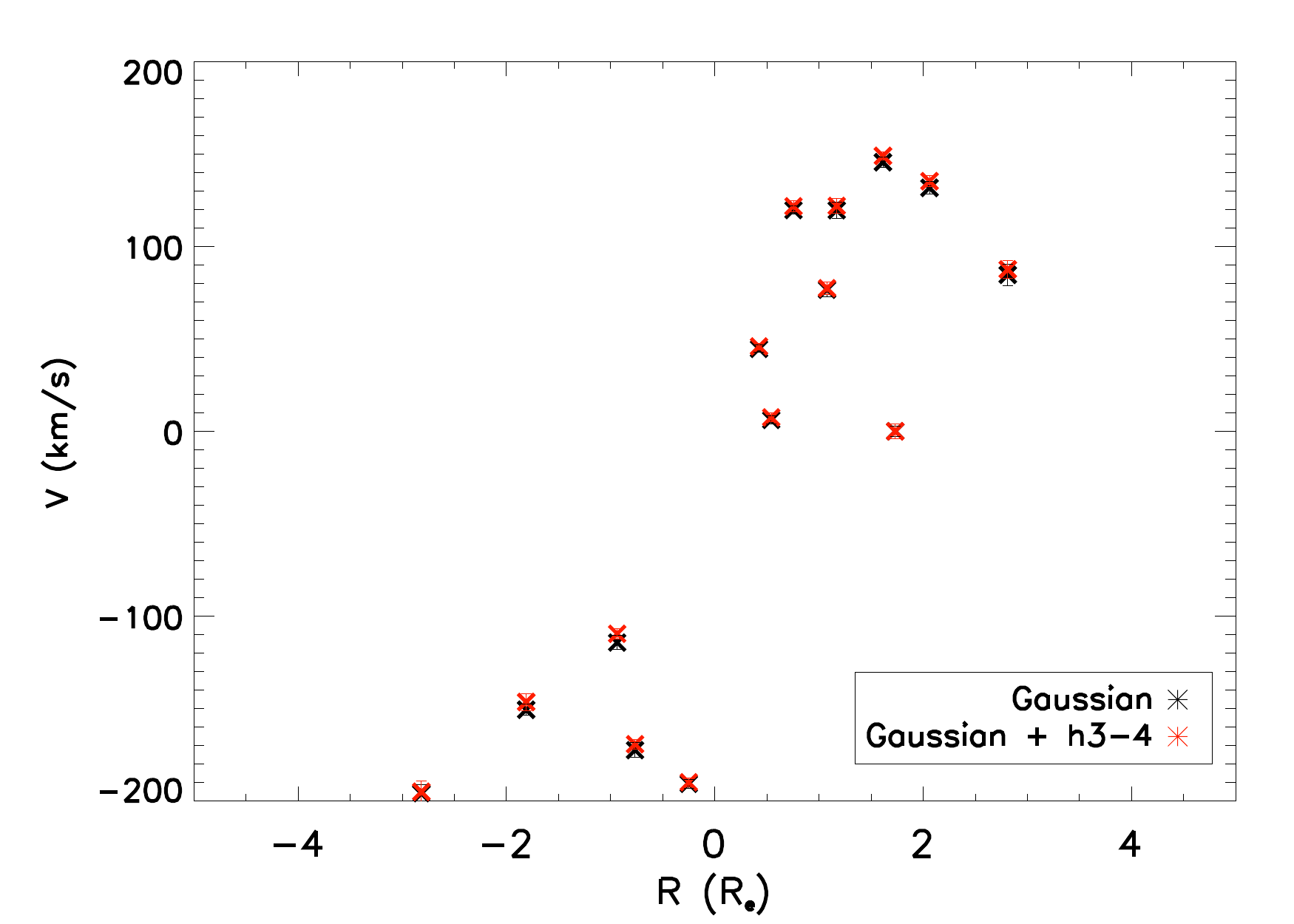}
\includegraphics[width=0.45\columnwidth,angle=0,clip]{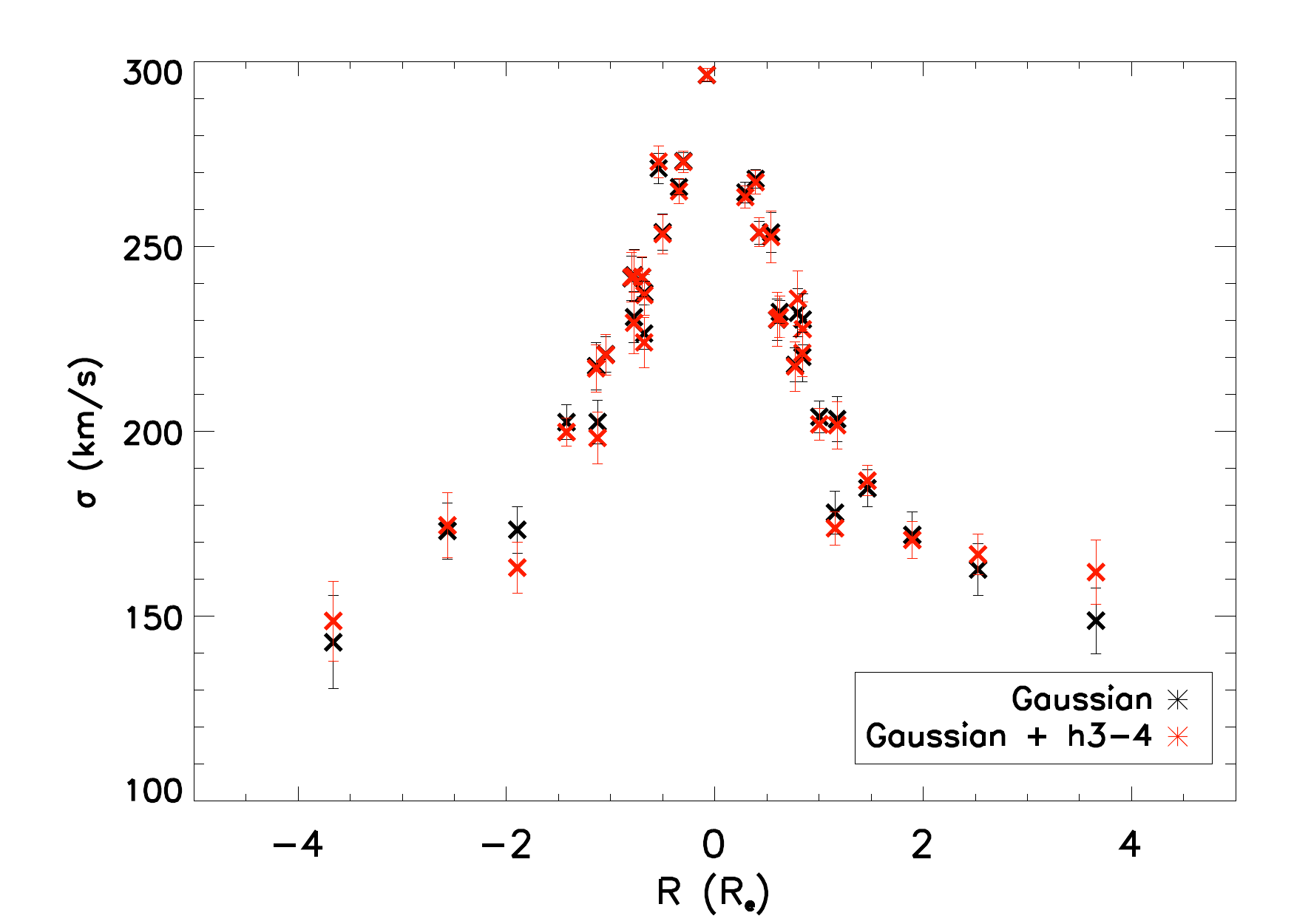}
\includegraphics[width=0.45\columnwidth,angle=0,clip]{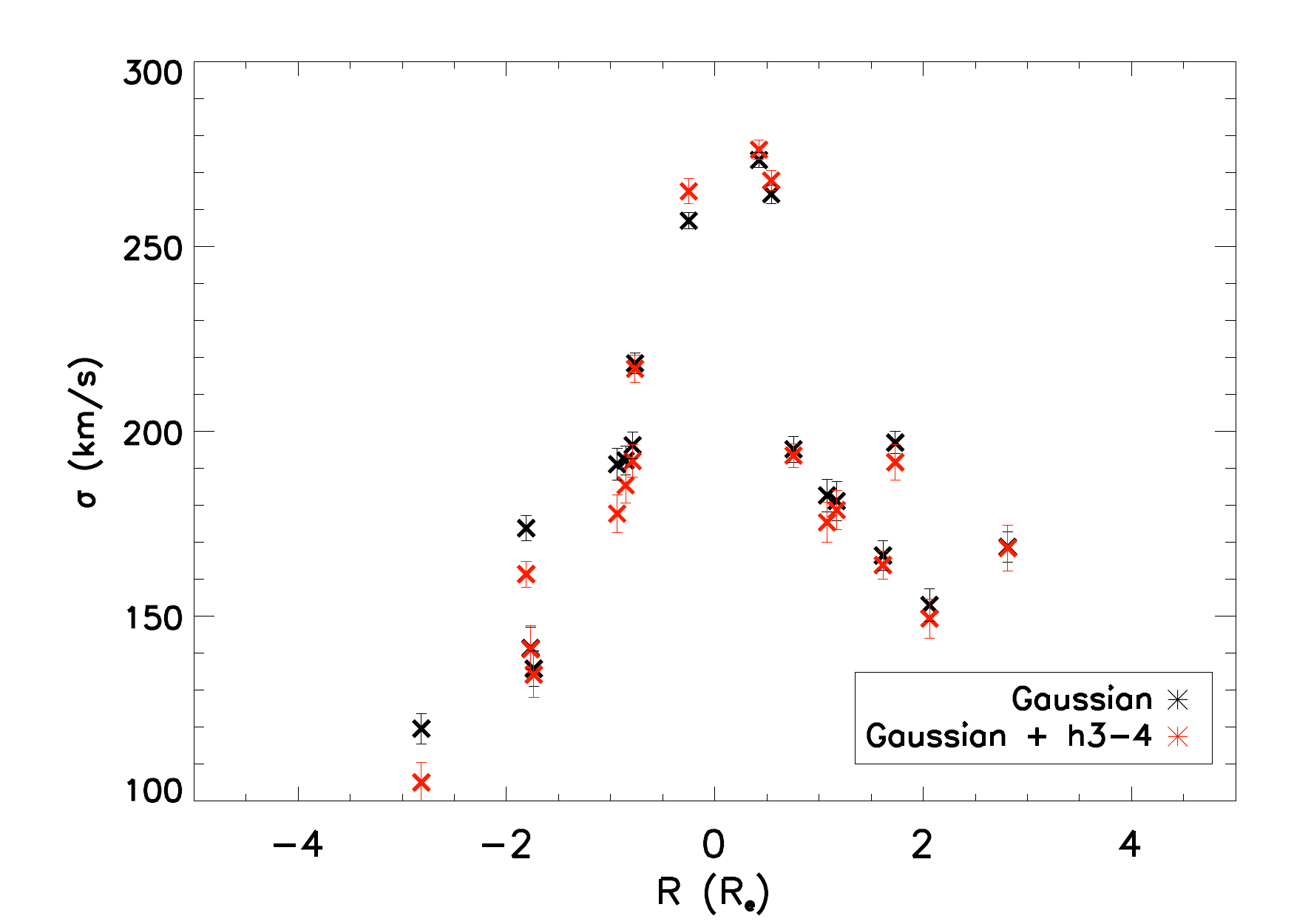}
\caption{
  Velocity (top) and dispersion (bottom) profiles with
  radius of galaxies NGC~4952 (left) and NGC~426 (right). We compare the
  profiles calculated using a Gaussian LOSVD (black) and a 4th order
  Gauss-Hermite series (red).}
\label{Fig:Hermite}
\end{center}
\end{figure}

The effect of including Hermite polynomials of differing degree to the
LOSVD function specified in pPXF has also been addressed in some
detail by \cite{Emsellem2004}. Since pPXF
selects against large deviations from Gaussianity, we do not
expect the exclusion of higher order hermite moments to create
significant systematic errors. We simply ask whether characteristic galaxies
in our sample show any systematic differences when we choose to model
the LOSVD as a 4th order Gauss-Hermite series.

Figure~\ref{Fig:Hermite} shows the velocity and dispersion profiles
for galaxies NGC~4952 and NGC~426 calculated both with and without 3rd
and 4th order Hermite terms. In both cases we find almost no
difference in the velocity profiles as expected.
More importantly, the dispersion profiles appear to show no
significant differences when we model $h_3$ and $h_4$. Arguably, there
is a trend towards slightly lower dispersion in the higher order case,
particularly at large radius or equivalently lower S/N. This again
makes sense since some of the LOSVD power is shifted from the Gaussian
term to the Hermite series, lowering the dispersion. However, any
deviations are well within 1-$\sigma$ and below $\sim 5$~\kms so we
may safely ignore them, and accurately model the LOSVD as a Gaussian
alone.

\subsection{Absorption Features}
\label{subsec: Absorption Features}

\begin{figure}[!htb]
\begin{center}
\includegraphics[width=0.4\columnwidth,angle=0,clip]{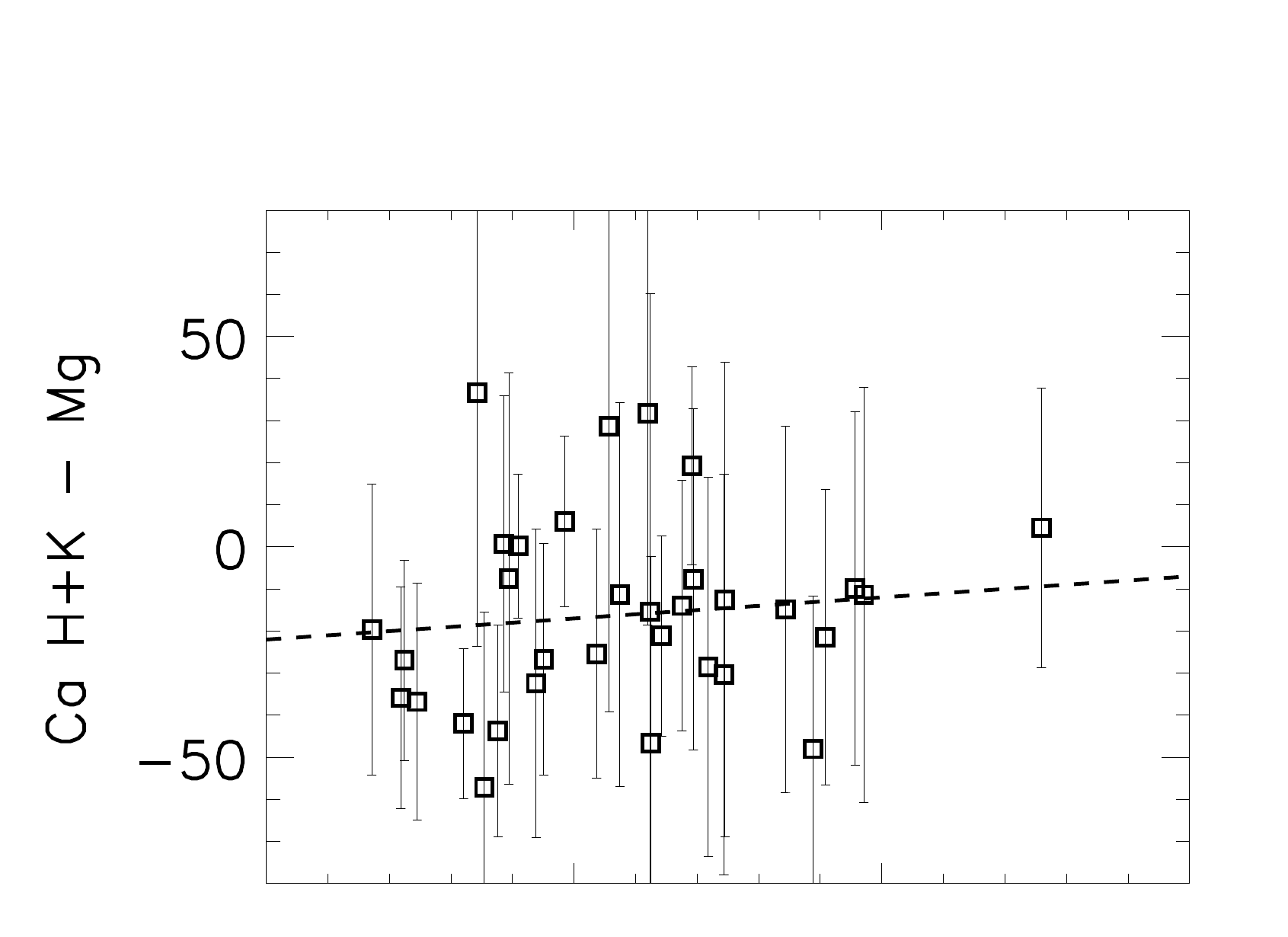}
\includegraphics[width=0.4\columnwidth,angle=0,clip]{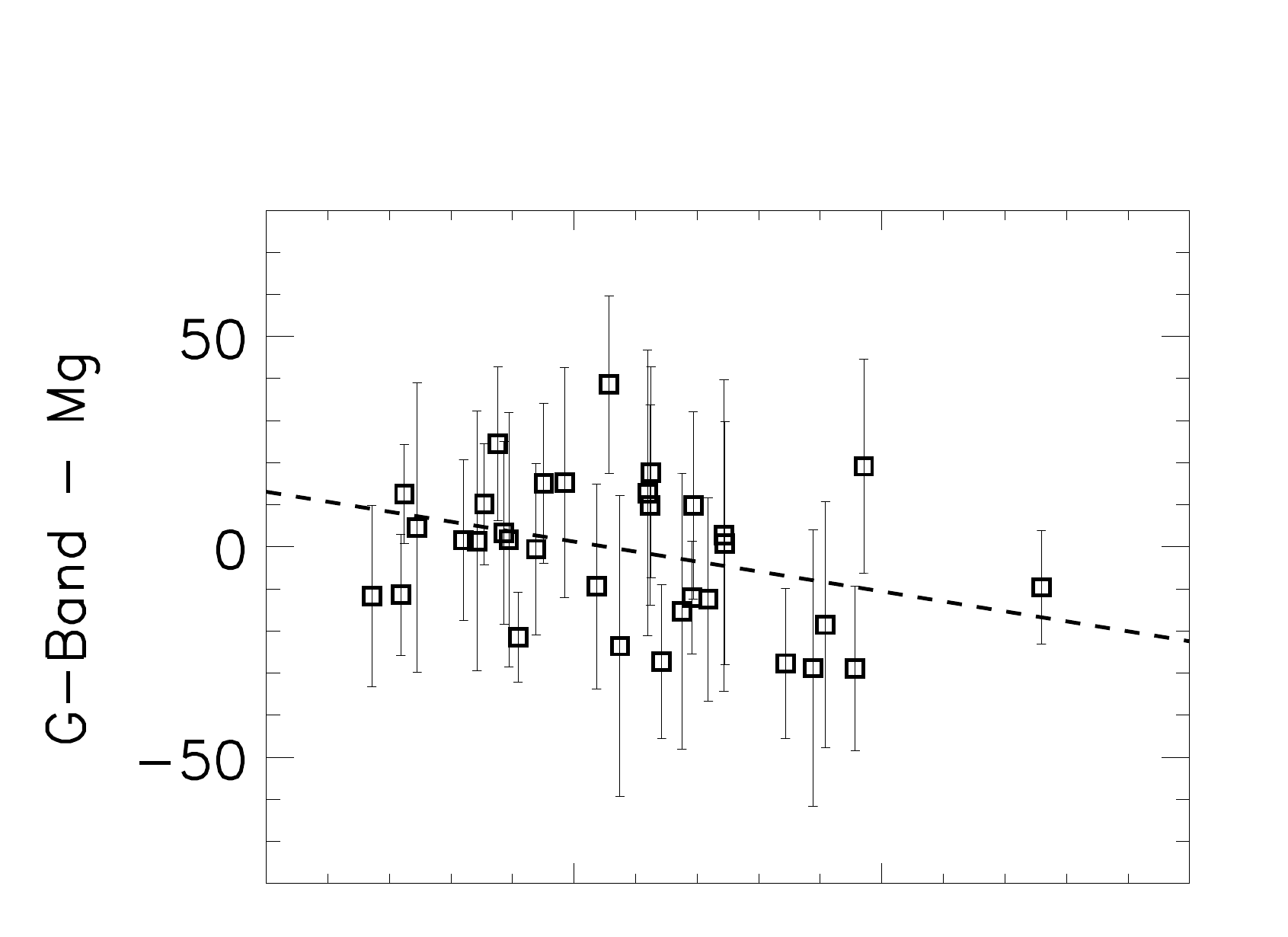}
\includegraphics[width=0.4\columnwidth,angle=0,clip]{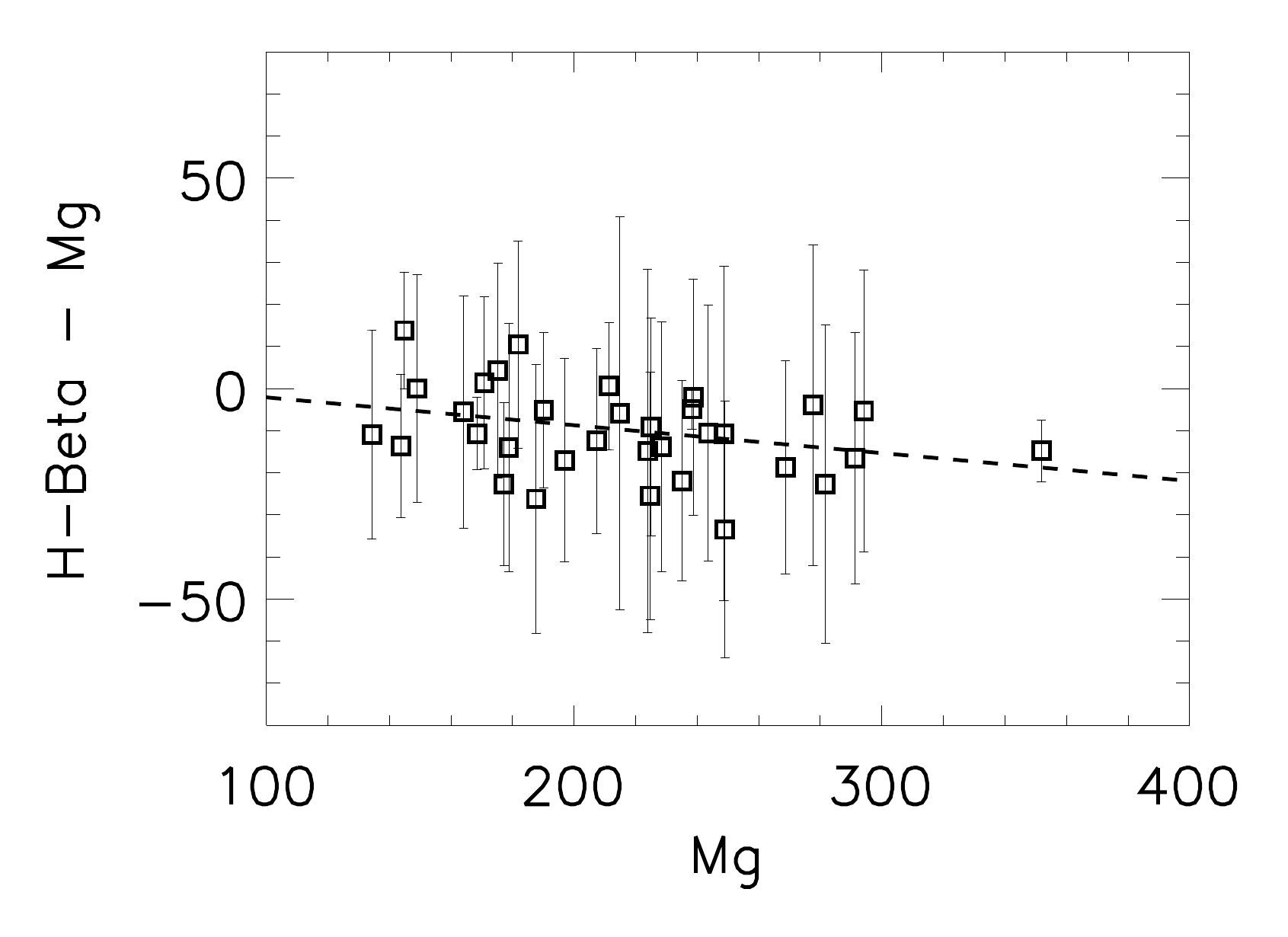}
\includegraphics[width=0.4\columnwidth,angle=0,clip]{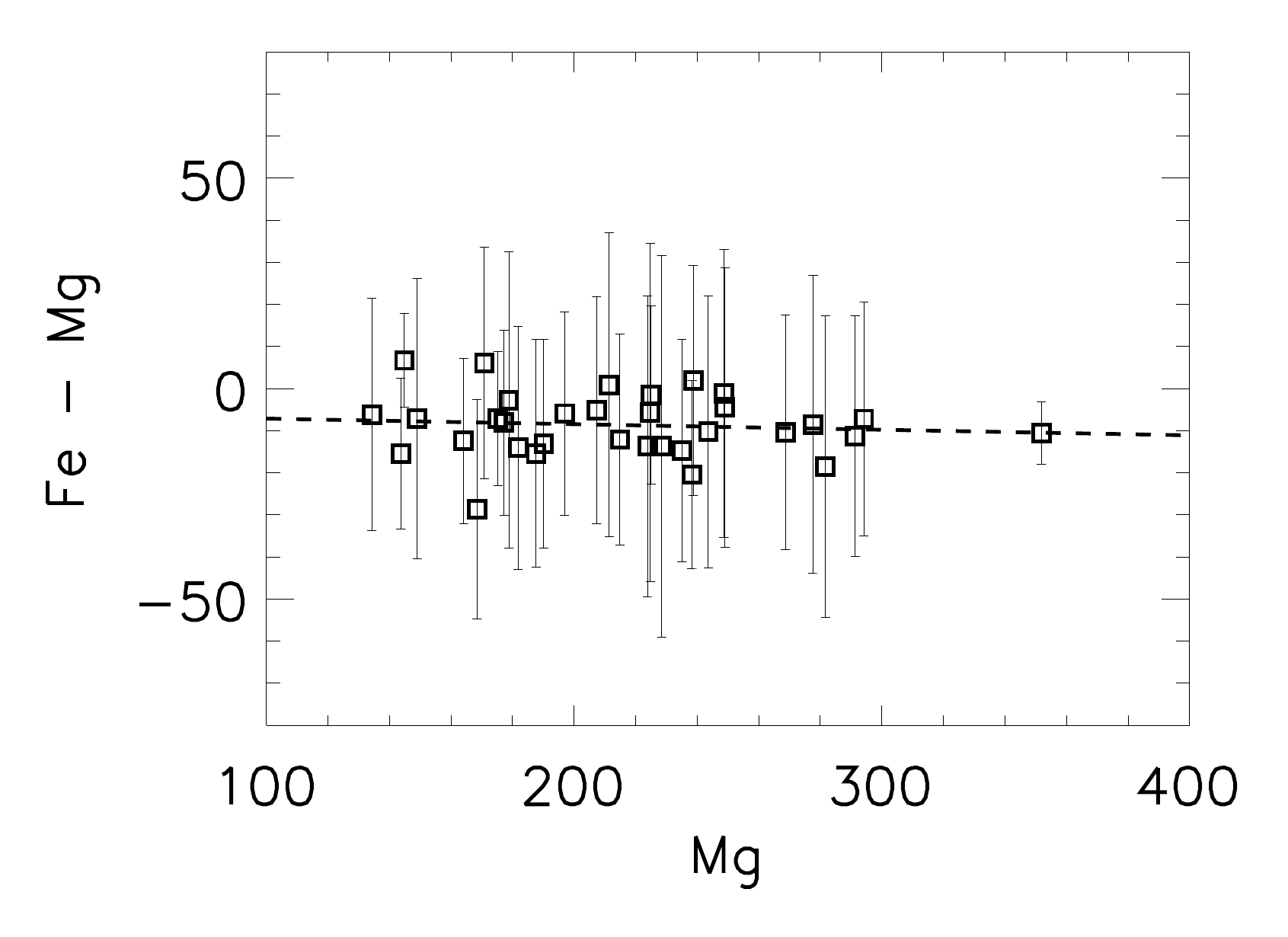}
\caption{
  A comparison between dispersions in the Ca H+K region $(3650$~\AA -
  $4050$~\AA), the G-Band (4215 \AA$ - 4575$~\AA), H$\beta$, (4445~\AA
$ - 4975$~\AA), the MgI{\it b} region (4900~\AA$ - 5420$~\AA) and the Fe
  lines redward of MgI{\it b} (5250~\AA$ - 5820$~\AA). We show the median
  offset from the MgI{\it b} measurement in all central fibers for each
  galaxy. Error bars represent standard deviations in the offset
  distribution and the line of best fit to the offset distribution is
  shown as a dashed line.}
\label{Fig:DispvsWError}
\end{center}
\end{figure}

\begin{figure}[!htb]
\begin{center}
\includegraphics[width=0.4\columnwidth,angle=0,clip]{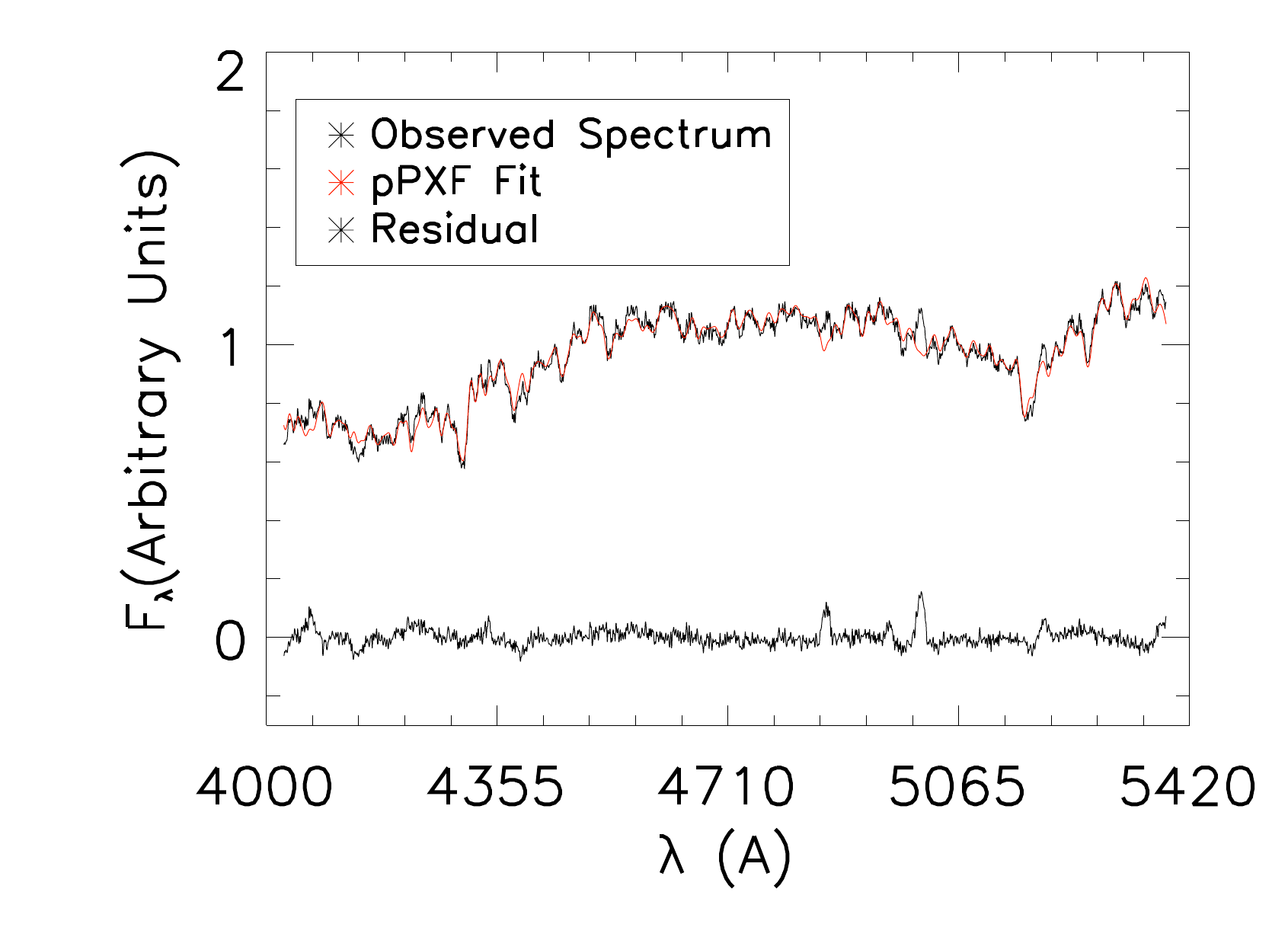}
\includegraphics[width=0.4\columnwidth,angle=0,clip]{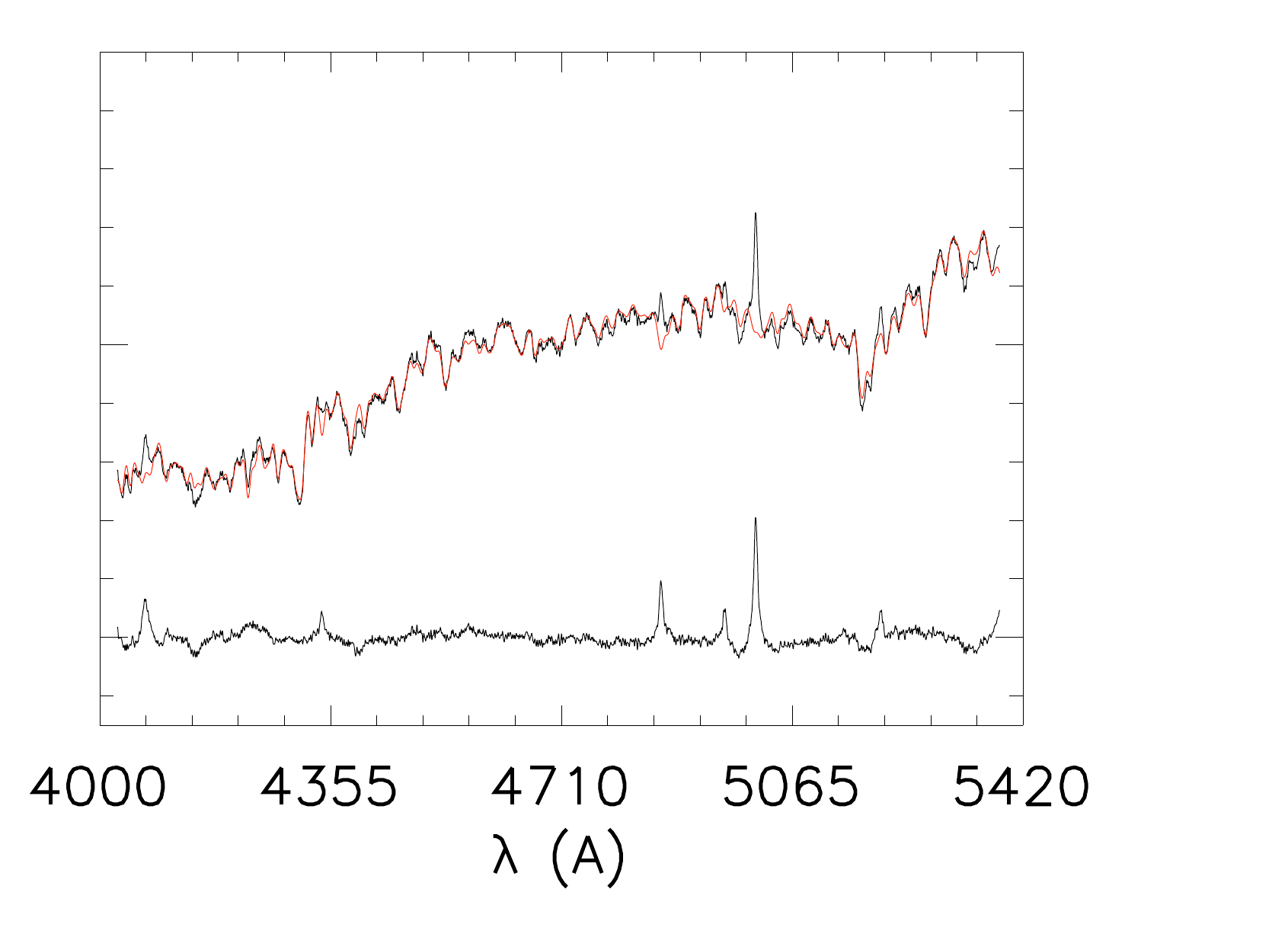}
\caption{Fits to the central fibers of galaxies NGC426 (left) and NGC7509 (right), both of which show 
strong emission lines characteristic of AGN.}
\label{Fig:CentralFibres}
\end{center}
\end{figure}

Although pixel-fitting is robust at relatively low S/N, it can be quite
sensitive to template mismatch, which introduces feature-dependent
systematics into our dispersion estimates. To investigate how robust
we are to this problem, we therefore compare results from a variety of
different wavelength regions, centered on different strong 
spectral features: Ca H+K (3650~\AA$ - 4050$~\AA), G-Band(4215~\AA$ -
4575$~\AA), H$\beta$ (4445 \AA$ - 4975$~\AA), MgI{\it b} 
(4900~\AA$ - 5420$~\AA) and Fe (5250~\AA$ - 5820$~\AA), 
where for the Fe region, we mask
out wavelengths between 5570 and 5610~\AA, which often shows strong
emission lines.  Since each of these regions is relatively small, the
continuum is fit well enough by a fourth order Legendre polynomial, and
uncertainties of less than $\sim 2 \%$ are introduced by fitting any
polynomial of order between $\approx 2 - 6$.

We compare regions by taking all of the central fibers from a given
galaxy, evaluating the fit in each region and then calculating offsets
from the MgI{\it b} measurements. Results showing the median and standard
deviation of the offset distribution for each galaxy are shown in
Figure~\ref{Fig:DispvsWError}.  We firstly note that, in general,
there doesn't appear to be a significant offset between any of the
different features in the sense that most galaxies show a median
offset consistent with zero.  Importantly, the average offset between
regions including both MgI{\it b} and Fe blends and Fe blends alone appears
to be less than $\sim 10$~\kms, significantly smaller than the
discrepancies found in e.g. \cite{Barth2002}. This is perhaps
unsurprising, as our set of templates include a range in $[\alpha /
{\rm Fe}]$ above solar, but is nevertheless reassuring.

The H$\beta$, MgI{\it b} and Fe regions align quite well over the range
$150 < \sigma < 300$~\kms, with discrepancies limited to about $\pm
30$~\kms and almost negligible mean offsets.  By contrast Ca H+K,
shows a mean offset of close to $20$~\kms\ alongside a much larger
scatter. This is relatively unsurprising as the continuum drops
sharply near $4000$\AA, and the line shape is quite sensitive to spectral 
type \citep[see discussion and further references in][]{Greene2005}.

The only possibly problematic region therefore appears to be the
G-Band, which shows a mild trend with dispersion, underestimating the
dispersion of the largest galaxies by up to $\sim 20$~\kms. The offset 
is largely due to the presence of a small number of
outlying galaxies with large positive offsets from the MgI{\it b} region.
In particular, NGC~426 and NGC~7509 show G-Band offsets of between $+30$
and $+40$~\kms.  Figure~\ref{Fig:CentralFibres} shows the central
fibers of these two galaxies, which both display strong emission
features characteristic of AGN. Our stellar templates cannot fit these
features, but their presence in the MgIb and CaH+K spectra has the
effect of artificially driving down the dispersion in these regions
and therefore creating a discrepancy with the relatively smooth
G-Band. For instance, the central fibers in NGC~7509 and NGC~426 yield
MgIb dispersions of $149.8 \pm 1.5$ and $207.5 \pm 5.9$~\kms
respectively. However, if we fit the same region with the strongest
lines between $4900$~\AA and $5120$~\AA masked out, these values
become $186.7 \pm 1.2$ and $230.5 \pm 3.0$~\kms, a shift of $\sim
30$~\kms in each case.

Given the presence of strong emission lines in these two galaxies at
least, it doesn't necessarily make sense to blindly fit over a
relatively small $500$~\AA\ window as these results can be
significantly skewed by one or two features. We therefore choose to
calculate dispersions over a much larger average region 4000 \AA$-$
5420~\AA, which is less affected by the presence of a small number of
lines.  The central fibers of NGC~426 and NGC~7509, for example, yield
dispersions of $190.0 \pm 1.1$ and $231.1 \pm 1.4$, broadly consistent
with the masked MgIb region. Our chosen region includes all of the
G-Band, H-$\beta$, MgIb and some Fe lines, which should not pose a
problem as these regions show no systematic offsets. However, we
exclude shorter wavelengths since, as discussed earlier, the S/N drops
precipitously near Ca H+K.

\subsection{Continuum Fitting}
\label{subsec: Continuum Fitting}

\begin{figure}[!htb]
\begin{center}
\includegraphics[width=0.27\columnwidth,angle=0,clip]{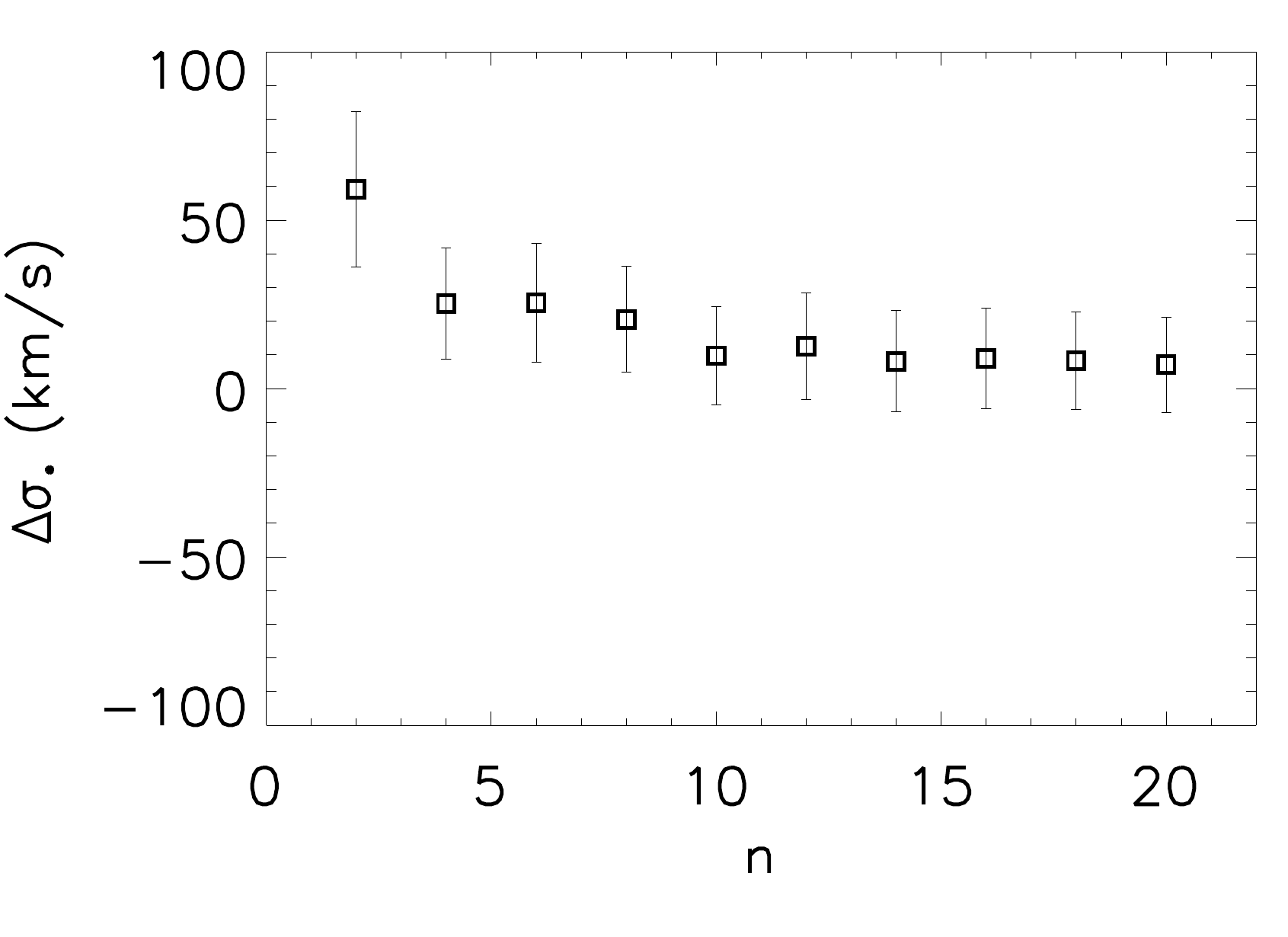}
\includegraphics[width=0.27\columnwidth,angle=0,clip]{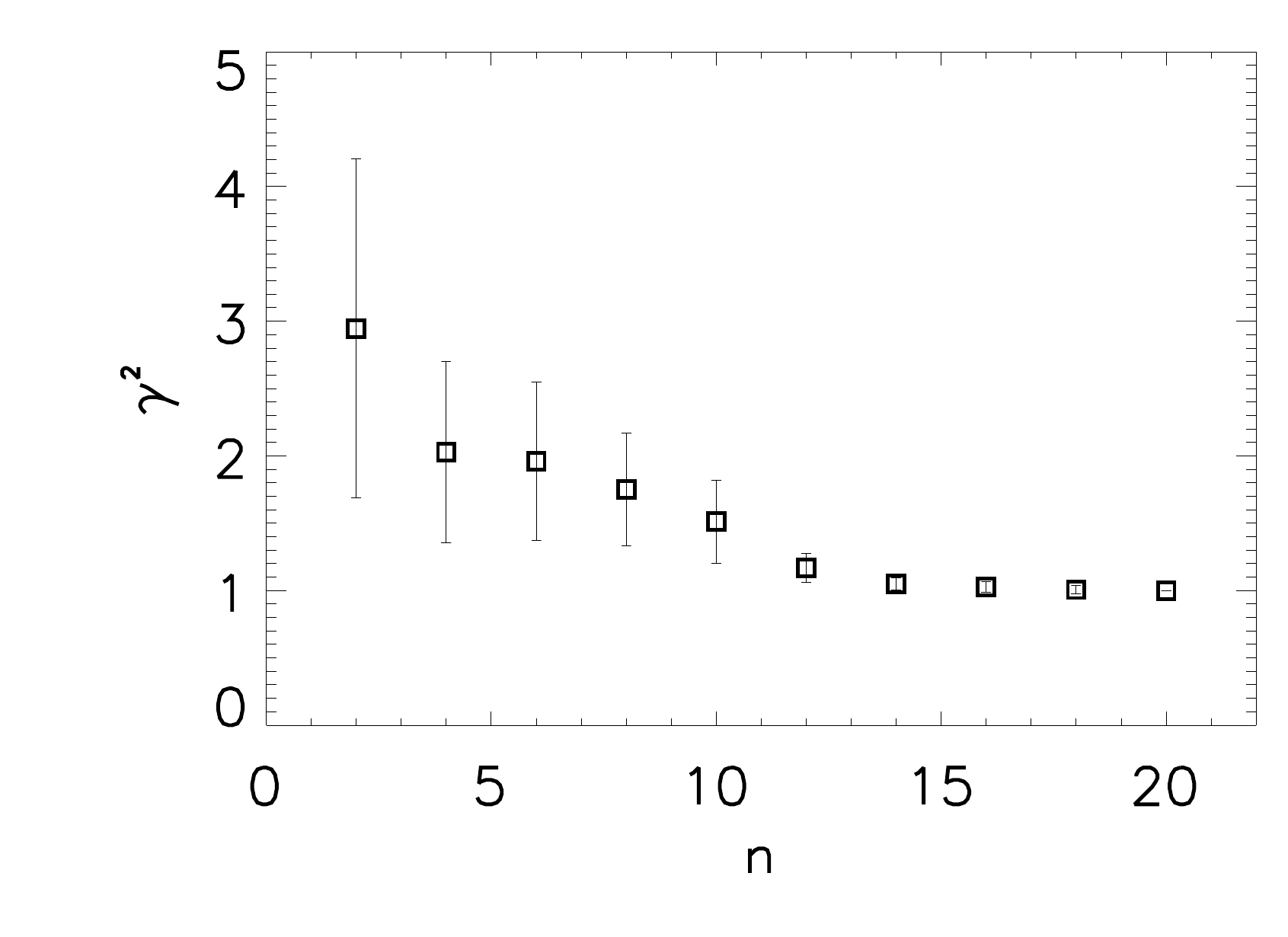}
\includegraphics[width=0.27\columnwidth,angle=0,clip]{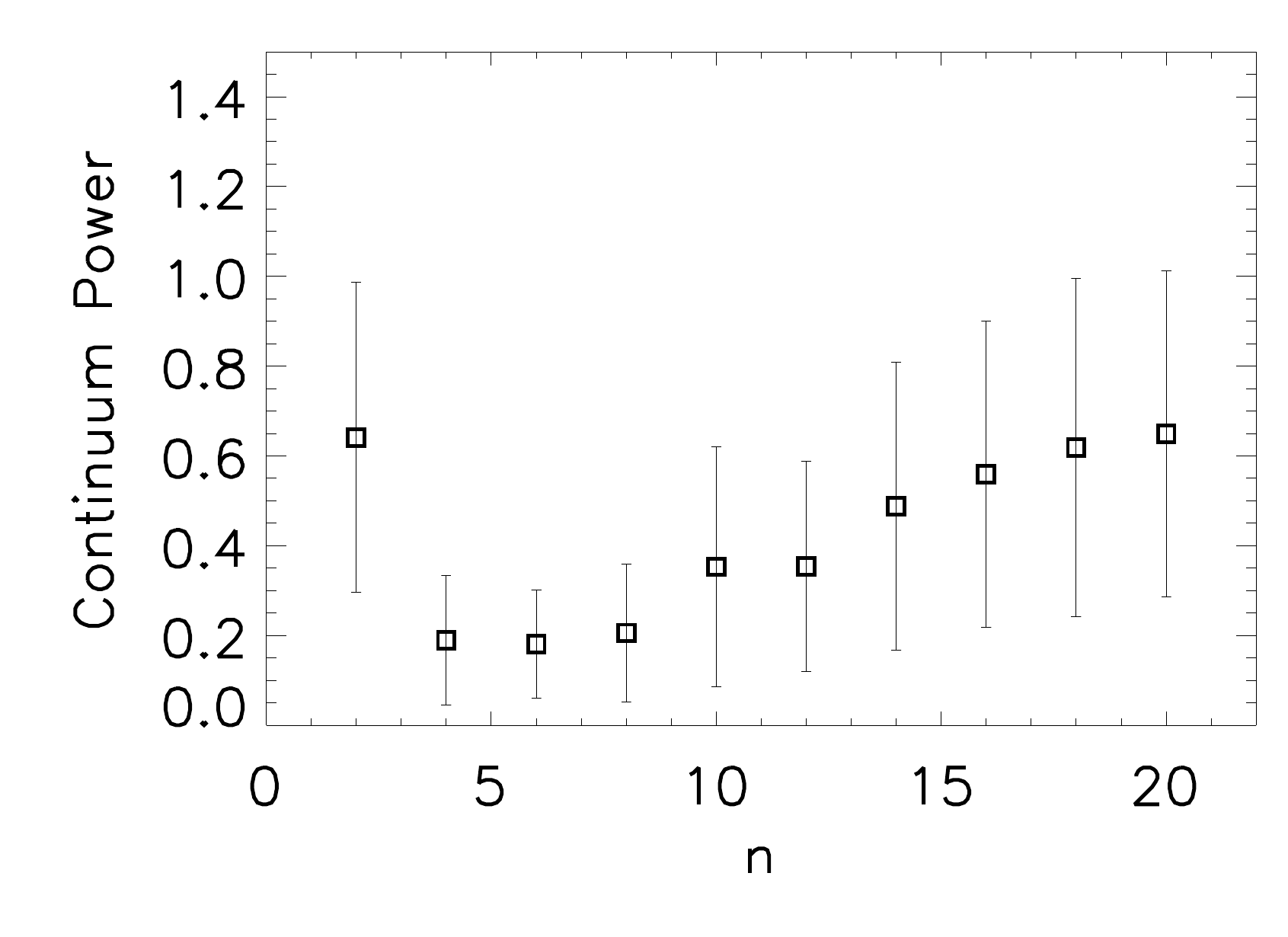}
\caption{
  Measures of goodness of fit as a function of continuum polynomial
  degree (n) for dispersions calculated over our fiducial region 
4000~\AA$- 5420$~\AA. We show systematic offsets from the median of
  dispersions evaluated over the G-Band (4215~\AA$ - 4575$~\AA),
  H$\beta$, (4445~\AA$ - 4975$~\AA), the MgI{\it b} (4900~\AA$ - 5420$~\AA)
  and the Fe lines redward of MgI{\it b} (5250~\AA$ - 5820$~\AA) on the
  left. In the middle we have the $\chi^2$ of the fit to each galaxy
  normalised by the minimum $\chi^2$. Finally, on the right we show the
  continuum power as a fraction of the total galaxy power again
  normalized to the maximum continuum power. Results shown are
  calculated as averages over the central fibers in all of our
  galaxies.}
\label{Fig:Continuum}
\end{center}
\end{figure}

Some care must be taken in fitting this larger region, since the
continuum is likely to vary a great deal more over a range of $\sim
1500$~\AA. This may not be captured accurately by stellar templates
and a 4th order polynomial alone. On the other hand, if we increase
the degree of the continuum polynomial too much, we may begin to fit
the absorption lines themselves We therefore need to further test the
response of pPXF to the degree of the continuum polynomial between
$4000$~\AA\ and $5420$~\AA.  In particular, we wish to find the right
continuum parameter range that strikes a balance between
over-simplifying the continuum and over-fitting it, which would tend
to draw power from the LOSVD.

We do this by successively fitting the central fibers of each galaxy
using a range of higher order polynomials with degree varying between
2 and 20. The derived dispersions are then compared to the dispersions
calculated from the G-Band, H$\beta$, MgI{\it b} and Fe regions
alone. Figure~\ref{Fig:Continuum} shows the resulting systematic
offsets from the median dispersion in these regions, the $\chi^2$ of each
fit, and the fraction of spectral power in the continuum.

We see that there are systematic offsets of close to $\sim 20$~\kms~
for continuum fits of degree less than $10$. Even above this value the fit
over our fiducial region slightly overestimates the dispersion
relative to the individual regions, but given that there is a small
systematic offset between Fe and MgI{\it b} regions, this is probably
expected. Above $10$ there is also a pronounced drop in the $\chi^2$ of
the overall fit, however, by this stage we are clearly over-fitting
the continuum since the continuum power rises dramatically. We
therefore use a $10^{th}$ order fit, which is found to best balance
the competition between underestimating the continuum and introducing
false structure on the dispersion scale.

\subsection{Template Mismatch}
\label{subsec: Template Mismatch}

\begin{figure}[!htb]
\begin{center}
\includegraphics[width=0.45\columnwidth,angle=0,clip]{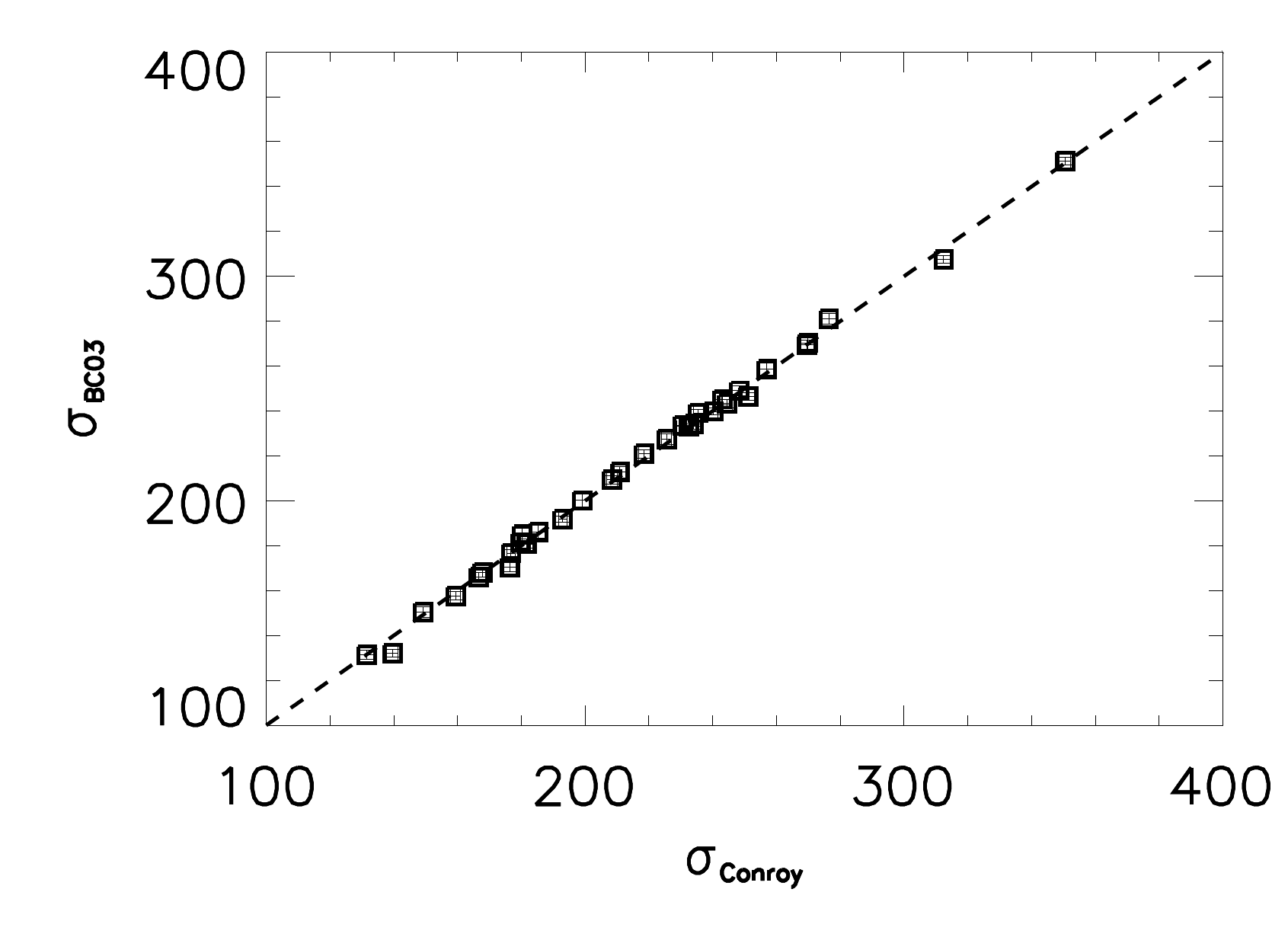}
\includegraphics[width=0.45\columnwidth,angle=0,clip]{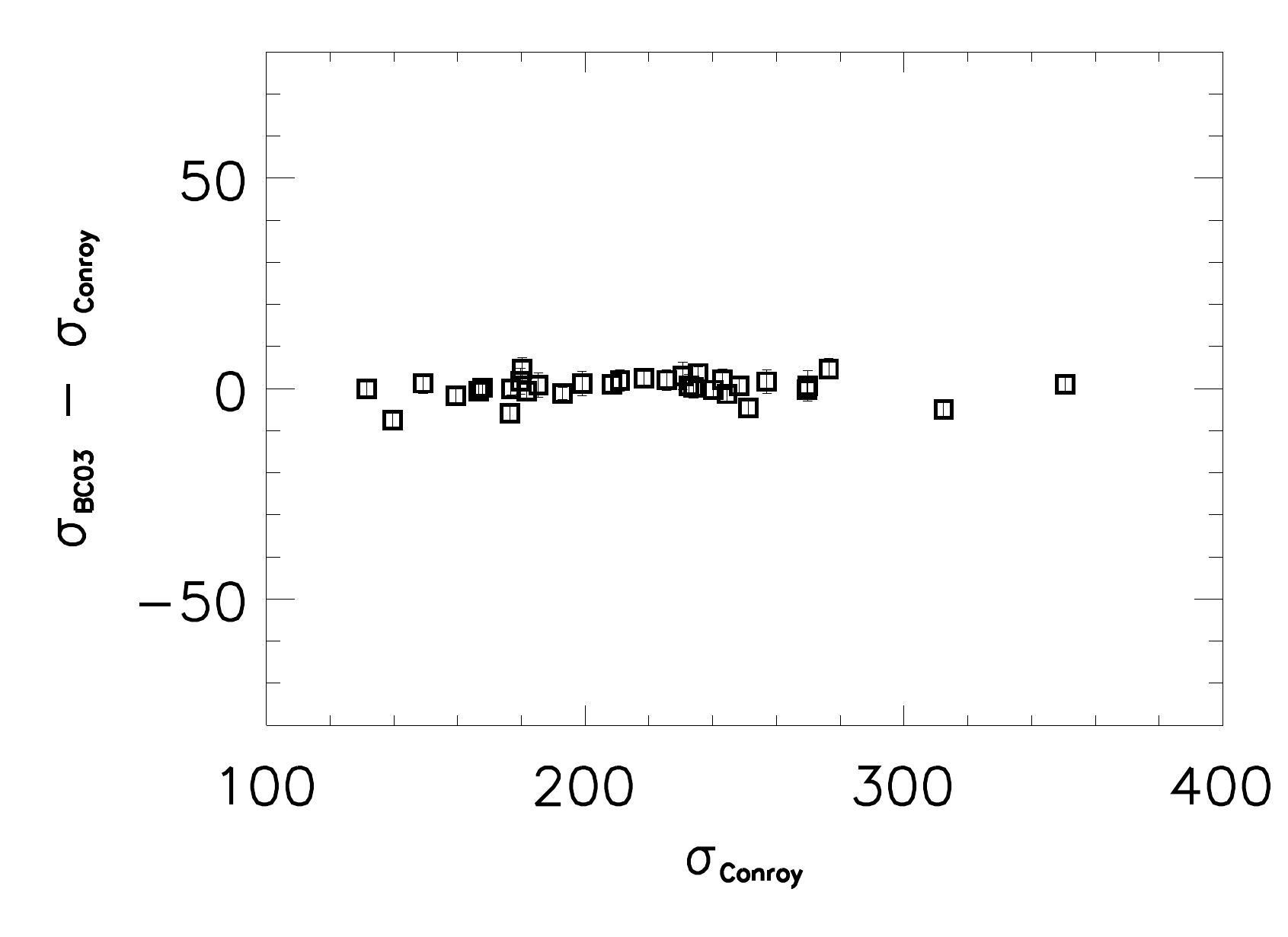}
\caption{A comparison between dispersions as calculated from the \cite{Conroy2009} and 
\cite{BruzualCharlot2003} templates. We show both the dispersions and the offsets between the two 
calculations. We note that the results are remarkably well correlated.}
\label{Fig:DispvsT}
\end{center}
\end{figure}

As a final sanity check to see how sensitive we are to the issue of template mismatch, we also 
compare our calculated dispersions to those extracted from a different set of 
templates. We use \cite{BruzualCharlot2003} single-age stellar population models with 
$\sigma \sim 70$~\kms resolution as our comparison set, and fit over our fiducial wavelength 
region. The results, in Figure~\ref{Fig:DispvsT} show a very tight correlation between the two sets, so 
either both sets of templates suffer from similar problems or they are both adequate for our set of 
galaxies.

\clearpage

\section{Observed Kinematics}
\label{ObservationsAppendix}

We show, in Figures~\ref{Fig:AllKinematicsa}-\ref{Fig:AllKinematicsl}, full 2D profiles of the velocity 
and velocity dispersion of all galaxies, as well as 1D kinemetric profiles.

\setcounter{figure}{0} \renewcommand{\thefigure}{B.\arabic{figure}}

\begin{figure}[!htb]
\begin{center}

\includegraphics[width=0.6\textwidth,angle=0,clip]{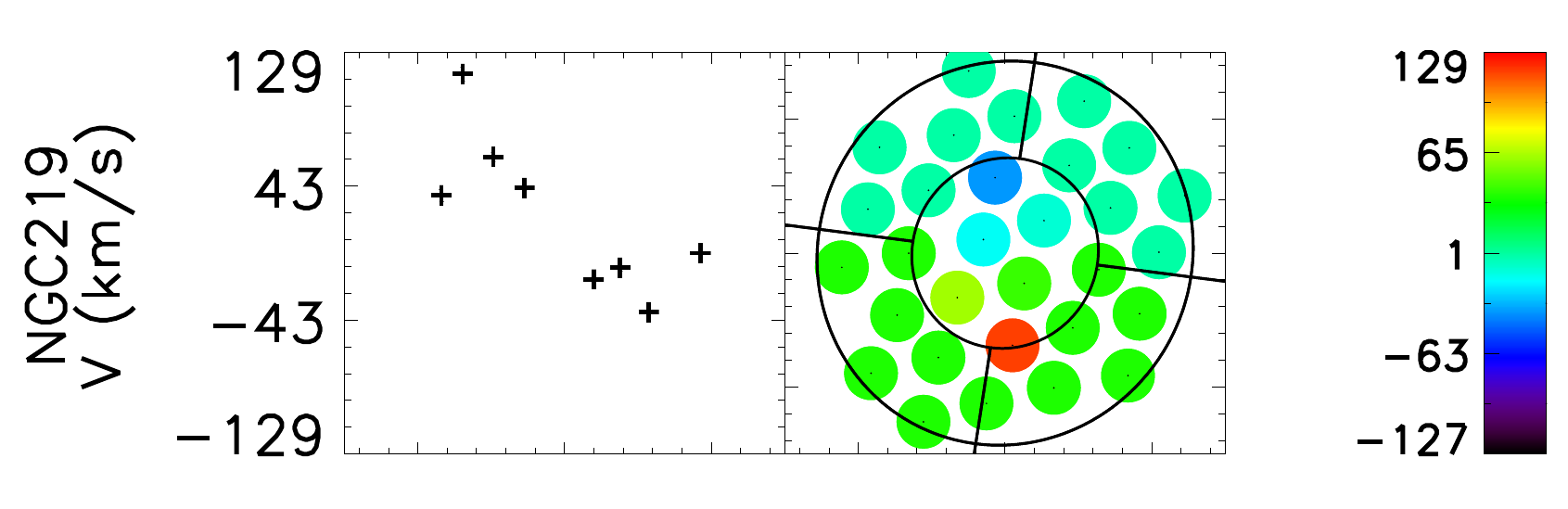}
\includegraphics[width=0.6\textwidth,angle=0,clip]{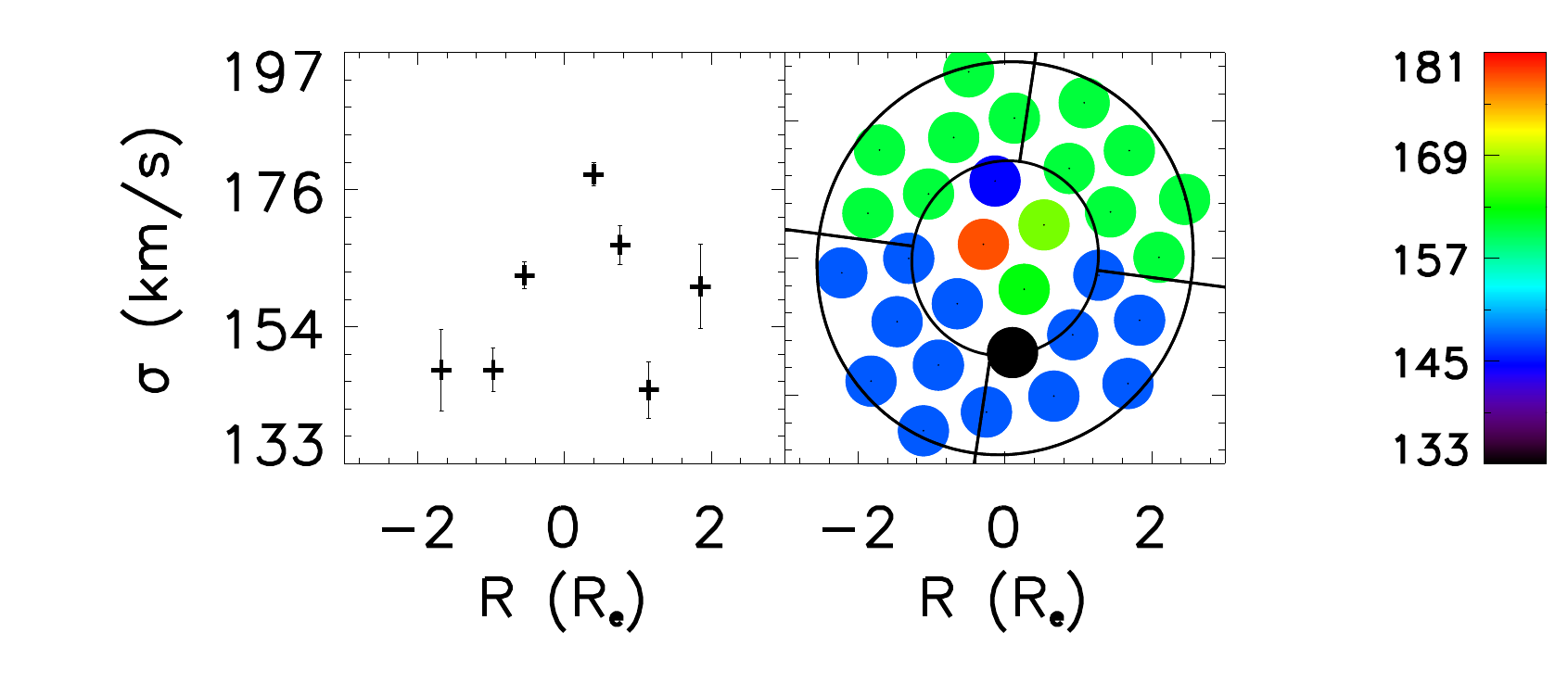}
\includegraphics[width=0.6\textwidth,angle=0,clip]{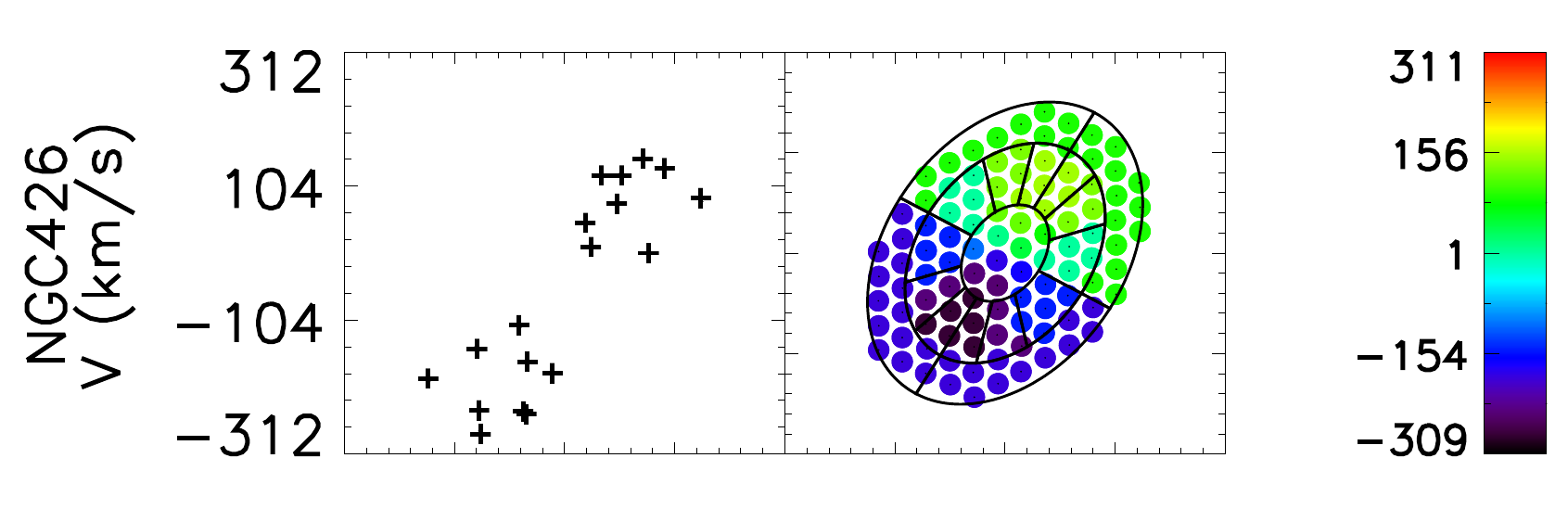}
\includegraphics[width=0.6\textwidth,angle=0,clip]{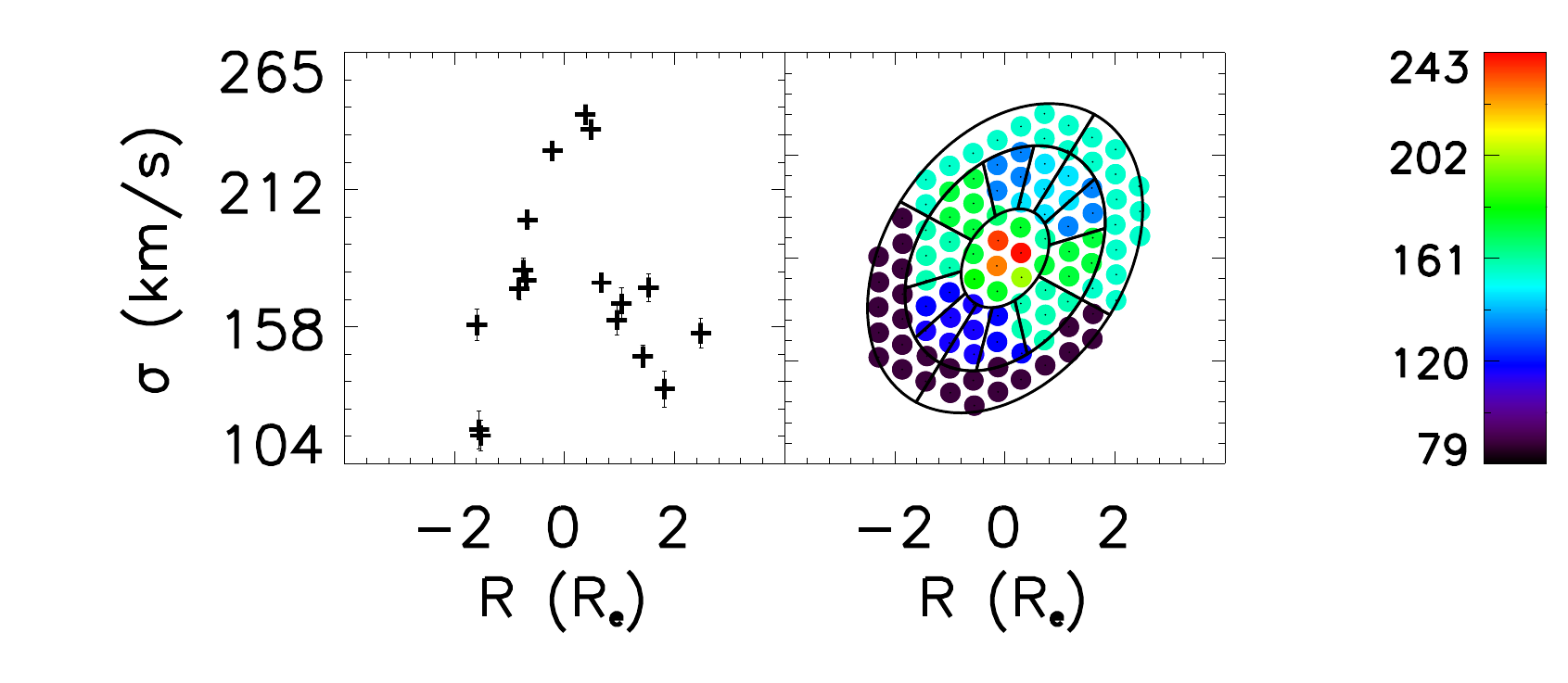}

\caption{Maps of the stellar kinematics of all galaxies in our sample. 
We show from left to right: (i) 1D radial map of stellar mean velocity, where each 
point corresponds to a different bin at possibly different angular positions (top left) (ii) 2D map of stellar 
mean velocity (top right), (iii) 1D radial map of stellar velocity dispersion $\sigma$ (bottom left), 
(iv) 2D map of $\sigma$ (bottom right). 
{In a small number of the galaxies, fibres are absent in certain regions due to masking of external 
sources in those areas. Several galaxies also show non-uniform binning (e.g. IC~1153) due to 
small astrometric shifts during observation \citep[see e.g.][for more detail]{Greene2013}. The 
outermost bin is shown in all cases, but as mentioned in the text, occasionally measurements here 
needed to be discarded due to excessive masking or limited field-of-view. Finally, low dispersions 
($\sigma \lesssim 100$~\kms, which are generally unreliable with errors $\gtrsim 20\%$, are shown 
in the 2D maps, but excluded from 1D plots and all calculations of $\lambda$ and other radial 
kinematic profiles.}}
\label{Fig:AllKinematicsa}
\end{center}
\end{figure}

\begin{figure}
\begin{center}

\includegraphics[width=0.6\textwidth,angle=0,clip]{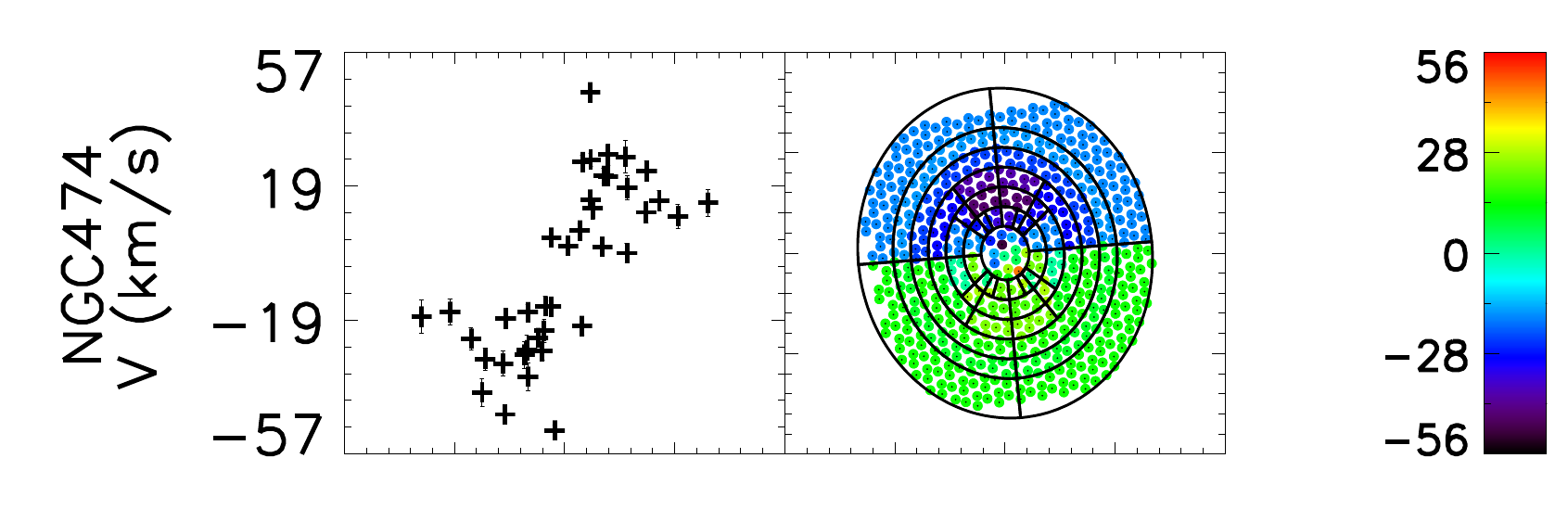}
\includegraphics[width=0.6\textwidth,angle=0,clip]{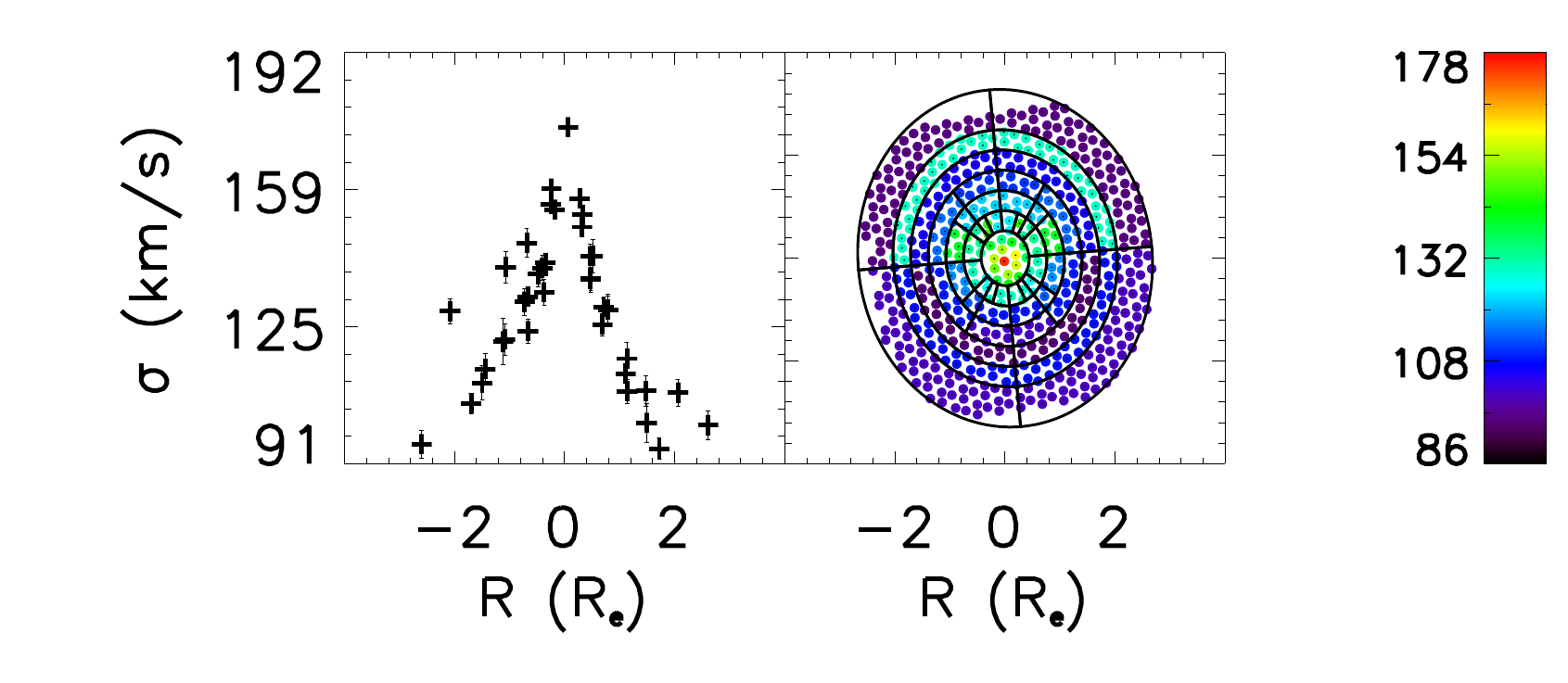}
\includegraphics[width=0.6\textwidth,angle=0,clip]{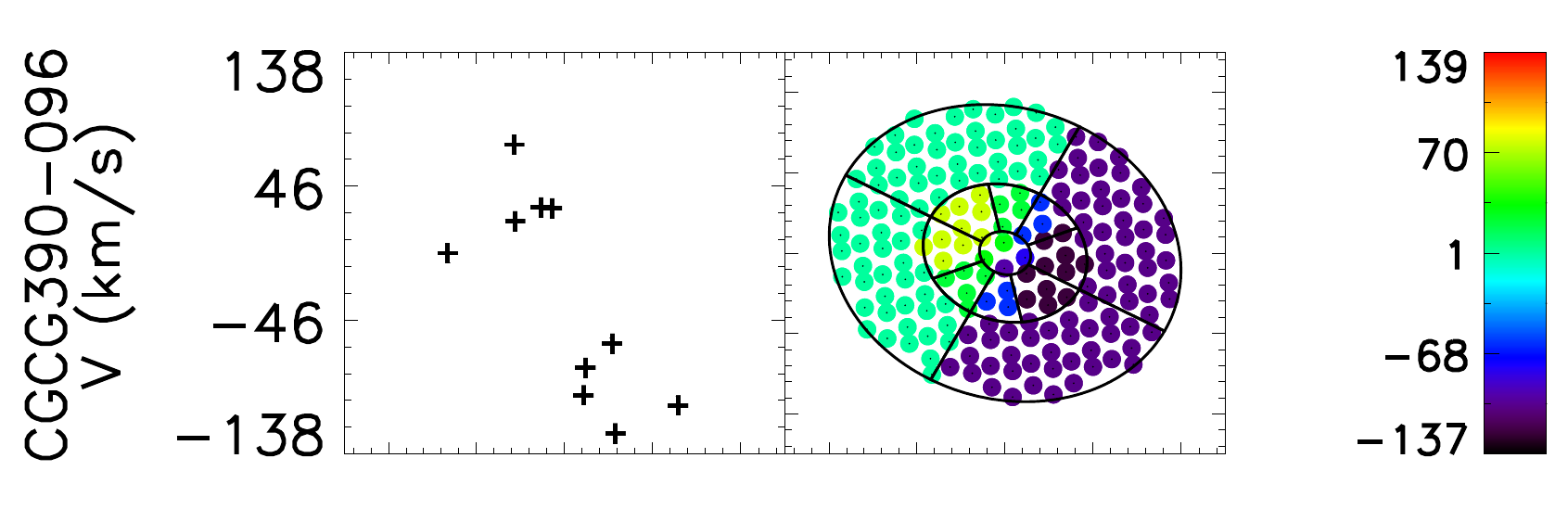}
\includegraphics[width=0.6\textwidth,angle=0,clip]{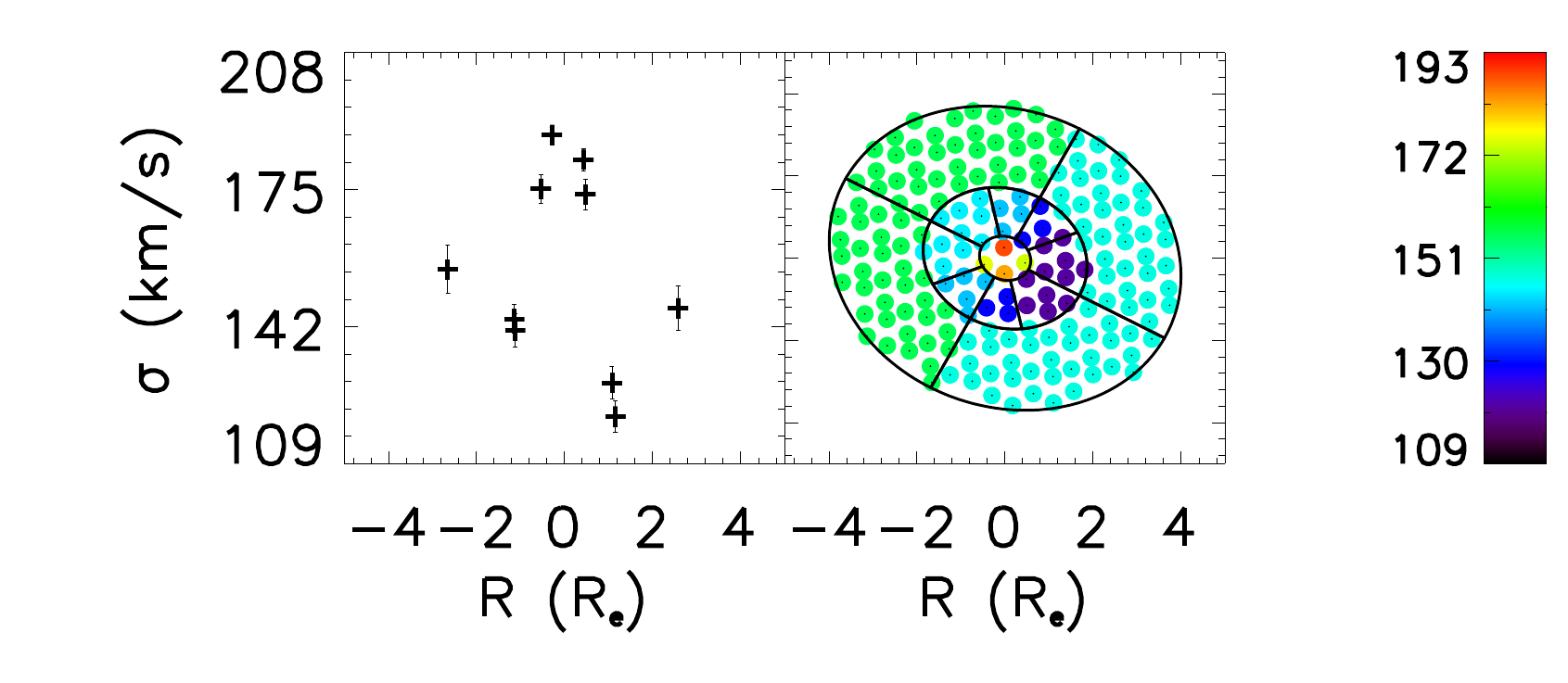}
\includegraphics[width=0.6\textwidth,angle=0,clip]{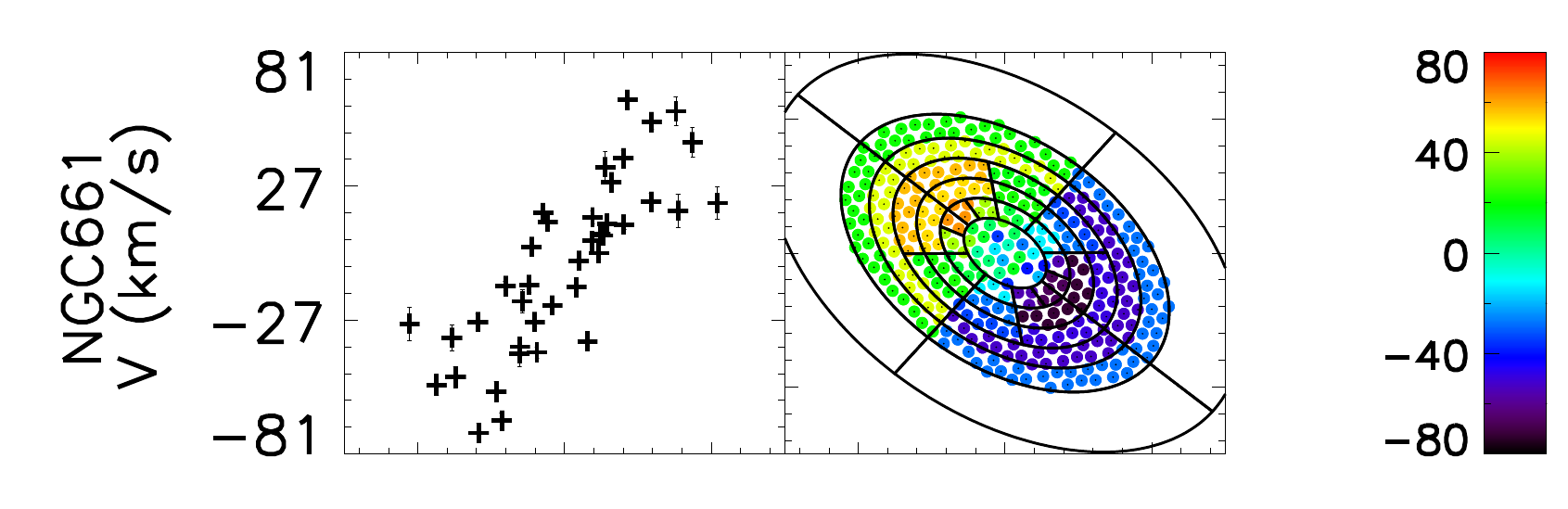}
\includegraphics[width=0.6\textwidth,angle=0,clip]{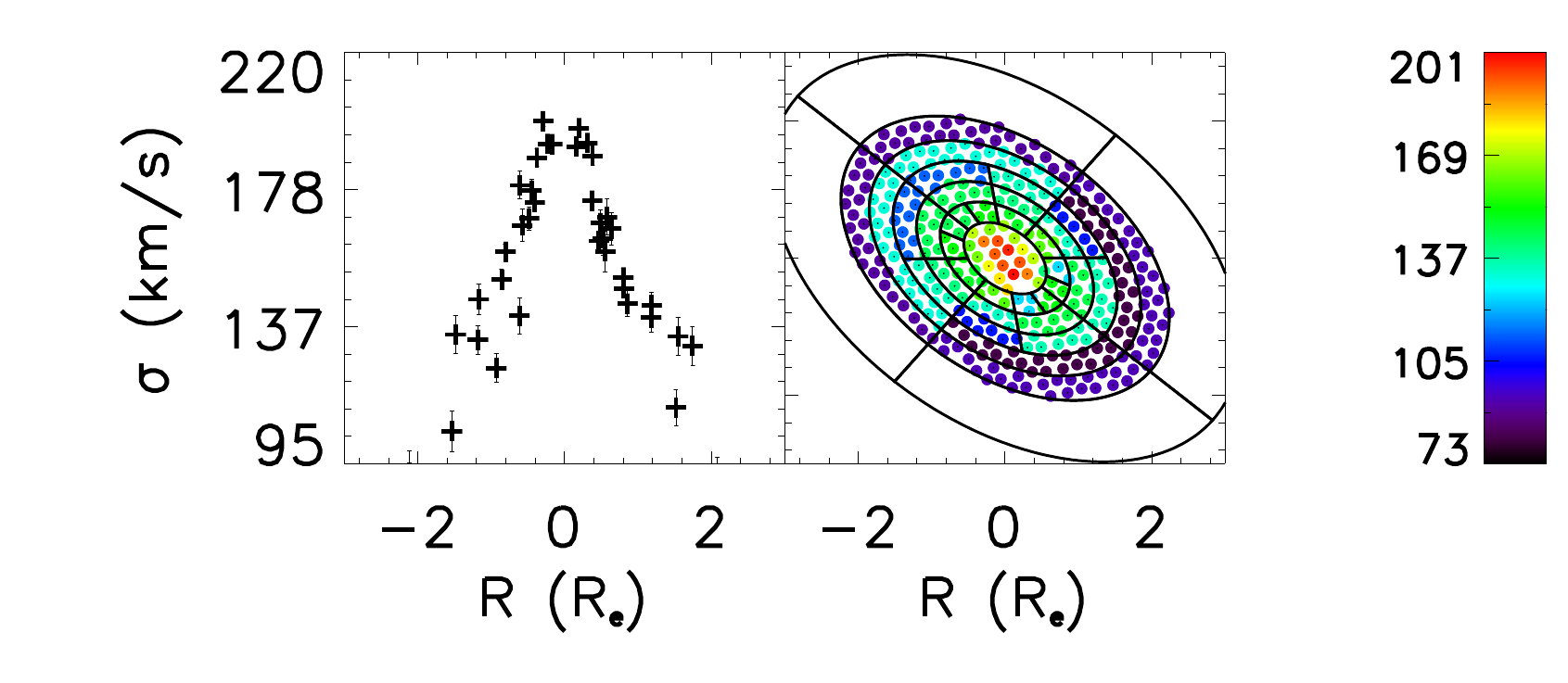}

\caption{Continued...}
\label{Fig:AllKinematicsb}
\end{center}
\end{figure}

\begin{figure}
\begin{center}

\includegraphics[width=0.6\textwidth,angle=0,clip]{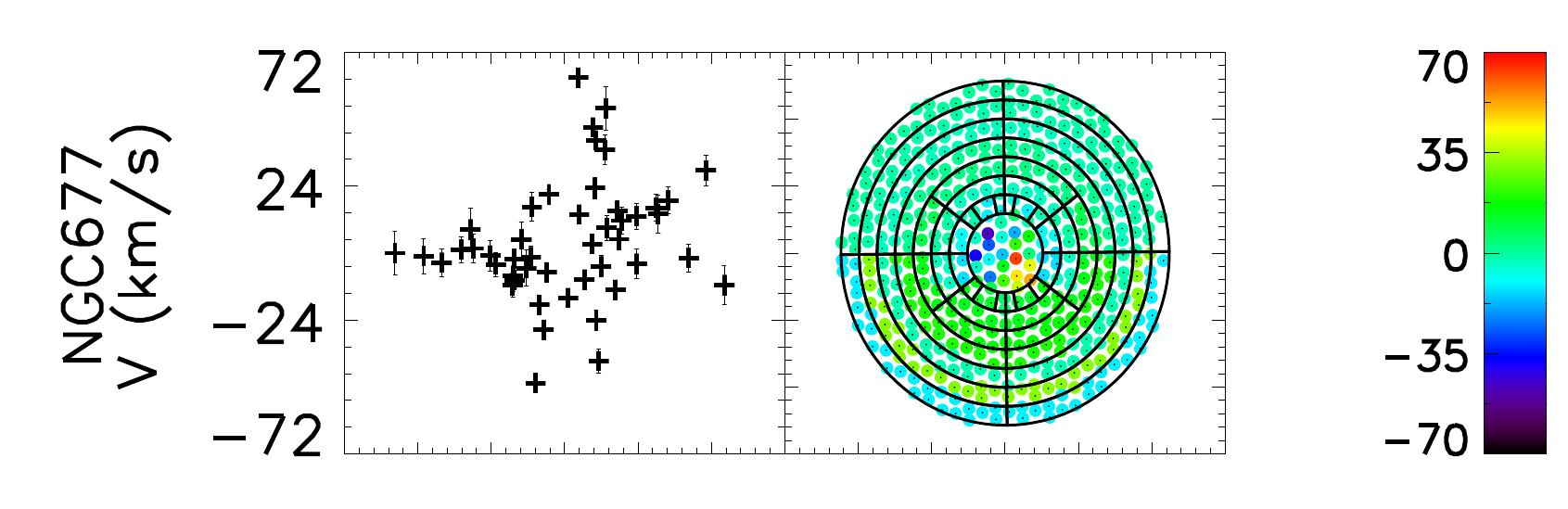}
\includegraphics[width=0.6\textwidth,angle=0,clip]{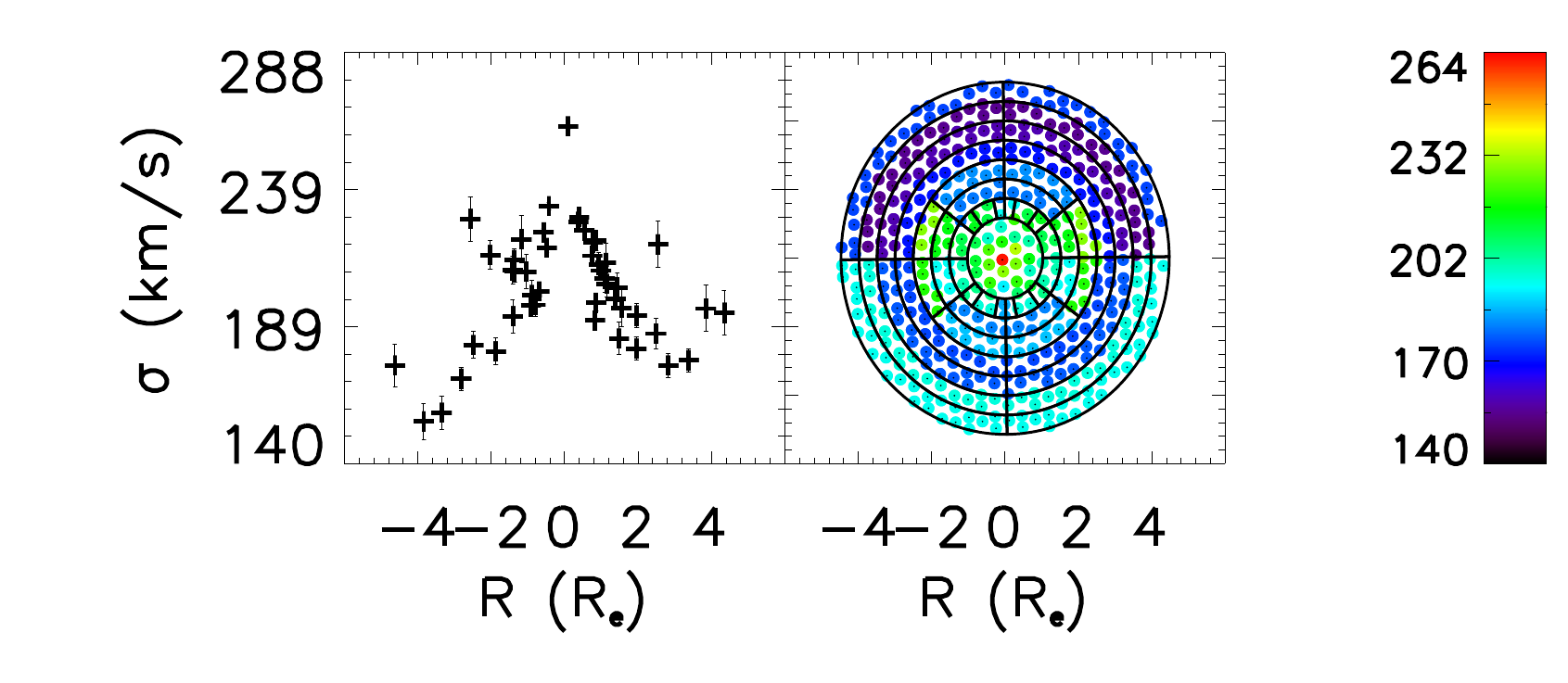}
\includegraphics[width=0.6\textwidth,angle=0,clip]{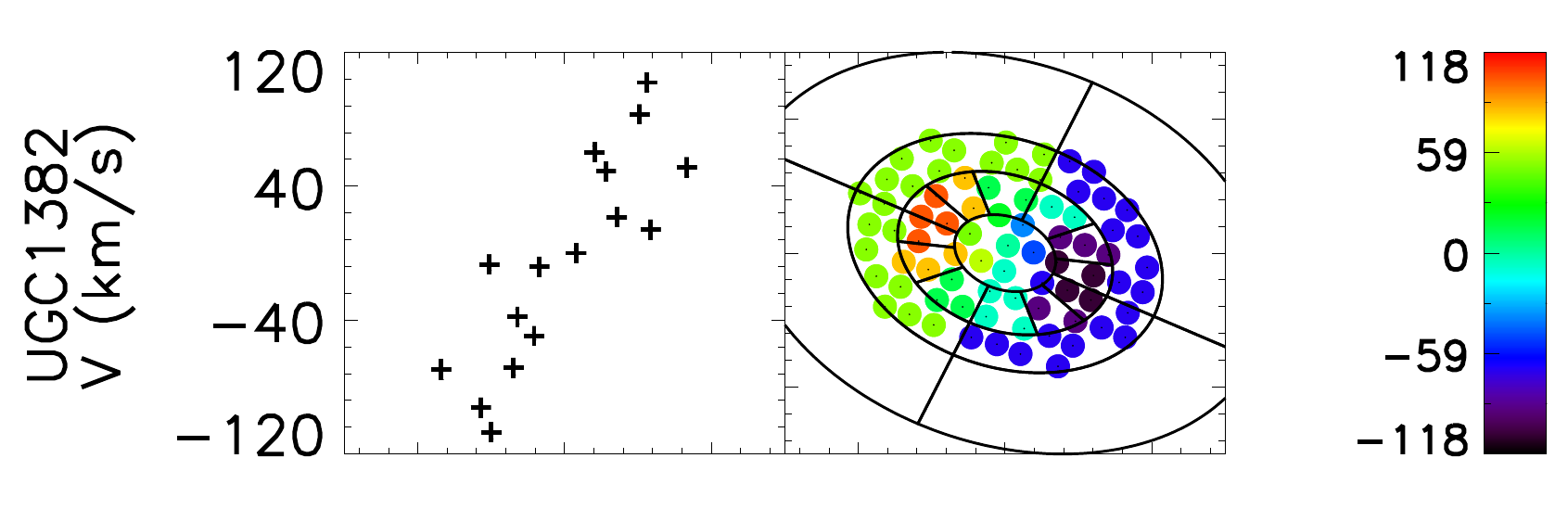}
\includegraphics[width=0.6\textwidth,angle=0,clip]{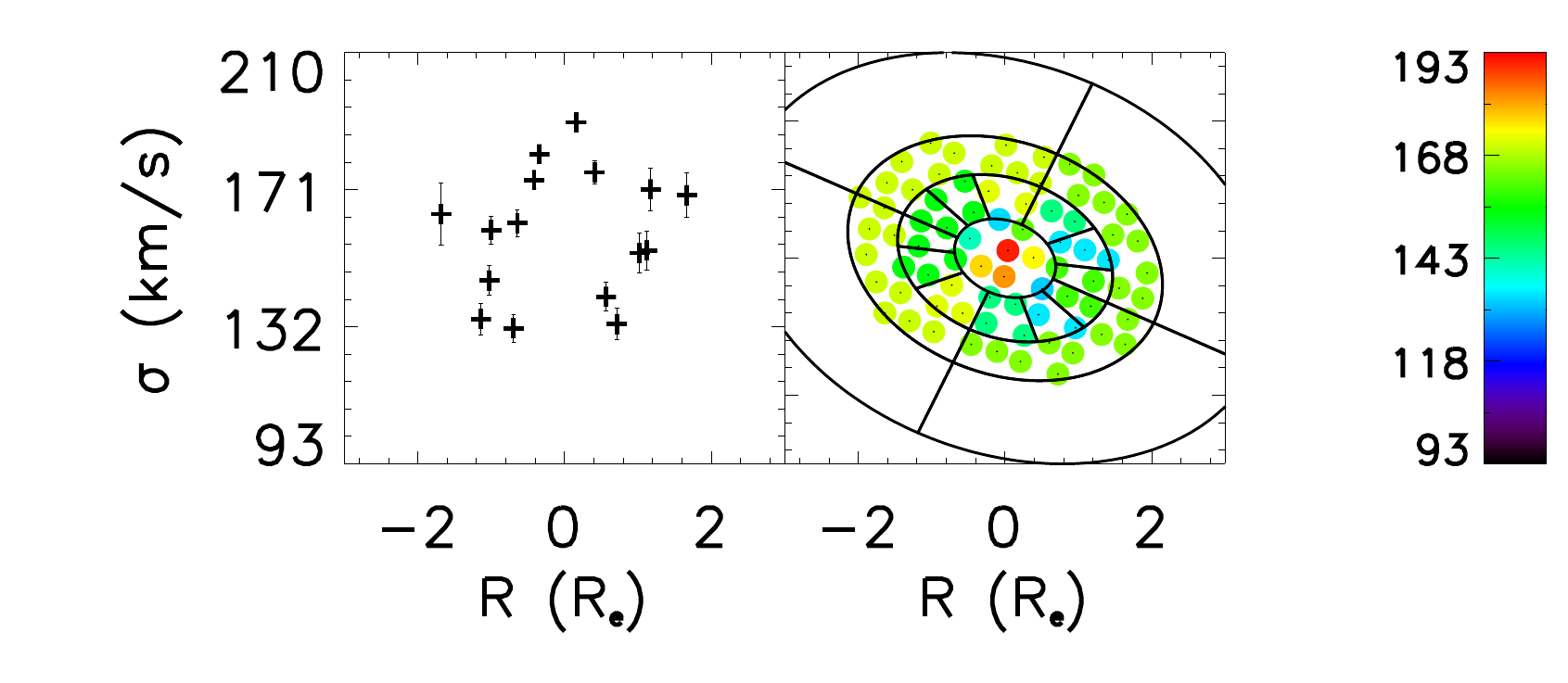}
\includegraphics[width=0.6\textwidth,angle=0,clip]{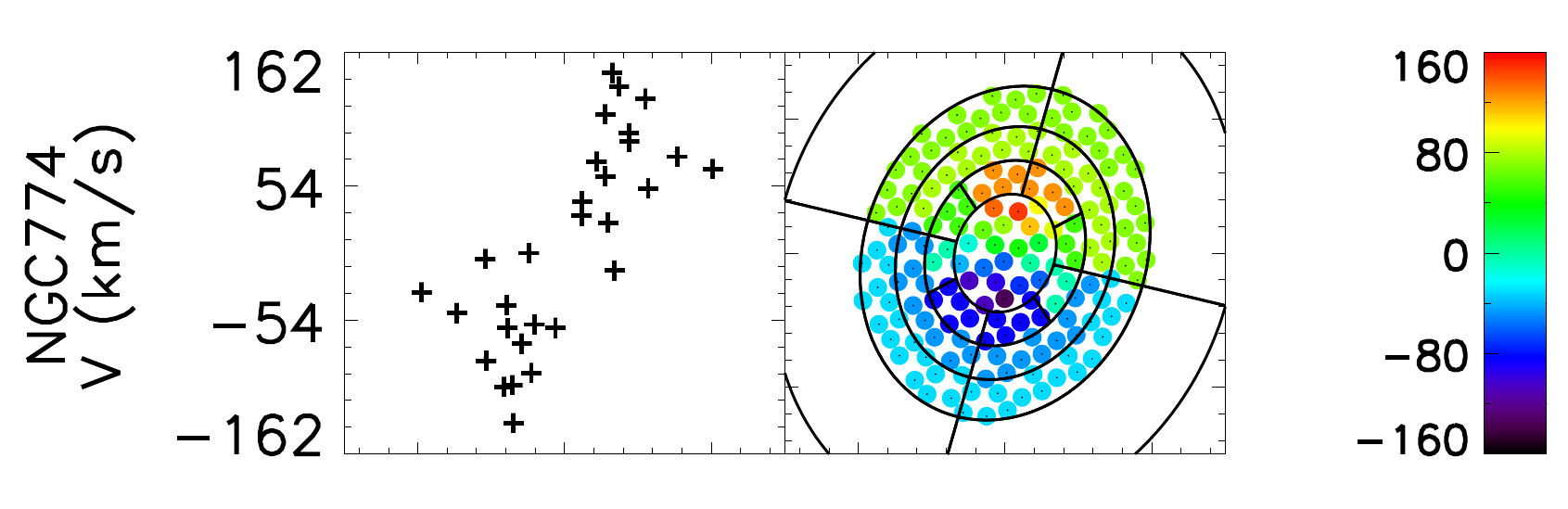}
\includegraphics[width=0.6\textwidth,angle=0,clip]{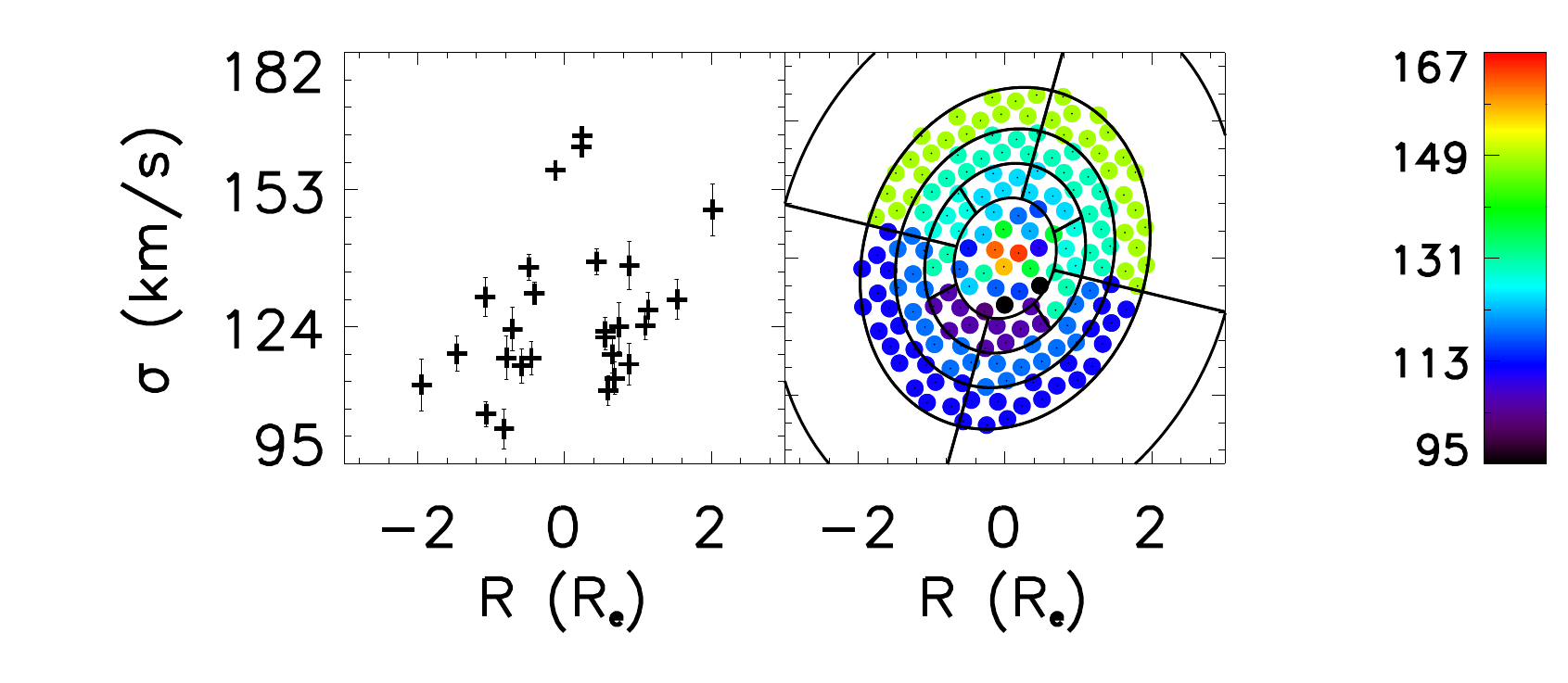}

\caption{Continued...}
\label{Fig:AllKinematicsc}
\end{center}
\end{figure}

\begin{figure}
\begin{center}

\includegraphics[width=0.6\textwidth,angle=0,clip]{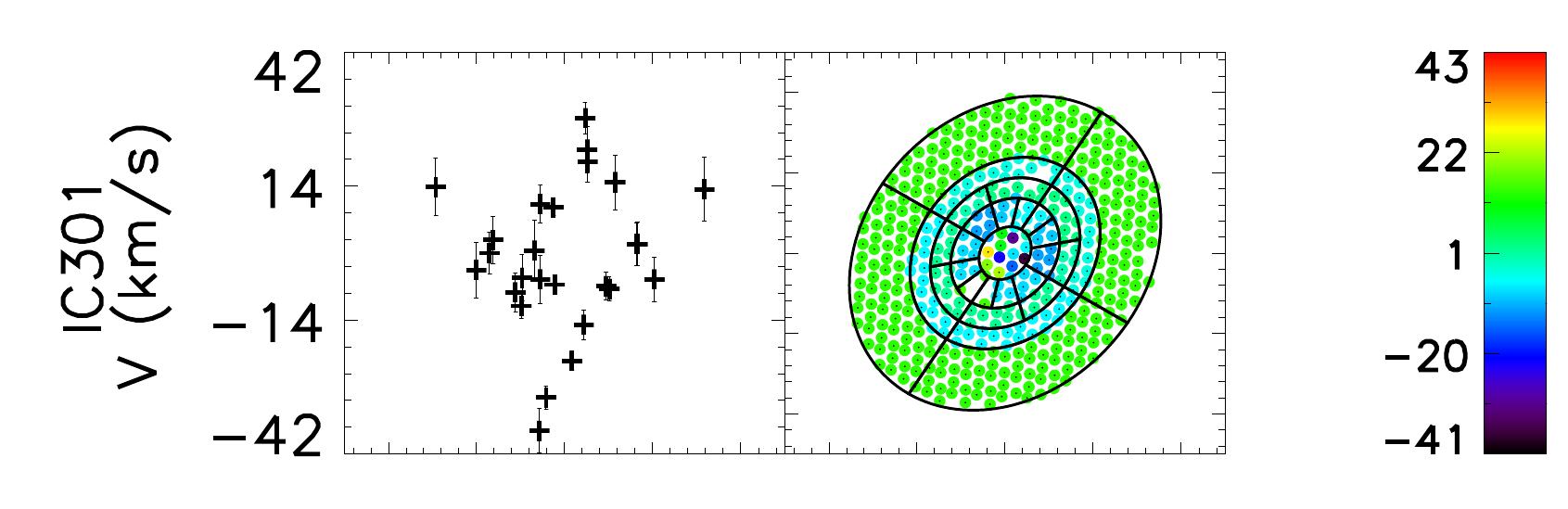}
\includegraphics[width=0.6\textwidth,angle=0,clip]{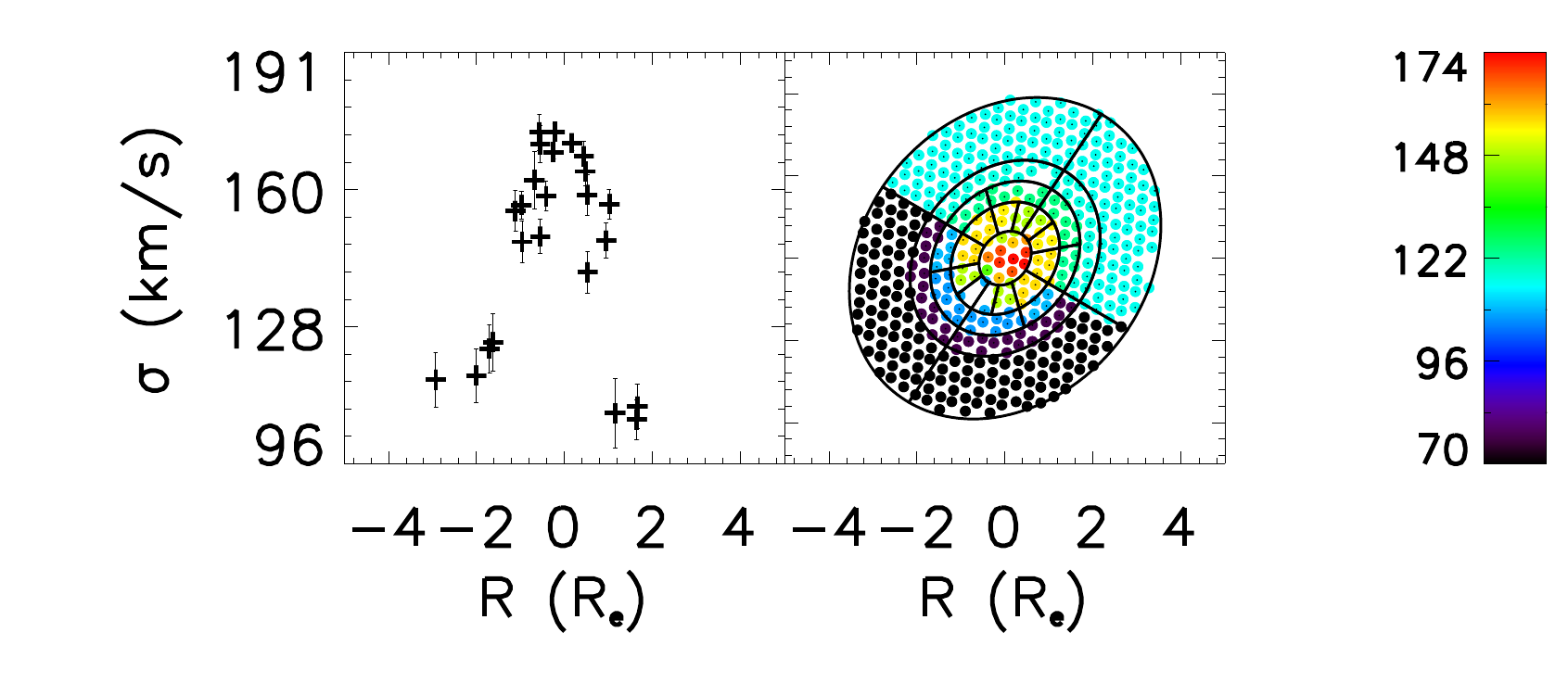}
\includegraphics[width=0.6\textwidth,angle=0,clip]{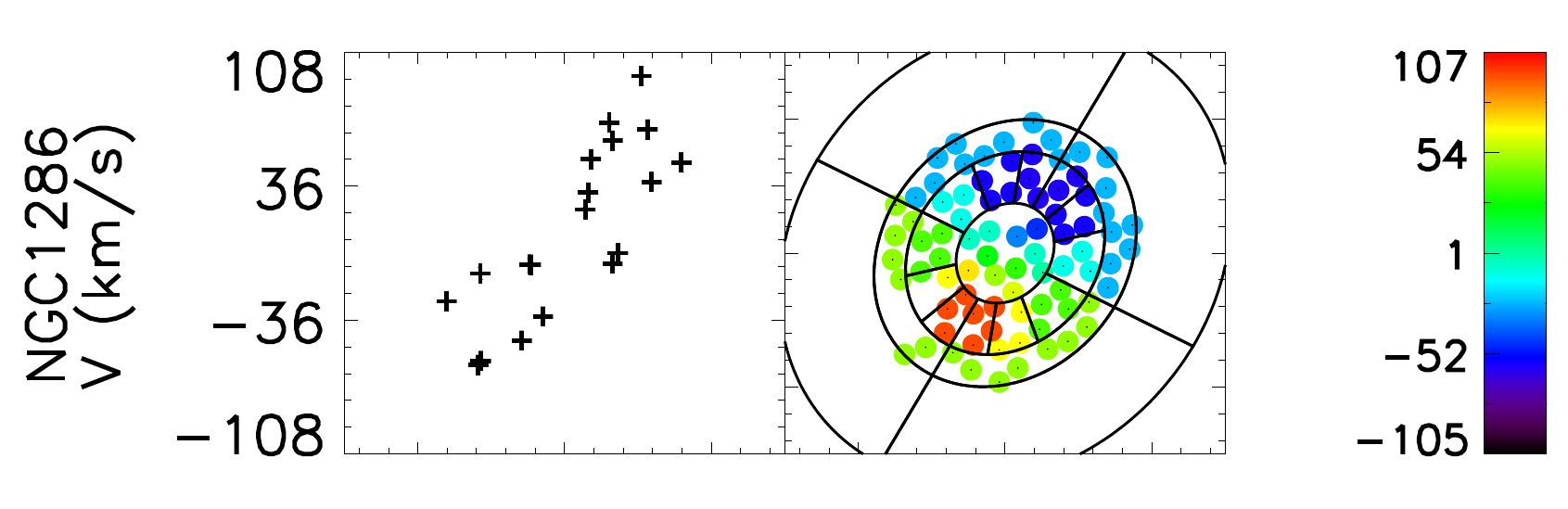}
\includegraphics[width=0.6\textwidth,angle=0,clip]{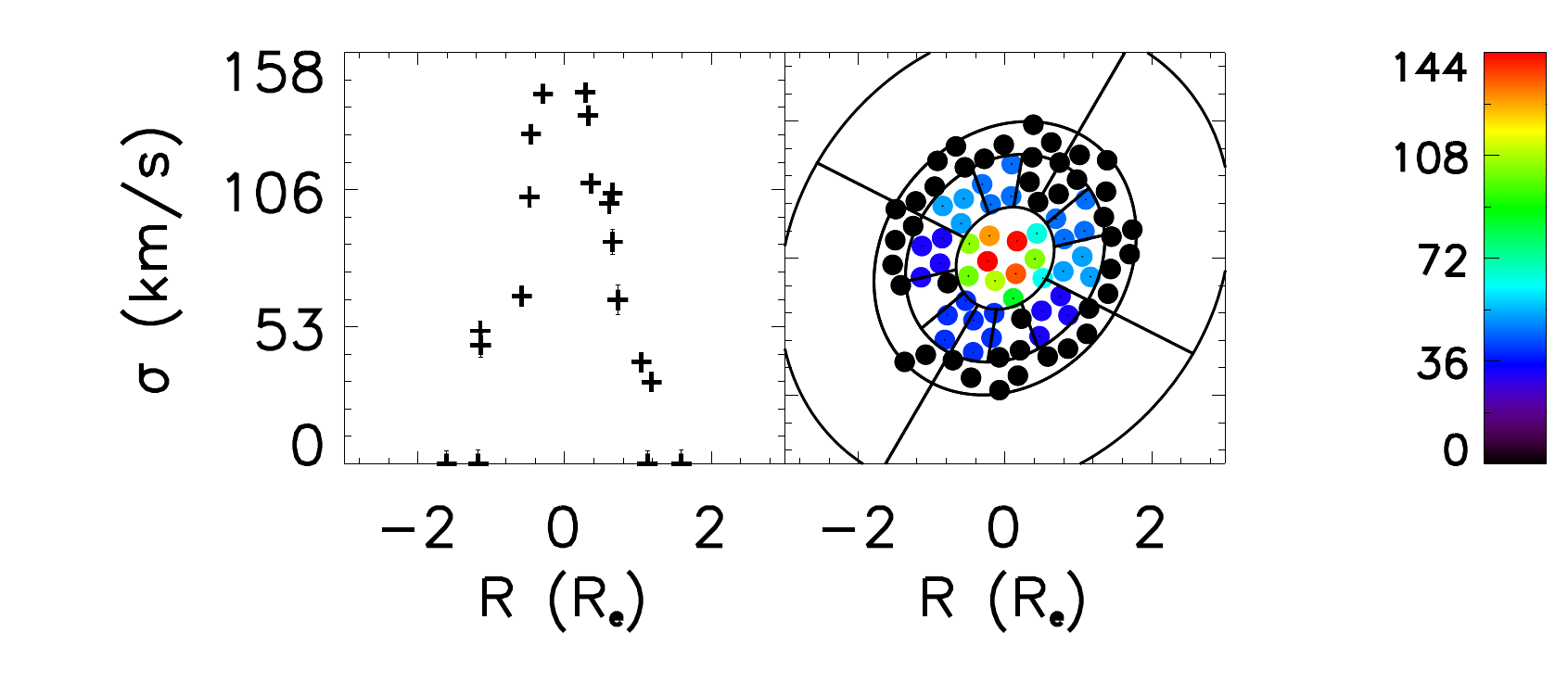}
\includegraphics[width=0.6\textwidth,angle=0,clip]{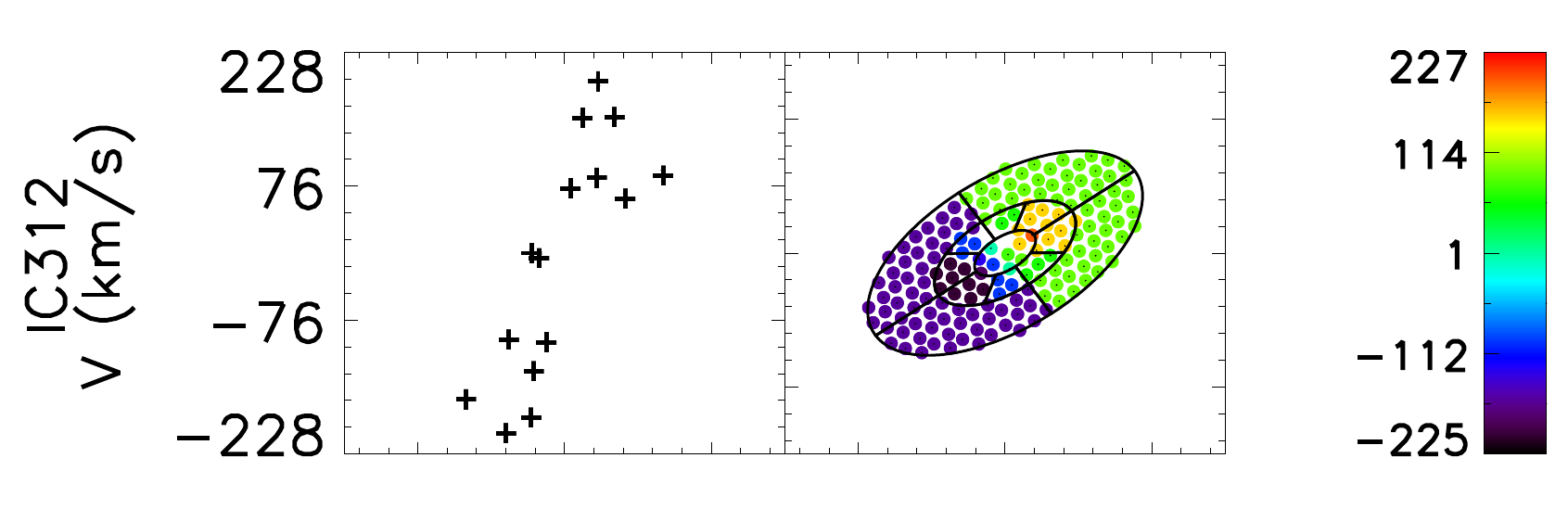}
\includegraphics[width=0.6\textwidth,angle=0,clip]{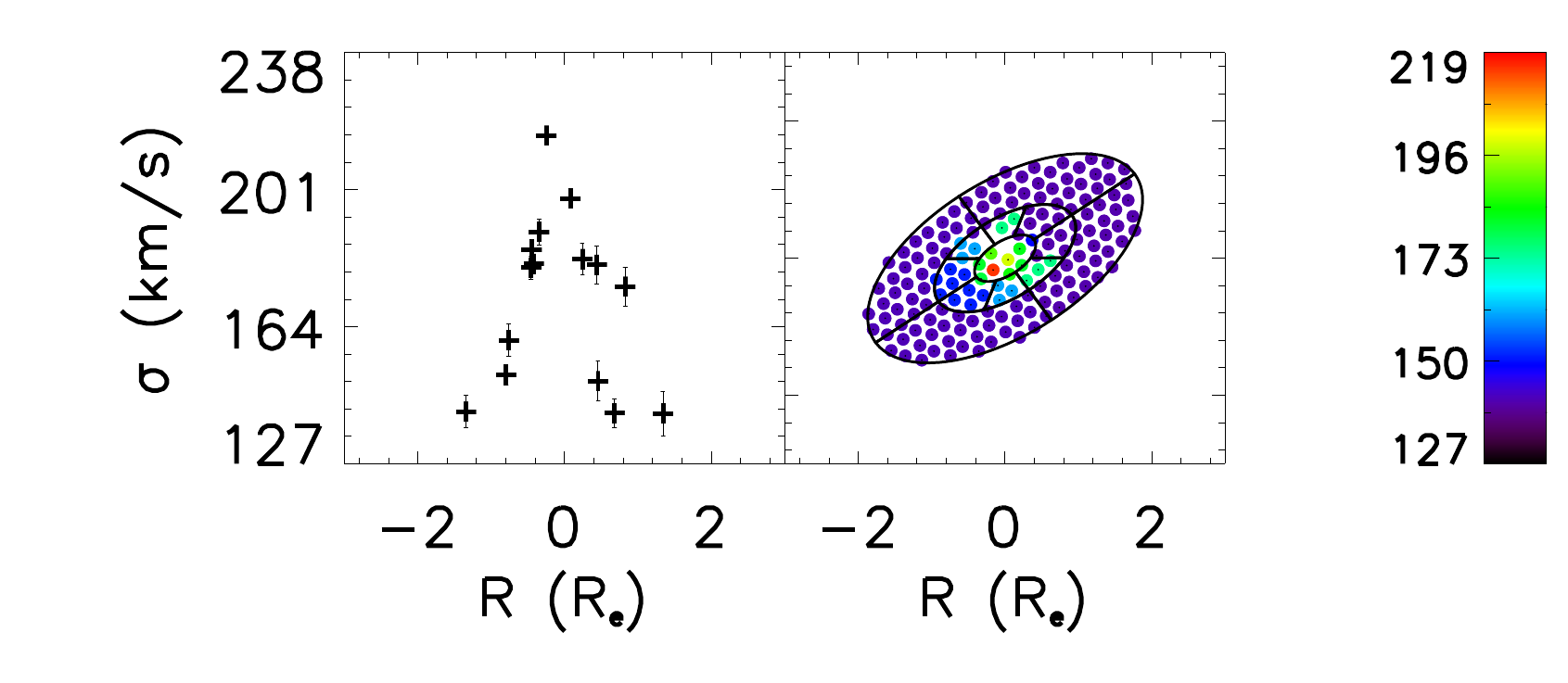}

\caption{Continued...}
\label{Fig:AllKinematicsd}
\end{center}
\end{figure}

\begin{figure}
\begin{center}

\includegraphics[width=0.6\textwidth,angle=0,clip]{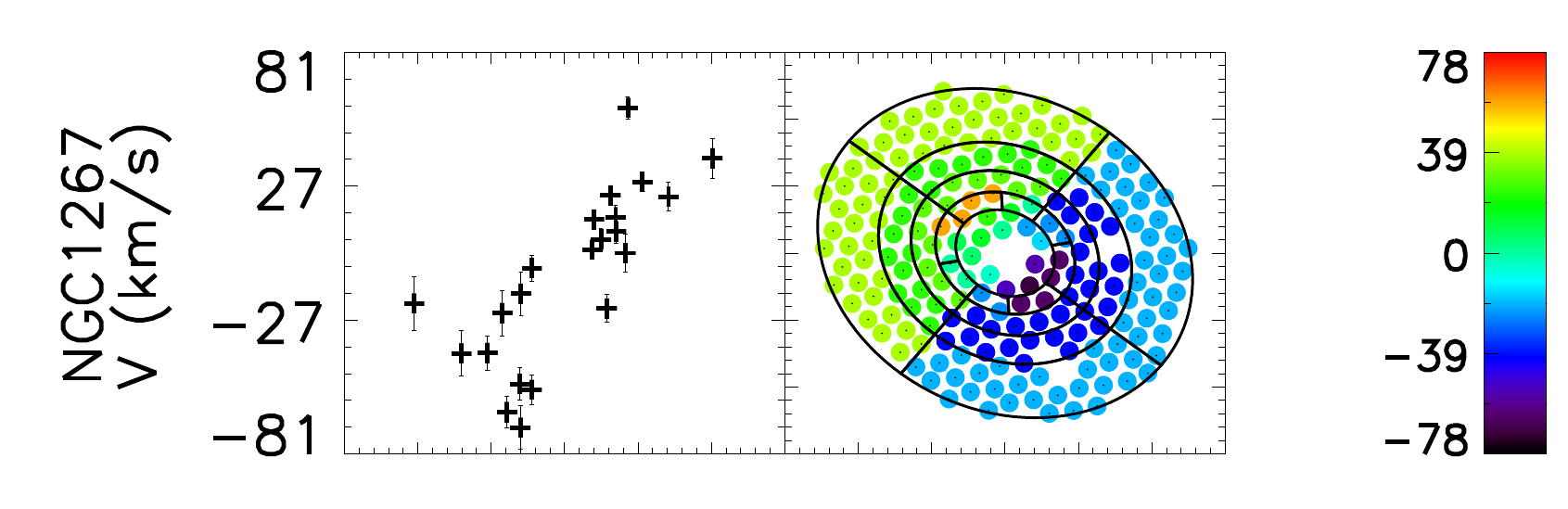}
\includegraphics[width=0.6\textwidth,angle=0,clip]{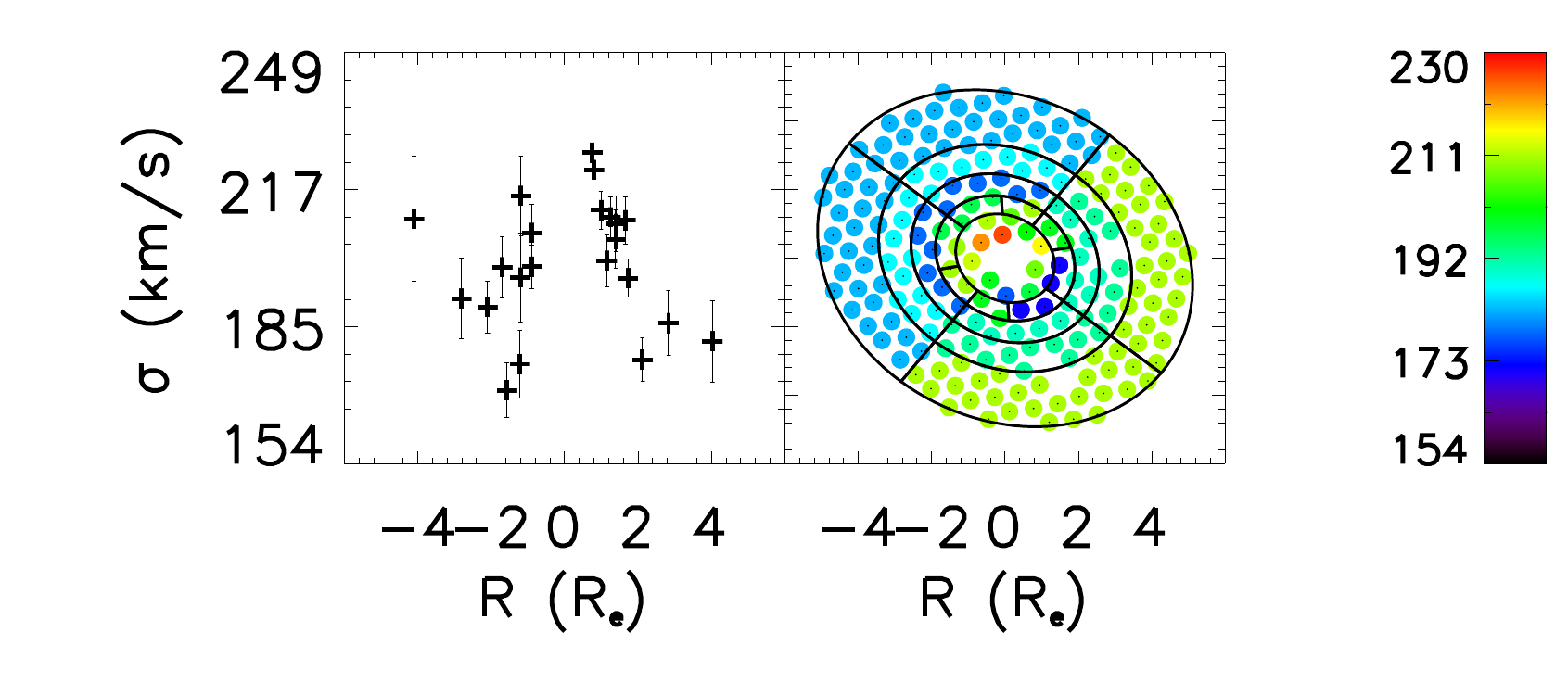}
\includegraphics[width=0.6\textwidth,angle=0,clip]{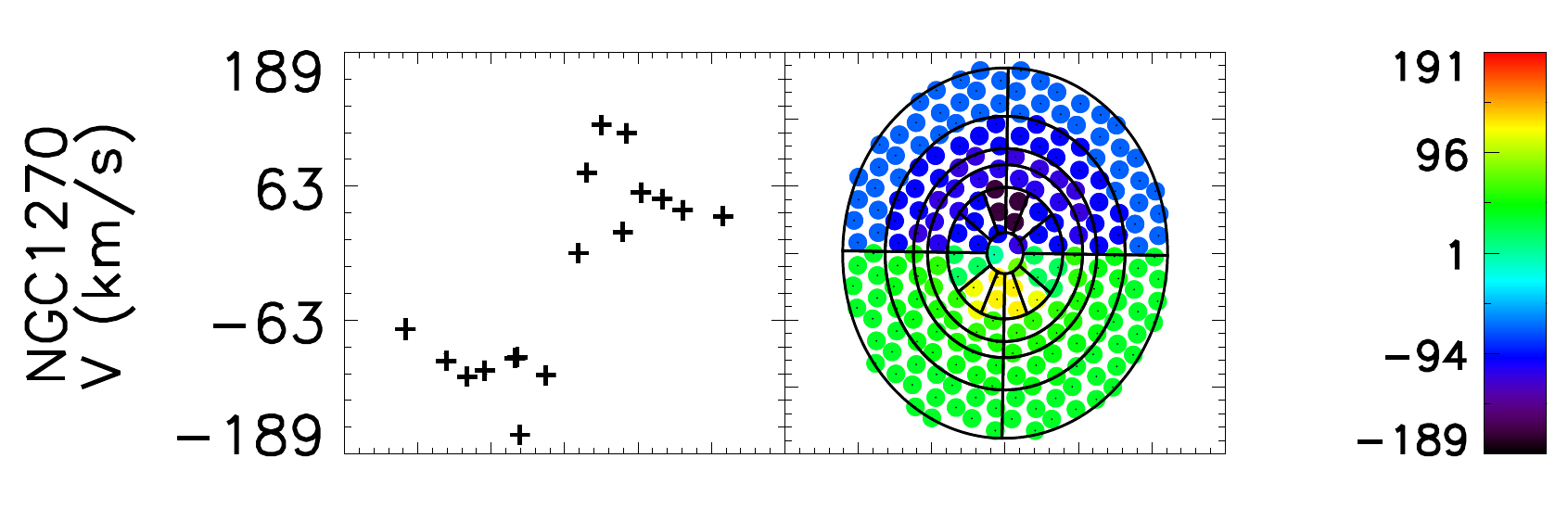}
\includegraphics[width=0.6\textwidth,angle=0,clip]{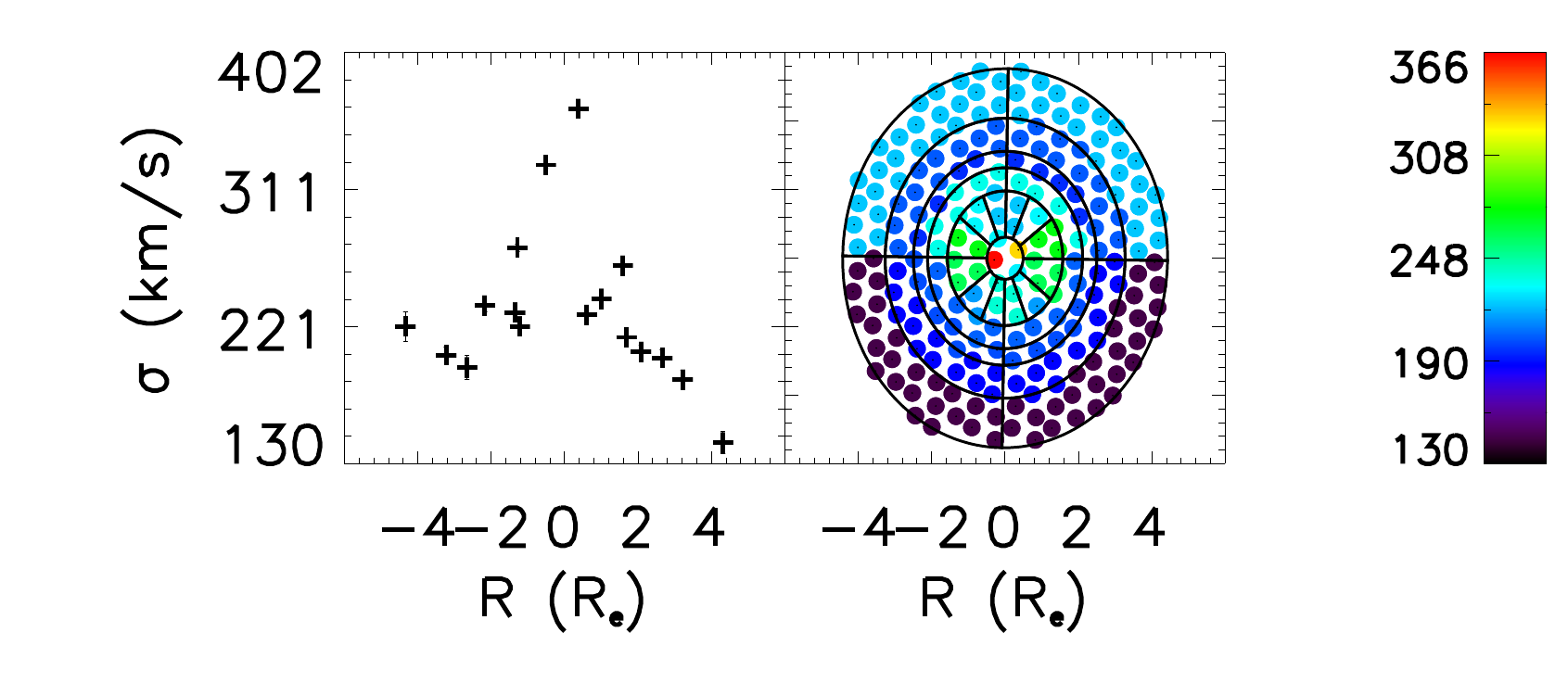}
\includegraphics[width=0.6\textwidth,angle=0,clip]{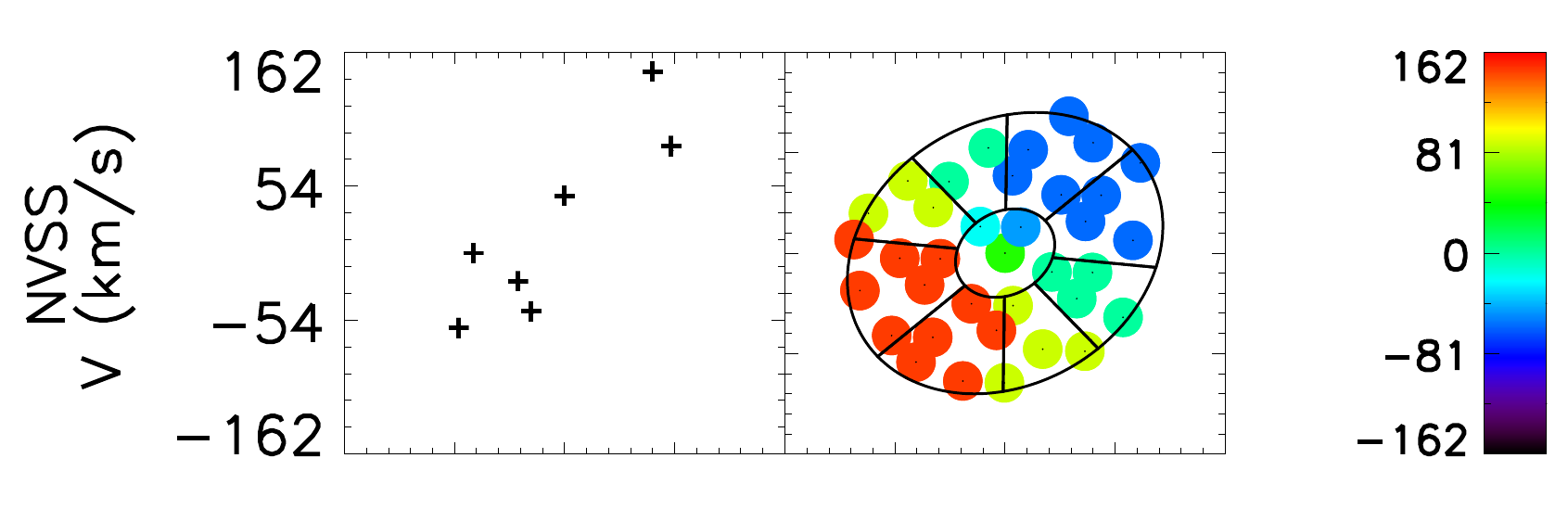}
\includegraphics[width=0.6\textwidth,angle=0,clip]{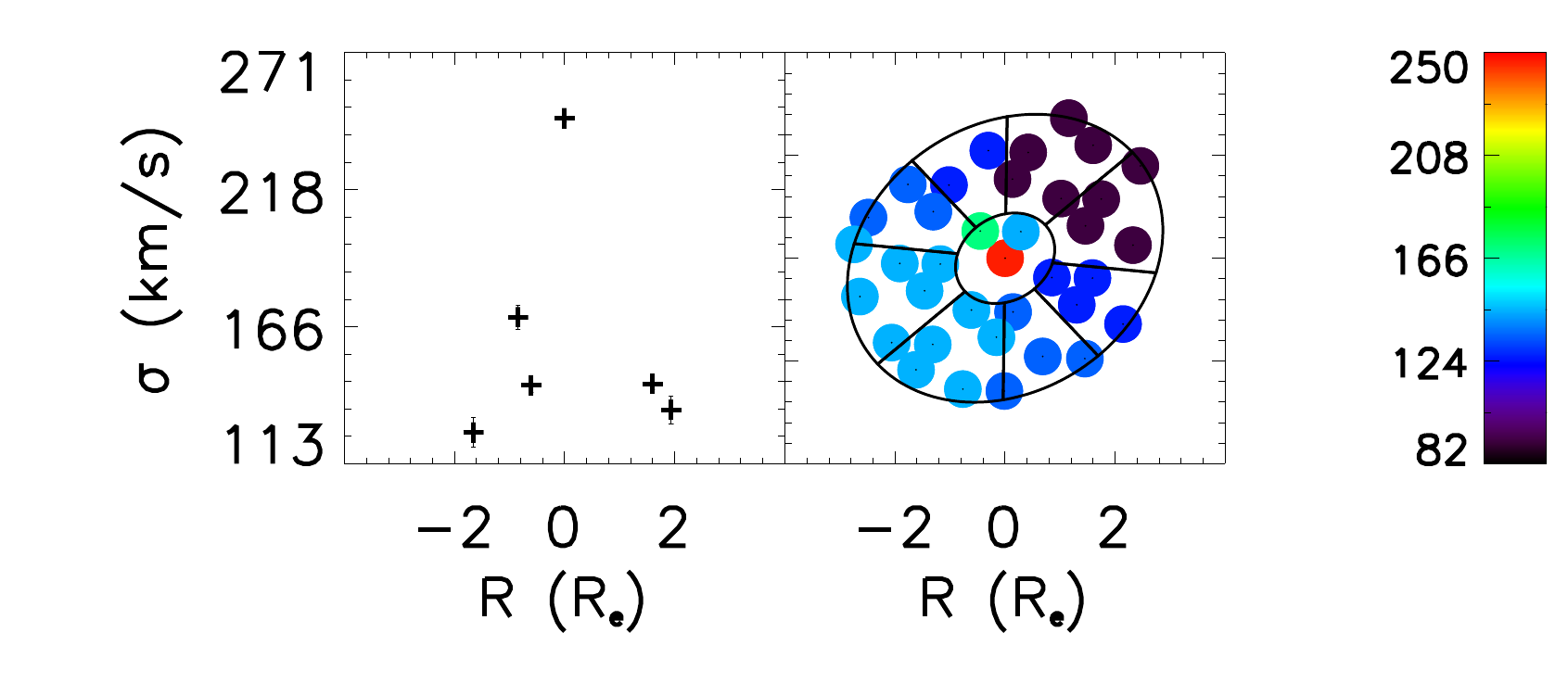}

\caption{Continued...}
\label{Fig:AllKinematicse}
\end{center}
\end{figure}

\begin{figure}
\begin{center}

\includegraphics[width=0.6\textwidth,angle=0,clip]{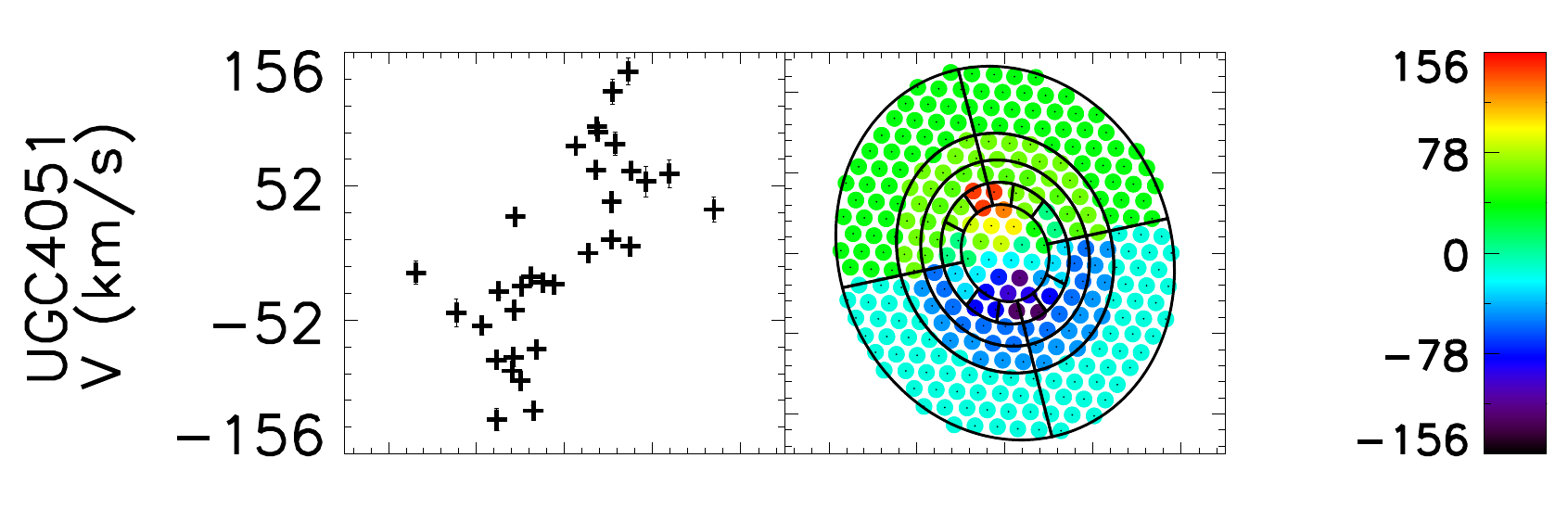}
\includegraphics[width=0.6\textwidth,angle=0,clip]{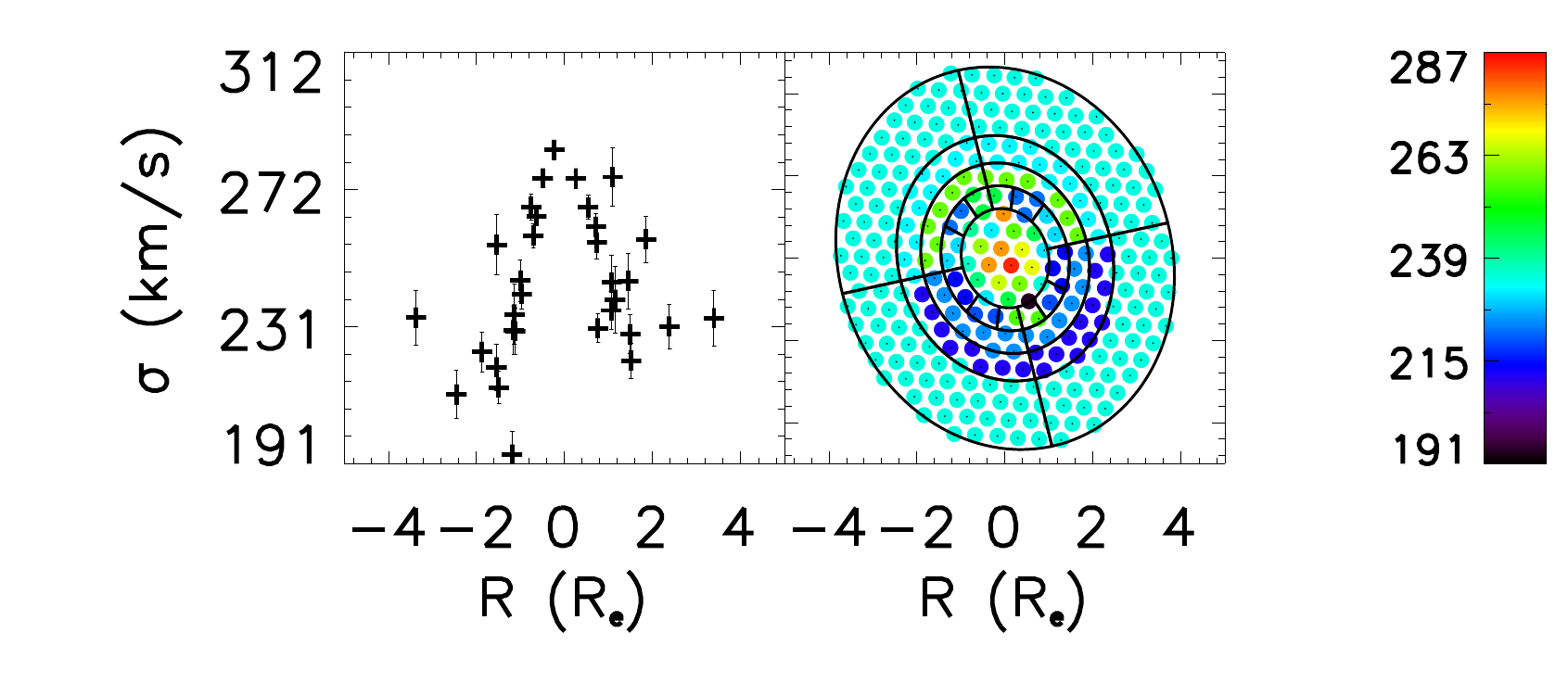}
\includegraphics[width=0.6\textwidth,angle=0,clip]{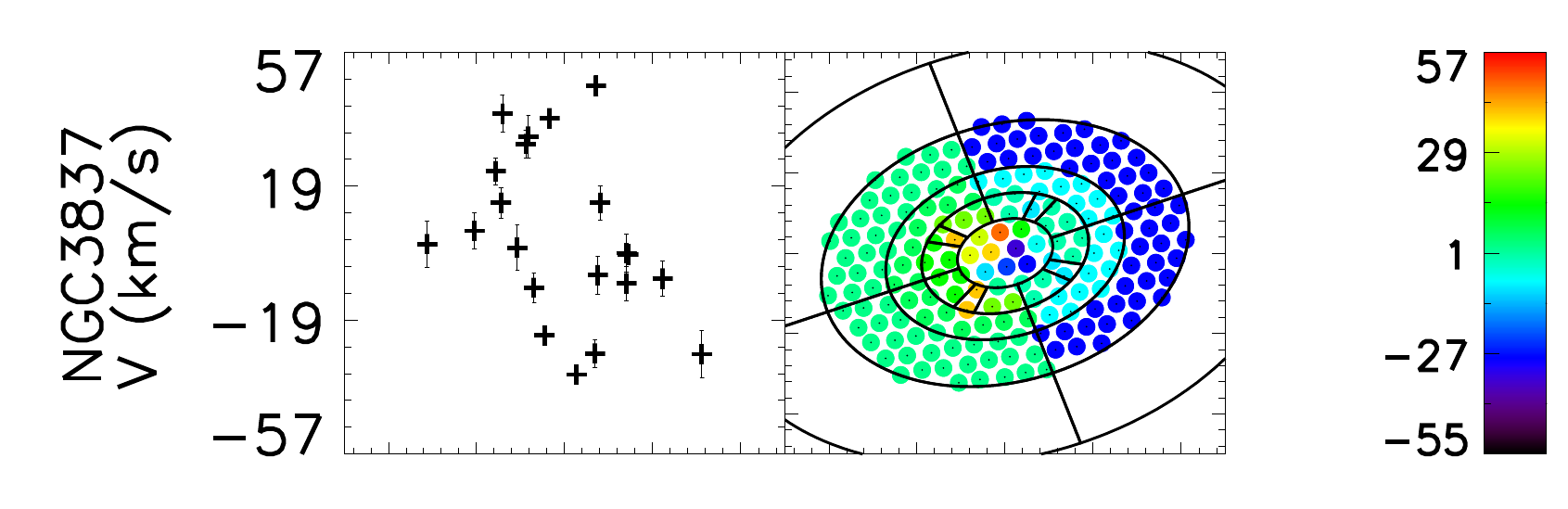}
\includegraphics[width=0.6\textwidth,angle=0,clip]{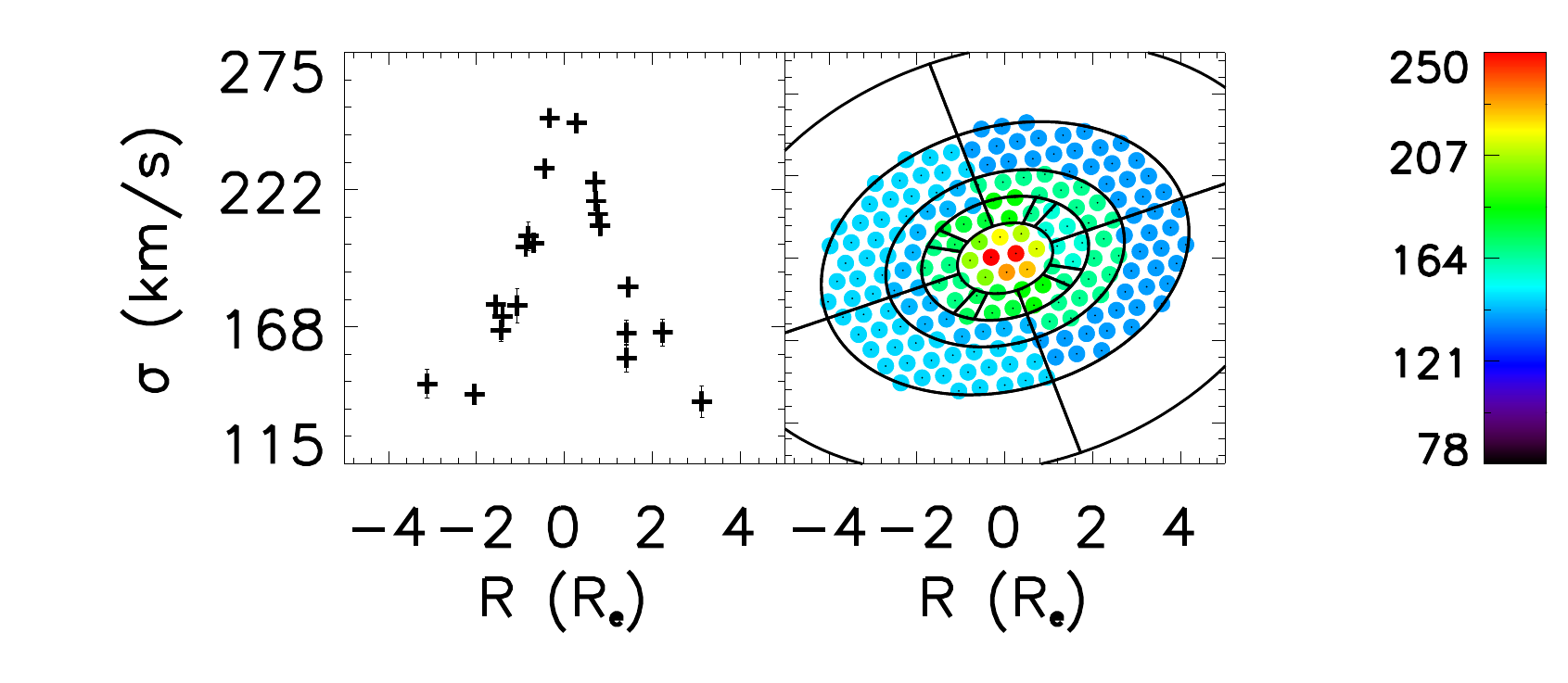}
\includegraphics[width=0.6\textwidth,angle=0,clip]{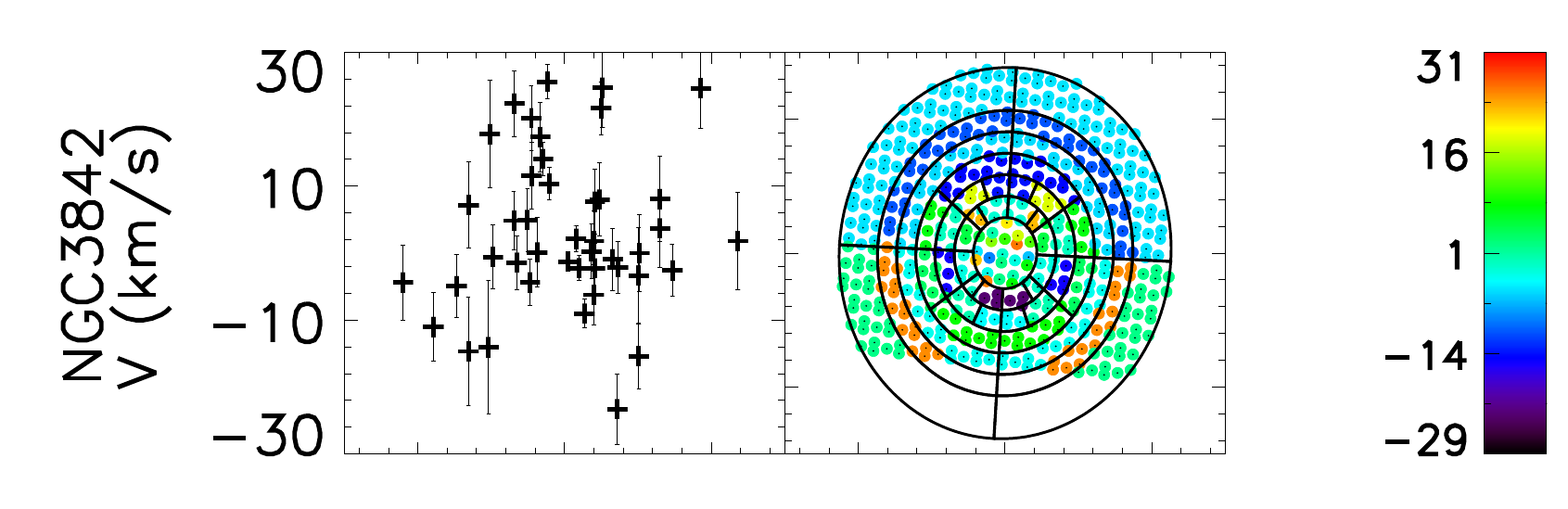}
\includegraphics[width=0.6\textwidth,angle=0,clip]{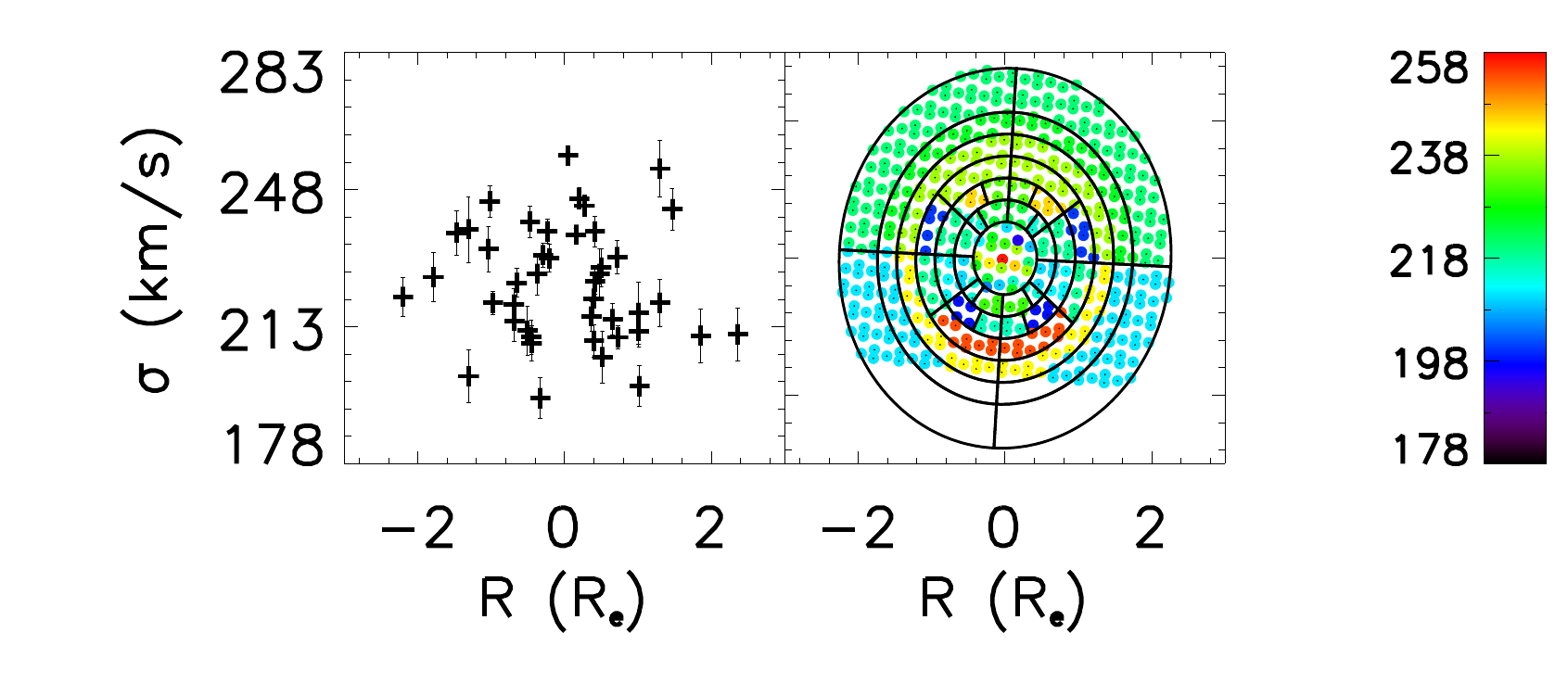}

\caption{Continued...}
\label{Fig:AllKinematicsf}
\end{center}
\end{figure}

\begin{figure}
\begin{center}

\includegraphics[width=0.6\textwidth,angle=0,clip]{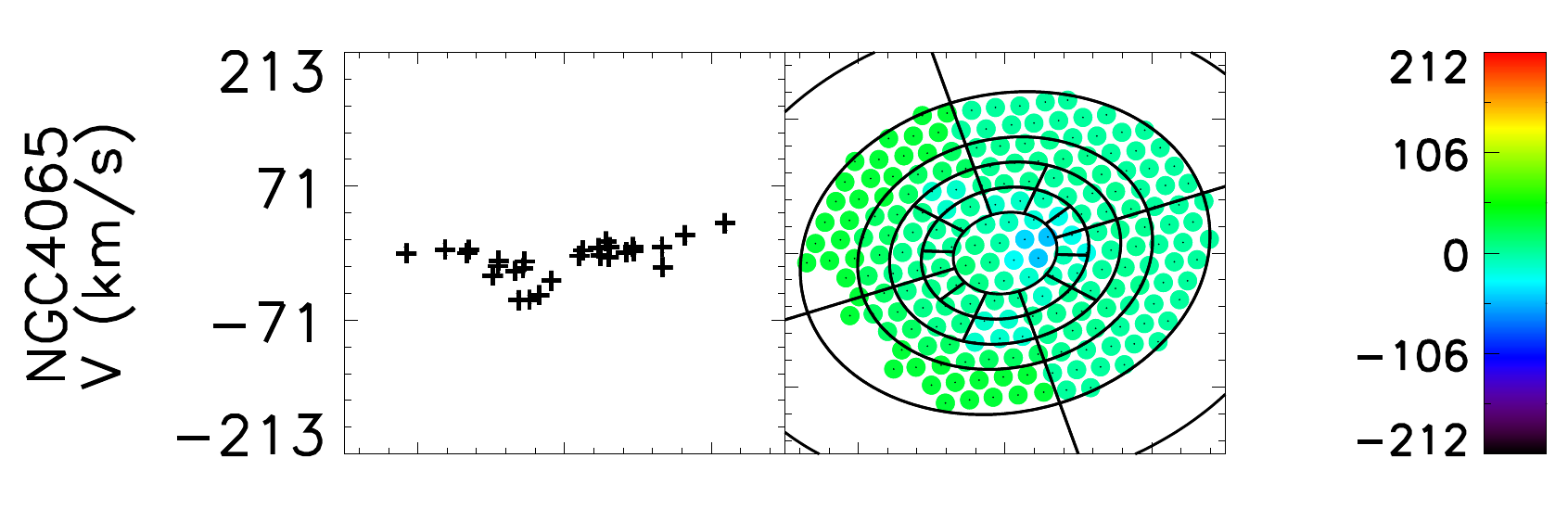}
\includegraphics[width=0.6\textwidth,angle=0,clip]{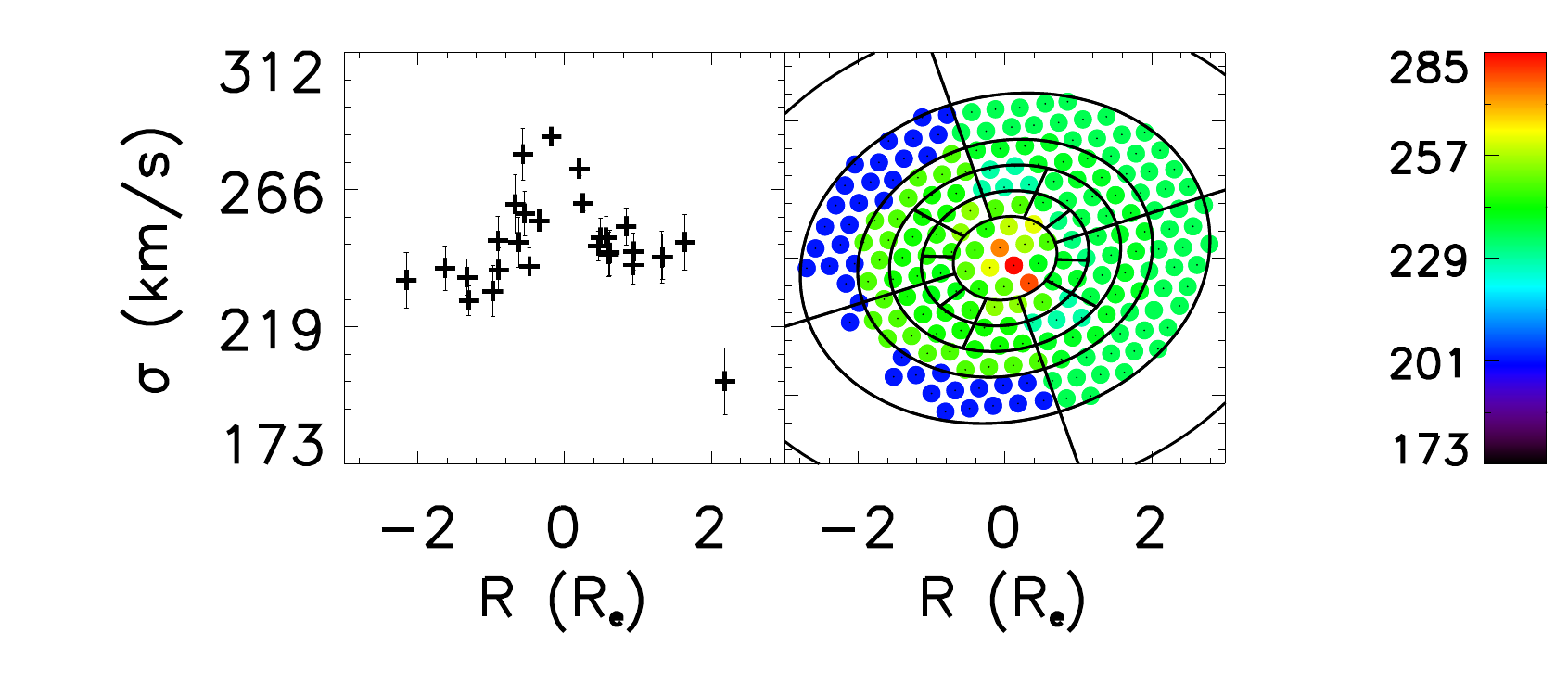}
\includegraphics[width=0.6\textwidth,angle=0,clip]{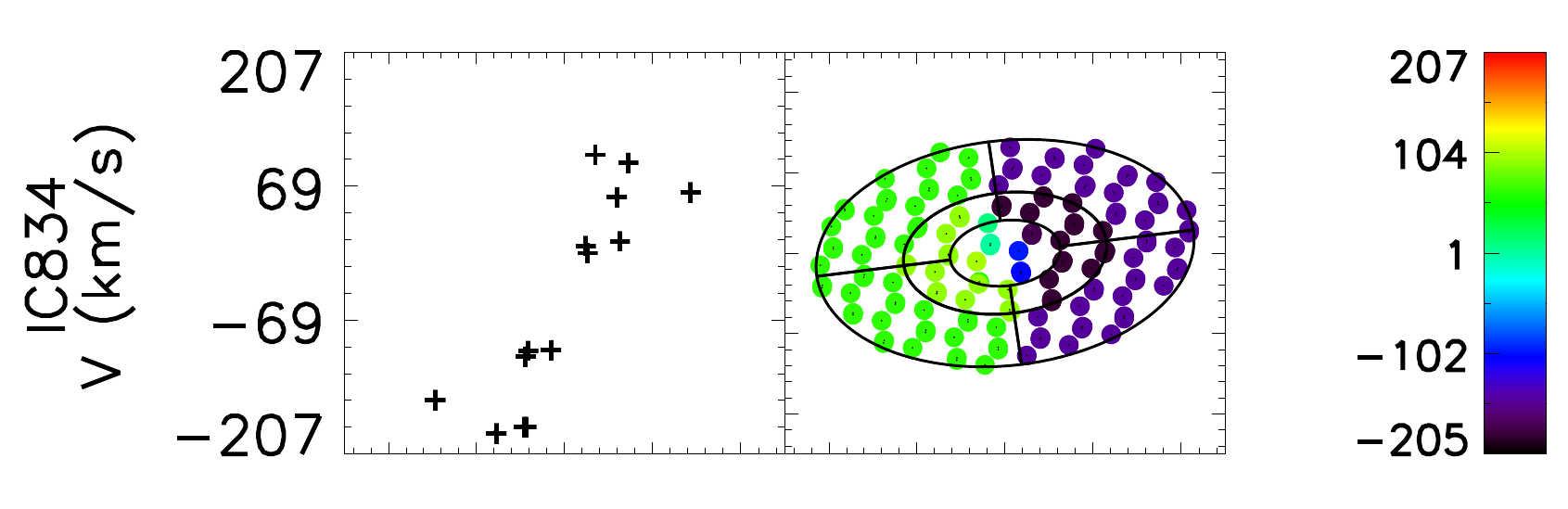}
\includegraphics[width=0.6\textwidth,angle=0,clip]{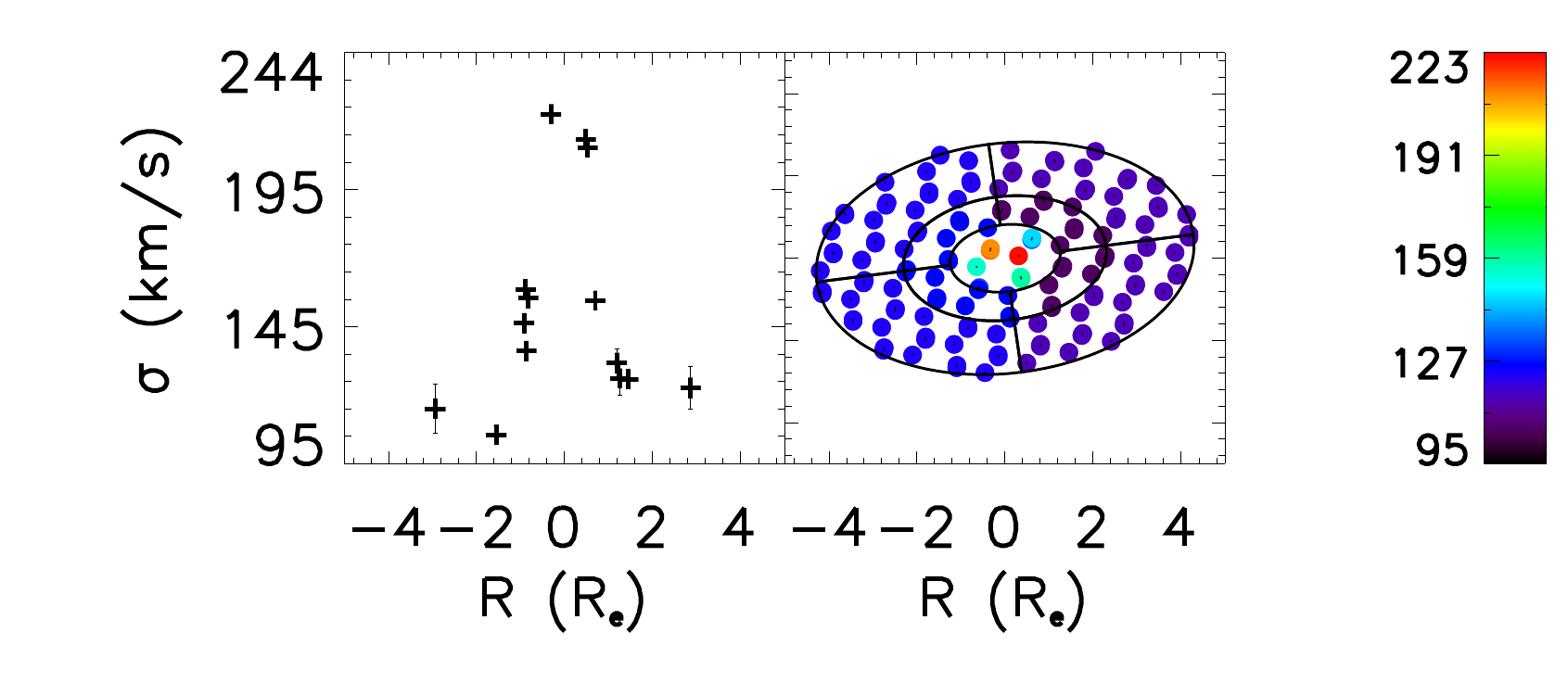}
\includegraphics[width=0.6\textwidth,angle=0,clip]{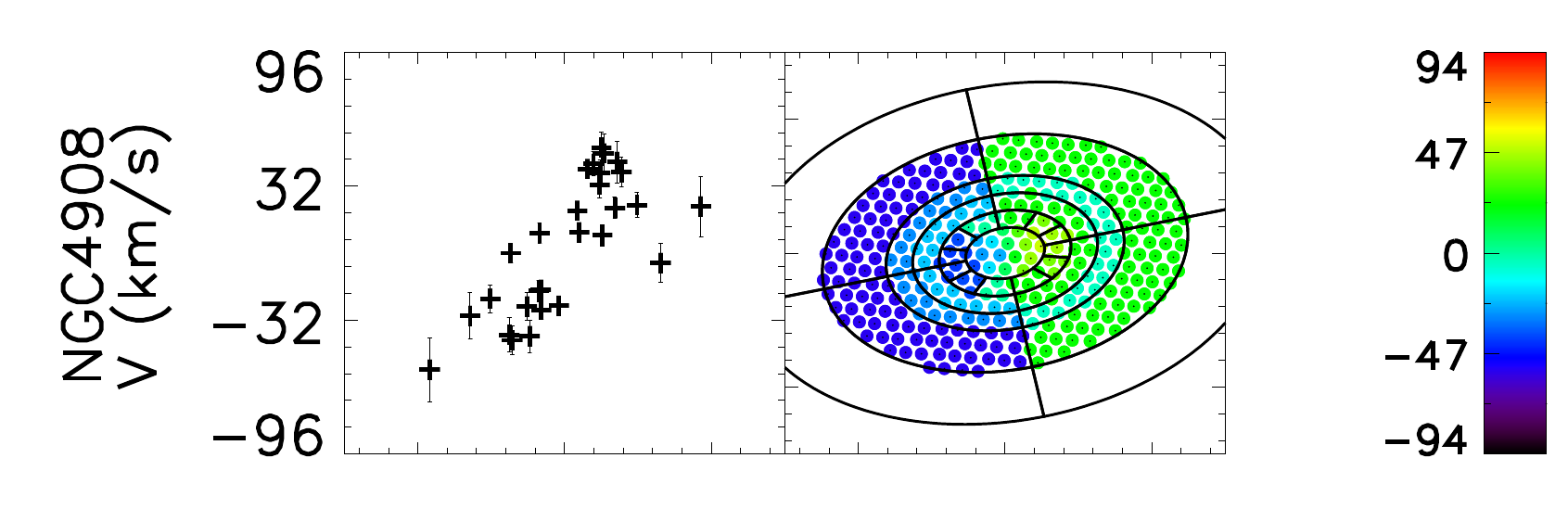}
\includegraphics[width=0.6\textwidth,angle=0,clip]{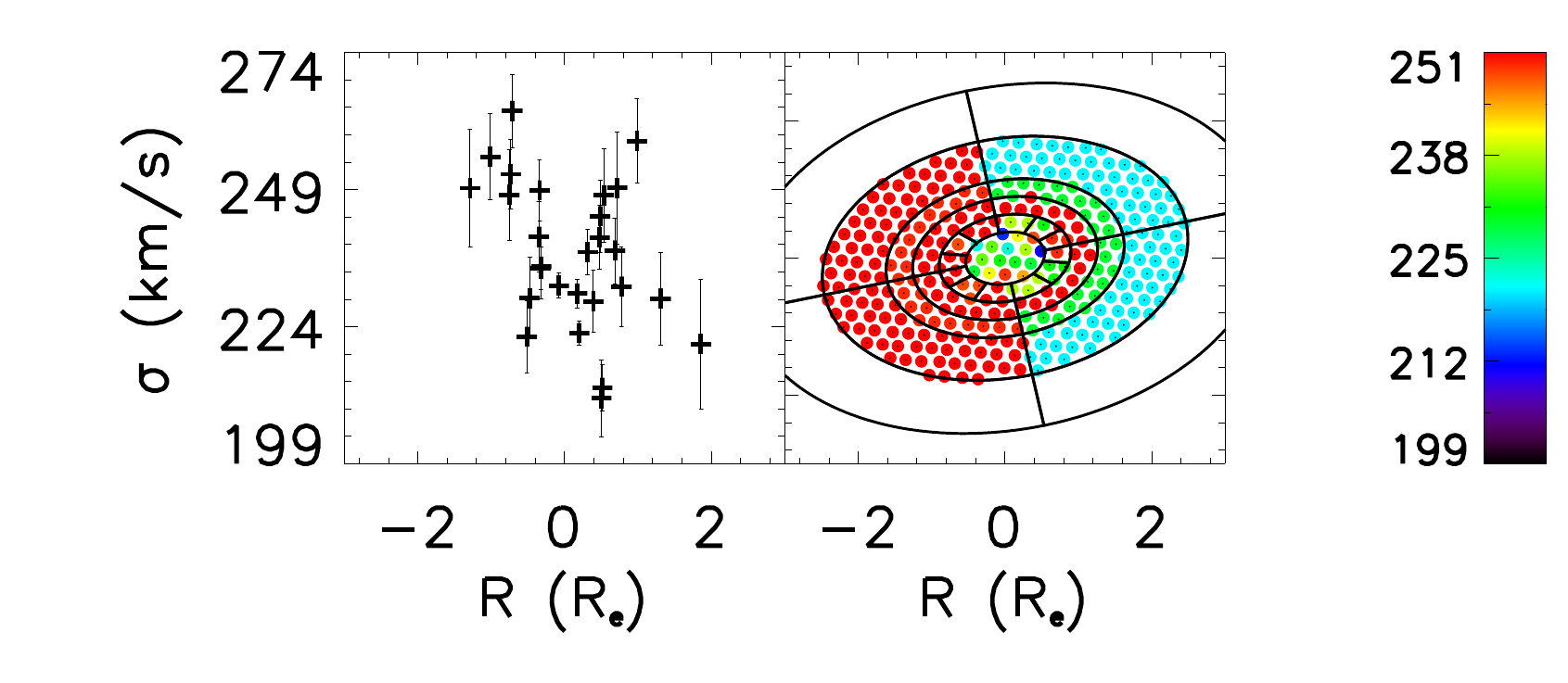}

\caption{Continued...}
\label{Fig:AllKinematicsg}
\end{center}
\end{figure}

\begin{figure}
\begin{center}

\includegraphics[width=0.6\textwidth,angle=0,clip]{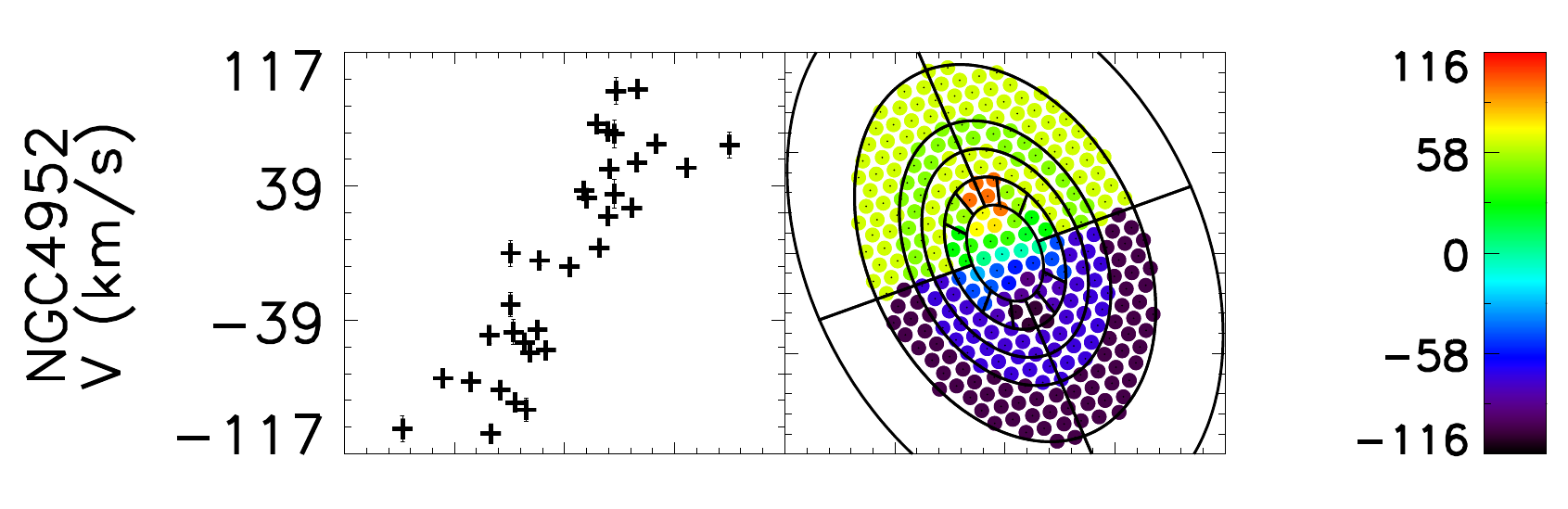}
\includegraphics[width=0.6\textwidth,angle=0,clip]{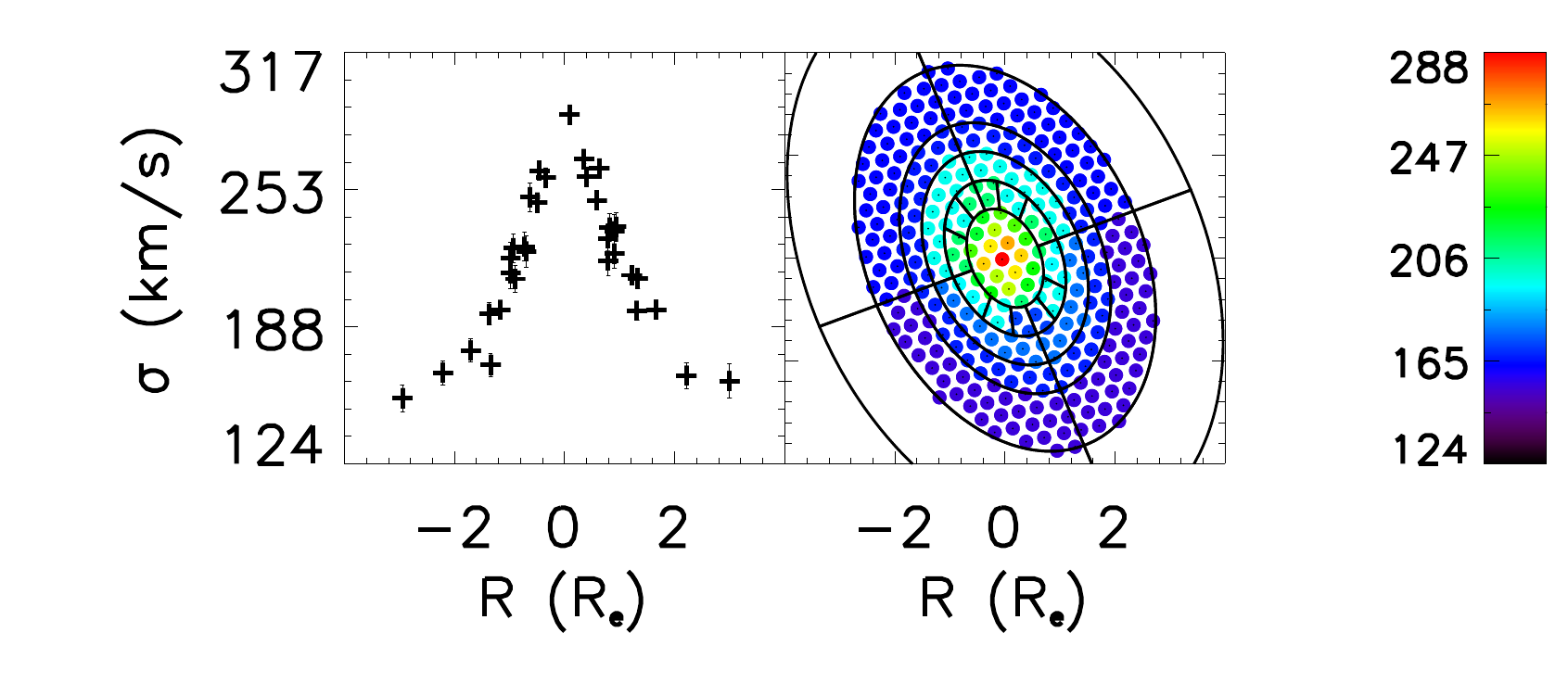}
\includegraphics[width=0.6\textwidth,angle=0,clip]{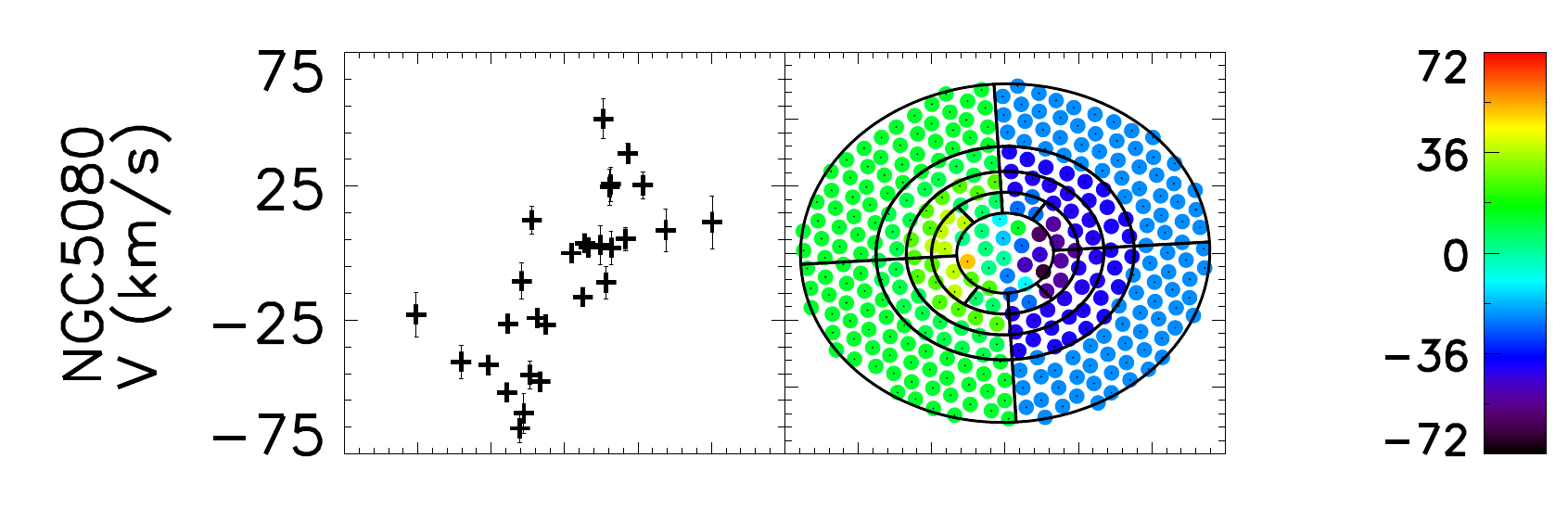}
\includegraphics[width=0.6\textwidth,angle=0,clip]{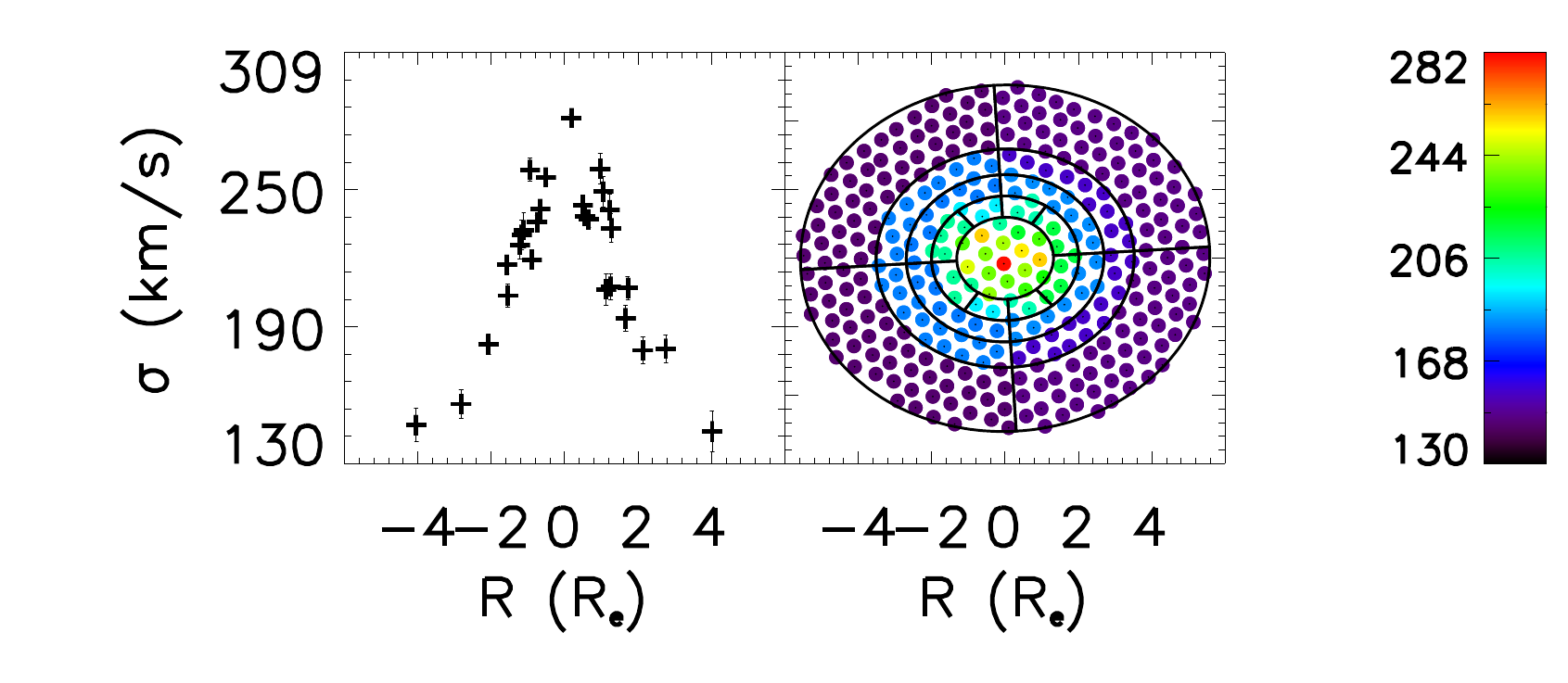}
\includegraphics[width=0.6\textwidth,angle=0,clip]{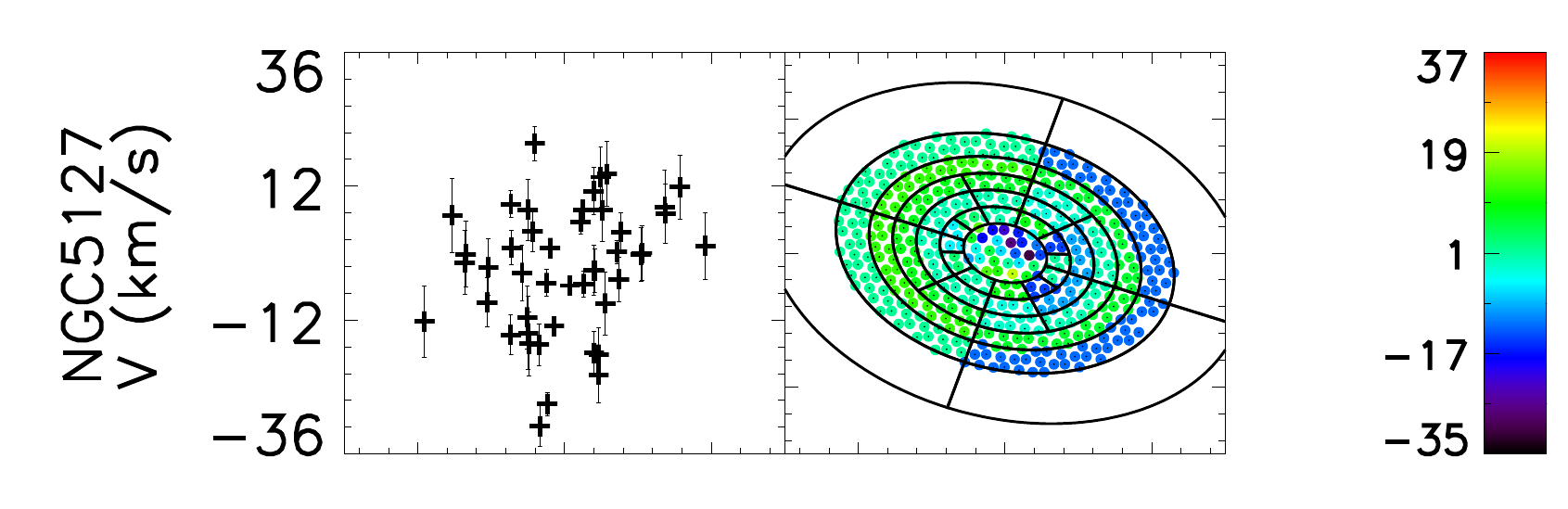}
\includegraphics[width=0.6\textwidth,angle=0,clip]{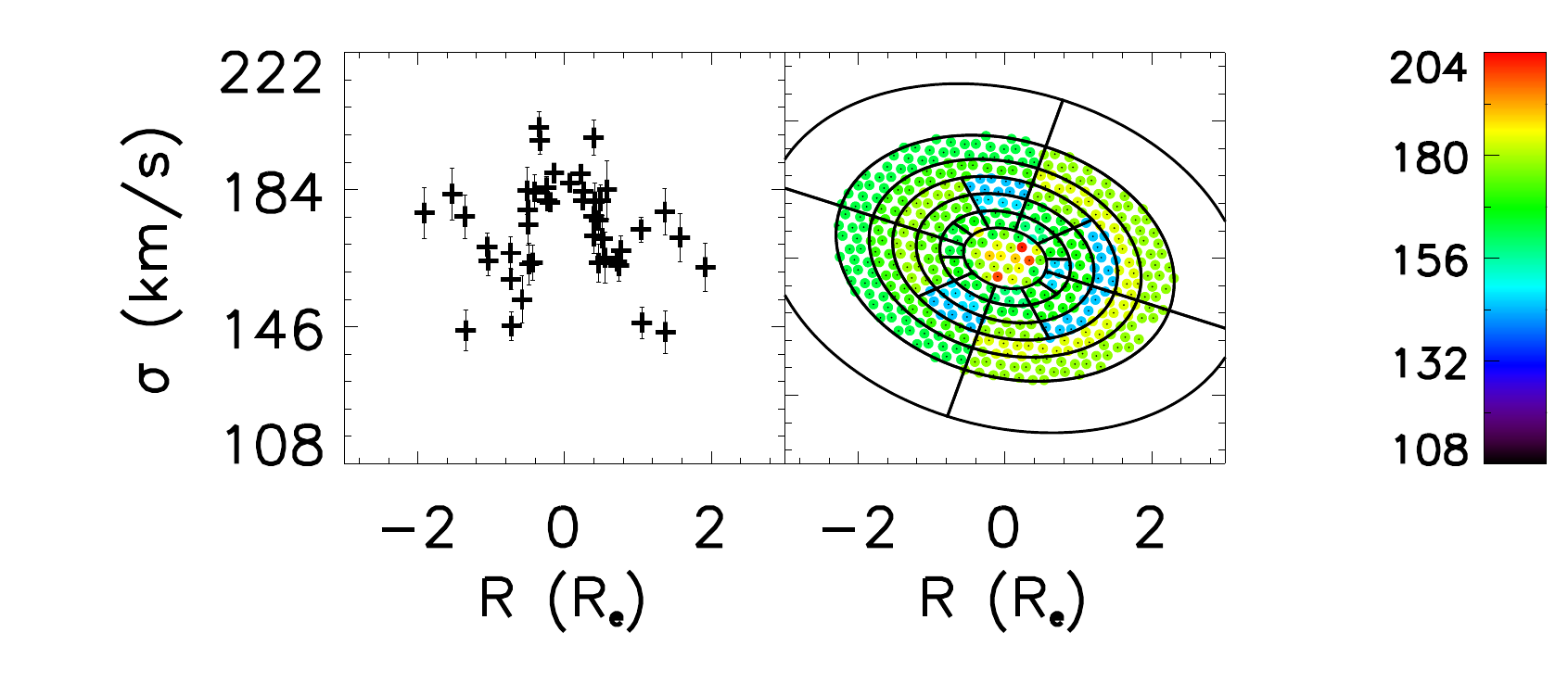}

\caption{Continued...}
\label{Fig:AllKinematicsh}
\end{center}
\end{figure}

\begin{figure}
\begin{center}

\includegraphics[width=0.6\textwidth,angle=0,clip]{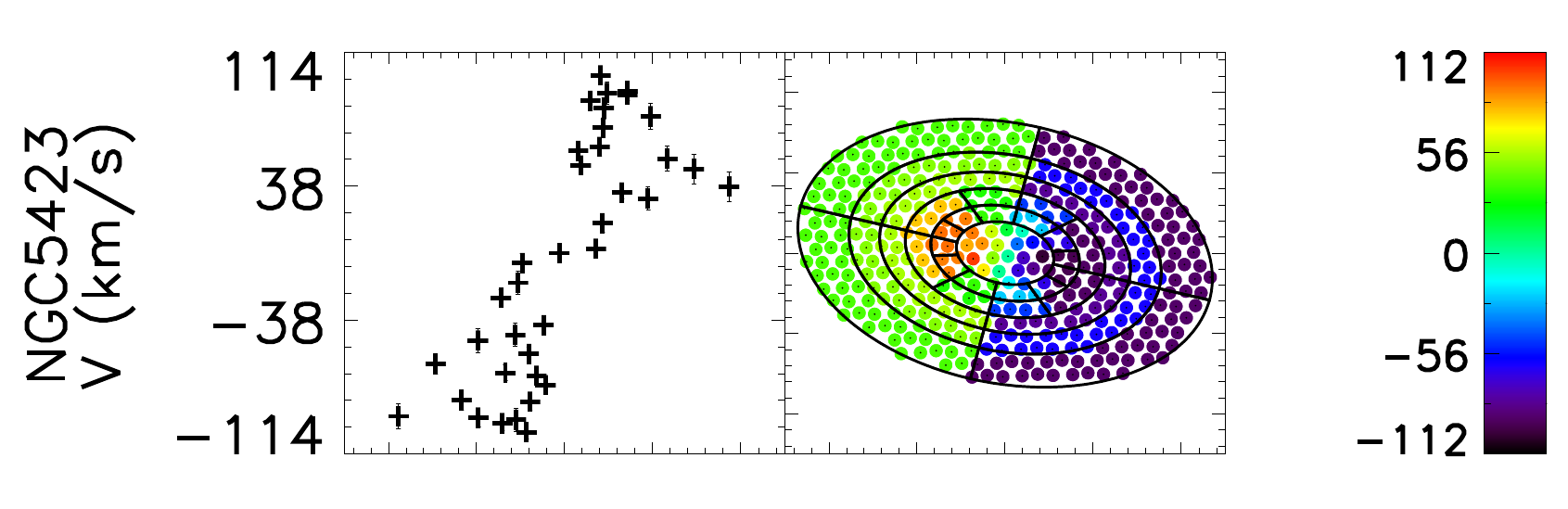}
\includegraphics[width=0.6\textwidth,angle=0,clip]{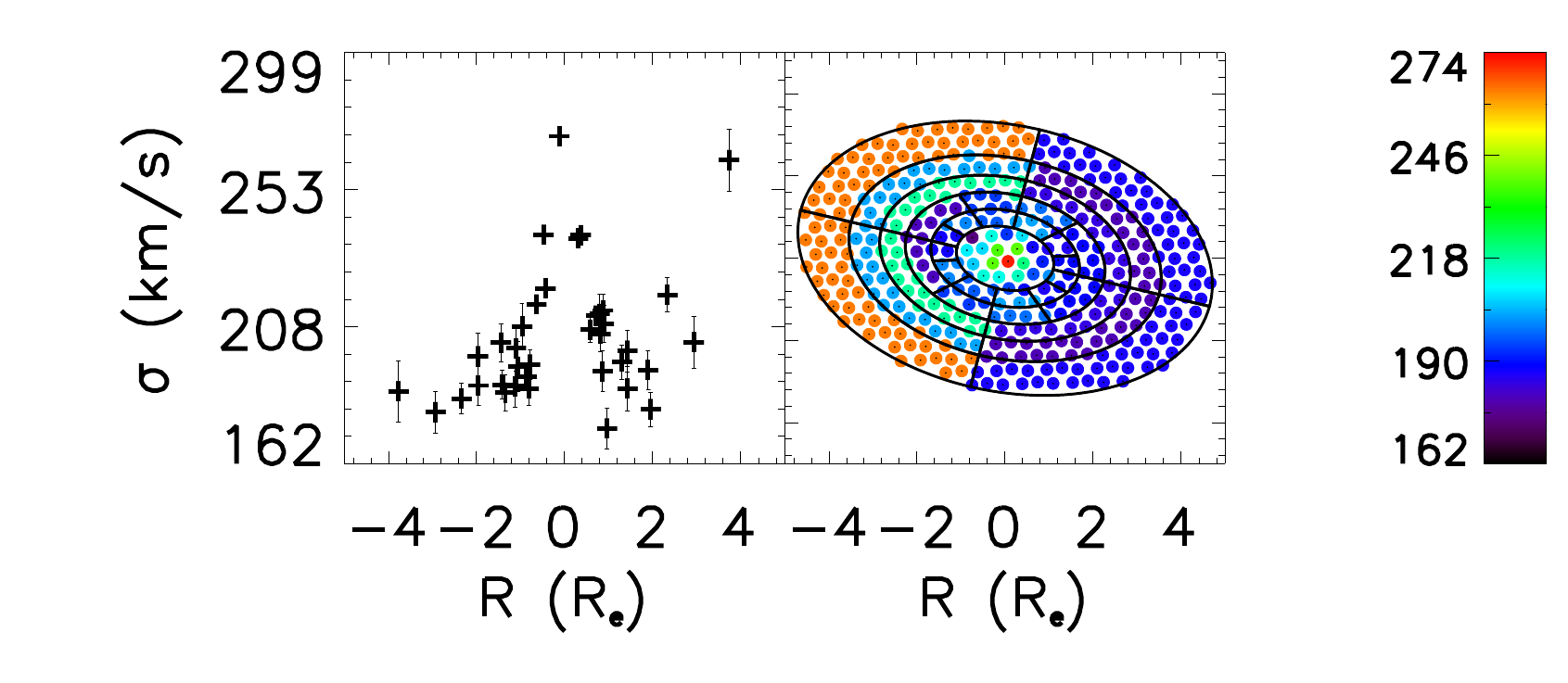}
\includegraphics[width=0.6\textwidth,angle=0,clip]{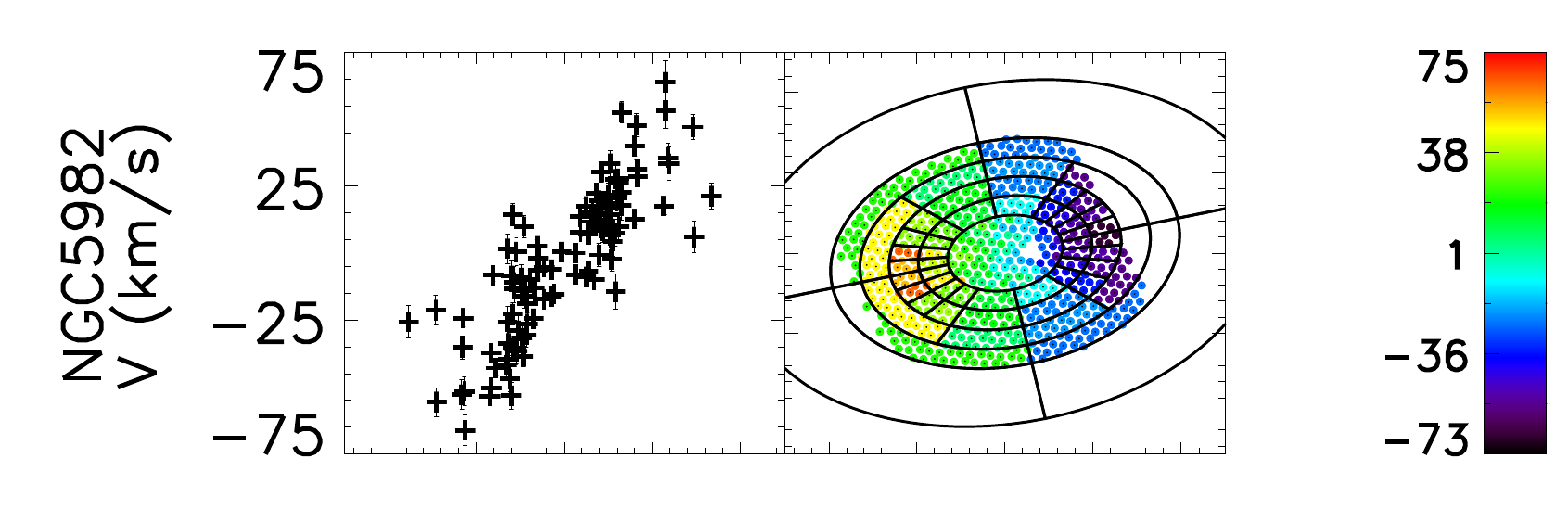}
\includegraphics[width=0.6\textwidth,angle=0,clip]{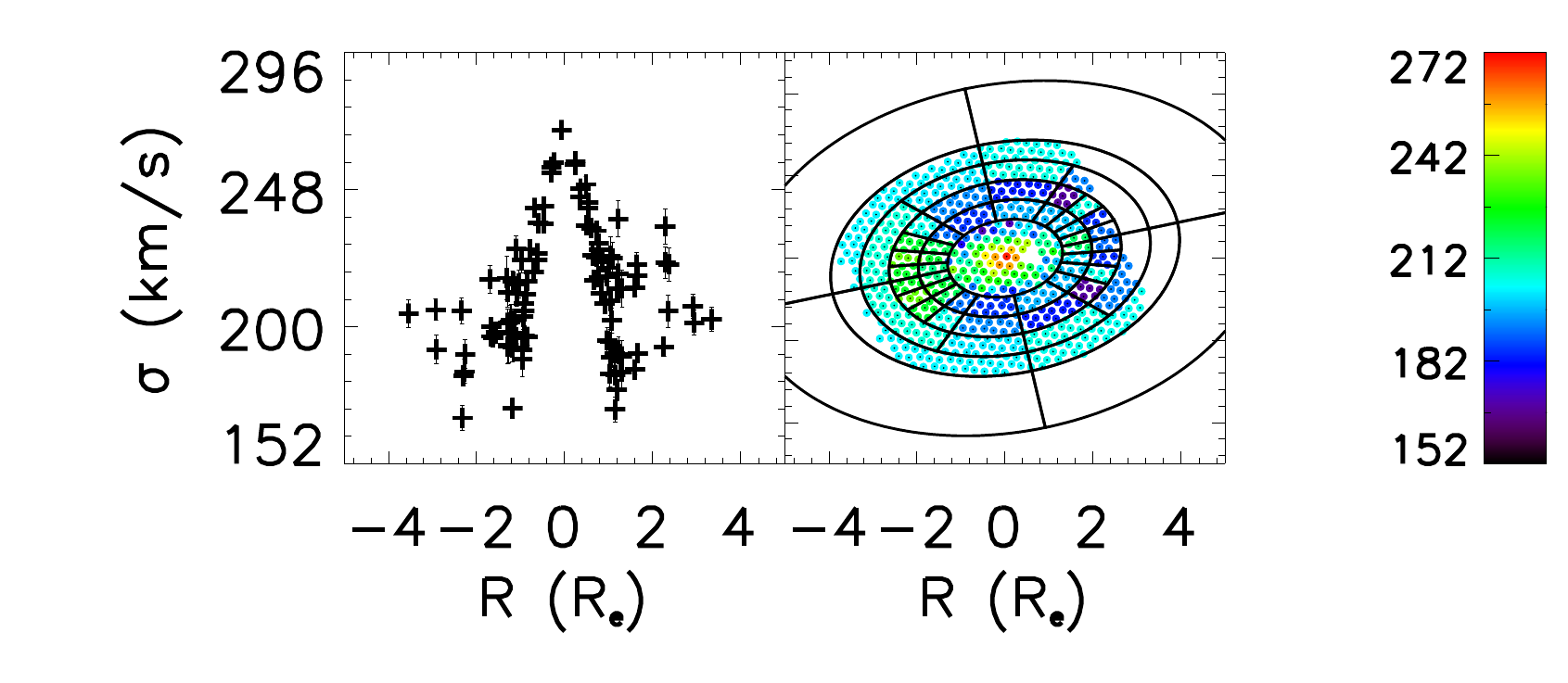}
\includegraphics[width=0.6\textwidth,angle=0,clip]{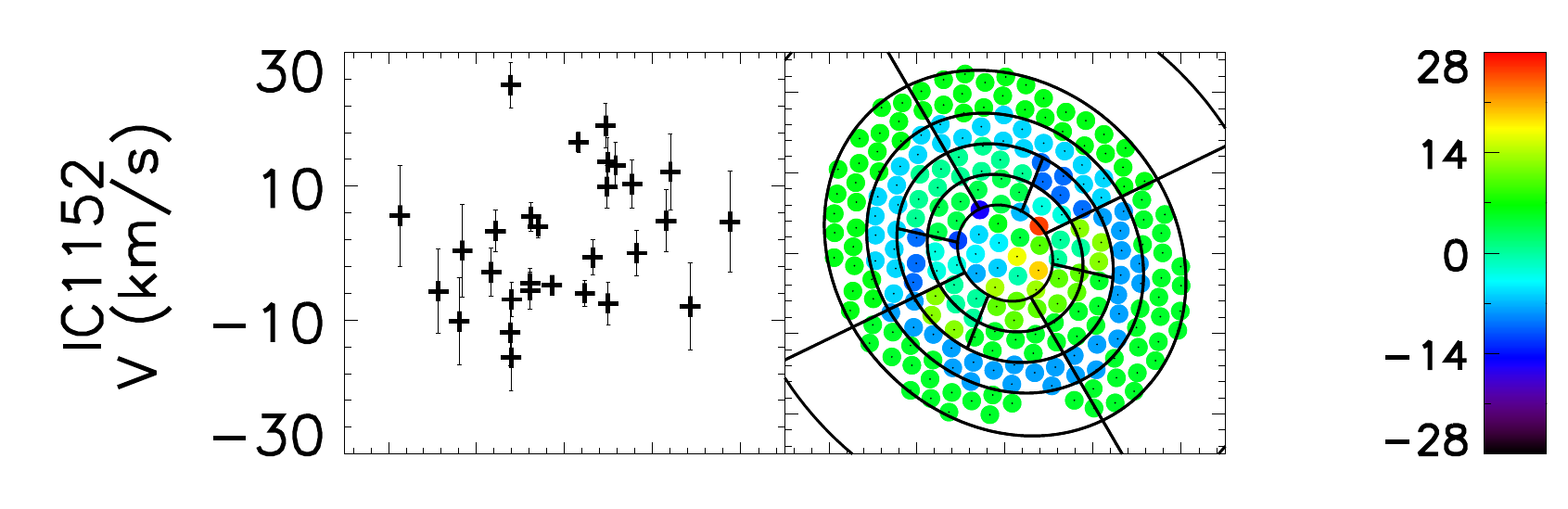}
\includegraphics[width=0.6\textwidth,angle=0,clip]{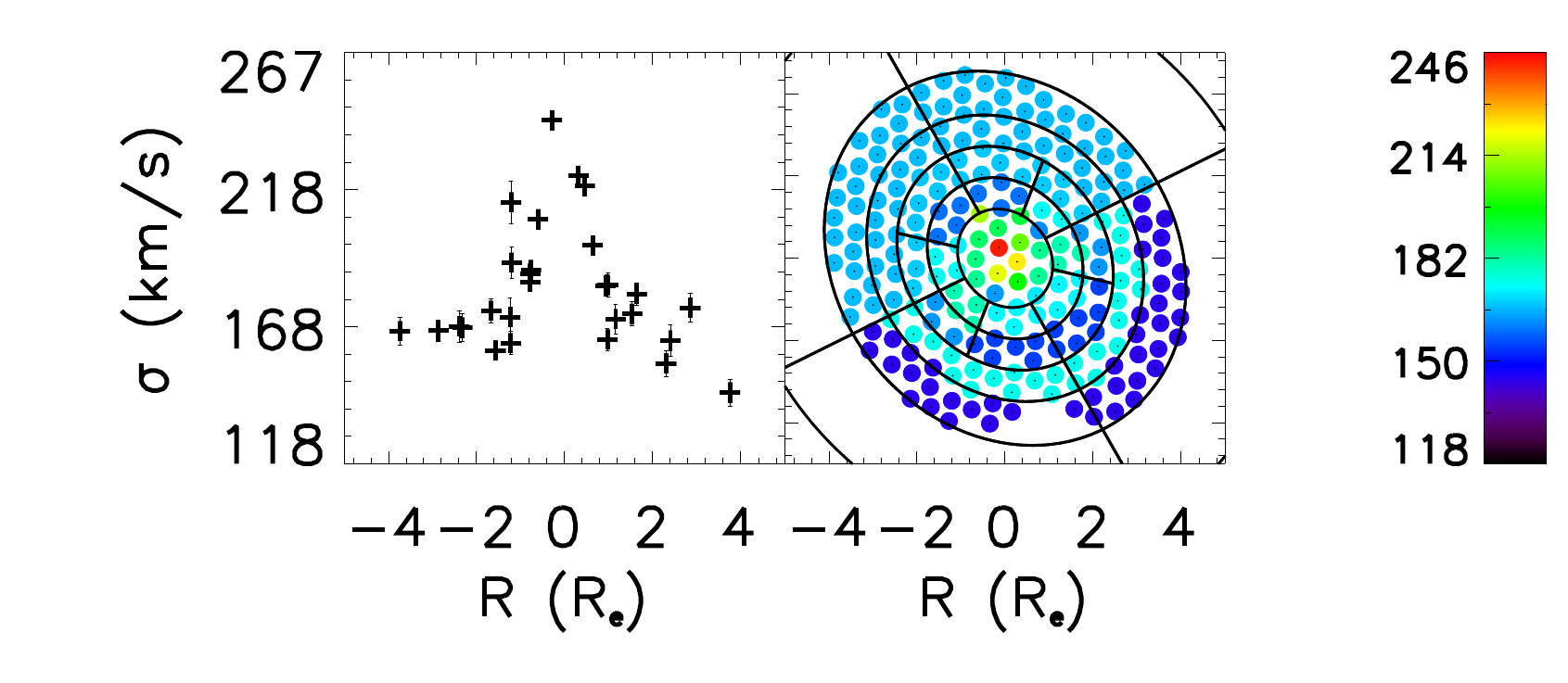}

\caption{Continued...}
\label{Fig:AllKinematicsi}
\end{center}
\end{figure}

\begin{figure}
\begin{center}

\includegraphics[width=0.6\textwidth,angle=0,clip]{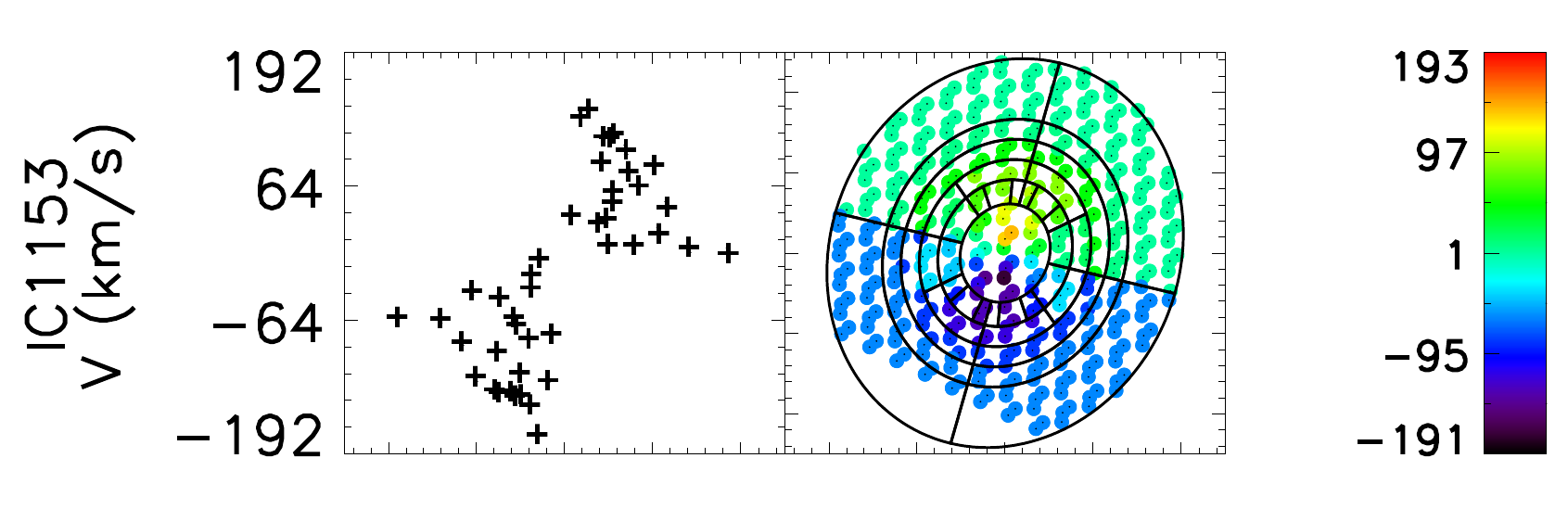}
\includegraphics[width=0.6\textwidth,angle=0,clip]{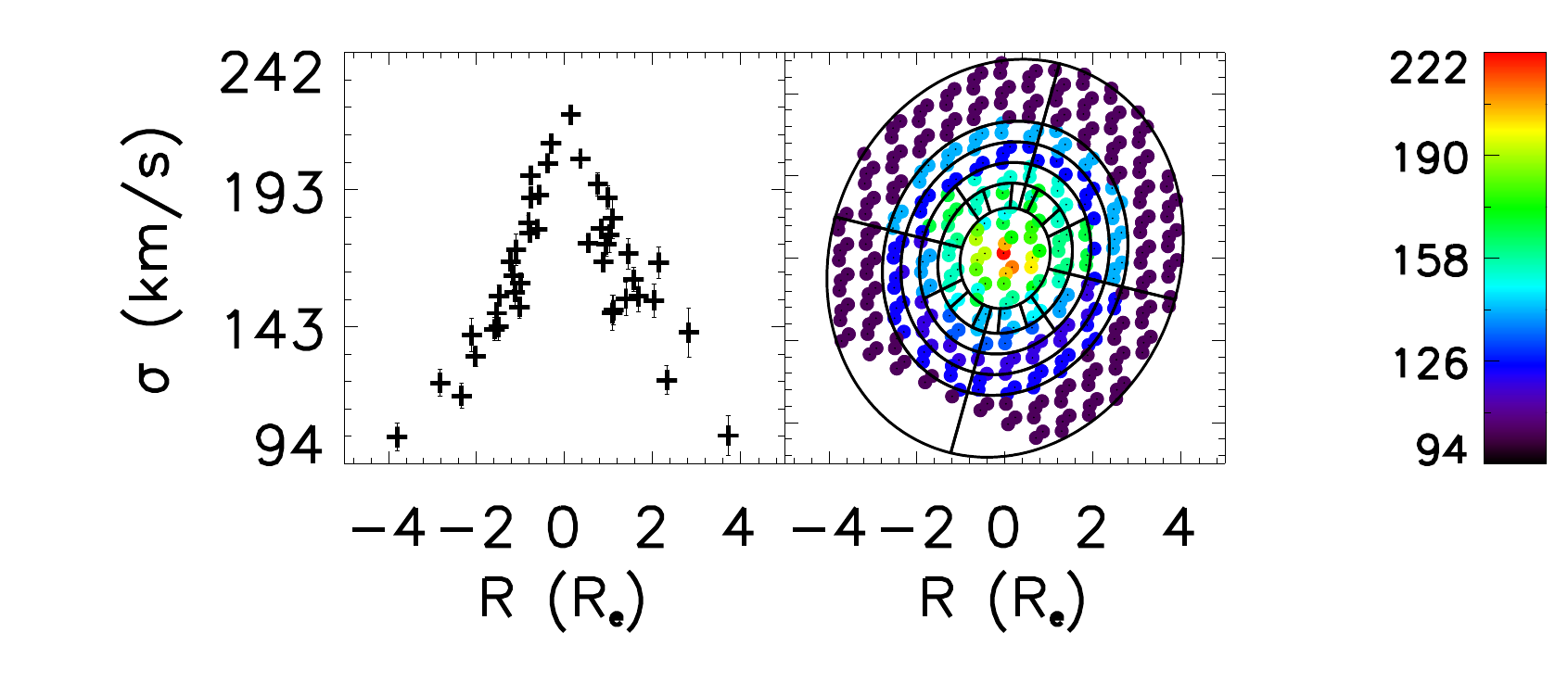}
\includegraphics[width=0.6\textwidth,angle=0,clip]{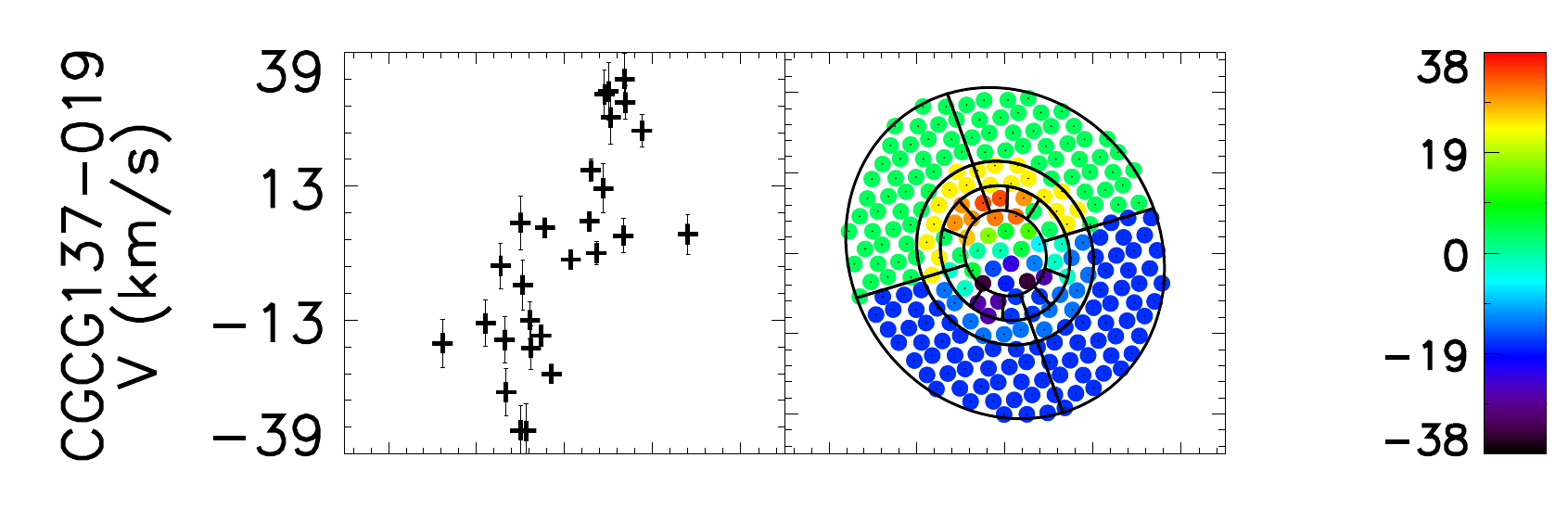}
\includegraphics[width=0.6\textwidth,angle=0,clip]{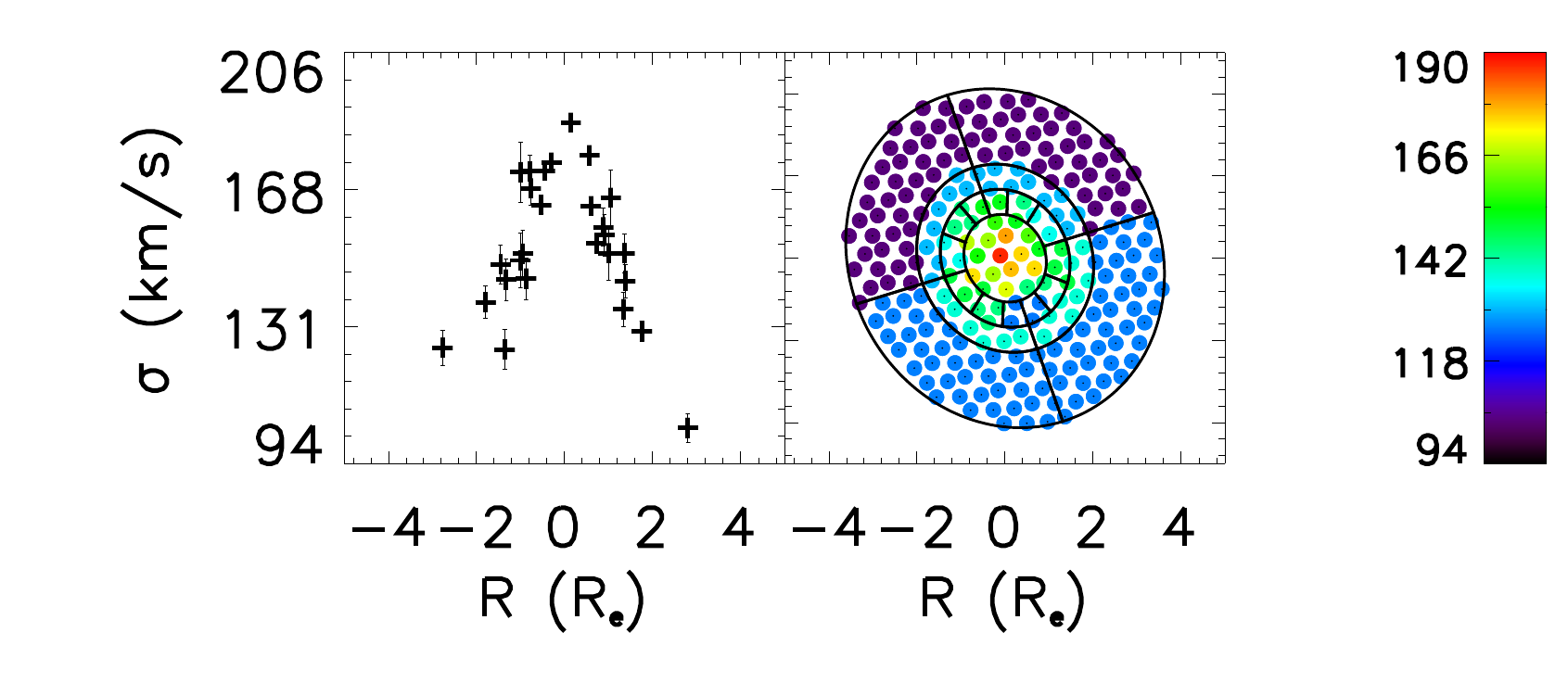}
\includegraphics[width=0.6\textwidth,angle=0,clip]{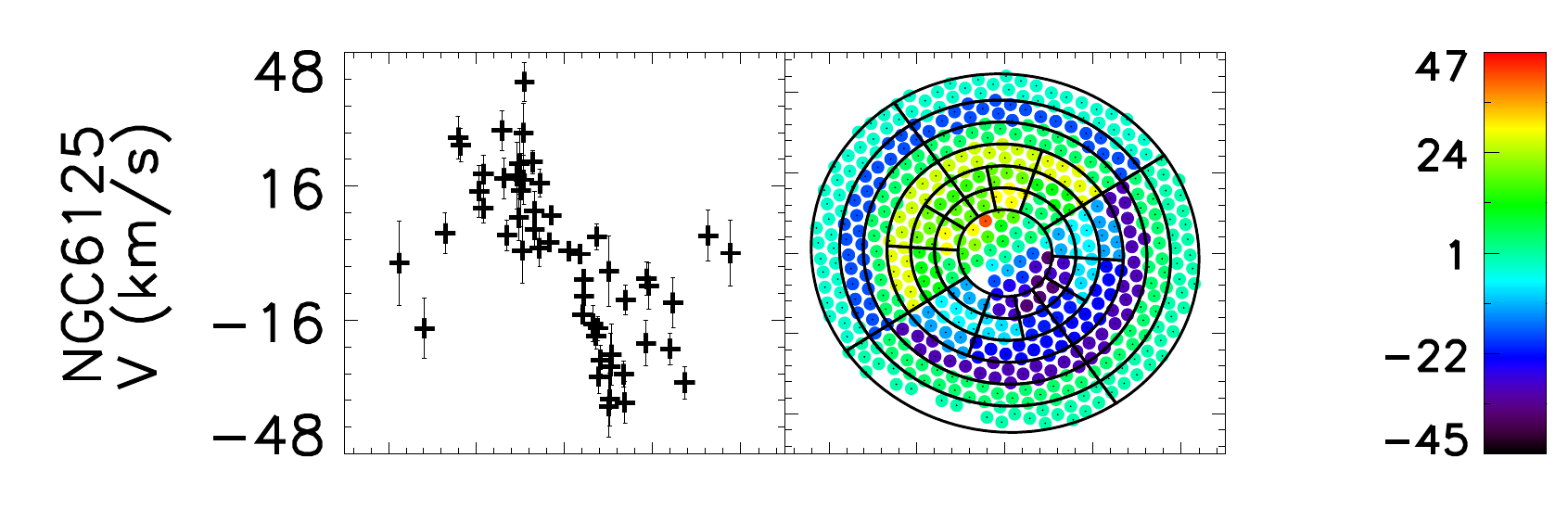}
\includegraphics[width=0.6\textwidth,angle=0,clip]{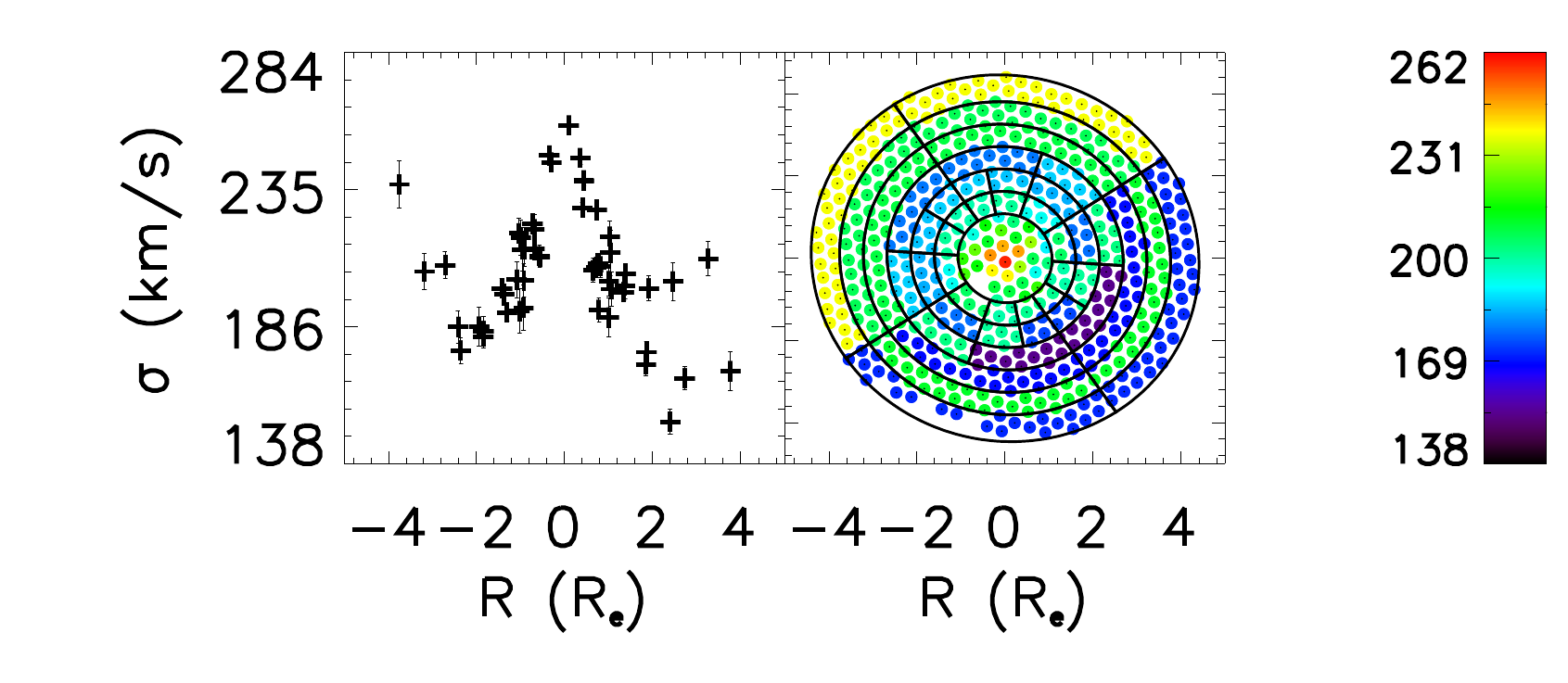}

\caption{Continued...}
\label{Fig:AllKinematicsj}
\end{center}
\end{figure}

\begin{figure}
\begin{center}

\includegraphics[width=0.6\textwidth,angle=0,clip]{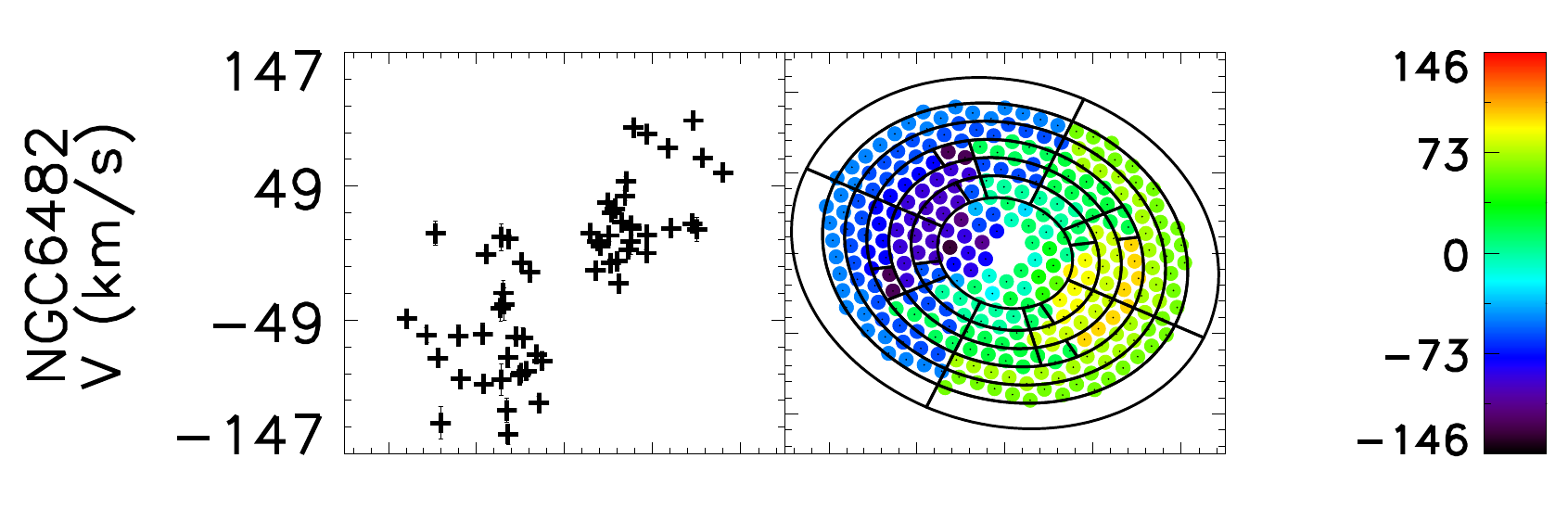}
\includegraphics[width=0.6\textwidth,angle=0,clip]{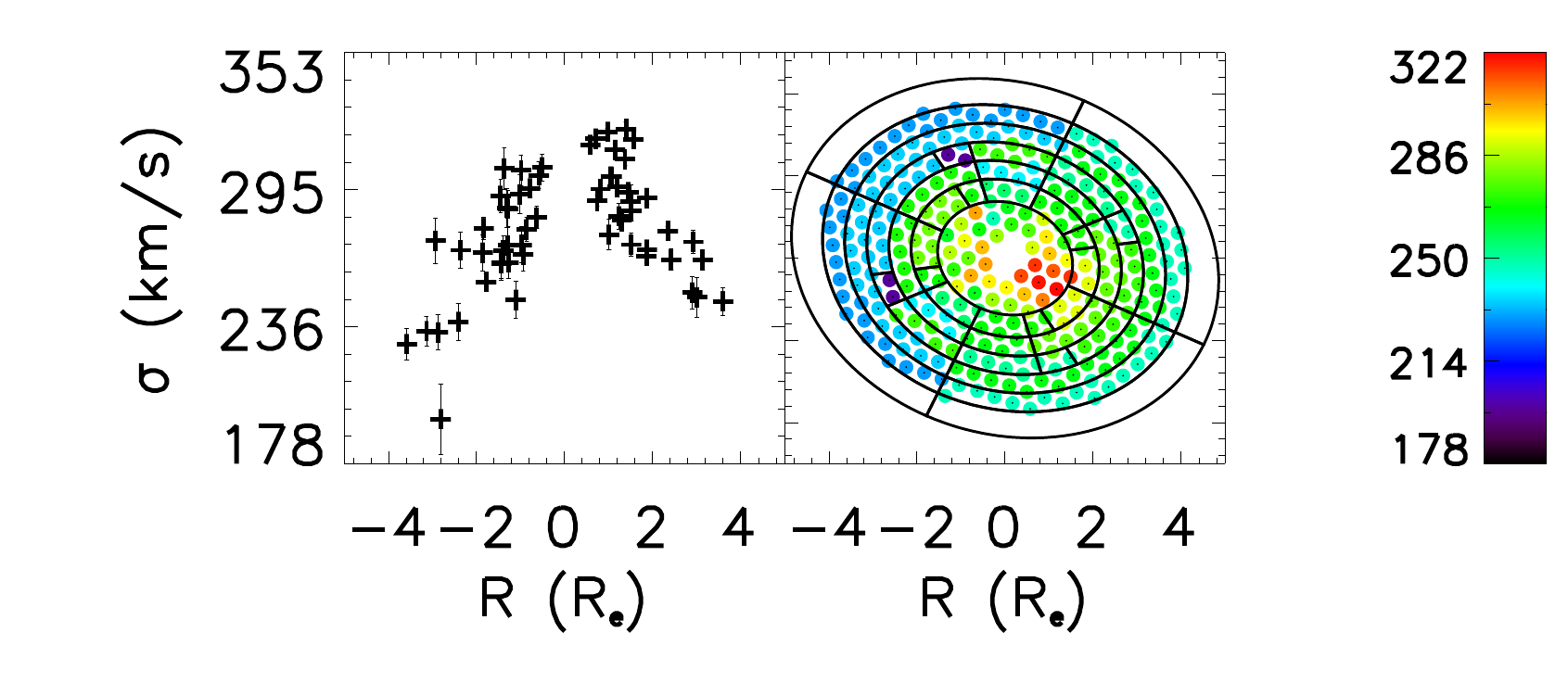}
\includegraphics[width=0.6\textwidth,angle=0,clip]{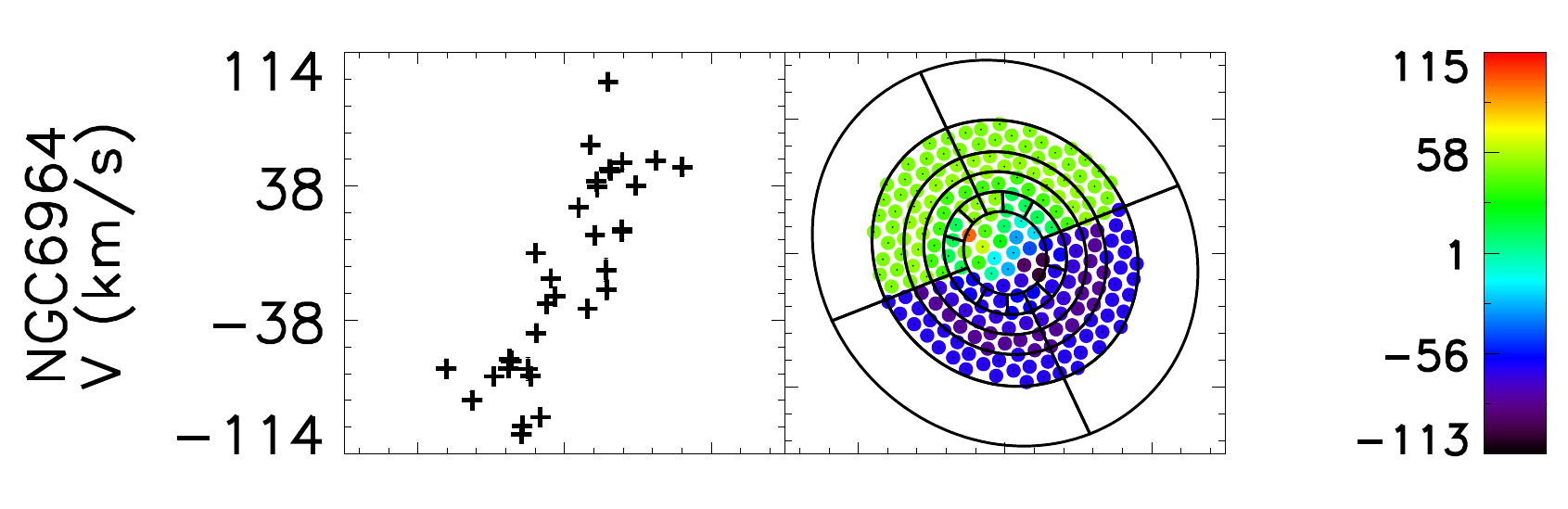}
\includegraphics[width=0.6\textwidth,angle=0,clip]{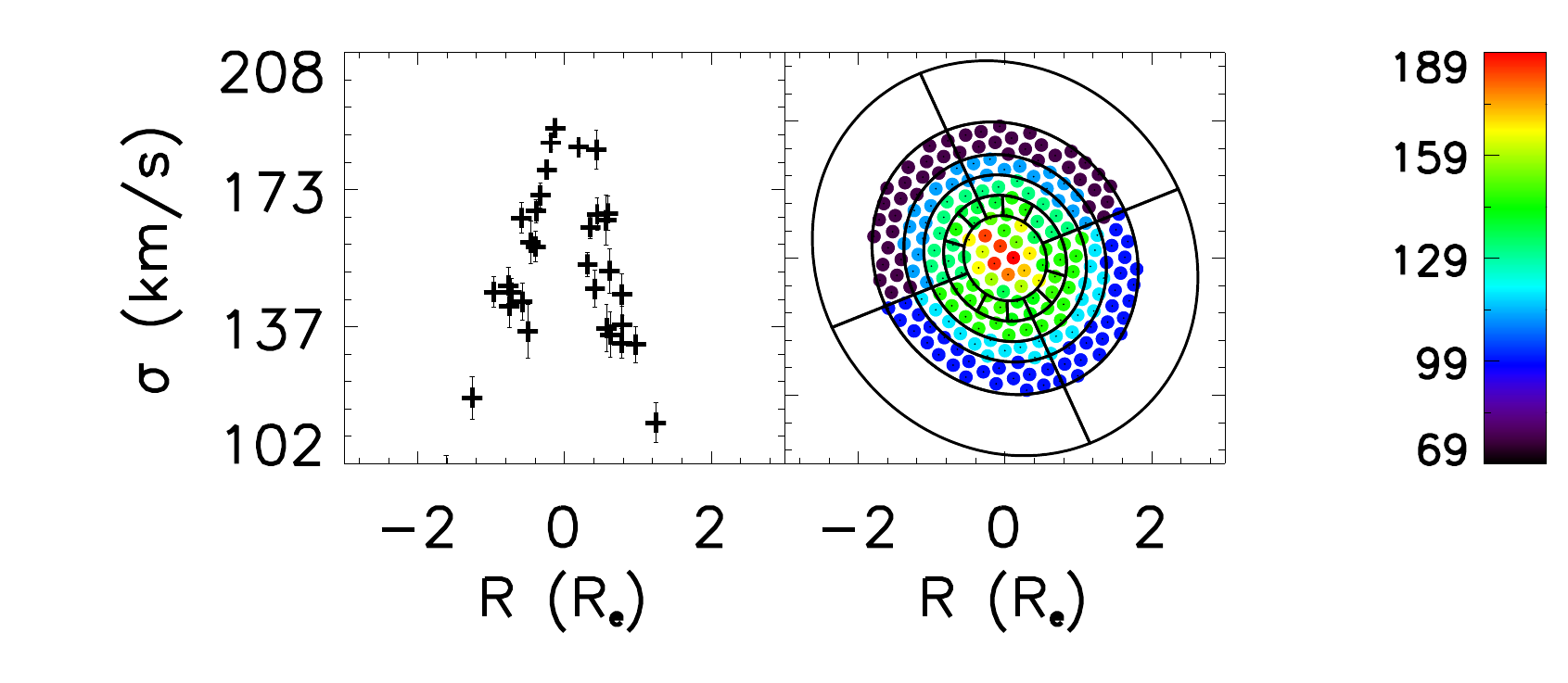}
\includegraphics[width=0.6\textwidth,angle=0,clip]{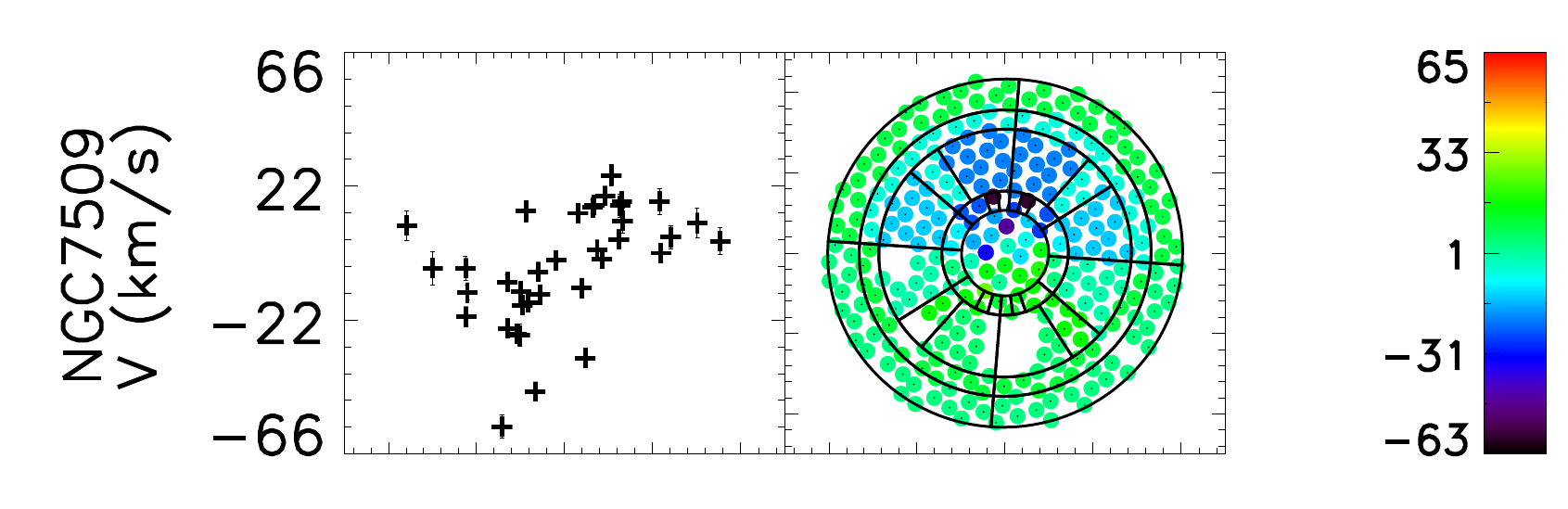}
\includegraphics[width=0.6\textwidth,angle=0,clip]{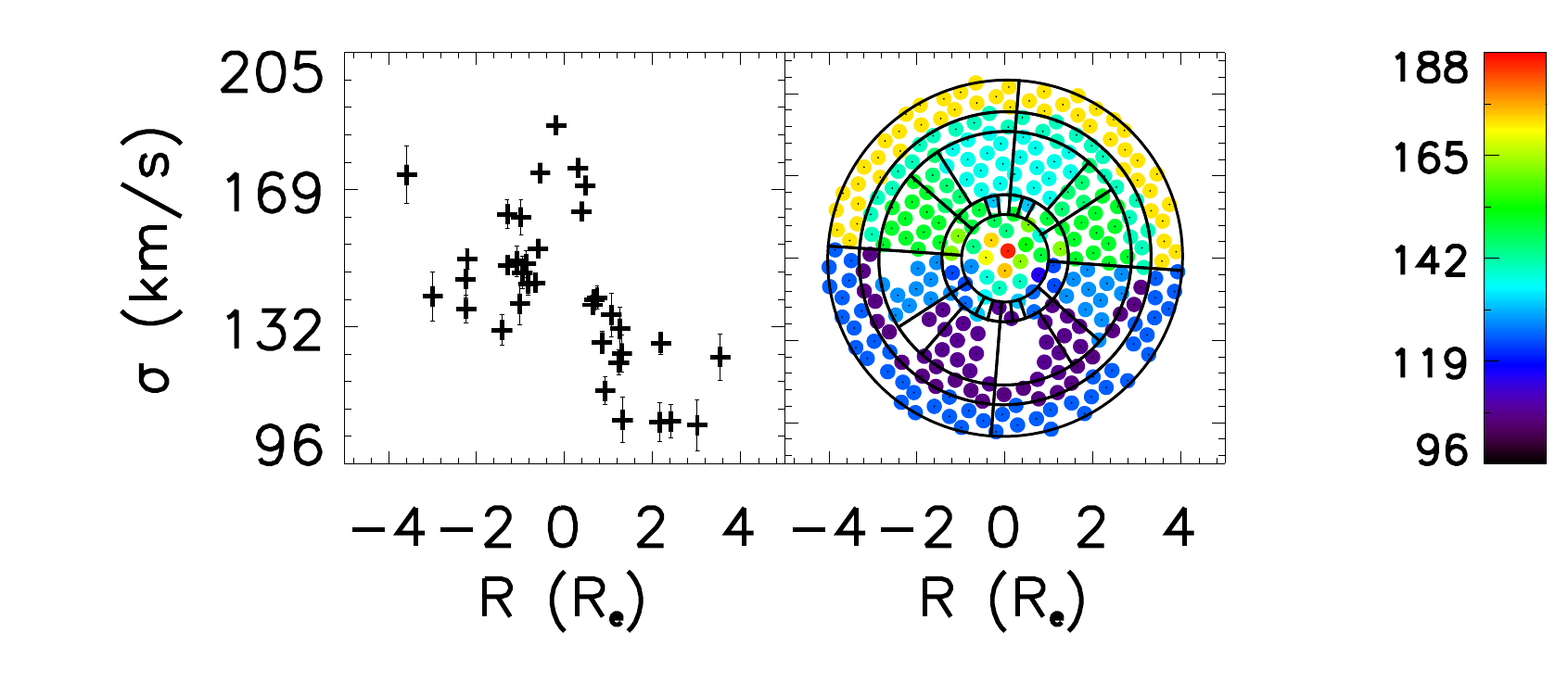}

\caption{Continued...}
\label{Fig:AllKinematicsk}
\end{center}
\end{figure}

\begin{figure}[!htb]
\begin{center}

\includegraphics[width=0.6\textwidth,angle=0,clip]{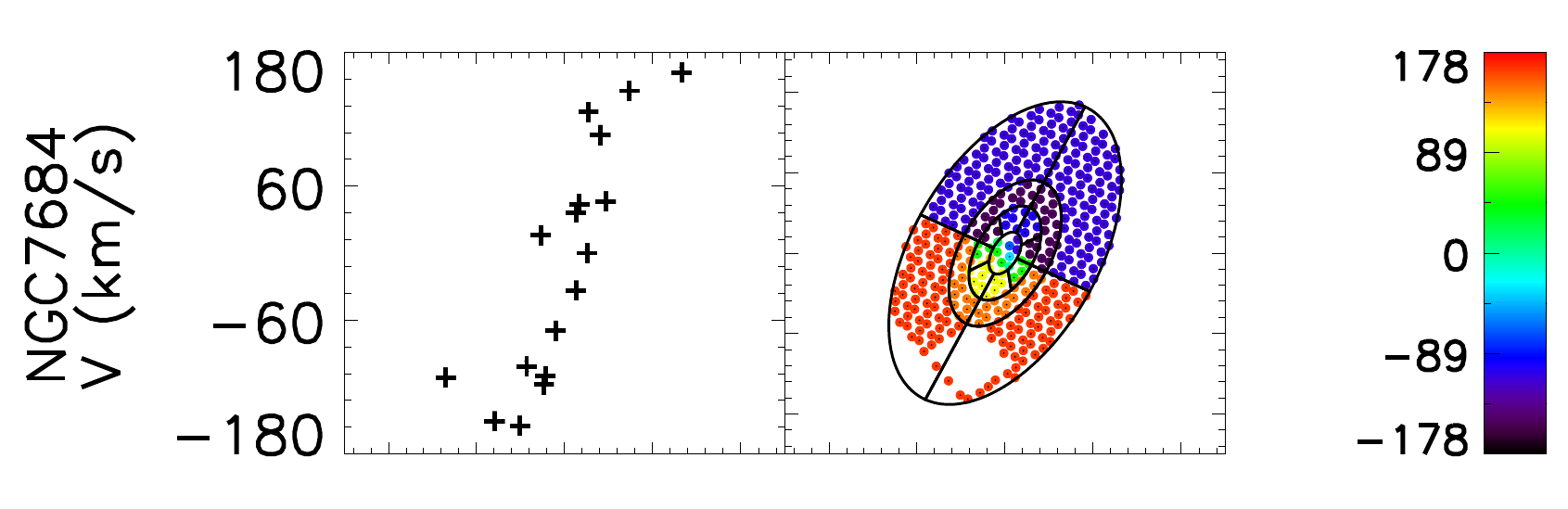}
\includegraphics[width=0.6\textwidth,angle=0,clip]{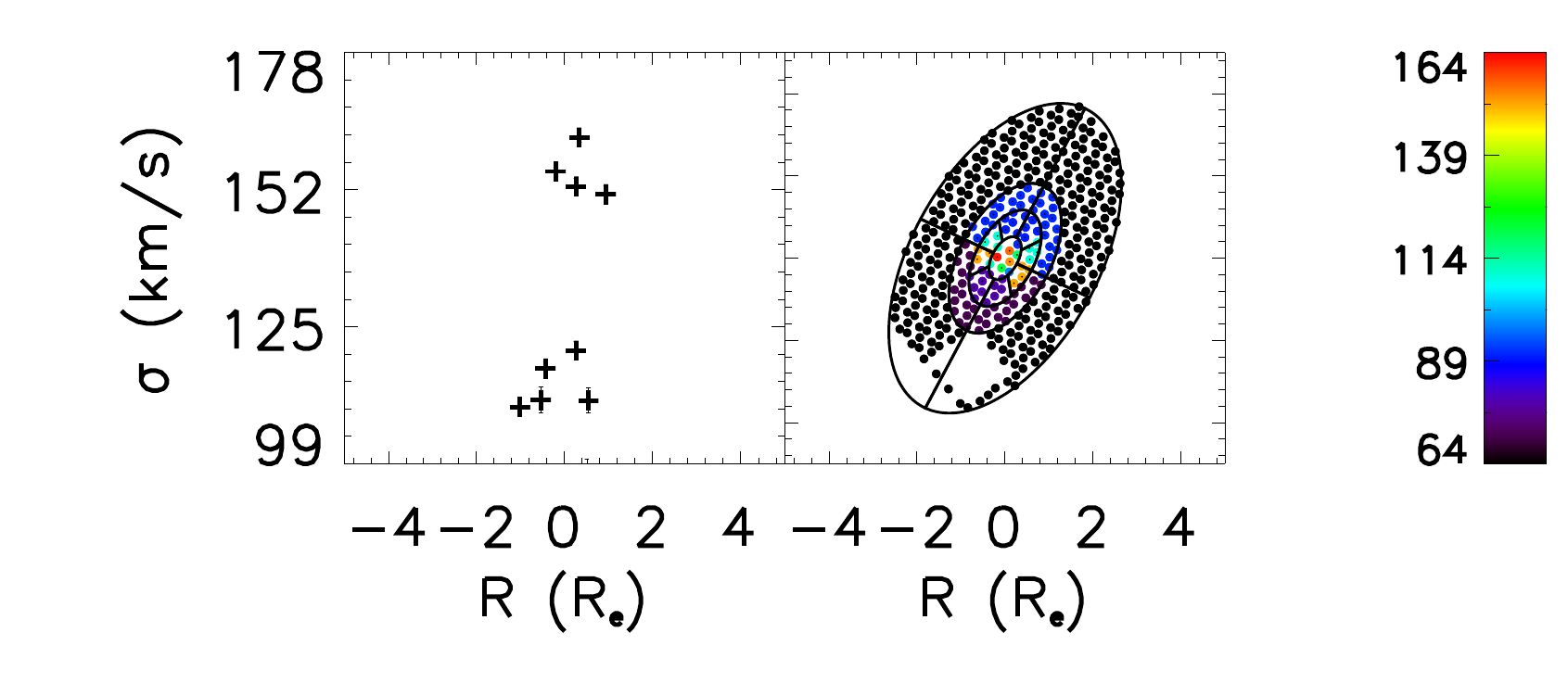}

\caption{Continued...}
\label{Fig:AllKinematicsl}
\end{center}
\end{figure}

\newpage

\end{document}